\documentclass[11pt,aps,prc,amsmath,superscriptaddress,amssymb,floats,nofootinbib]{revtex4-1}
\usepackage{graphicx} % Include figure files
\usepackage{dcolumn} % Align table columns on decimal point  
\usepackage{hyperref}
\usepackage{comment}
\oddsidemargin=\evensidemargin
\begin{document}  
%\newcommand{\url}[1]{URL #1}
%-----------------------------------------------------------------  

\title{
{\Large \bf Technical Design Report}\\
{\large \bf for the}\\
{\Large \bf Paul Scherrer Institute Experiment R-12-01.1:}\\
{\Large  \bf Studying the Proton ``Radius'' Puzzle with \boldmath{$\mu p$}
  Elastic Scattering}\\
{June 1, 2017}\\ %\today\\
\vspace{0.3cm}  
{\Large  The MUon Scattering Experiment collaboration (MUSE):}\\
} 
\author{R.~Gilman (Contact person)} 
\affiliation{Rutgers University, New Brunswick, New Jersey, USA}
\author{E.J.~Downie (Spokesperson)}
\affiliation{George Washington University, Washington, DC, USA} 
\author{G.~Ron (Spokesperson)}
\affiliation{Hebrew University of Jerusalem, Jerusalem, Israel}
\author{S.~Strauch (Spokesperson)}
\affiliation{University of South Carolina, Columbia, South Carolina, USA}
\author{A.~Afanasev}
\affiliation{George Washington University, Washington, DC, USA}
\author{A.~Akmal}
\affiliation{Montgomery College, Rockville, MD, USA}
\author{J.~Arrington}
\affiliation{Argonne National Lab, Argonne, IL, USA} 
\author{H.~Atac}
\affiliation{Temple University, Philadelphia, Pennsylvania, USA}
\author{C.~Ayerbe-Gayoso}
\affiliation{College of William \& Mary, Williamsburg, Virginia, USA}
\author{F.~Benmokhtar}
\affiliation{Duquesne University, Pittsburgh, PA, USA} 
\author{N.~Benmouna}
\affiliation{Montgomery College, Rockville, MD, USA}
\author{J.~Bernauer}
\affiliation{Massachusetts Institute of Technology, Cambridge,
  Massachusetts, USA}
\author{A.~Blomberg}
\affiliation{Temple University, Philadelphia, Pennsylvania, USA}
\author{W.~J.~Briscoe}
\affiliation{George Washington University, Washington, DC, USA} 
\author{D. Cioffi}
\affiliation{George Washington University, Washington, DC, USA} 
\author{E.~Cline}
\affiliation{Rutgers University, New Brunswick, New Jersey, USA}
\author{D.~Cohen}
\affiliation{Hebrew University of Jerusalem, Jerusalem, Israel}
\author{E.~O.~Cohen}
\affiliation{Tel Aviv University, Tel Aviv, Israel}
\author{C.~Collicott}
\affiliation{Johannes Gutenberg-Universit\"at, Mainz, Germany}
\author{K.~Deiters}
\affiliation{Paul Scherrer Institut, CH-5232 Villigen, Switzerland}
\author{J.~Diefenbach}
\affiliation{Johannes Gutenberg-Universit\"at, Mainz, Germany}
\author{B.~Dongwi}
\affiliation{Hampton University, Hampton, Virginia, USA}
\author{D.~Ghosal}
\affiliation{University of Basel, Switzerland}
\author{A.~Golossanov}
\affiliation{George Washington University, Washington, DC, USA} 
\author{R.~Gothe}
\affiliation{University of South Carolina, Columbia, South Carolina, USA}
\author{D.~Higinbotham}   
\affiliation{Jefferson Lab, Newport News, Viginia, USA}
\author{D.~Hornidge}
\affiliation{Mount Alison University, New Brunswick, Canada}
\author{Y.~Ilieva}
\affiliation{University of South Carolina, Columbia, South Carolina, USA}
\author{N.~Kalantarians}
\affiliation{Hampton University, Hampton, Virginia, USA}
\author{M.~Kohl}
\affiliation{Hampton University, Hampton, Virginia, USA}
\author{B.~Krusche}
\affiliation{University of Basel, Switzerland}
\author{G.~Kumbartzki} 
\affiliation{Rutgers University, New Brunswick, New Jersey, USA}
\author{I.~Lavrukhin}
\affiliation{George Washington University, Washington, DC, USA} 
\author{L.~Li}
\affiliation{University of South Carolina, Columbia, South Carolina, USA}
\author{J. Lichtenstadt}
\affiliation{Tel Aviv University, Tel Aviv, Israel}
\author{W.~Lin}
\affiliation{Rutgers University, New Brunswick, New Jersey, USA}
\author{A.~Liyanage}
\affiliation{Hampton University, Hampton, Virginia, USA}
\author{W.~Lorenzon}
\affiliation{University of Michigan, Ann Arbor,  Michigan, USA}
\author{K.~E.~Mesick}
\affiliation{Los Alamos National Laboratory, Los Alamos, NM, USA}
\author{Z.-E.~Meziani}
\affiliation{Temple University, Philadelphia, Pennsylvania, USA}
\author{P.~Mohanmurthy}
\affiliation{Massachusetts Institute of Technology, Cambridge,
  Massachusetts, USA}
\author{P.~Moran}
\affiliation{Temple University, Philadelphia, Pennsylvania, USA}
\author{J.~Nazeer}
\affiliation{Hampton University, Hampton, Virginia, USA}
\author{E.~Piasetzsky}
\affiliation{Tel Aviv University, Tel Aviv, Israel}
\author{R.~Ransome}
\affiliation{Rutgers University, New Brunswick, New Jersey, USA}
\author{R.~Raymond}
\affiliation{University of Michigan, Ann Arbor,  Michigan, USA}
\author{D.~Reggiani}
\affiliation{Paul Scherrer Institut, CH-5232 Villigen, Switzerland}
\author{P.E.~Reimer}   
\affiliation{Argonne National Lab, Argonne, IL, USA}
\author{A.~Richter}
\affiliation{Technical University of Darmstadt, Darmstadt, Germany}
\author{T.~Rostomyan}
\affiliation{Rutgers University, New Brunswick, New Jersey, USA}
\author{P.~Roy}
\affiliation{University of Michigan, Ann Arbor,  Michigan, USA}
\author{A.~Sarty}
\affiliation{St.~Mary's University, Halifax, Nova Scotia, Canada}
\author{Y.~Shamai}
\affiliation{Soreq Nuclear Research Center, Israel}
\author{N.~Sparveris}
\affiliation{Temple University, Philadelphia, Pennsylvania, USA}
\author{N.~Steinberg}
\affiliation{University of Michigan, Ann Arbor,  Michigan, USA}
\author{I. Strakovsky}
\affiliation{George Washington University, Washington, DC, USA} 
\author{V.~Sulkosky}
\affiliation{University of Virginia, Charlottesville, Virginia, USA}
\author{A.S.~Tadepalli}
\affiliation{Rutgers University, New Brunswick, New Jersey, USA}
 \author{M.~Taragin}
\affiliation{Weizmann Institute, Rehovot, Israel}

\begin{abstract}
The difference in proton radii measured with $\mu p$ atoms and
with $ep$ atoms and scattering remains an unexplained puzzle.
The PSI MUSE proposal is to measure $\mu p$ and $e p$ 
scattering in the same experiment at the same time.
The experiment will determine cross sections, two-photon effects, 
form factors, and radii independently for the two reactions, and 
will allow $\mu p$ and $ep$ results to be compared with reduced
systematic uncertainties.
These data should provide the best test of lepton
universality in a scattering experiment to date, about an order of
magnitude improvement over previous tests.
Measuring scattering with both particle polarities will allow a test of
two-photon exchange at the sub-percent level, about a factor
of four improvement on uncertainties and over an order of magnitude
more data points than
previous low momentum transfer determinations,
and similar to the current generation of higher momentum transfer
electron experiments.
The experiment has the potential to demonstrate whether the $\mu p$ and
$ep$ interactions are consistent or different, and whether any
difference results from novel physics or two-photon exchange. 
The uncertainties are such that if the discrepancy is real it should be
confirmed with $\approx$5$\sigma$ significance, similar to that 
already established between the 
regular and muonic hydrogen Lamb shift. 
\end{abstract}
\maketitle 

\tableofcontents   
\section{Motivation}
In 2010, a Paul Scherrer Institute
(PSI) experiment \cite{Pohl:2010zza} reported that the proton radius determined 
from  muonic hydrogen level transitions is $0.84184 \pm 0.00067$~fm, about 
5$\sigma$ off from the nearly order-of-magnitude less precise, non-muonic
measurements. This ``proton radius puzzle'' was confirmed in 2013 by a
second measurement of muonic hydrogen \cite{Antognini:2013} that determined the radius
to be $0.84087 \pm 0.00039$~fm.

New electronic results of 0.879 $\pm$ 0.008 fm \cite{Bernauer:2010wm}
and 0.875 $\pm$ 0.010 fm \cite{,Zhan:2011ji}, all from
scattering measurements, confirmed the puzzle, and
a new CODATA analysis \cite{2012arXiv1203.5425M} 
increased the significance of the discrepancy between electronic
and muonic measurements to $>$7$\sigma$.
The situation has
been discussed extensively in a number of papers - here we point
out a review paper in Annual Review of Nuclear and Particle Science
\cite{Pohlreview:2012} - and in many talks and two dedicated workshops
\cite{prpw:2012,prpw:2014}.
It is generally agreed that new data are needed to resolve the puzzle,
and the MUon Scattering Experiment (MUSE) discussed here uniquely
attempts to resolve the puzzle.
Our intent is
\begin{itemize}
\item to directly compare $ep$ to $\mu p$ elastic scattering at the sub-percent
level, in simultaneous measurements, more precisely than done before,
and at much lower $Q^2$,
\item  to compare cross sections of positive vs.\ negative charged particles to test 
two-photon exchange effects in both $ep$ to $\mu p$  elastic scattering at the sub-percent level, 
more precisely than done before,
\item and to extract the proton radius from the slope of the
elastic  electric form factor at $Q^2$ $=$ 0, the first significant $\mu p$ scattering
radius determination, at roughly the same level as done in
previous electron scattering experiments.
\end{itemize}

\begin{figure}[h]
\centerline{\includegraphics[height=2.5in]{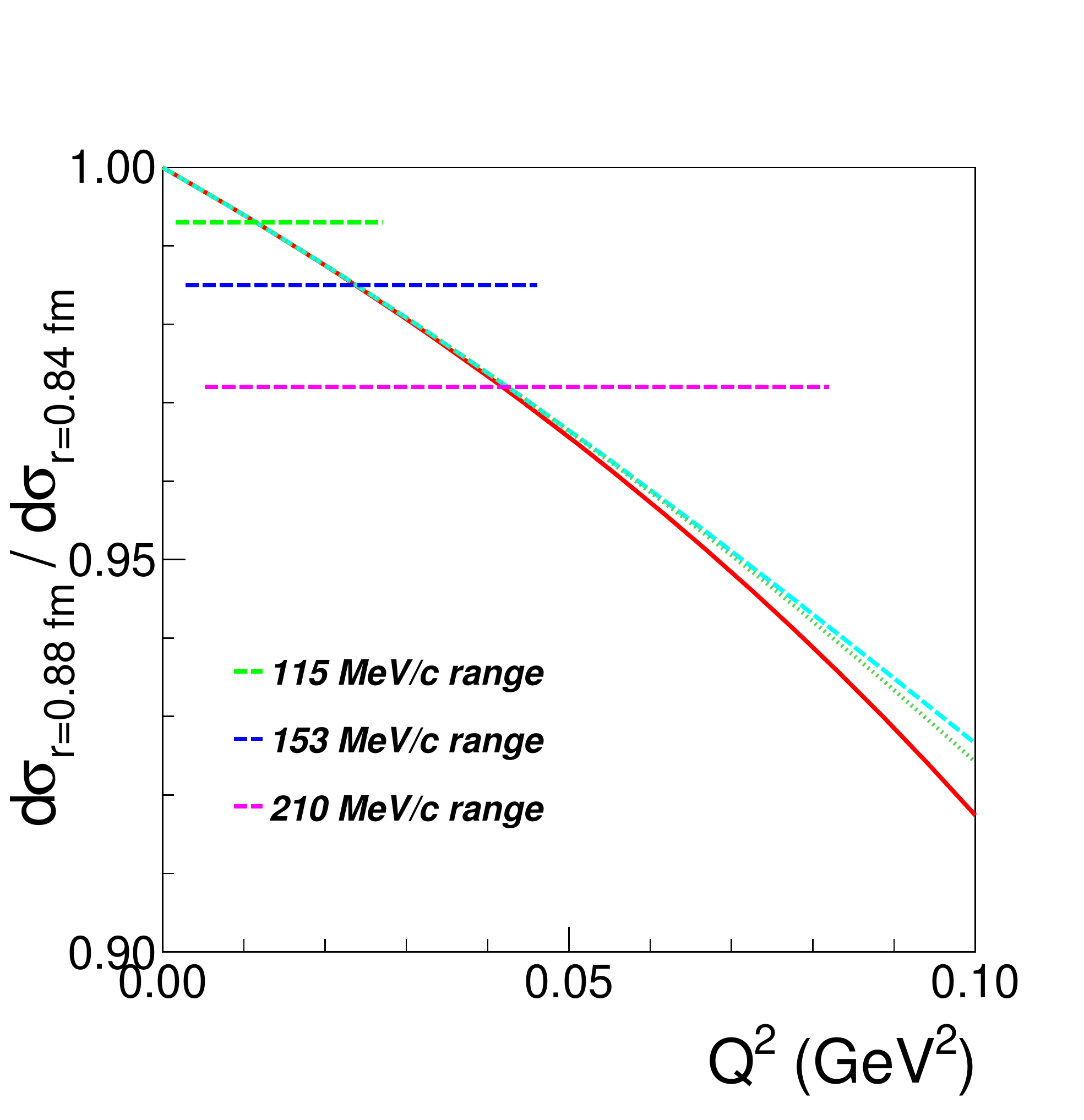}
\includegraphics[height=2.5in]{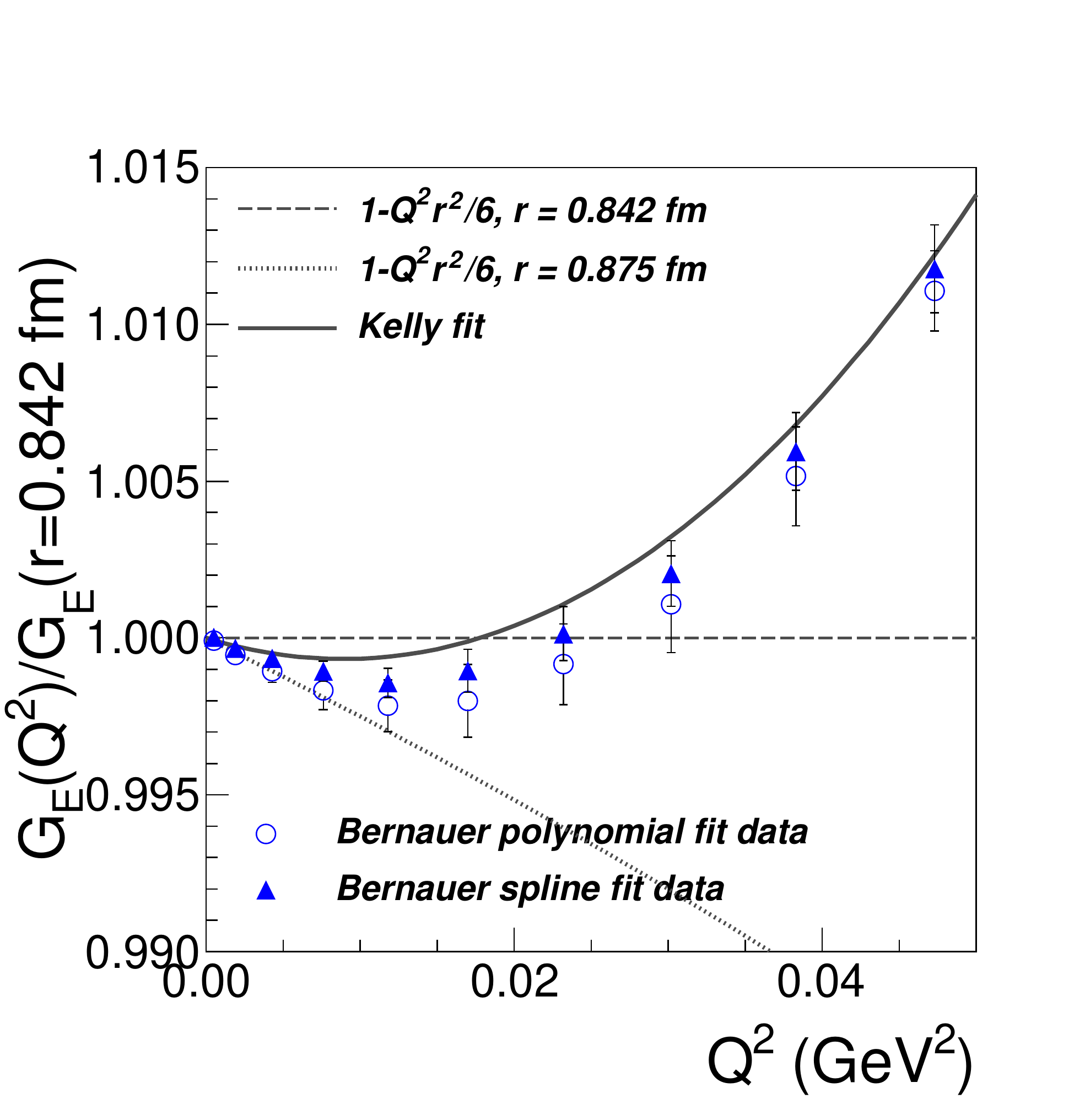}}
\caption{Left: Effect on the cross section from the proton radius
  being 0.88 fm vs.\ 0.84 fm. 
The solid red line uses a linear approximation, $G_E = 1 - Q^2 r^2
/6$.
The cyan dashed and green dotted curves include $Q^4$ and $Q^6$ 
terms taken from the Kelly parameterization, respectively.
The horizontal green, blue, and magenta dashed lines indicate the
kinematic range of the proposed MUSE data.
Right: Mainz results for the proton
  electric form factor determined by spline and polynomial fit
  analyses of the cross sections, over a low $Q^2$ range to show how
  terms beyond the linear term quickly become important.
  Also shown are the Kelly parameterization and a linear fit assuming
  the radius determined by $ep$ measurements, all relative to expectations
 from a linear fit using the radius determined from $\mu p$ atoms.
 Note the greatly expanded vertical scale compared to the left panel.
} 
\label{fig:mainz_ana}
\end{figure}

Figure~\ref{fig:mainz_ana} gives a quick indication of the minimum
requirements for the experiment.
The left panel shows that a radius of 0.88 fm vs.\ 0.84 fm results
in a cross section that falls off about 6\% faster over the range of
the MUSE kinematics.
Thus a determination of the radius at a 5$\sigma$, $<$ 0.01 fm, level
requires cross sections at at least the percent level. 
The right panel shows on an expanded $Q^2$ scale some parameterizations compared to the
Mainz data for $G_E^p(Q^2)$, extracted from the cross sections
using spline and polynomial fit functions to the data, all relative to
the expected behavior if the form factor is linear with a radius of
0.842 fm.
In $ep$ scattering, the charge radius is determined from the slope of the
form factor $G_E$ at $Q^2$ = 0.
The lowest $Q^2$ Mainz data points are more consistent with the larger
radius found in $ep$ experiments, but even before $Q^2 = 0.02$ GeV$^2$
the form factor is starting to show nonlinearities. 
The curvature at low $Q^2$ indicates the importance of measuring at low 
$Q^2$ to be sensitive to the radius, and over a range of $Q^2$ to have
sensitivity to higher-order terms.
In this report, we discuss many aspects of the MUSE experiment,
including the backgrounds and systematic effects that drive the key 
performance specifications for the equipment, 
factors that lead to the experimental statistics, 
and the data taking and analysis, 
leading to the extraction of the proton radius at a level of $<$ 0.01 fm, 
in Section~\ref{sec:radiusextraction}.
The Kelly parameterization \cite{Kelly:2004hm} shown generally
predicts the trends of the data -- as do several other standard data 
parameterizations -- and will be used for estimating rates and systematics.

\section{Experiment Overview}
\label{sec:overviewintro}

\begin{table}[h]
\caption{\label{tab:kinoverview} 
MUSE kinematic coverage.
}
\begin{tabular}{|l|c|}
\hline
Quantity & Coverage \\
\hline
Beam momenta & 115, 153, 210 MeV/$c$ \\
Scattering angle range & 20$^{\circ}$ - 100$^{\circ}$ \\
Azimuthal coverage & 30\% of 2$\pi$ typical \\
$Q^2$ range for electrons & 0.0016 GeV$^2$ - 0.0820 GeV$^2$ \\
$Q^2$ range for muons &  0.0016 GeV$^2$ - 0.0799 GeV$^2$ \\
\hline
\end{tabular}
\end{table}

MUSE will measure
cross sections for elastic $\mu^{\pm} p$  and $e^{\pm}p$ 
scattering in the PSI $\pi$M1 beam line.
Table~\ref{tab:kinoverview} summarizes the kinematics.

\subsection{Beam Properties}
\label{sec:beamproperties}

\begin{figure}[h]
\centerline{\includegraphics[width=0.6\textwidth]{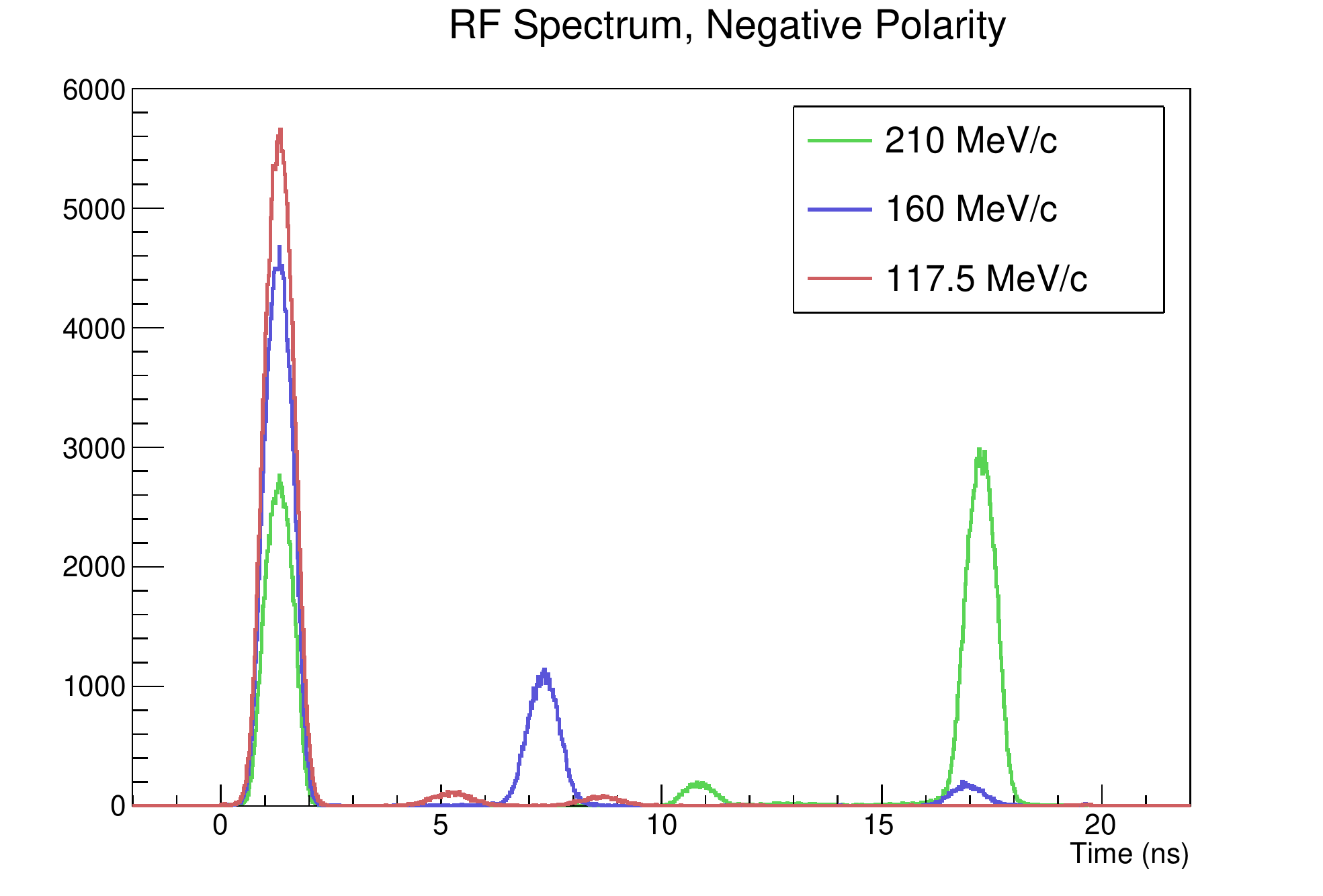}}
\caption{Measured RF time spectrum for negative charge beams at 
$117.5$ MeV$/c$, $160$ MeV$/c$, and $210$ MeV$/c$,
at a distance of $\approx$23.5 m from the production target.
The peaks from left to right are $e$, $\pi$, $\mu$ for $117.5$ MeV$/c$ and $160$ MeV$/c$, 
and $e$, $\mu$, $\pi$ for $210$ MeV$/c$. 
The absolute scale is arbitrary.
The 3 -- 6 ns separation is large compared to the intrinsic width in
time of the particle peaks of $\approx$300 ps.
The spectrum wraps around every $\approx$20 ns.}
\label{fig:rftimeall}
\end{figure}

The $\pi$M1 channel transports mixed secondary beams of
$e$'s, $\mu$'s, $\pi$'s  and protons (for high momenta) generated 
by interactions of the primary proton beam at the M1 production target.
The accelerator time structure of $\approx$50 MHz leads to
$\approx$300~ps ($\sigma$) wide pulses of each particle type every
$\approx$20 ns.
MUSE uses beam momenta of 115, 153, and 210 MeV/$c$ to provide good $e$ and $\mu$
fluxes at momenta where the particles arrive separated in time at the
MUSE scattering target - see Fig.~\ref{fig:rftimeall}.
The beam spot size is $\approx$1 cm ($\sigma$) at the scattering
target, with angular divergences $\approx$1$^{\circ}$ - 3$^{\circ}$.
More details regarding the properties of the beam
are given in Section~\ref{sec:beamline}.

The $\pi$M1 beam properties coupled with the need to measure precise cross
sections make it necessary to measure incoming beam particles to
identify each particle type and determine its trajectory. 
To accomplish this, we limit the beam flux to $\approx$3.3 MHz
with a collimator at the intermediate focal point of the beam line.
At this point the beam is $\approx$5~cm high and momentum dispersed to
a width of $\approx$21~cm, for the 3\% channel momentum acceptance.
We reduce the momentum acceptance.

\subsection{Experimental Equipment}
\label{sec:equipmentoverview}

\begin{figure}
\centerline{\includegraphics[width=0.7\textwidth]{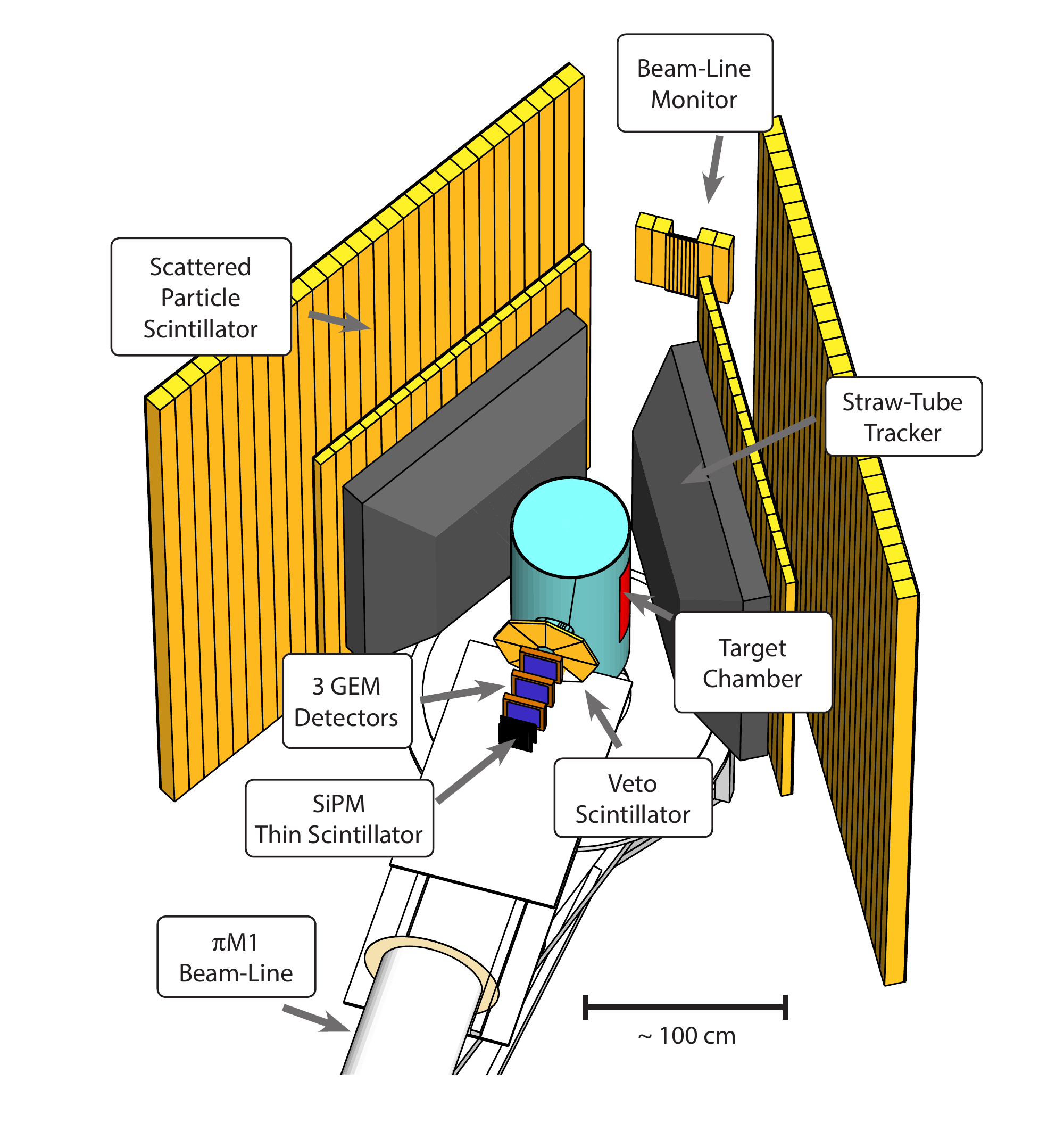}}
\caption{Implementation of detectors in the $\pi$M1 area in a Geant4 \cite{Agostinelli:2002hh} simulation.
The beam strikes the thin scintillator beam hodoscope and three GEM chambers, 
passes through a hole in the annular veto scintillator, 
enters the cryotarget vacuum chamber and strikes one of the targets, 
then exits the vacuum chamber and goes through the beam monitor.
Scattered particles are detected by two symmetric spectrometers, each
with two straw chambers wrapped in RF shielding and two planes of scintillator paddles.
}
\label{fig:layout}
\end{figure}

\begin{figure}
\centerline{\includegraphics[width=0.7\textwidth]{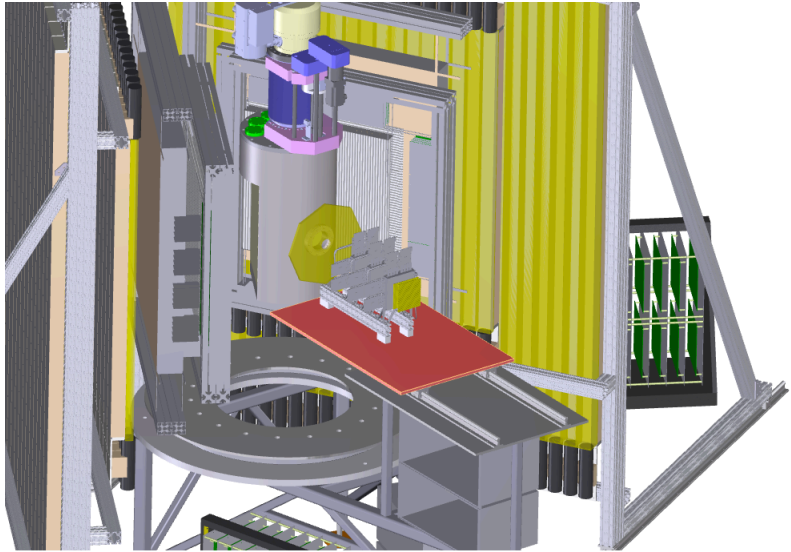}}
\caption{CAD drawing of the experiment, showing some of the detectors,
support structures, and electronics, and the cryotarget vacuum chamber.
}
\label{fig:layoutmd}
\end{figure}

Figures~\ref{fig:layout} and \ref{fig:layoutmd} shows the MUSE
experimental equipment in the target region.
After exiting the channel, the beam passes through the beam hodoscope,
a fast scintillator array that determines particle type through time 
measurements, relative to the accelerator RF.
The beam next passes through GEM chambers to measure the trajectories
going into the target.
An annular veto detector surrounds the beam as it enters the vacuum chamber
thin entrance window. 
Particles going through the veto detector are mainly muons from pion
decay,\footnote{At MUSE beam momenta, $\pi$'s decay at a rate of $\approx$
10\%/m, while $\mu$'s decay at a rate of $\approx$ 0.1\%/m.} 
which would hit the thick vacuum chamber wall, and might otherwise trigger the system.
A target ladder inside the vacuum chamber holds the liquid hydrogen 
cryotarget, an empty cell for background measurements, and a solid 
alignment target. 
The ladder can be moved so that no target is in the beam.
Downstream of the target is a beam monitor that monitors beam
particles that do not scatter in the cryotarget.
The beam monitor is also used in conjunction with the beam hodoscope
to determine the $\mu$ and $\pi$ beam momenta through time of flight
techniques.

Symmetric left and right spectrometers detect particles that scatter
in the cryotarget at angles of 20$^{\circ}$ - 100$^{\circ}$.
The spectrometers consist of straw chambers for tracking and
scintillator walls for precision timing measurements.
The two symmetric spectrometers are intended both to double the 
experimental statistics, and give simultaneous but independent
cross section measurements that provide an overlap and check
of some of the experiment systematics.

The pion scattering rate from the strong interaction is much greater 
than the rates of interest, the electron and muon scattering rates 
from the electromagnetic interaction.
Thus the experiment uses a trigger that includes beam particle
identification, scattered particle detection, and no veto signal. 

Detector performance requirements arise from several different
considerations.
Detector time resolutions need to be at the $\approx$100 ps / plane
level, so that time of flight between detectors can measured to
$\leq$ 100 ps. This is needed for beam momentum measurements
and for identifying reaction types.
Trigger detector efficiencies need to be at the $\approx$ 99\%
level, to not lose statistics, to reject pion induced backgrounds,
and to control any angle-dependent inefficiencies that might change
the angular distribution shape, and the extracted proton radius.
Chamber position resolutions need to be at the 100 - 150 $\mu$m
level, with relative orientations at the sub-mr level, so that
calibrations can precisely determine the chamber coordinates.
This is because the measured cross sections are sensitive to 
systematic offsets in the scattering angle, particularly at forward
angles where the electromagnetic cross section varies rapidly
with angle.
More details on the requirements are given in the detector
sections.

\subsection{Rates, Accidentals, Statistics}

Singles and trigger rates are presented in Table~\ref{tab:actratesest} 
for all kinematic settings.
Elastic scattering is calculated from measured form factors, and
singles and trigger rates from all processes are estimated with
Geant4 simulations.
We also consider backgrounds, discussed in Section~\ref{sec:physreacback}.
The estimates use
beam fluxes given in Table~\ref{tab:beamflux} in Section~\ref{sec:beamline},
the detector configuration shown in Fig.~\ref{fig:layout},
and a 6 cm thick LH$_2$ target, with 0.125 mm thick kapton entrance and exit windows.
The trigger rates are based on sufficient energy being deposited
in two planes of scintillator paddles, no hit detected in the veto scintillator, 
and the efficiency of 
the beam PID system at rejecting $\pi$ events at the trigger level -- see  
Section~\ref{sec:triggerdaq}.
The singles rate is the integrated rate for all scintillator paddles
in one wall, which is dominated by forward-angle particles, with the 
most forward scintillator paddle having up to about one-third of the total rate quoted.
Background rates that were cross-checked with standalone estimates include 
$\pi^{\pm} p$ scattering, evaluated using the SAID partial wave analysis, 
available online at \url{http://gwdac.phys.gwu.edu/}, and
particle decays in flight.

\begin{table}[h]
\caption{\label{tab:actratesest} 
Rates for both detector arms combined
for various processes in Hz
with the estimated beam fluxes totaling 3.3 MHz for all particle types.
The ``+(-)'' momenta indicate positive (negative) polarity particles.
For elastic processes from the target the singles and trigger rates
are basically equal, but for particles from decays in flight or
low-energy particles knocked out of the target this is not the case.
The rates are for both detector arms combined.
}
\begin{tabular}{|r|c|c|c|c|c|c|}
\hline
Momentum (MeV/$c$) & ~~+115~~ & ~~+153~~ & ~~+210~~ & ~~-115~~ & ~~-153~~ & ~~-210~~ \\
\hline
{\bf $\mu+p$ elastic scattering }  & {\bf 0.6} & {\bf 2.6} & {\bf 1.0} & {\bf 0.3} & {\bf 0.7} & {\bf 0.5}\\
$\mu$+kapton elastic scattering     & 0.8   & 2.0   & 0.4 & 0.4 & 0.5 & 0.2\\
Geant4: $\mu$ singles & 590 & 1452 & 680 & 278 & 387 & 352 \\
Geant4: $\mu$ triggers & 96 & 384 & 136 & 55 & 96 & 76 \\
\hline
{\bf $e+p$ elastic scattering}  & {\bf 54} & {\bf 20} & {\bf 1.9}  & {\bf 55} & {\bf 28} & {\bf 7.5} \\
$e$+kapton elastic scattering  & 21 & 6.6 & 0.5  & 22 & 9.5 & 2.0 \\
Geant4: $e$ singles  & 77309  & 47691  & 8820 & 89036 & 73080  & 36942\\
Geant4: $e$ triggers & 2619 & 1386 & 288 & 2465 & 2070 & 1128 \\
\hline 
Geant4: $\pi$ singles  & 8442  & 126750  & 274960 & 5176 & 34660  & 152341 \\
Geant4: $\pi$ triggers & 2340 & 43725 & 85360 & 1393 & 11948 & 42630\\
Geant4: $\pi$ triggers + beam PID & 0 & 0 & 0 & 0 & 0 & 0 \\
\hline
Total singles rate & 86341 & 175893 & 284460 & 94489 & 108126 & 189635 \\
Total Geant4 triggers + beam PID & 2715 & 1774 & 433 & 2520 & 2167 & 1208 \\
\hline
\end{tabular}
\end{table}

The Geant4 triggers in Table~\ref{tab:actratesest} exceed the
elastic triggers, mainly from interactions with the upstream beam line
detectors scattering particles straight into the scintillators to
generate a trigger.  
These events are removed at the analysis stage through the
reconstructed target vertex, which is far from the actual target.
The highest DAQ rates of nearly 4 kHz come from electrons at negative
polarity.
Our goal is for the DAQ system to be able to read out an event in 0.1
ms, which would lead to a 20\% dead time at 2 kHz trigger rate.
Given the ratio of electron scattering to muon scattering events,
the DAQ rate can easily be reduced to a more manageable level
by prescaling the electron triggers.
There are additional triggers described in Section~\ref{sec:triggerdaq} 
that are not in Table~\ref{tab:actratesest} but which will be read out by the DAQ.

The highest total background singles rate of about 280 kHz
contaminates $\approx$3\% of the event data, assuming a 100 ns time 
window, the scale of the straw chamber drift times. 
A more significant random background is the $\approx$6.5\% probability of a second
beam particle in the same RF bucket as the triggering beam particle. 
The potential statistical precision gained from including these events
is outweighed by the potential that their inclusion changes the shape
of the angular distribution, and the extracted radius.

The number of counts measured is related to the cross section by
\begin{equation}
N_{counts} = N_{beam} \times (x\rho)_{target} \times {{d\sigma}\over{d\Omega}}
    \times \Delta\Omega \times \epsilon~,
\label{eq:cseq}
\end{equation}
where $N_{counts}$ is the number of elastic events counted,
$ N_{beam}$ is the number of beam particles,
$(x\rho)_{target}$ is the target areal density,
${{d\sigma}\over{d\Omega}}$ is the elastic differential cross section,
$\Delta\Omega$ is the detector solid angle,
and $\epsilon$ accounts for all efficiency factors
(detection, electronic, data acquisition, and analysis efficiencies)
and radiative corrections. 
We estimate that we typically lose about 30\%
of potential events due to a combination of
detector inefficiencies (hardware and reconstruction),
accidental secondary beam particles, 
target fiducial cuts, and
DAQ live time.

With the planned MUSE system and a two-calendar-year production run,
the planned statistical precision of MUSE is $\approx$1\% for muons
in our lowest precision bins, well below 1\% in most of our kinematic
range, and typically several times better for electrons.
The statistical uncertainty has a small increase from subtraction of
the target wall background.
For muons, there is an additional small increase in the uncertainties
from the subtraction of the muon decay background.
These and other factors are considered in leading to the projected
data shown in Section~\ref{sec:dataprojections}.

\subsection{Systematics}

To obtain precise cross sections and a precise proton radius,
systematic precision is required as well as statistical precision.
The main systematic issue in electromagnetic scattering is that the
cross sections changes rapidly with angle for low $Q^2$,
forward-angle data, and with energy, due to the kinematic
factors in the cross section.
These issues are analyzed in 
Appendix~\ref{sec:uncertaintiesoverview}.
The main result of these studies is that 
we can control changing the shape of the angular distributions
significantly 
by determining detector orientations at the sub-mr level,
by determining beam momentum at the $\approx$0.3\% level, and 
by correcting for the effects of multiple scattering through Monte Carlo.
The necessary calibration procedures are discussed elsewhere in this report.

%\section{The \boldmath{$\pi$}M1 Beam Line}
\section{The $\pi$M1 Beam Line and Detectors}

\subsection{Beam Properties}
\label{sec:beamline}
\subsubsection{Measured Beam Properties}

The $\pi$M1 beam line views the proton beam spot on the M1 target
at an angle of 22.5$^{\circ}$ relative to the proton beam.
The pion production region viewed by $\pi$M1 has a size about 
2 mm horizontal (full width) by 
2.9 mm vertical (2$\sigma$) by 
13 mm along the $\pi$M1 $z$ axis (6$\sigma$).
The  $\pi$M1 channel nominal acceptance is about 6 msR solid angle
and with a 3\% momentum bite.
The channel includes focusing
quads, two dipoles which each bend the beam 75$^{\circ}$ in the
horizontal direction, and two sets of jaws.
The default tune is point-to-point, producing an image
of the production target at a path length distance of about 23.5 m.
Detailed properties of the channel magnets (field maps) do not exist,
and it is understood that the fields of the upstream quads are
distorted by the steel enclosures.

Except for a Hall probe is mounted in the downstream
dipole, channel magnet stability is only monitored through the
power supply currents. 
The dipole field appears to generally have been stable to better 
than $10^{-4}$ during our test runs.

The $\pi$M1 channel has previously mainly been used and studied for $\pi$ 
beam properties, we have now studied the $\mu$ and $e$ beam properties as well.
Here we present several examples. 

\begin{figure}[h]
\centerline{\includegraphics[width=0.95\textwidth]{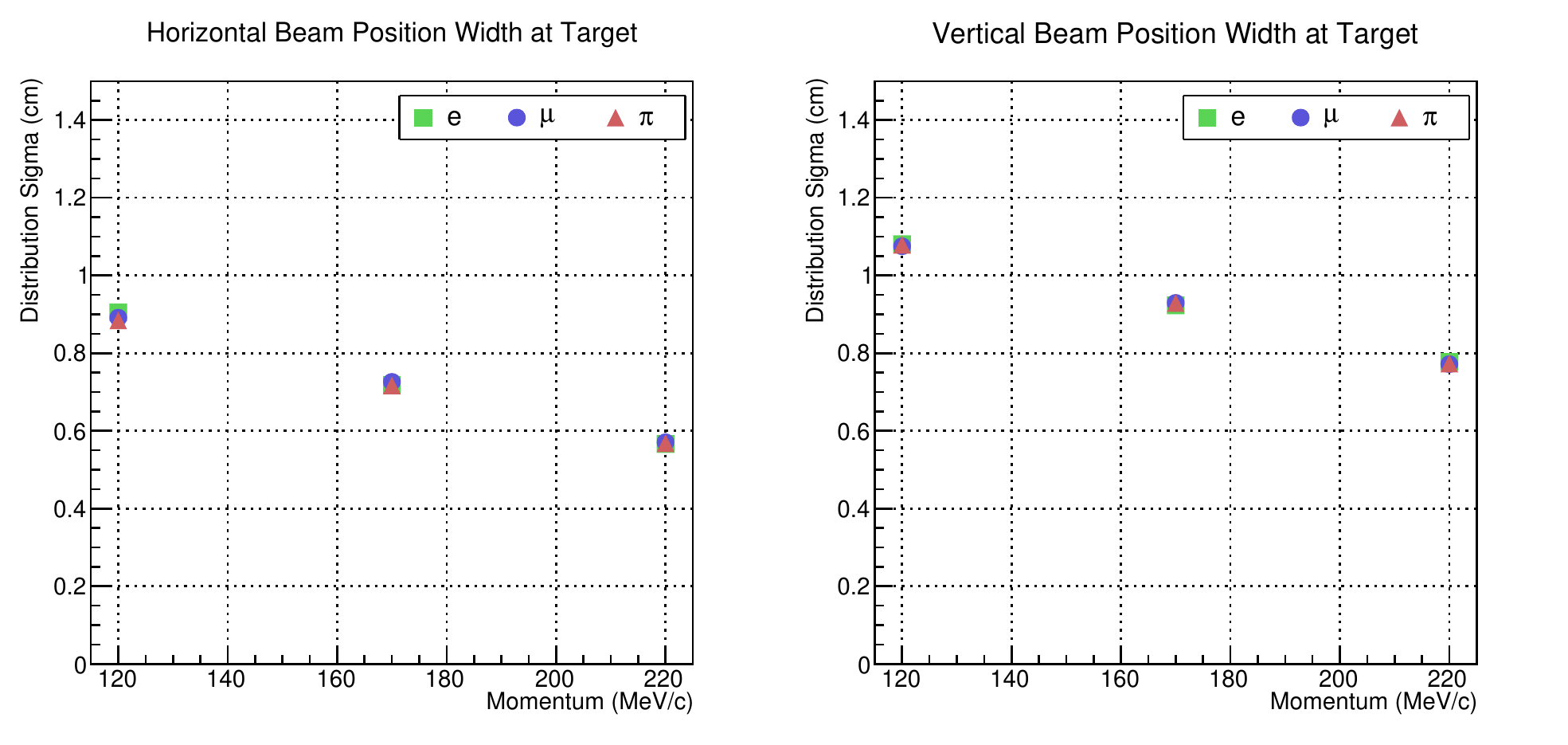}}
\caption{Width ($\sigma$) of the beam as measured in the June 2014 test setup with GEM chambers as a function of momentum
and particle type.}
\label{fig:beamdist}
\end{figure}

Figure~\ref{fig:beamdist} shows the beam distribution measured with
a GEM telescope (see Sec.~\ref{sec:gems}), projected to the target position.
The beam width obtained varies with momentum but is independent of particle type.
At the central momentum setting, the nominal beam spot ($\sigma$) is $\approx$0.8$\times$0.9~cm$^2$.
The angular divergence of the beam 
was found to be at worst (lowest momentum) $\sim$24 mr ($\sigma$) in the horizontal
direction and $\sim$15 mr ($\sigma$) in the vertical direction.
These results are similar to the nominal $\pi$M1 characteristics at
\url{http://aea.web.psi.ch/beam2lines/beam_pim1.html}.
In the MUSE experiment, the detectors in the beam cause multiple
scattering, which makes the beam spots on target slightly different
for different particle types.

\begin{figure}[h]
\centering{
\includegraphics[width=0.48\textwidth]{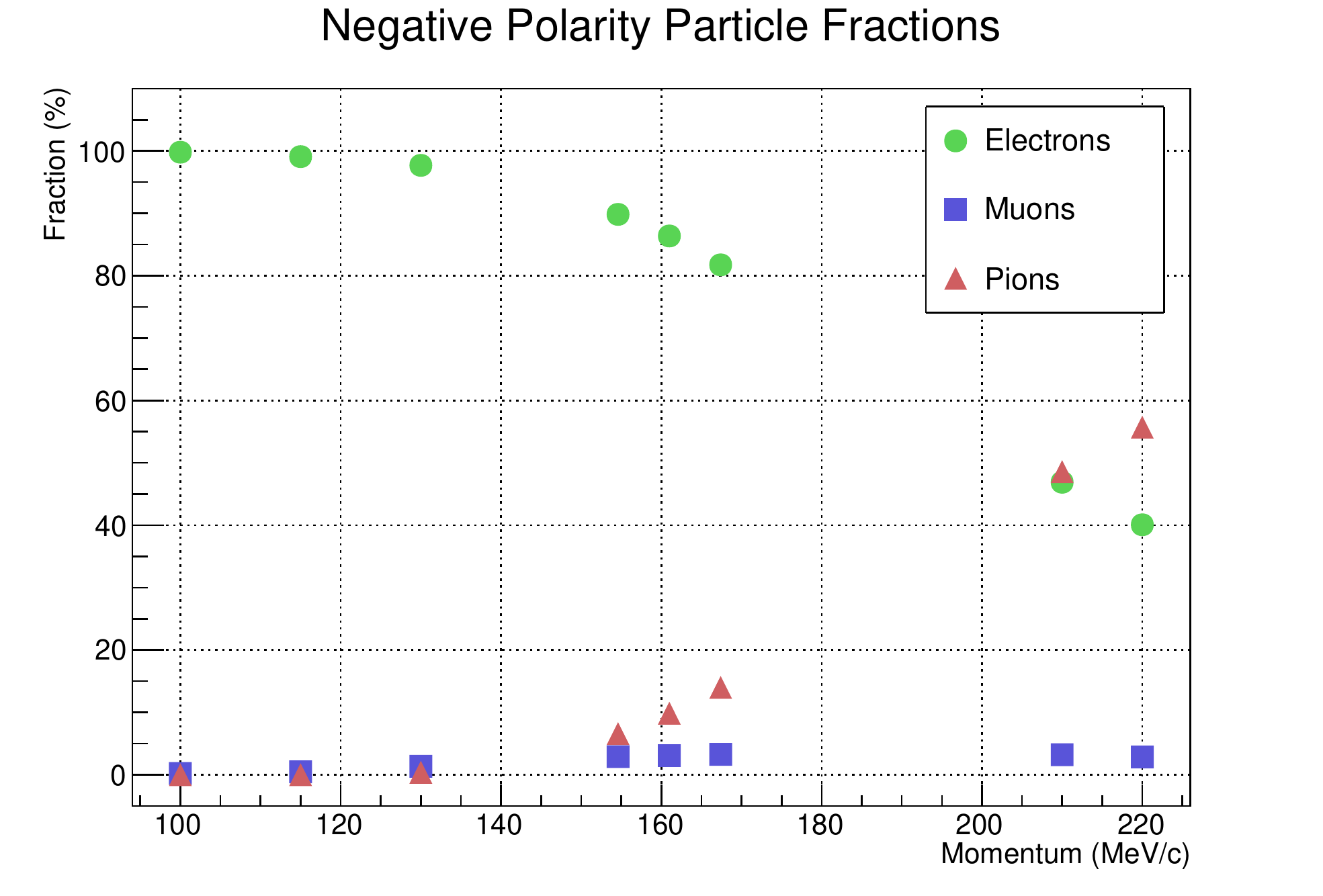}
\includegraphics[width=0.48\textwidth]{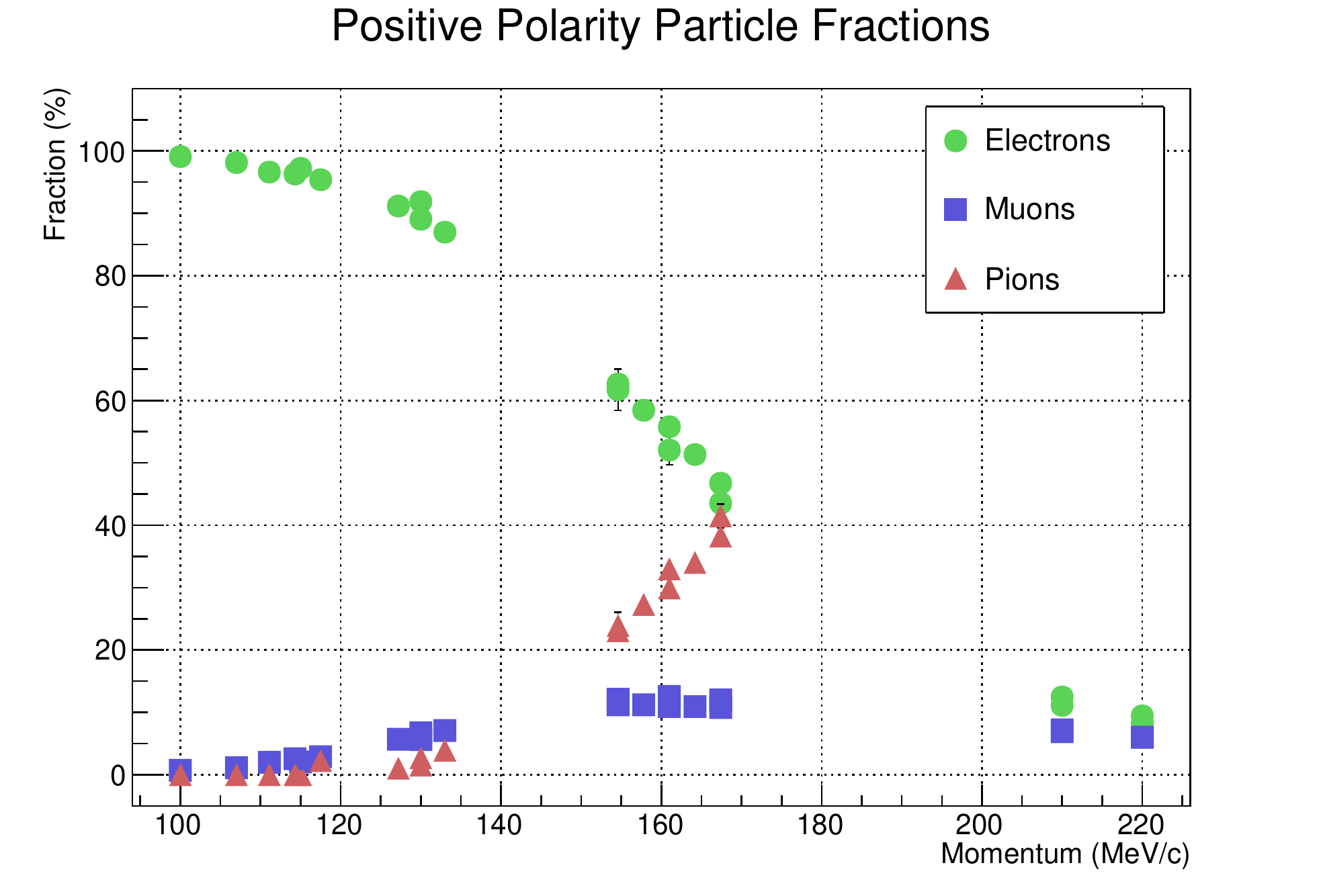}}
\caption{Measured particle fractions versus beam momentum as measured in June 2013.}
\label{fig:beamfractions}
\end{figure}

The $\pi$M1 $\pi$ and $e$, but not $\mu$, fluxes were measured by Schumacher and Sennhauser
in 1987 \cite{schumacher1987}. The RF time spectrum -- see Fig.~\ref{fig:rftimeall} -- 
was measured at several beam momenta to determine
particle fractions for each polarity, shown in
Fig.~\ref{fig:beamfractions}. 
(We have not had the appropriate equipment and readout to measure
the absolute beam flux.)
Combining the absolute measurements of \cite{schumacher1987}
scaled to a 2.2 mA primary proton current with our fractional
measurements results in the fluxes given in Table~\ref{tab:beamflux}.
No protons were observed in the channel in our momentum range.

\begin{table}[h]
\caption{\label{tab:beamflux} Beam flux at the target for 
full $\pi$M1 channel acceptance with 2.2 mA primary proton current.
The total flux is based on previous measurements, while the relative
fluxes of each particle types are based on MUSE measurements.
Also shown in parentheses is the flux of each particle type when the
combined flux is limited to the MUSE planned total flux of 3.3 MHz.}
\begin{tabular}{|c|c|c|c|c|c|}
\hline
Momentum & Polarity & Total Flux & $e$ Flux & $\mu$ Flux & $\pi$ Flux \\
(MeV/$c$) & & (MHz)  & (MHz)  & (MHz)  & (MHz) \\
\hline
\hline
115 & + & 8.3 & 8.05 (3.20) & 0.17 (0.07) & 0.08 (0.02) \\
153 & + & 16.9 & 10.65 (2.08) & 2.03 (0.40) & 4.23 (0.83) \\
210 & + & 79.2 & 9.50 (0.40) & 6.34 (0.26) & 63.36 (2.64)  \\
115 & $-$ & 7.4 & 7.29 (3.25) & 0.07 (0.03) & 0.04 (0.02) \\
153 & $-$ & 11.9 & 10.71 (2.97) & 0.38 (0.11) & 0.81 (0.22) \\
210 & $-$ & 24.0 & 11.28 (1.55)  & 0.96 (0.13)  & 11.76 (1.62)  \\
\hline
\end{tabular}
\end{table}

We studied beam properties at the IFP, using a SciFi 
detector, and compared the results to TURTLE simulations.
The simulated $\pi$ beam envelope at the IFP is
22.5 cm wide (full width at 10\% maximum), with sharp edges
and a momentum dispersion of 7 cm/\%,
and roughly Gaussian in the vertical direction with width
$\sigma$ = 0.60 cm, and no visible tails outside $\pm$2.25 cm.
The measured beam that reached the scattering target
came through the IFP in a region about 20 cm wide by 5 cm high,
with significant uncertainty in the horizontal direction.

We checked the momentum dispersion of $\pi$'s and $\mu$'s 
in the channel by using a collimator slot at the IFP and measuring 
both the shift in the RF time at the target and time of flight between two
scintillators.
All data to date are consistent within uncertainties with the 
expected 0.14\%/cm.

\subsubsection{Beam Simulations}
\label{sec:beamsim}

We have implemented in our Geant4 \cite{Agostinelli:2002hh} simulation
a realistic beam parameterization based on our measured beam
properties.
The beam properties are similar to TURTLE predictions of the parent
particle distributions.

\begin{figure}[ht]
\centerline{
\includegraphics[width=3.0in]{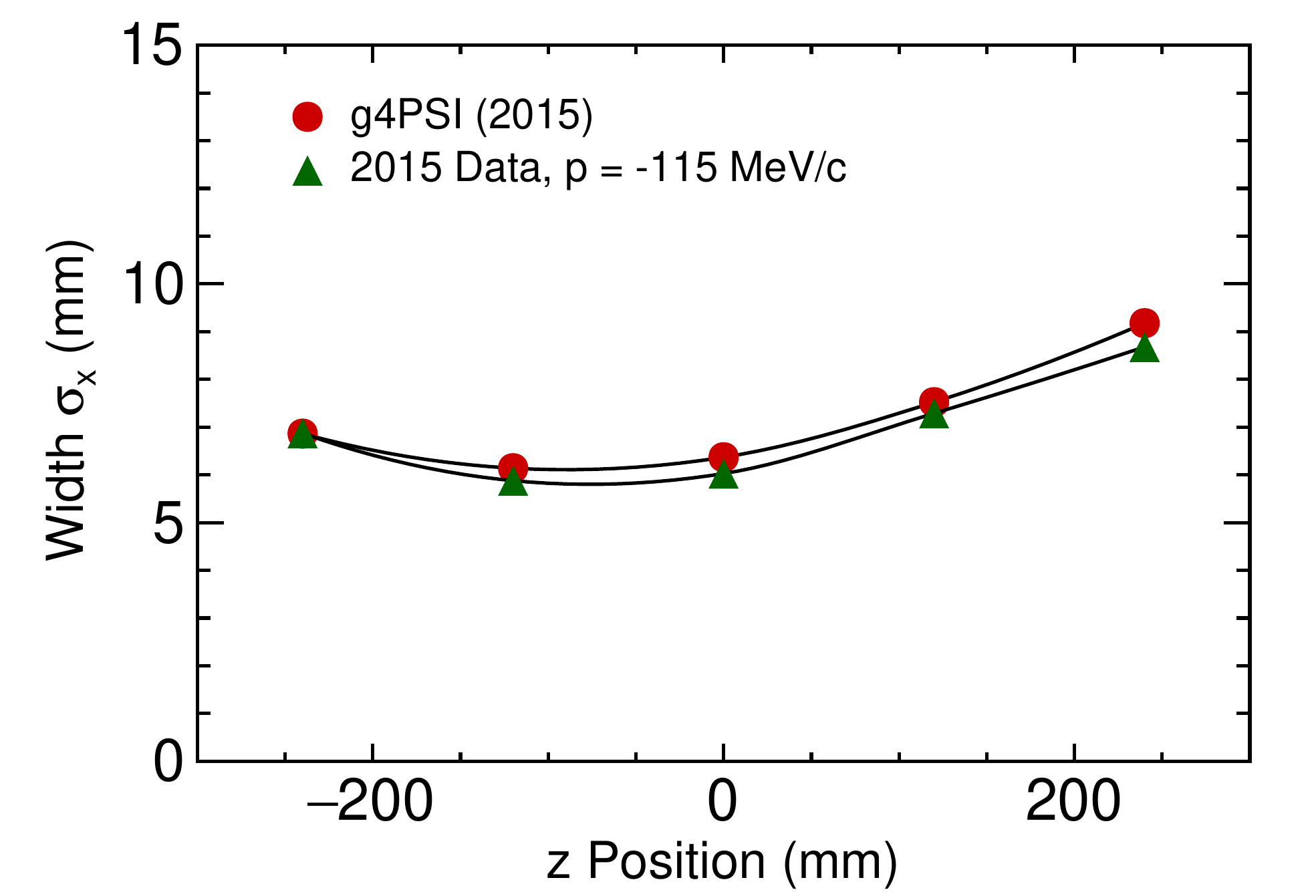}\hfill
\includegraphics[width=3.0in]{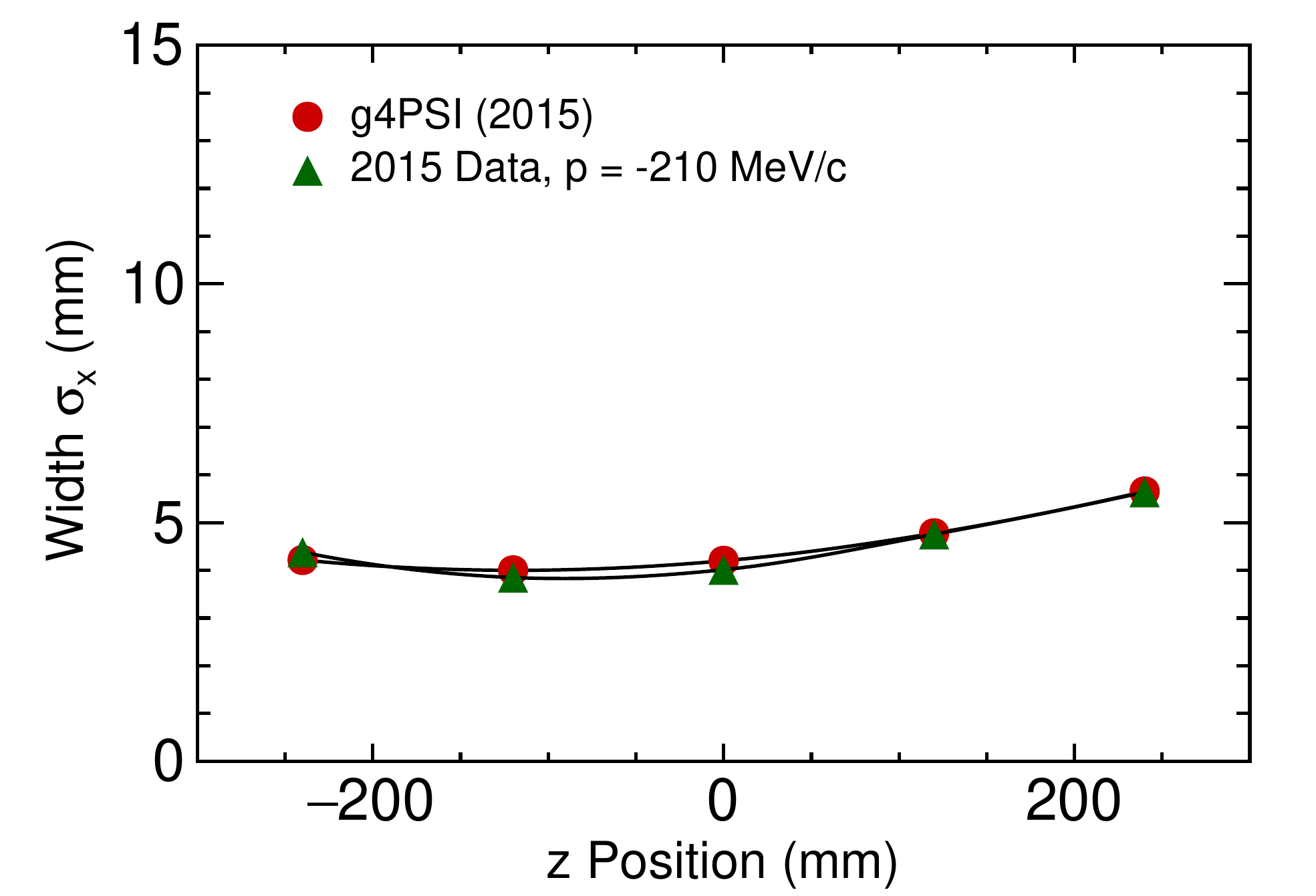}
}
\caption{Comparison of measured (triangles) beam widths $\sigma_x$ at
  various positions with simulated beam distributions (circles).  The
  simulation is based on beam parameters which were tuned to fit the
  experiment.  The examples are for electrons at beam momenta of
  115~MeV/c and 210~MeV/c, respectively.}
\label{fig:sim_beam_fit}
\end{figure}

Figure~\ref{fig:sim_beam_fit} shows the horizontal beam width for the
measured data (triangles) and for the simulated beam distribution
(circles) for electrons at 115~MeV/c and 210~MeV/c, respectively.
Similar results are obtained for the vertical beam width and for other
beam momenta and polarities. The simulation adequately 
reproduces the measured beam properties.

\begin{figure}[ht]
\centerline{
\includegraphics[width=3.0in]{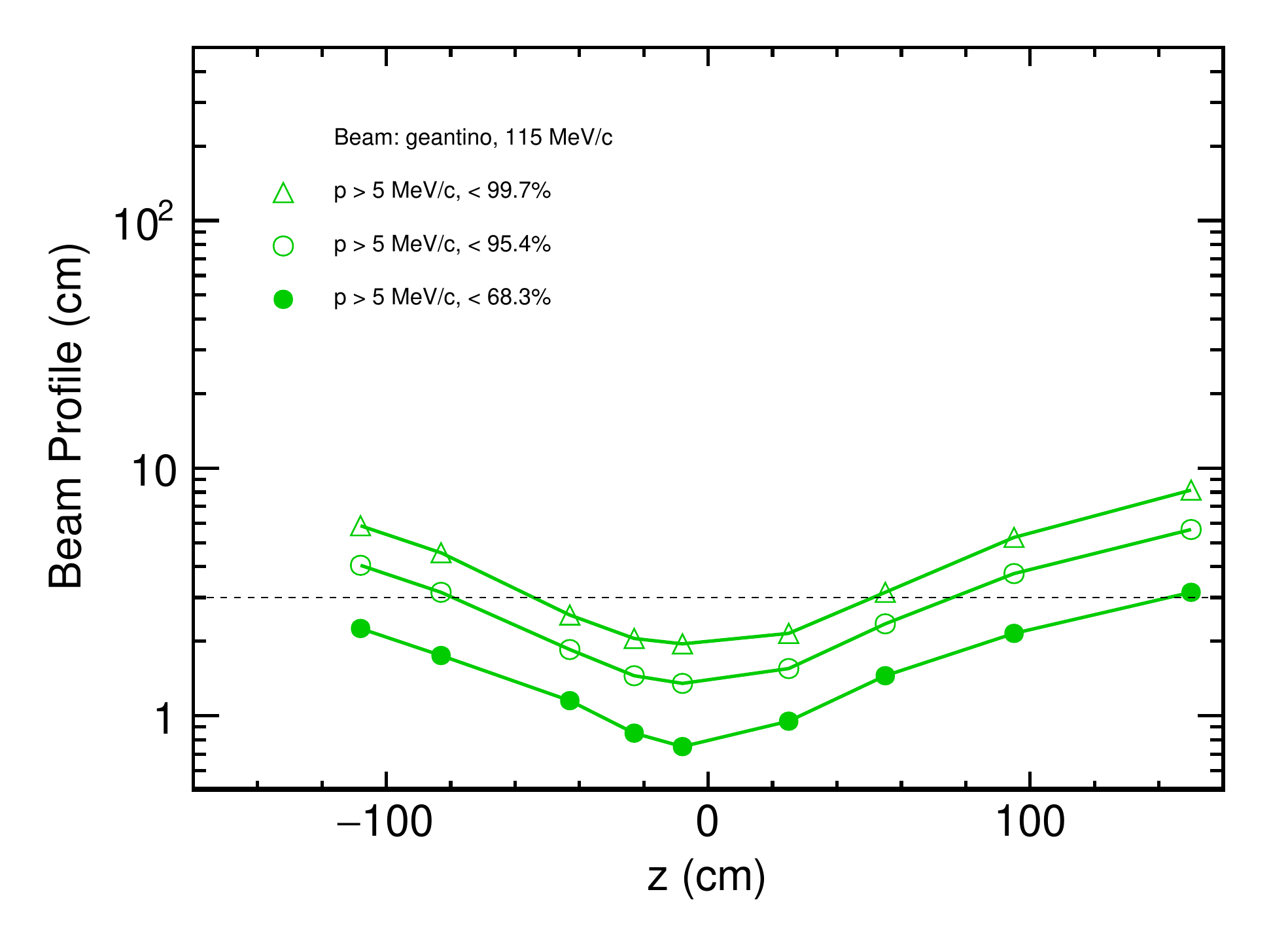}\hfill
\includegraphics[width=3.0in]{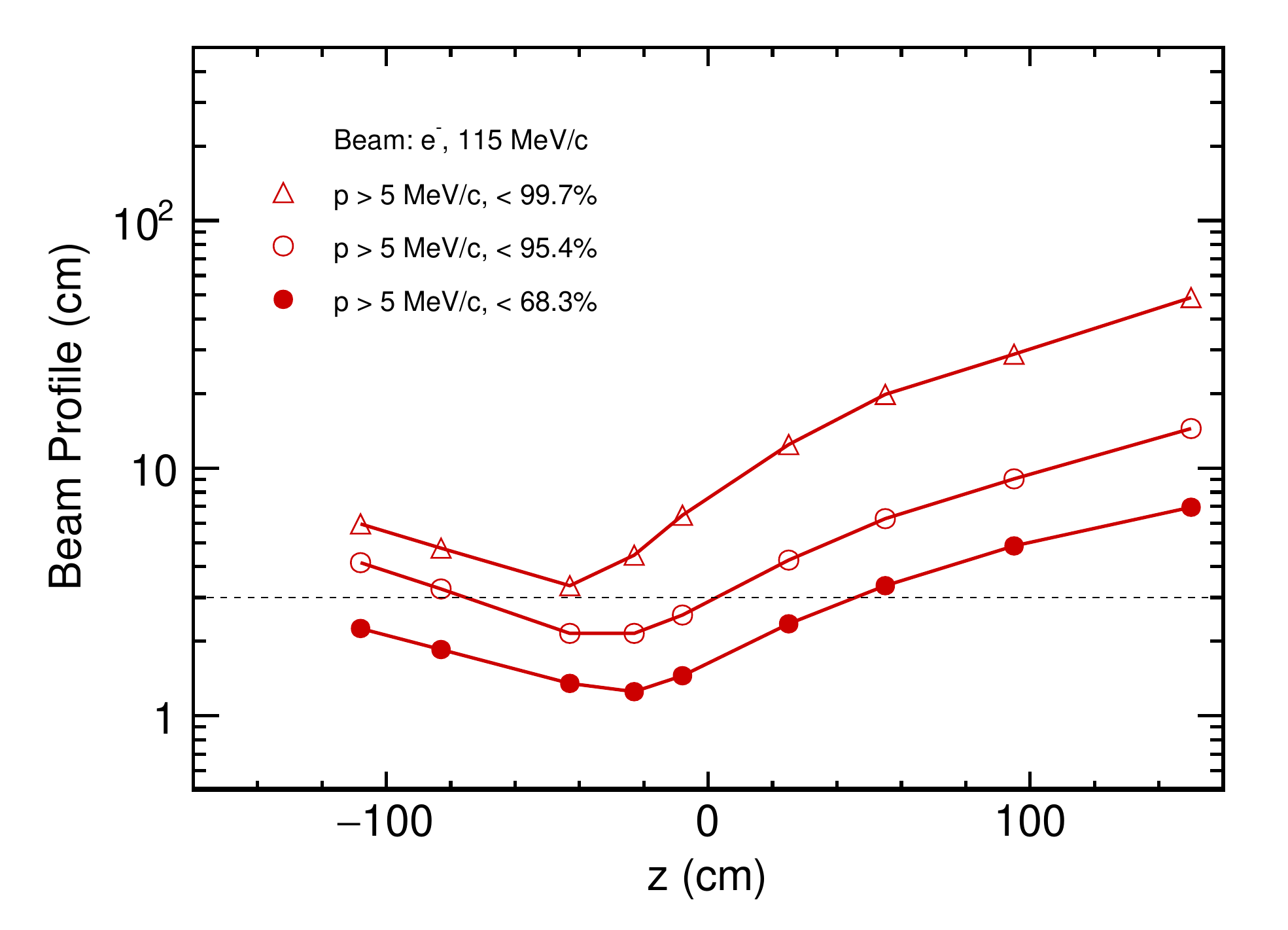}
}
\centerline{
\includegraphics[width=3.0in]{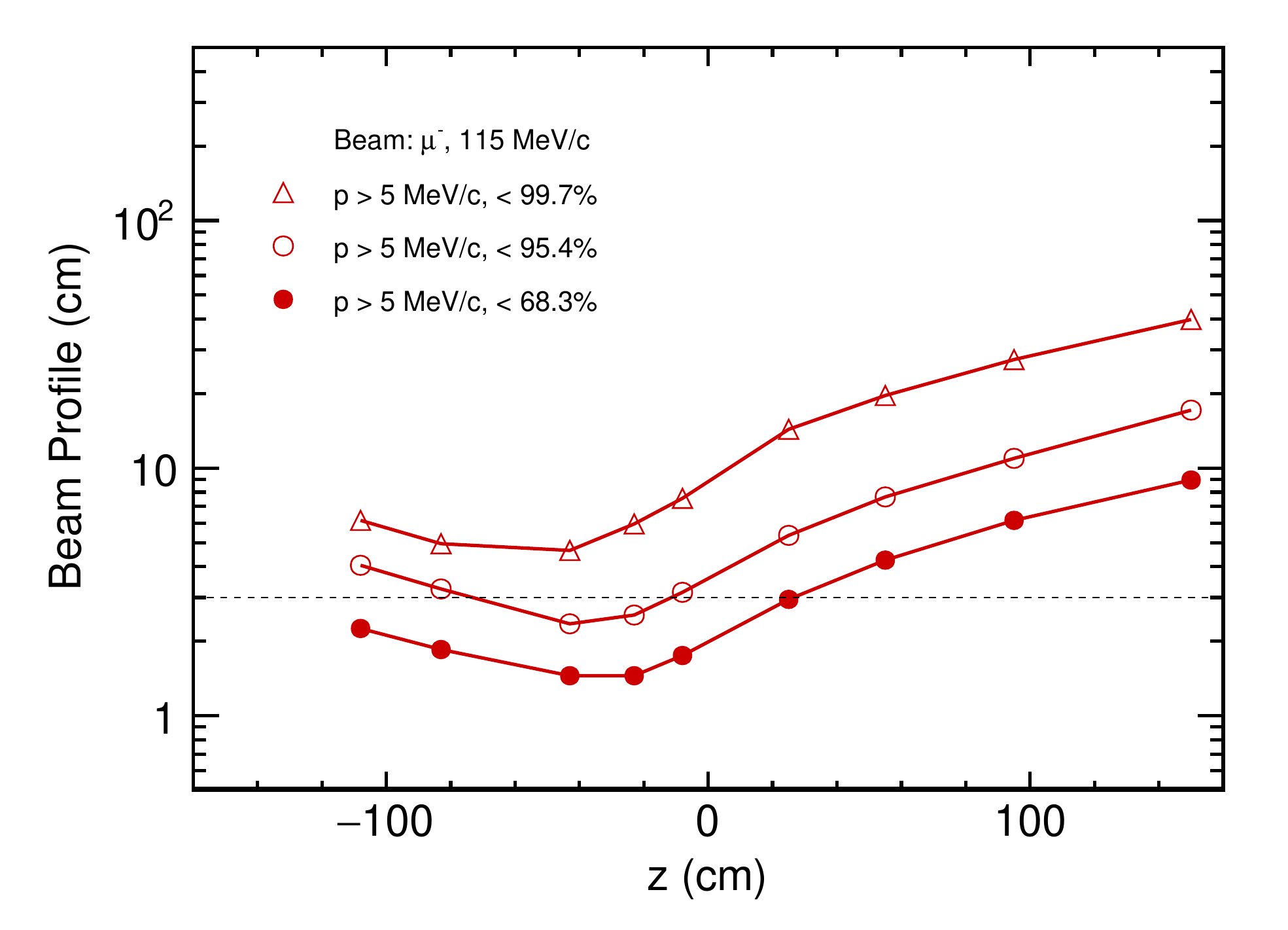}\hfill
\includegraphics[width=3.0in]{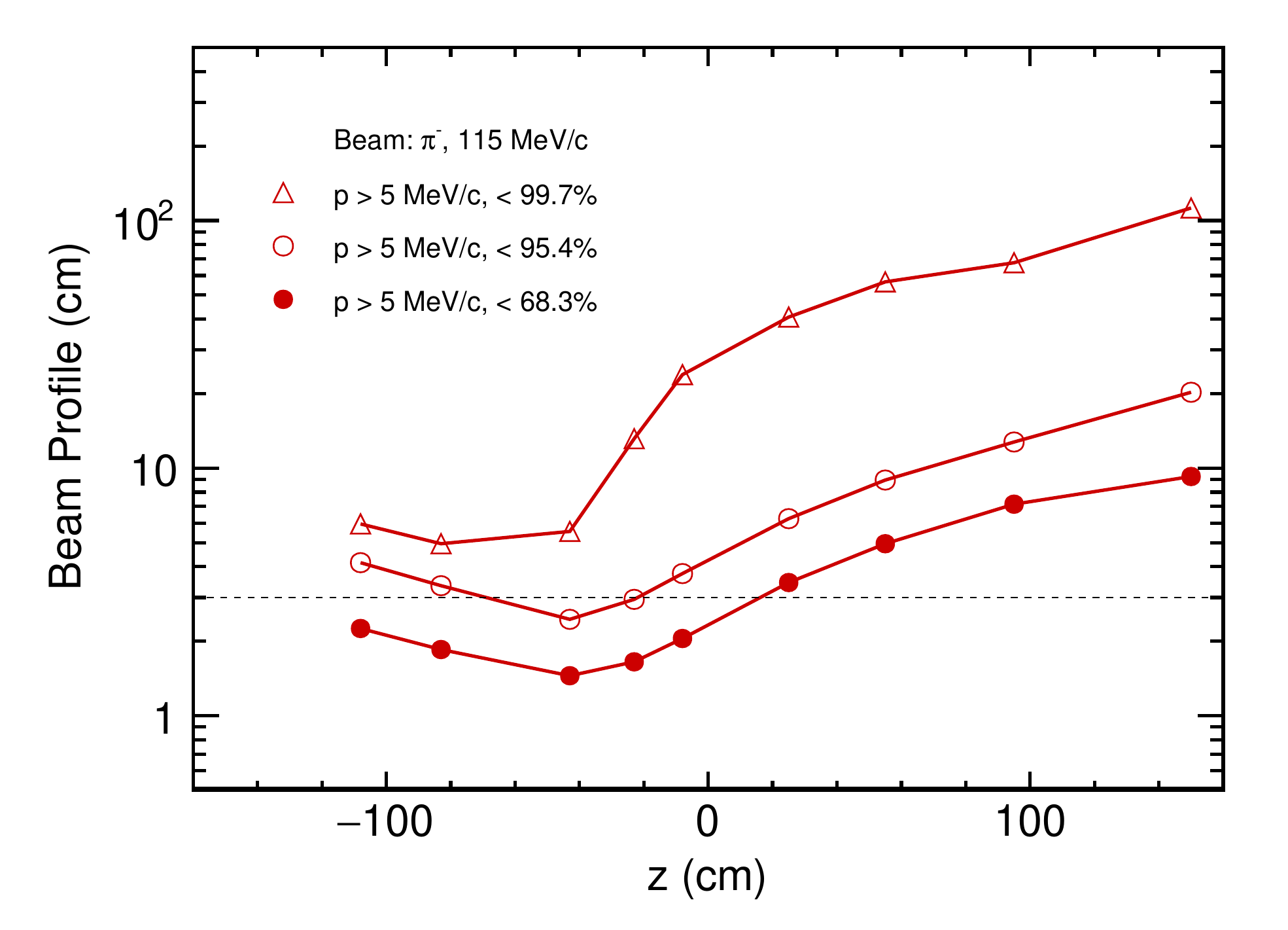}
}
\caption{Evolution of the beam profile along the beam line from
  simulation for various particle types at 115~MeV/$c$ incident particle
  momentum.  The curves show distances from the beam axis which
  enclose the given fractions of beam particles with at least 5 MeV/$c$
  momentum.  The effect of multiple scattering in the MUSE setup is
  included for $e^-$, $\mu^-$, and $\pi^-$ (red); the
  \textit{geantino} particles (green) do not interact and reflect the
  original beam distribution. The dashed line marks the size of the
  target with $r = 3$~cm.}
\label{fig:sim_beam_profile}
\end{figure}

The realistic beam parameterization allows the prediction of beam
properties in the presence of the MUSE beam line detectors.
Figure~\ref{fig:sim_beam_profile} shows the minimum radial distance
from the beam axis that encloses a certain fraction of primary beam
particles.  The panels are for different beam particles:
\textit{geantinos} (green), which do not interact in the simulation,
electrons, muons, and pions in red.  The effect of multiple scattering
in air and, especially, in detector components starting at about
$z = -40$~cm, is evident.  
The distributions in
Fig.~\ref{fig:sim_beam_profile} are for a beam momentum of
115~MeV/$c$, where the effect of
multiple scattering is largest.
The target radius of 3~cm is indicated as horizontal dotted line.  The
electron and muon beams almost fully pass through the target
at about $z = 0$~cm.  A more quantitative measure of useful beam
particles is the fraction of primary particles which enter the target
volume within $r < 2.5$~cm and $5^\circ$ in the forward direction.  An
overview of these fractions is given in
Table~\ref{tab:target_fraction} for two setup options: one with two
and one with four beam hodoscope detectors just upstream of the three GEM
detectors.
\begin{table}[h!]
\begin{center}
\renewcommand{\tabcolsep}{2mm}
\renewcommand{\arraystretch}{1.1}
\caption{Fraction of selected beam particles entering the target volume for the
  MUSE setup including 2 or 4 SiPM detectors and 3 GEM planes.
  \label{tab:target_fraction}}
\begin{tabular}{c|ccc|ccc}
  \hline\hline
& \multicolumn{3}{c|}{two SiPM}& \multicolumn{3}{c}{four SiPM} \\
 Particle & 115~MeV/c & 153~MeV/c & 210~MeV/c & 115~MeV/c & 153~MeV/c & 210~MeV/c  \\
\hline
  $e^+$      & 0.943 &  0.976 &  0.989   &     0.922 &  0.968 &  0.986    \\
  $\mu^+$ & 0.881 &  0.969 &  0.989   &    0.832  &  0.955 &  0.989   \\  
  $e^-$      & 0.950  & 0.977  & 0.989    &    0.929  &  0.972  & 0.987    \\
 $\mu^-$  & 0.887 &  0.969 &  0.991    &    0.843 &  0.959  &  0.990  \\  
\hline\hline
\end{tabular}
\end{center}
\end{table}

\subsection{Beam Line Detectors}

\subsubsection{Beam Cerenkov}
\label{sec:quartz}
\paragraph{Purpose:}
The beam Cerenkov provides a high-resolution timing measurement
at the IFP for RF time and time-of-flight (TOF) determinations.
In order to minimize material in the beam line for regular experimental 
running, the beam Cerenkov detector will only be used for calibration and background
measurements, not for regular data taking.

\paragraph{Requirements:}
The requirements for the Cerenkov detector are laid out in table \ref{table:WBS3_requirements}.
\begin{table}
\caption{Cerenkov detector requirements}
\label{table:WBS3_requirements}
\begin{tabular}{|c|c|c|}
\hline 
Parameter & Performance Requirement & Achieved\tabularnewline
\hline 
\hline 
Time Resolution & $<$80~ps & $\checkmark$ $<$60~ps\tabularnewline
\hline 
Efficiency & 99\% for e, $\mu$ & 99\% $\pm$2\%\tabularnewline
\hline 
Positioning & $\approx$1~mm, $\approx$20~mr & not attempted, easy\tabularnewline
\hline 
Rate Capability & 2.5~MHz & $\checkmark$ 3~MHz\tabularnewline
\hline 
\end{tabular}
\end{table}

\paragraph{Detector design:}
The design is based on the work of Albrow {\it et al.}
\cite{Albrow:2012ha}, who obtained timing resolution better 
than 10 ps ($\sigma$) with a beam Cerenkov, using thick quartz bars 
read out through a Photek PMT240 multichannel plate (MCP) with an
Ortec 9327 preamp/CFD.
The time resolution extrapolates to $\approx$50 ps in MUSE
experimental conditions, as we use thinner radiators to minimize
effects on our lower energy beam.
We use the same MCP and preamp/CFD as Albrow {\it et al.}
We tilt the Cerenkov plate at close to the muon
Cerenkov angle, to provide good efficiency for all particle types.

\begin{figure}
\centerline{\includegraphics[width=2.6in]{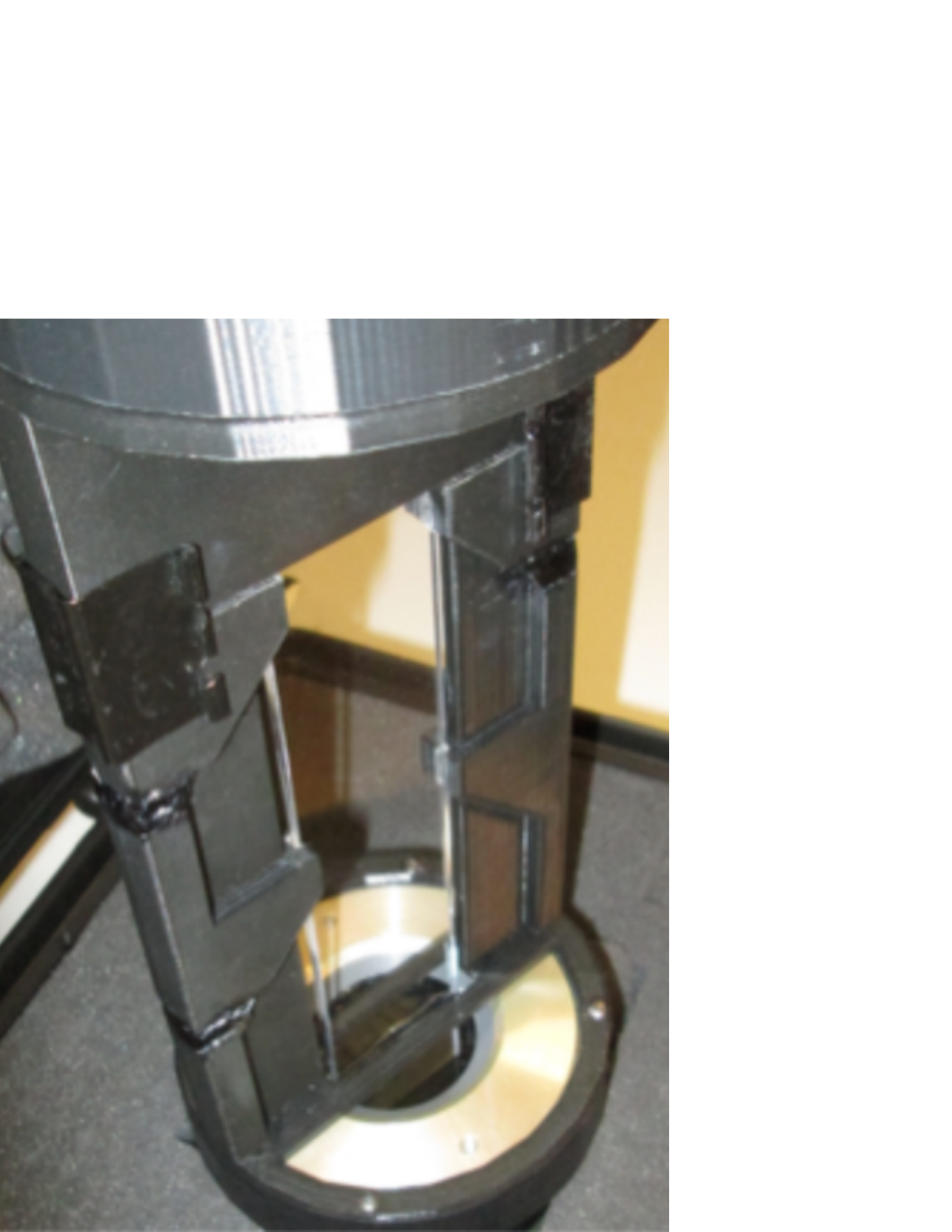}}
\caption{
Quartz plate mounted on a PMT240 with a 3d printed
mounting fixture, during the June 2015 test run.
The assembly is wrapped in tedlar for light tightness.
}
\label{fig:qsonmcp}
\end{figure}

\begin{figure}
\centerline{\includegraphics[width=3.0in]{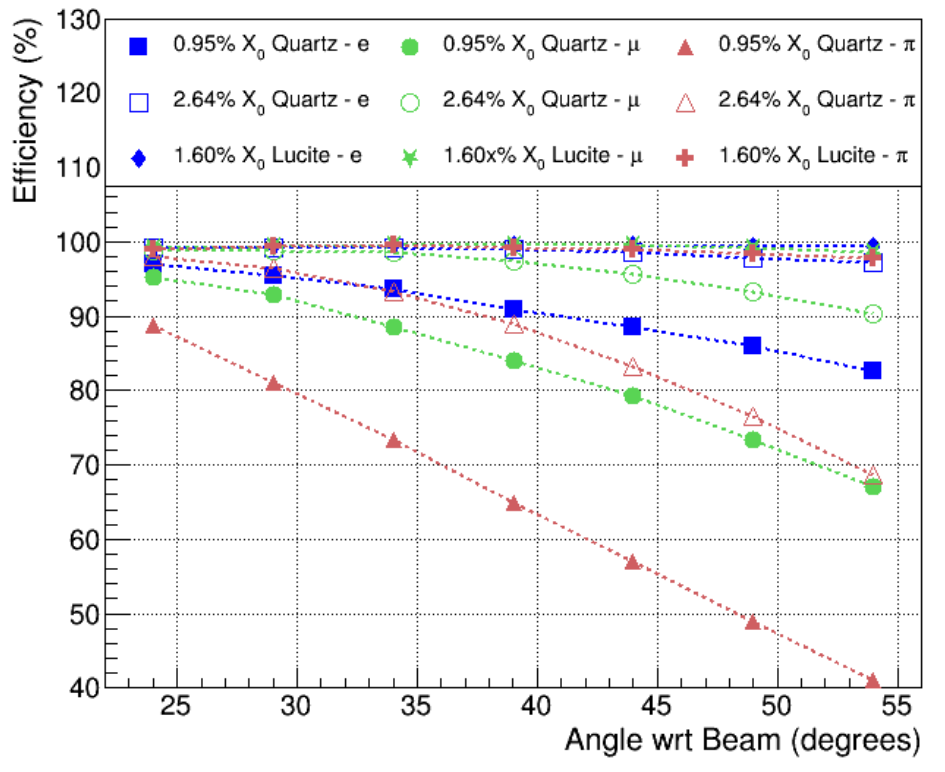}
\includegraphics[width=3.0in]{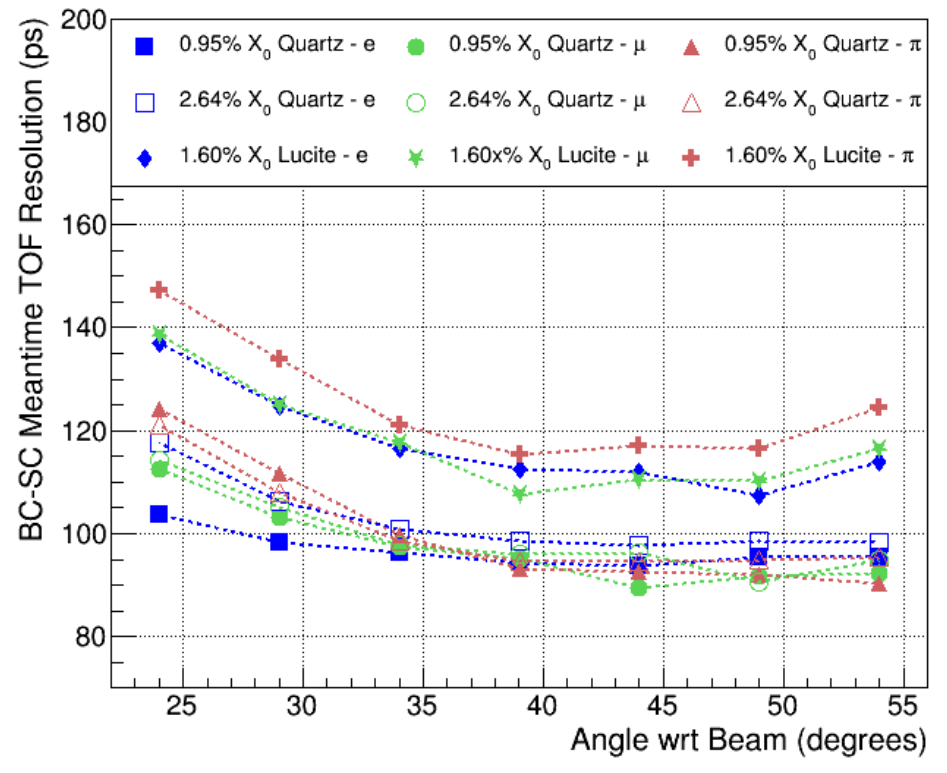}}
\caption{
Left: efficiency data for the beam Cerenkov prototype for 3 radiators.
Systematic uncertainties on the efficiencies are a few percent.
Right: time of flight resolution between the beam Cerenkov and
a fast scintillator.
}
\label{fig:bctestdata}
\end{figure}

\paragraph{Current status:}
We have achieved better than 90 ps time of flight resolution between our
prototype beam Cerenkov -- see Fig.~\ref{fig:qsonmcp} --
and a fast scintillator in tests in $\pi$M1,
corresponding to a Cerenkov resolution of 50 - 60 ps.
Examples of the test data taken are shown in Fig. ~\ref{fig:bctestdata}.
Rate dependence tests of the beam Cerenkov
response up to 3 MHz of incident particles show no effects on the 
analog output pulse shape.

\paragraph{Path to completion:}
We descoped further beam Cerenkov development and construction
after adopting the beam hodoscope technology.
The existing detector remains in PiM1, ready for
installation and use in MUSE as needed.
We will retest the detector before the experiment to ensure it is
ready if needed.

\subsubsection{Beam Hodoscope}
\label{sec:beamsipm}
\paragraph{Purpose:}
The beam hodoscope provides precise timing information 
for beam particles, along with position information.
The beam hodoscope timing information will be used in conjunction with the RF signal to
provide beam particle identification at low precision for the trigger and
at high precision for the event analysis.
Time of flight from the beam hodoscope to the scattered particle
scintillators, determines the reaction type, in particular separating muon decay
from muon scattering.
Time of flight from the beam hodoscope to the beam monitor allows
an RF time vs.\ time of flight comparison that confirms particle 
identification, helps to identify backgrounds, and determines the muon
and pion beam momenta.
The beam hodoscope also counts the total incident beam flux, and
through sampling randomly coincident beam particles determines the
flux of each particle type.

\paragraph{Requirements:}
The requirements for the beam hodoscope detector are laid out in Table
\ref{table:WBS2_requirements}.
The time resolution requirement allows the needed time of flight
resolution from the combination of multiple planes of beam hodoscope
to the scintillator walls.
The efficiency allows efficient collection of statistics and rejection
of pion induced events.
The positioning requirement is to align the beam hodoscope to the
GEM chambers, allowing efficient analysis when there are randomly coincident 
additional beam particles within the detector time windows.
The rate capability allows the full planned beam flux of $\approx$3.3
MHz to be observed at high efficiency.

\begin{table}
\caption{Beam hodoscope detector requirements}
\label{table:WBS2_requirements}
\begin{tabular}{|c|c|c|}
\hline 
Parameter & Performance Requirement & Achieved\tabularnewline
\hline 
\hline 
Time Resolution & $<$100~ps / plane & $\checkmark$ 80~ps\tabularnewline
\hline 
Efficiency & 99\%  & $\checkmark$ 99.8\% \tabularnewline
\hline 
Positioning & $\approx$1~mm, $\approx$1~mr & not attempted; easy --
                                             calibrated by data\tabularnewline
\hline 
Rate Capability & 3.3~MHz / plane & $\checkmark$ $>$10~MHz / plane\tabularnewline
\hline 
\end{tabular}
\end{table}

\paragraph{Detector design:}

\begin{figure}[h]
\centerline{
\includegraphics[height=2.5in]{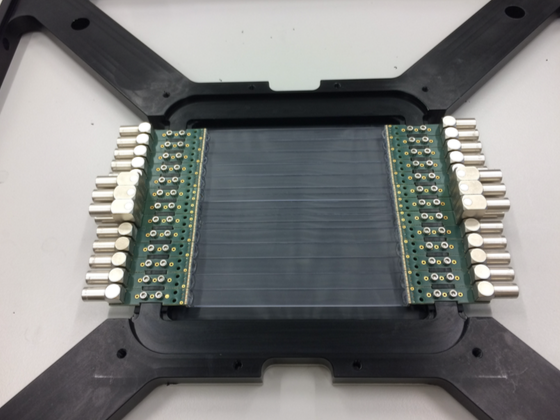}}
\caption{
A beam hodoscope plane of thin scintillators with SiPM readout
under construction.
}
\label{fig:SiPMProto}
\end{figure}

A beam hodoscope plane under construction is shown in
Fig.~\ref{fig:SiPMProto}.
We use BC-404 scintillator, with 6 central paddles of size
10 cm long $\times$ 4 mm wide $\times$ 2 mm thick, flanked
by 5 outer paddles on each side of size 
10 cm long $\times$ 8 mm wide $\times$ 2 mm thick.
This gives 16 paddles, and 32 readout channels per plane.
The paddles are glued to SiPMs which are mounted on custom printed circuit
boards (PCBs).
A foil is used during assembly to position the paddles 6 $\mu$m
apart, to provide an air gap and total internal reflection to capture
much of the light within the paddles, to be directed to the SiPMs
at each end.
The active area is $\approx$ 10 cm $\times$ 10.4 cm.

The Hamamatsu S13360-3075PE SiPMs have an active area of 
3 mm $\times$ 3 mm in a housing about 4 mm $\times$ 4 mm,
a peak quantum efficiency near 50\% at $\approx$450 nm, and
a gain of about 4$\times$10$^6$ at operating voltage.
For the central 6 paddles we use PCBs that house 2 SiPMs 
operating independently to read out 2 4-mm paddles - the PCBs
with double LEMO connectors in Fig.~\ref{fig:SiPMProto}.
For the outer 10 flanking paddles we use PCBs that house 2 SiPMs
connected in series and reading out the same paddle - the PCBs
with single LEMO connectors in Fig.~\ref{fig:SiPMProto}.
We operate the single SiPMs at $\approx$55 V, and the two SiPMs in
series at $\approx$110 V.

\begin{figure}[h]
\centerline{
\includegraphics[height=2.5in]{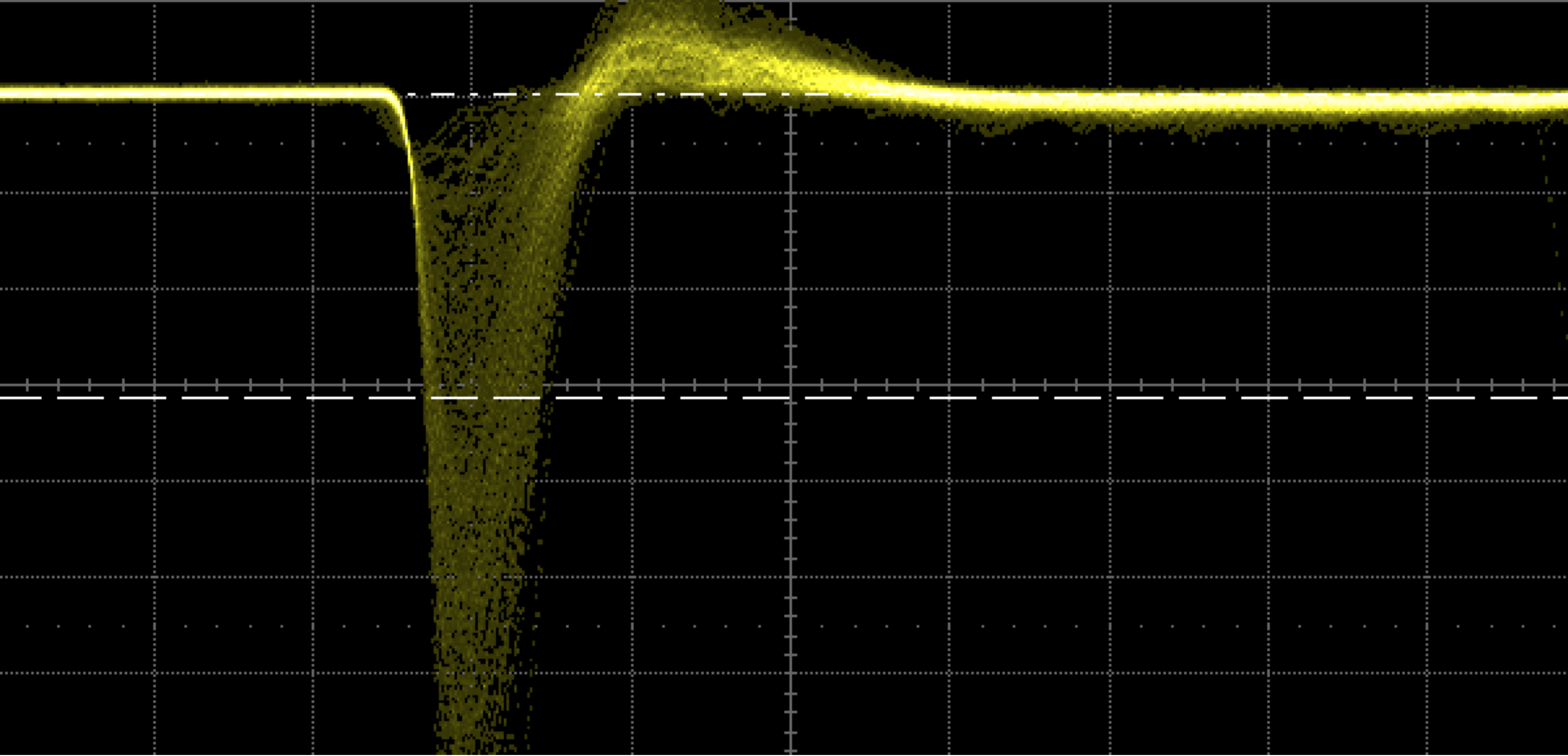}}
\caption{
A scope trace of amplified beam hodoscope prototype paddle signals.
The horizontal scale is 5~ns / division. 
The vertical scale is 100~mV / division.
}
\label{fig:SiPMSignal}
\end{figure}

The beam hodoscope analog signal is amplified to produce a fast
signal with a 1.2 ns rise time, a 3.0 ns fall time, and typically a 
few hundred mV peak.
See Fig.~\ref{fig:SiPMSignal}.
For read out, the amplified analog signal is sent to a Mesytec CFD,
which, in turn, sends  an analog copy of the signal 
to the Mesytec QDC input, and the discriminated signal to the TRB3 TDC, to the trigger, 
and (an OR of 16 such signals) to the Mesytec QDC gate.

We will build 4 planes of the beam hodoscope detector.
All 4 planes will be used at the planned highest beam momentum of 210 MeV/$c$.
Due to the effects of the beam hodoscope on the beam, and the less
precise timing requirements at lower beam momentum, the number of
planes installed will be reduced to 3 at 153 MeV/$c$, and 2 at 115 MeV/$c$.

\paragraph{Current status:}
The beam hodoscope design results from extensive prototyping of 
scintillator materials (BC-404, BC-418, BC-420, BC-422),
SIPMs (various Hamamatsu and AdvanSiD models),
geometries (lengths of 10 and 16 cm, widths of 4 mm, 5 mm, 8 mm, and
12 mm).
Both individual paddles and few-element hodoscopes were tested, to
also study cross talk.
Radiation damage of the SiPMs and recovery was also studied, to ensure
adequate lifetime of the detector.
SiPM performance was measured with them in the beam, for a total
particle flux exceeding Monte Carlo estimates of annual dosage.

The first beam hodoscope plane has now been constructed, but not yet
tested. 
An initial set of amplifier PCBs have been built and are
undergoing testing, to verify performance before the full set of PCBs
are constructed.

\begin{figure}[h]
\centerline{
\includegraphics[height=1.9in]{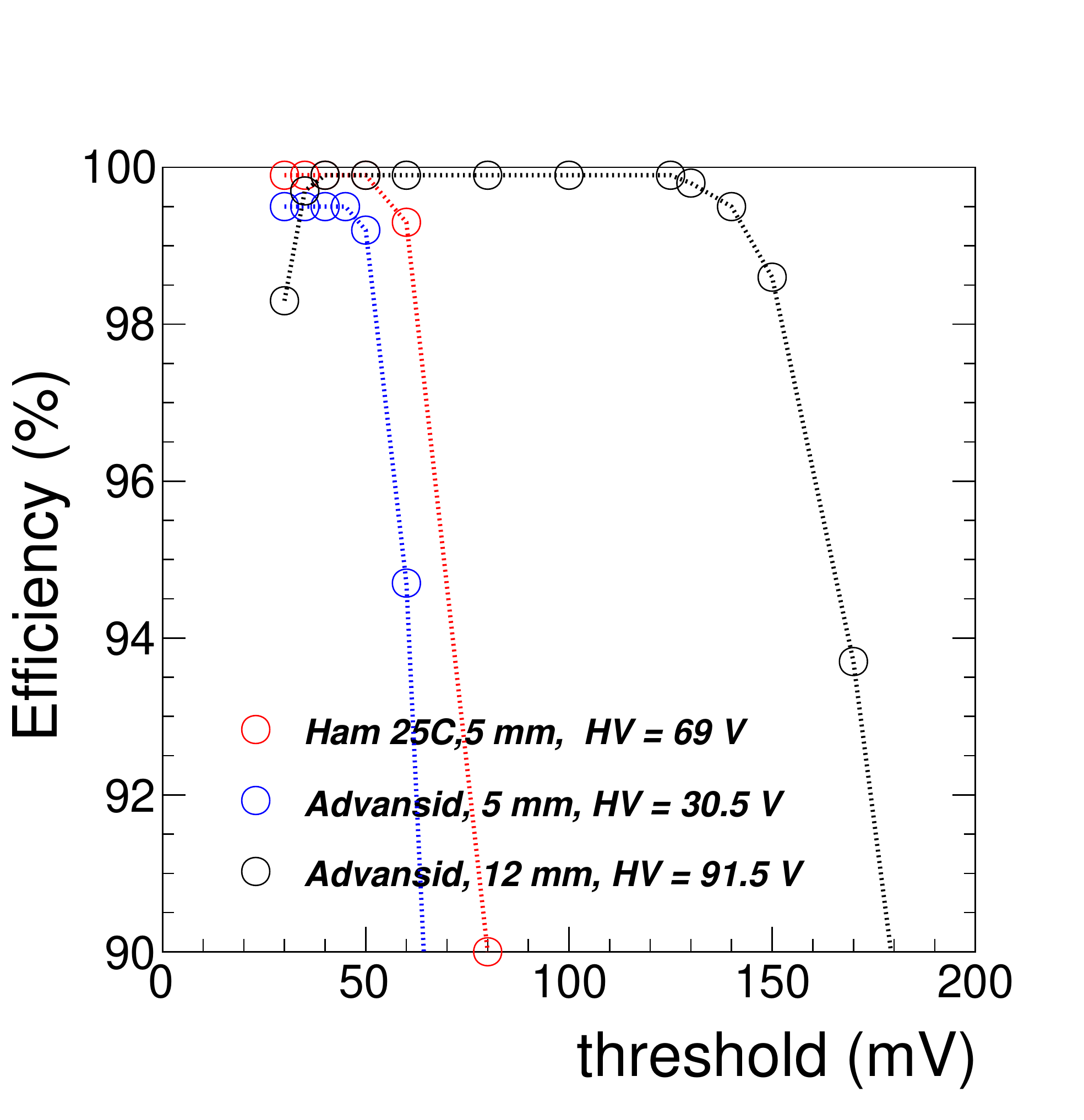}
\includegraphics[height=1.9in]{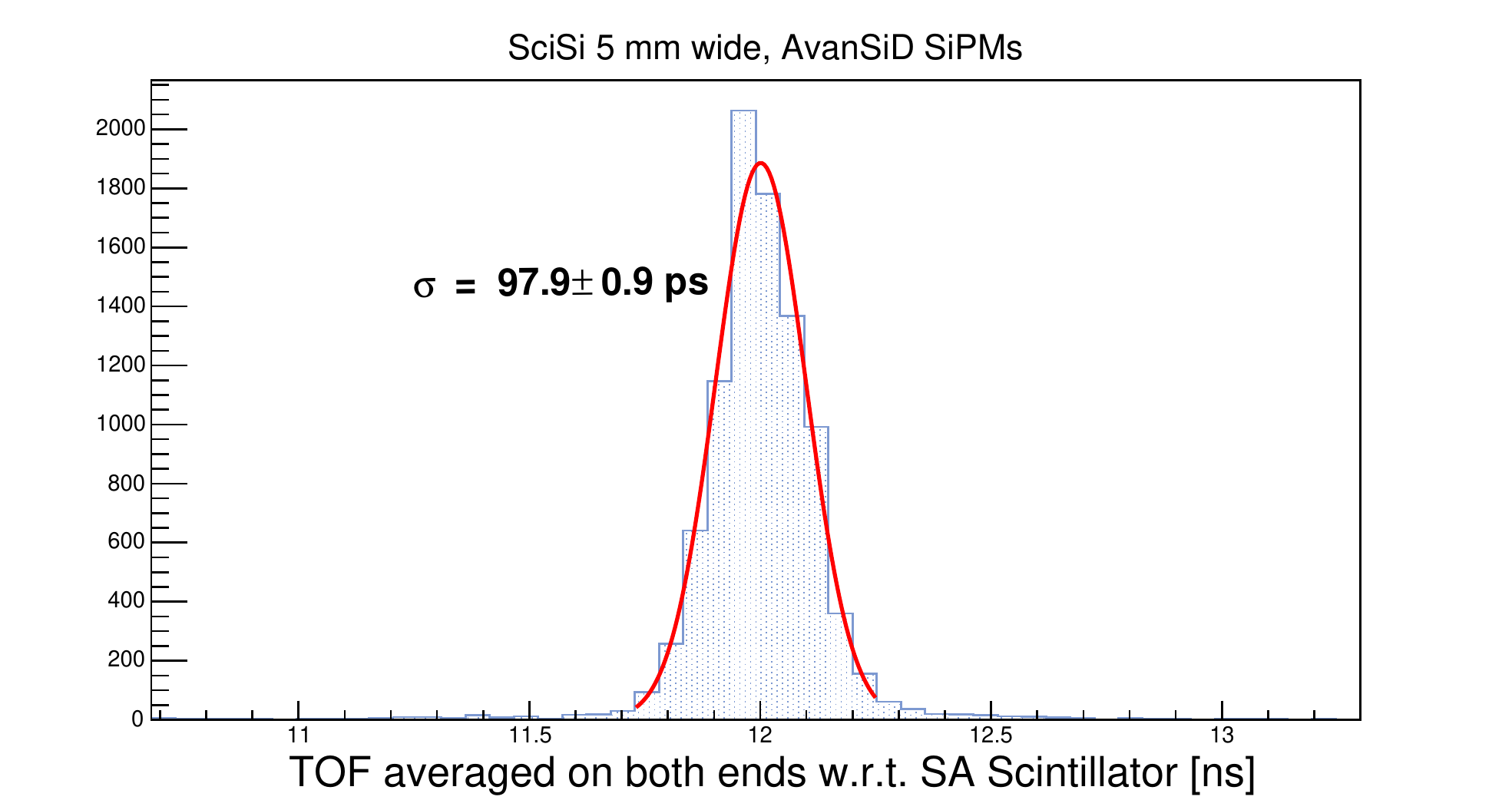}
}
\caption{
Left: Efficiency for several paddles as a function of threshold.
Right: Time resolution of a SiPM paddle, using the average of 
both ends, for events with muon RF time in fast scintillator.
}
\label{fig:SiPMProtodata}
\end{figure}

Two sample results from prototype tests follow.
 Figure~\ref{fig:SiPMProtodata} shows efficiency measurements and 
a timing spectrum from one of the SiPM detectors. 
Both 5- and 12-mm wide paddles had 99.9\% $\pm$ 0.1\%
efficiencies with properly-set thresholds at the recommended operating
voltages.
The 12-mm Advansid paddle had 3 APDs connected in series,
so it required 3$\times$ the operating voltage and threshold of the
single APD 5-mm paddle.
Resolutions for bars, averaging both ends, were less
then 100 ps for most configurations.

\paragraph{Path to completion:}
All components necessary to construct the beam hodoscope detector
have been purchased and are on hand, with the exception of four 
detector frames, which are now being machined.
The additional planes will be assembled in 2017 and early 2018.
The amplifier PCBs will be assembled after the performance of the
initial set of PCBs is verified.

\subsubsection{GEM Chambers}
\label{sec:gems}

\paragraph{Purpose:}
Measuring high-precision cross sections requires knowledge
of the scattering angle on an event-by-event basis at the level
of several mr, but the divergence of the beam is $\approx$15 -- 25 mr 
($\sigma$).
Thus high-resolution tracking detectors
are needed to measure trajectories into the target
to reconstruct the scattering kinematics.

The most effective solution for tracking a several MHz beam with $<100$ $\mu$m 
resolution is the use of GEM (Gas Electron Multiplier) detectors. 
GEMs have been demonstrated to withstand harsh radiation environments while 
maintaining high resolution and efficiency for single events and show little 
to no aging effects. 
GEMs have been successfully operated at intense high-energy muon beams at the
COMPASS experiment at CERN, which has served as a role model for the 
development of GEMs in many other experiments and applications. They are
low-mass detectors of order 0.5\% of a radiation length, thus keeping multiple 
scattering at a minimum.
Resolutions of 50 -- 100 $\mu$m are typically achieved with a two-dimensional 
strip readout at 
some 400 $\mu$m pitch. This way the amplified charge is distributed over 
several readout strips as a few-mm wide cluster, which allows for an improved 
resolution smaller than the pitch by using a centroid weighting technique. 
The two-dimensional hit information from several GEM detectors is combined to
determine the beam trajectory.
The reduced number of electronics channels and 
a rather simple construction scheme makes GEM detectors very cost-effective.

\paragraph{Requirements:}
The GEMs for beam particle tracking are required to provide 100 $\mu$m 
resolution and $>$98\% efficiency per element at operation of 3.3 MHz beam flux.
The requirements are presented in Table~\ref{table:gem_requirements}.

\begin{table}
\caption{GEM detector requirements}
\label{table:gem_requirements}
\begin{tabular}{|c|c|c|}
\hline 
Parameter & Performance Requirement & Achieved\tabularnewline
\hline 
\hline 
Resolution & 100 $\mu$m / element & $\checkmark$ 70 $\mu$m\tabularnewline
\hline 
Efficiency & 98\%  & $\checkmark$ 98\% \tabularnewline
\hline 
Positioning & $\approx$0.1~mm, $\approx$0.2~mr & not attempted; easy\tabularnewline
\hline 
Rate Capability & 3.3~MHz / plane & $\checkmark$ 5~MHz\tabularnewline
\hline 
Readout Speed & 2~kHz / 20\% deadtime & 1~kHz / 100\% deadtime\tabularnewline
\hline 
\end{tabular}
\end{table}

\paragraph{Detector design:}
The Hampton group developed, built, and successfully 
operated a set of $10 \times 10$ cm$^2$ GEM detectors at the OLYMPUS 
experiment at DESY~\cite{Milner:2014, Henderson:2017}.
At OLYMPUS, these GEM detectors were used for monitoring of the luminosity by 
determining the forward-angle
elastic $ep$ scattering rate on an event-by-event basis. 
These GEM detectors became available for the proposed MUSE experiment at PSI 
in the course of 2013, after OLYMPUS data taking was completed. 
The GEM elements were identified as US (upstream), MI (middle), and DS 
(downstream), left and right sector in OLYMPUS and this nomenclature 
will also be used for MUSE.

The $10 \times 10$ cm$^2$ OLYMPUS GEMs are operated with a 
70\% Ar / 30\% CO$_2$ 
gas mixture and are read out with strips in two 
dimensions with a pitch of 400 $\mu$m. 
The design of the GEM stack parameters such as the drift gap and gaps
between the three GEM layers and the readout plane follow that of the COMPASS
design, which has been demonstrated to provide reliable detection of
hit locations at routine rate densities of 2.5 MHz/cm$^2$ and of up to 
25--100 MHz/cm$^2$ in dedicated tests. The expected rate density for the
nominal $\pi$M1 tune at the final GEM just upstream of the target
is about 3.3 MHz / 5 cm$^2$ = 0.66 MHz/cm$^2$, 
with a single-track probability of over 90\%. 
Because the beam is coming to a focus, the upstream GEMs will have a
smaller rate density.
The OLYMPUS GEMs are therefore 
very well suited to provide event-by-event beam particle tracking under these 
conditions.

The GEMs are read out using FPGA-controlled frontend electronics based on the 
APV-25 chip developed for CMS and digitized with the Multi-Purpose Digitizer 
(MPD). The readout hardware was developed by INFN Rome and Genova for the
Hall A SBS spectrometer in the framework of the 12 GeV upgrade of Jefferson 
Lab, and was used for the first time in a realistic setting at OLYMPUS.
It consists of a frontend card hosting the APV-25 chip, which is 
directly attached to the GEM detector, and a VME based controller board (MPD)
hosting an FPGA located several meters away. 
Each APV processes 128 readout channels and pipelines both analog and digital 
information of 128 channels on a single cable. Raw signals on all strips
are sampled with either 20 or 40 MHz frequency. After adjusting the latency, 
``snapshots'' of the analog signal are taken and sent as frames to the VME 
based controller. 
\begin{figure}[h]
\centering
\includegraphics[angle=0,width=0.7\textwidth]{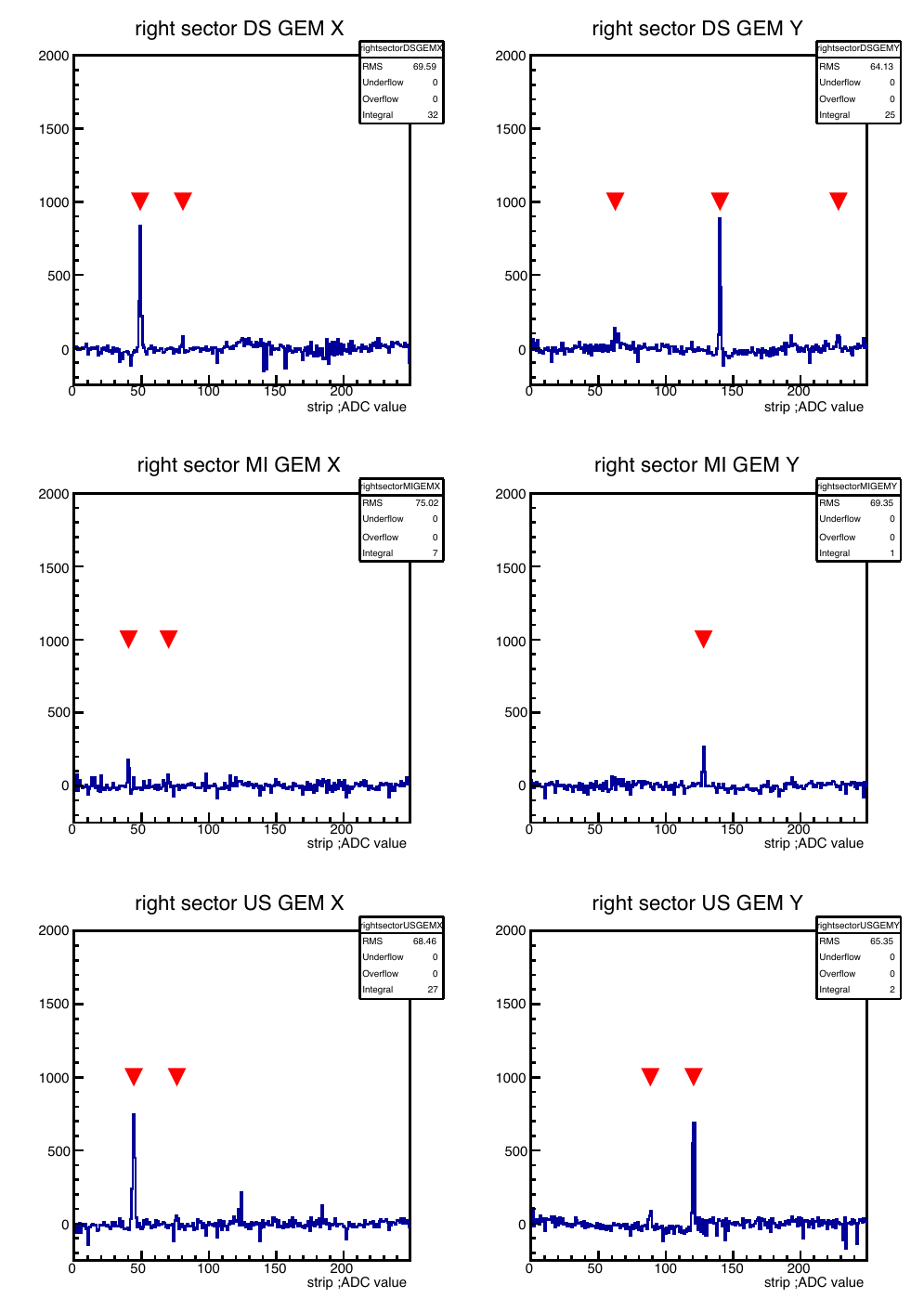}
\caption{ADC channel versus strip number in $x$ and $y$ direction for the 
US, MI, and DS GEM elements. The red triangles mark the location where the
cluster finding algorithm yields a candidate cluster location.}
\label{fig:gem_cluster} 
\end{figure}

The MPD controller provides clock, trigger and configuration via I2C to the APV,
and receives and digitizes the raw data into on-board ADCs. The DAQ 
frontend software was realized for OLYMPUS by the Hampton group and has been
running on a VME based single-board computer that controls the VME bus to 
read out and write the data to disk or to send it to the event builder.
As each APV chip reads out 128 channels, a $10 \times 10$ cm$^2$
chamber corresponds to $2 \times 250$ channels, which are read out with four frontend 
chips. One MPD can operate up to 16 APVs, \textit{i.e.} one such controller
can operate up to four GEMs, hence one telescope of three GEMS can be read out 
with a single MPD. 
The strip numbers and digitized pulse heights of the hit clusters in $x$ and 
$y$ give the spatial information for the track.
Figure~\ref{fig:gem_cluster} shows the digitized pulse height after pedestal
subtraction of a single event versus the strip number, of the US, MI, and DS 
GEM in both $x$ and $y$ direction (250 channels each). 
The red triangles indicate the candidate cluster locations returned by the
cluster finding algorithm.

\begin{figure}[h]
\centering
\includegraphics[angle=0,width=0.48\textwidth]{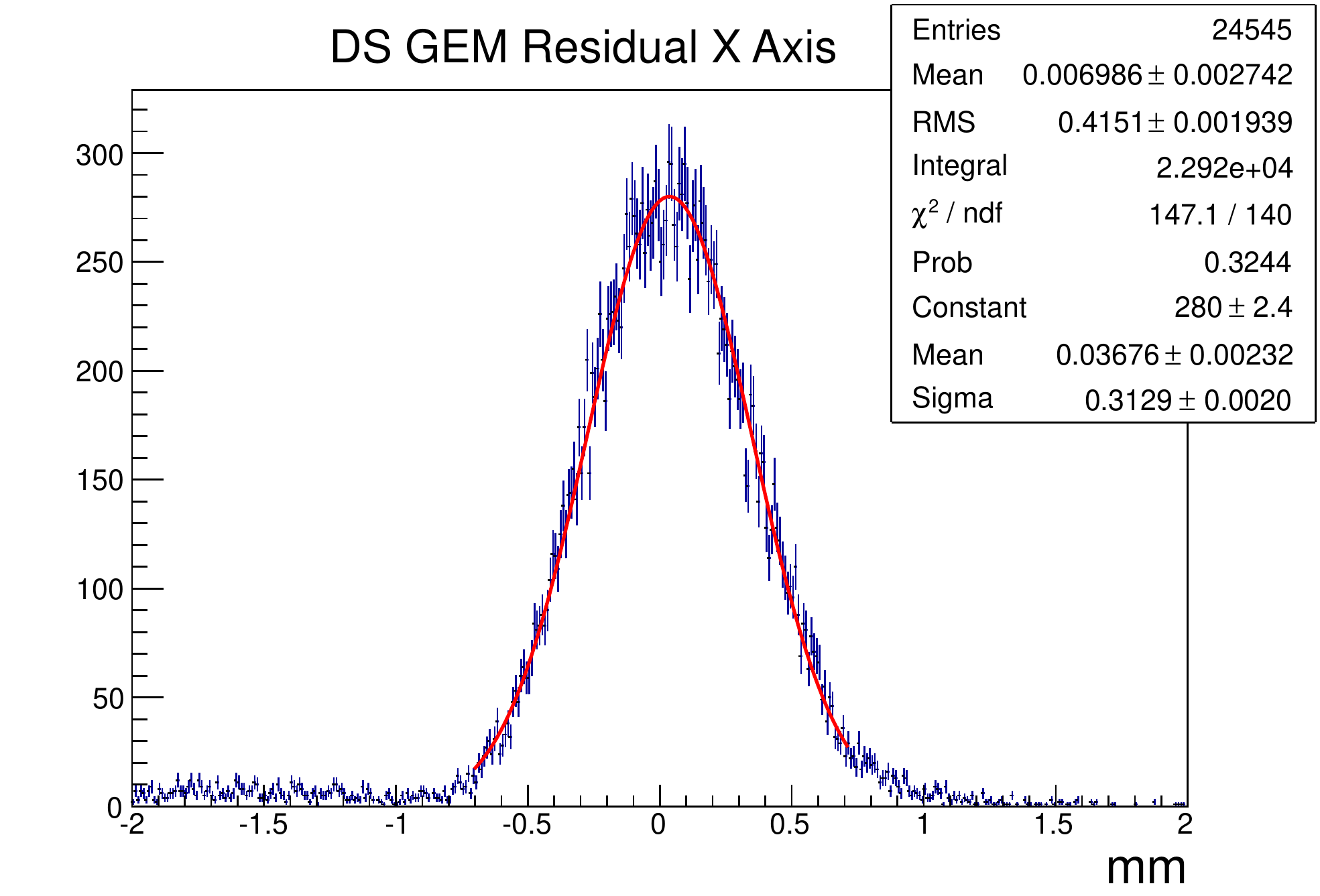}
\hfill
\includegraphics[angle=0,width=0.48\textwidth]{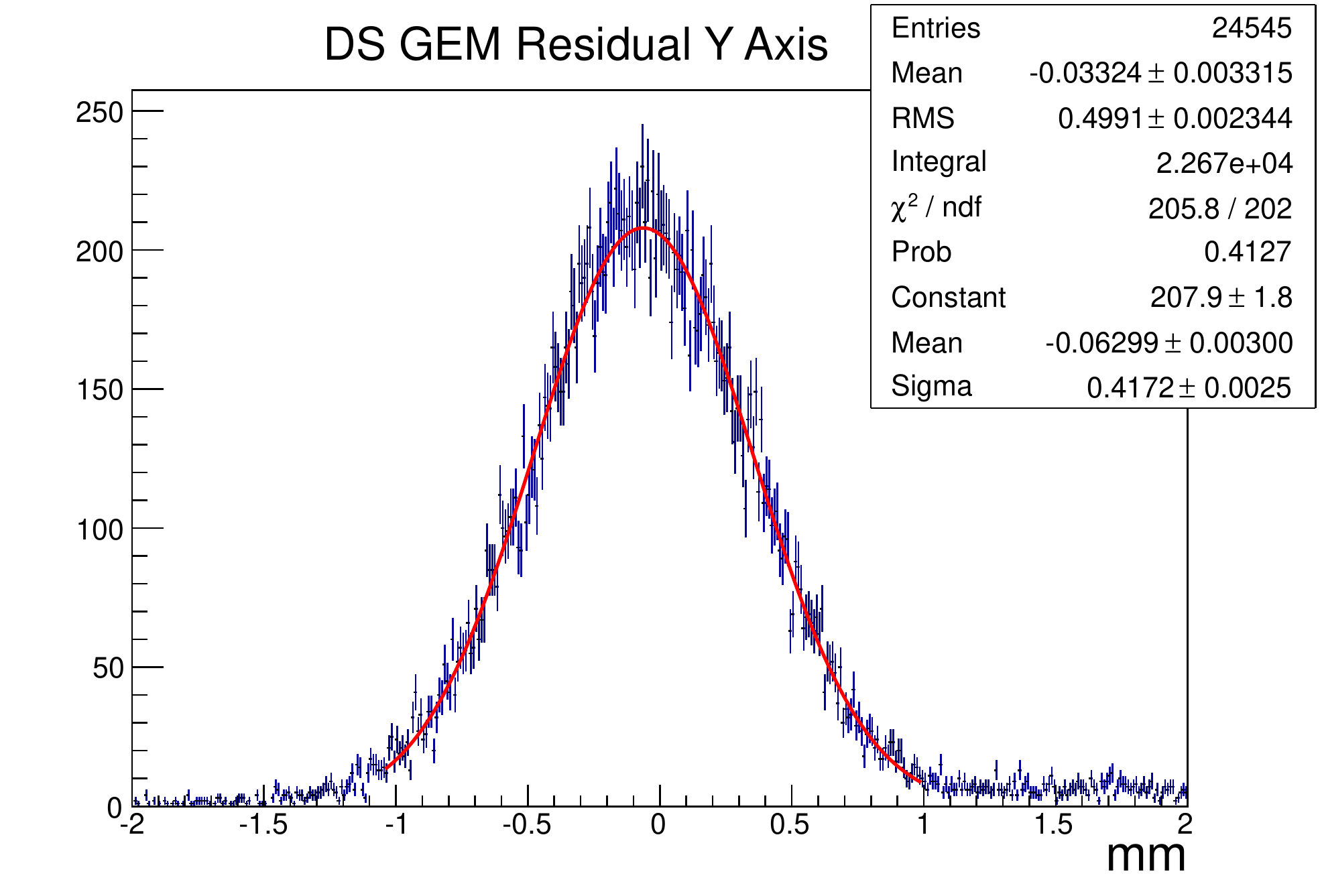}
\includegraphics[angle=0,width=0.48\textwidth]{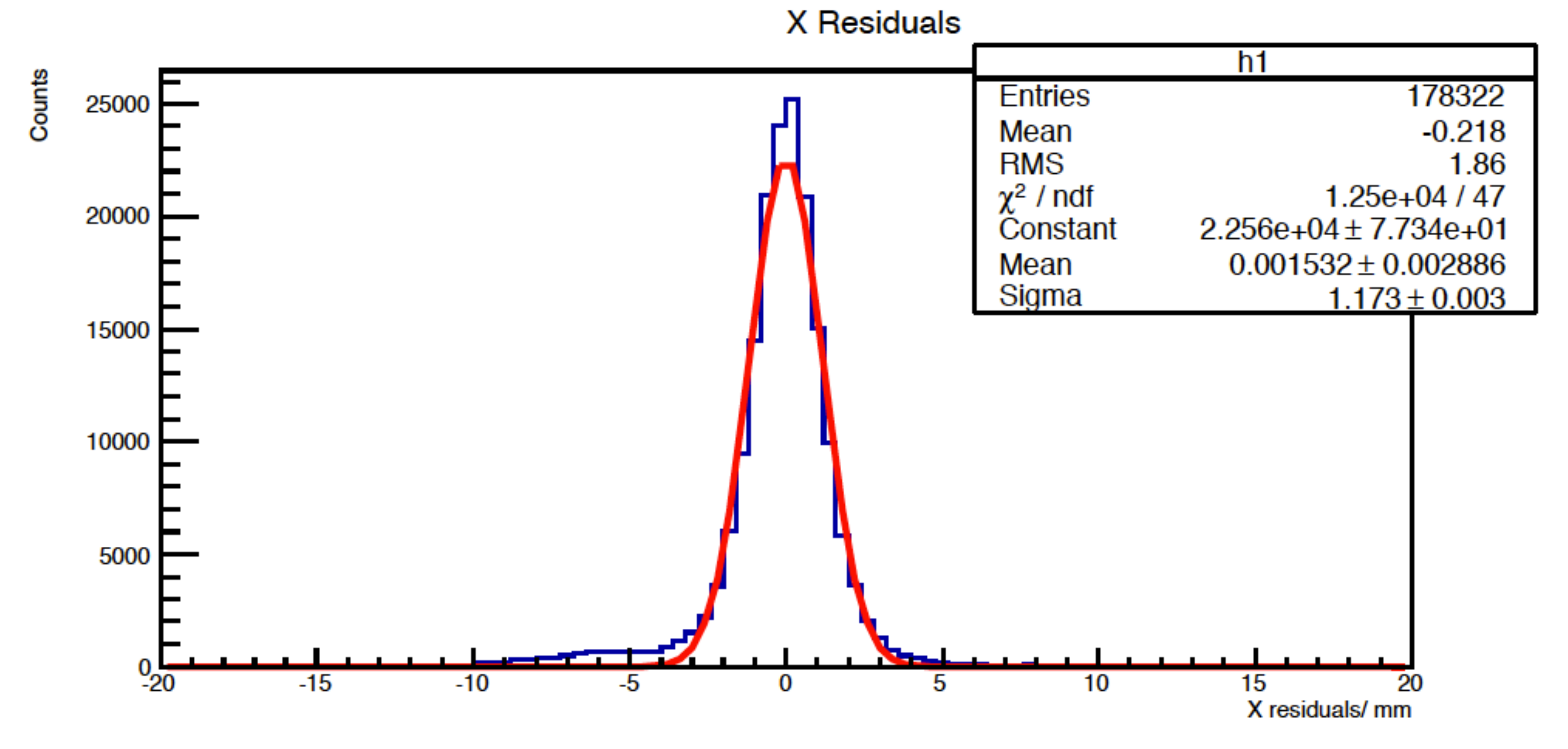}
\hfill
\includegraphics[angle=0,width=0.48\textwidth]{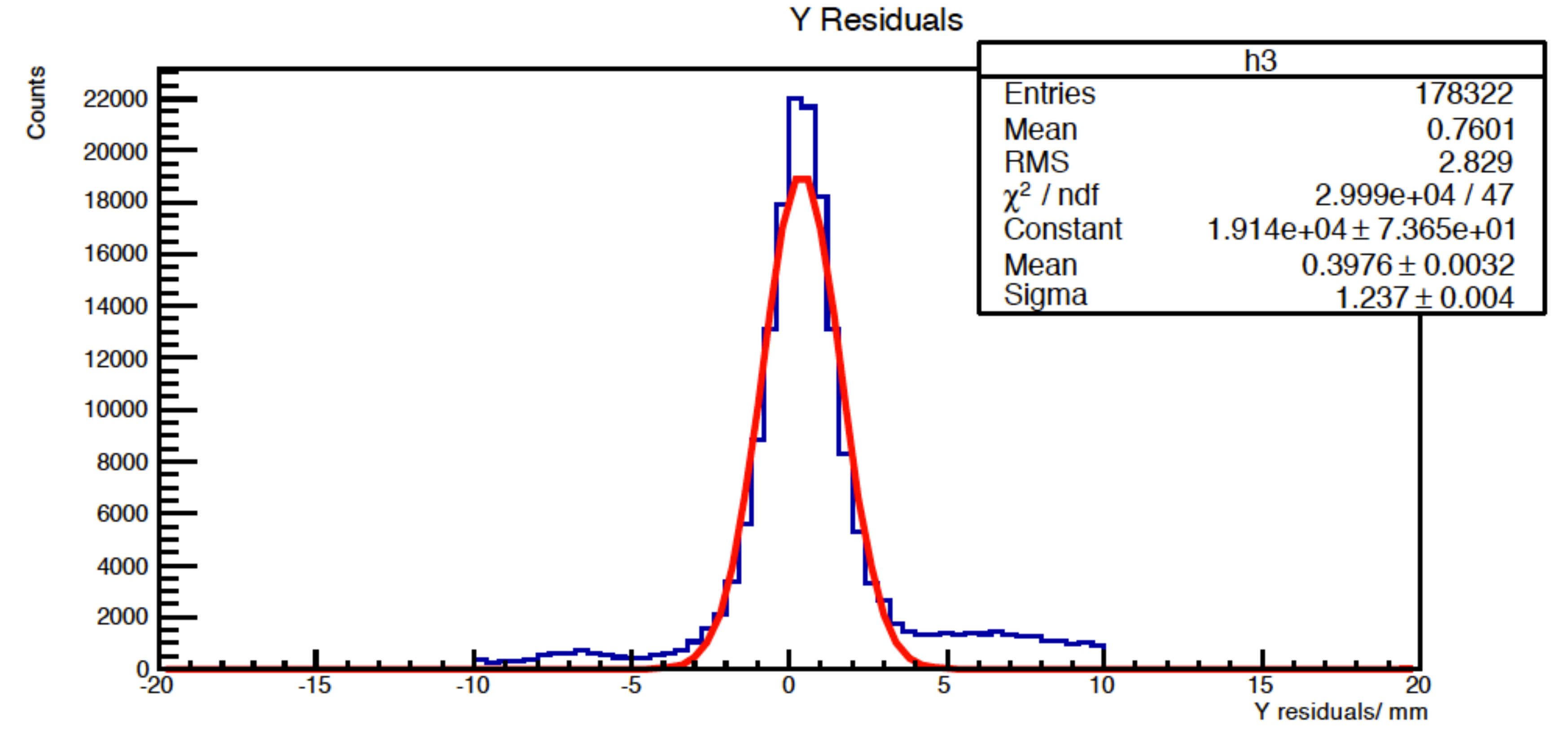}
\caption{Top: Track residuals from OLYMPUS fitted with 3 MWPC + 2 GEM elements. 
The residual width is composed of the intrinsic resolution and the track 
uncertainty. Intrinsic resolutions of around 80 $\mu$m have been achieved for 
each GEM element.
Bottom: Track residuals from the December 2013 test beam for MUSE beam 
trajectories in the GEM telescope fitted with 2 of 3 GEM elements. 
Residuals are bigger due to the less constrained tracks.
}
\label{fig:resolution} 
\end{figure}

\begin{figure}[h]
\centering
\includegraphics[angle=0,width=0.28\textwidth]{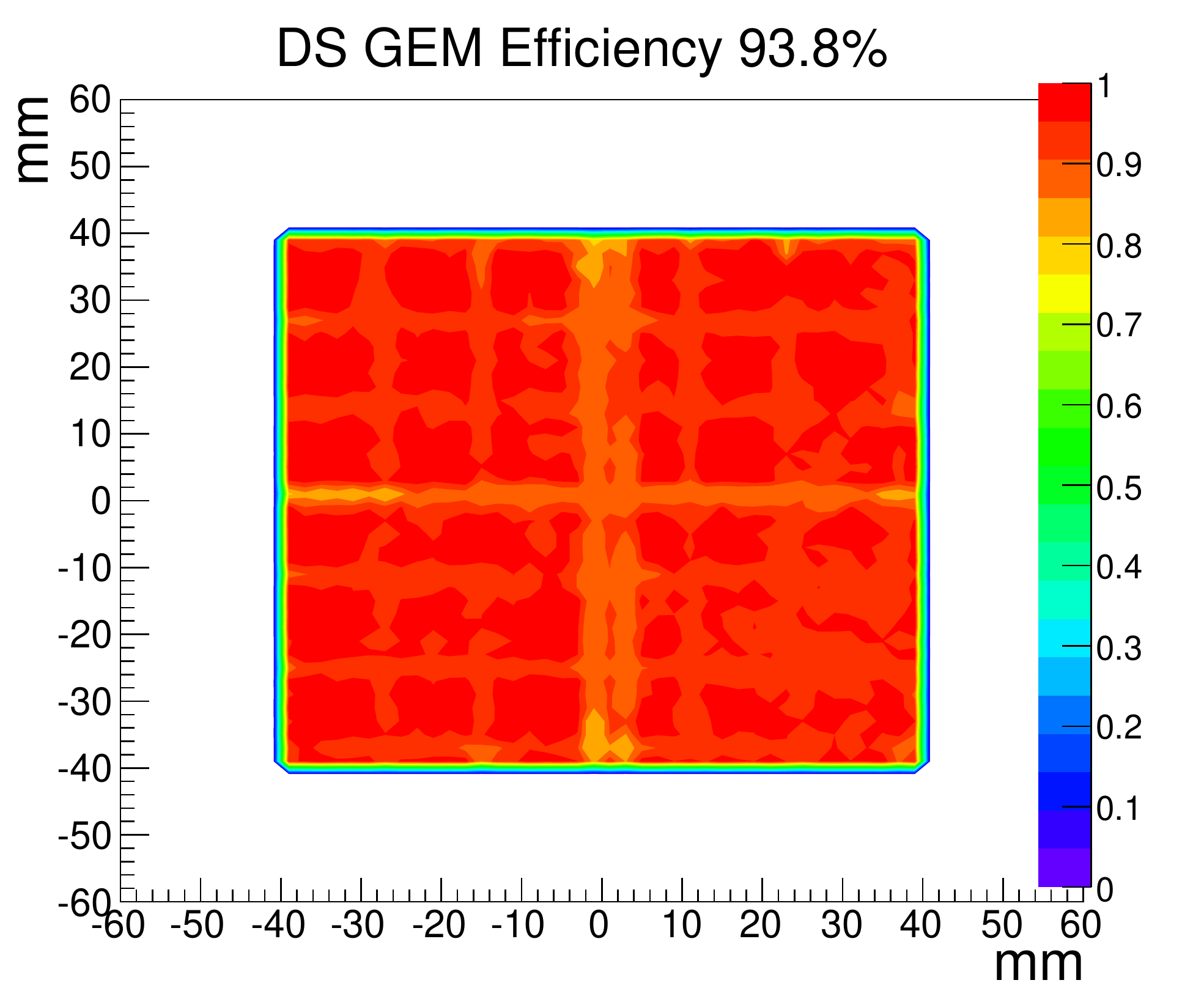}
%\hfill
\includegraphics[angle=0,width=0.35\textwidth]{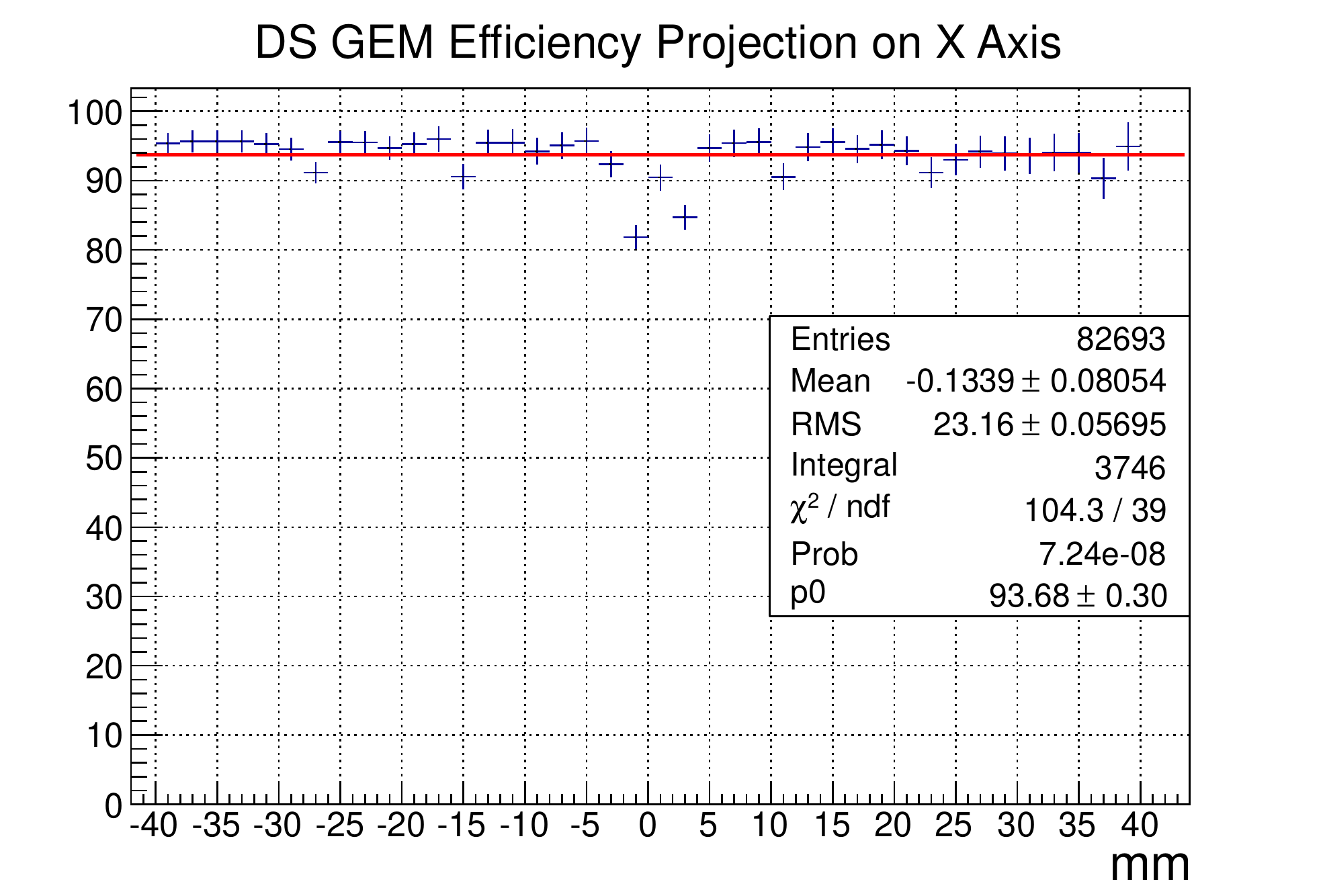}
%\hfill
\includegraphics[angle=0,width=0.35\textwidth]{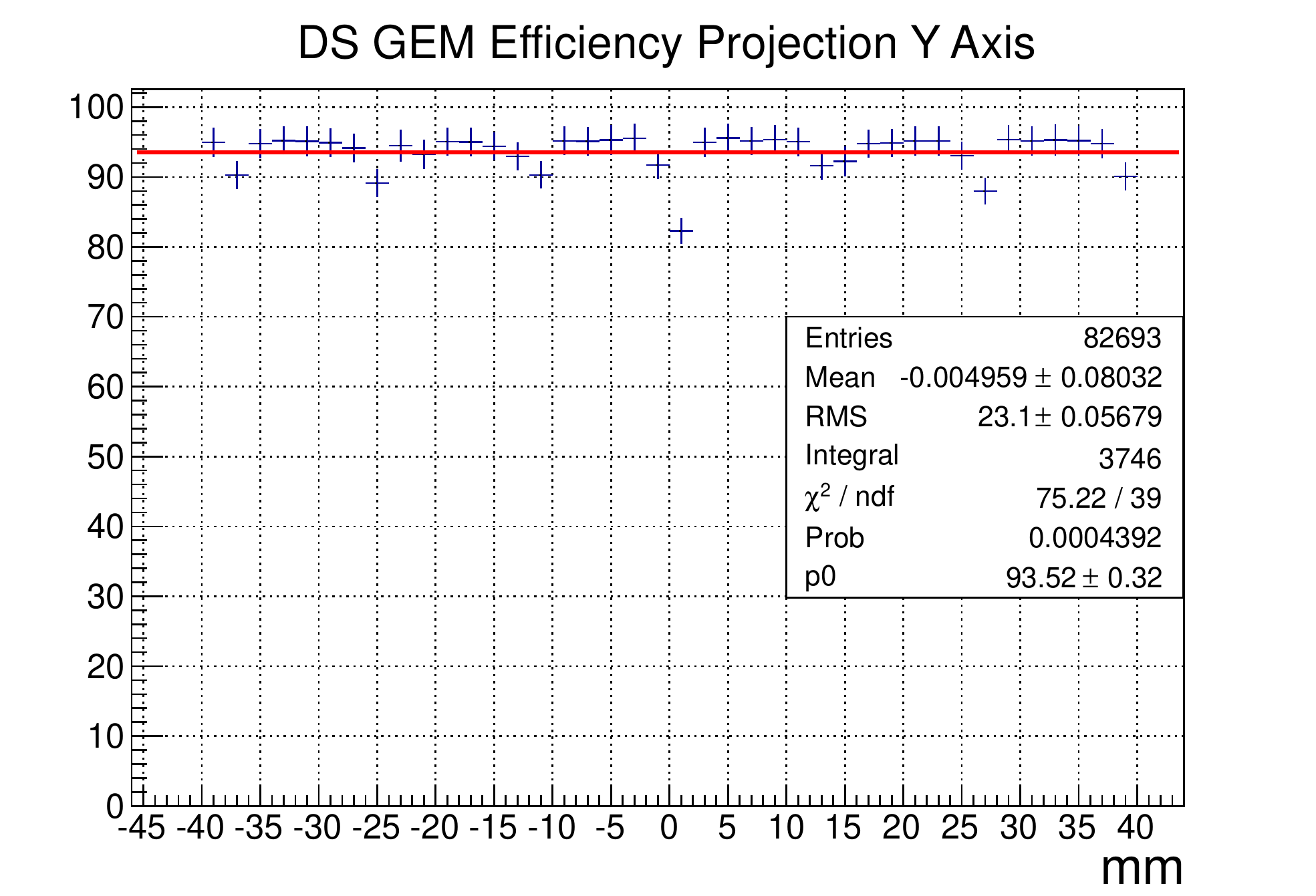}
\caption{Efficiency of the DS GEM element as a function
of $x$ and $y$ at OLYMPUS. Tracks were identified and fitted with 
3 MWPC + 2 GEM elements,
in order to verify if the respective third GEM element shows a hit at the 
expected location. Some localized structures are visible related to weaker
strips, which is under study. Efficiencies are generally around 93-95\%.}
\label{fig:efficiency} 
\end{figure}

The GEM telescopes at OLYMPUS worked very well. Operation was 
very stable, noise levels were very low. At OLYMPUS, intrinsic resolutions 
were found to be around 80 $\mu$m, and efficiencies around 93-95\%,
as shown in Figs.~\ref{fig:resolution} and ~\ref{fig:efficiency}.

\paragraph{Current status:}
The GEM readout was successfully added into the MUSE MIDAS DAQ and used
in beam tests starting in December 2013.

\begin{figure}[!t]
\centering
\includegraphics[angle=0,width=\textwidth]{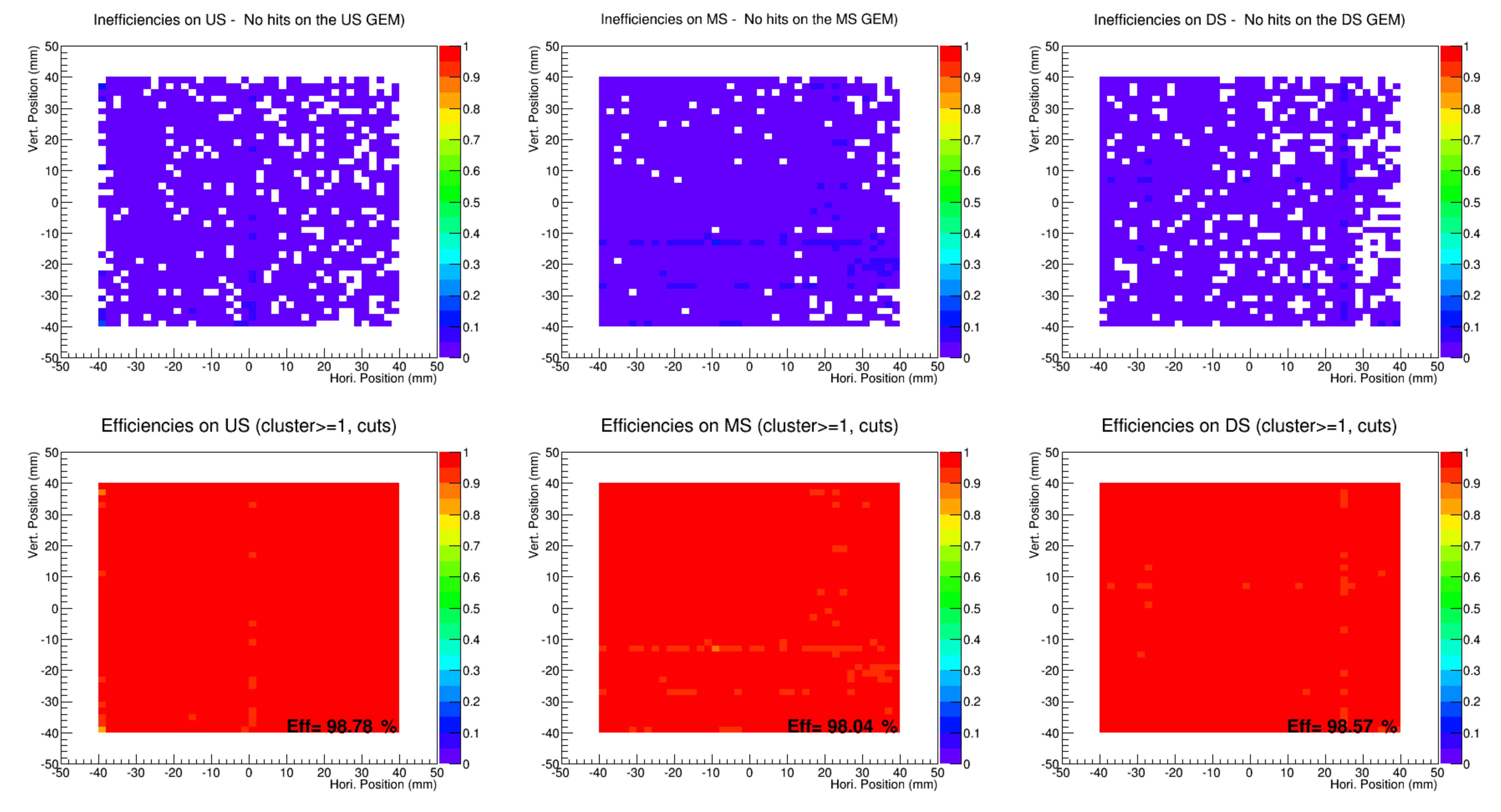}
\caption{Inefficiency maps (top row) and efficiency maps (bottom row) of the 
US, MI, and DS GEM element as a function of $x$ and $y$, obtained in test 
beamtimes at PSI in summer 2015. Tracks were determined with 2 GEM elements, 
in order to verify if the respective third GEM element shows a hit at the 
expected location within 10 mm radius. 
Some localized structures are still visible related to weaker
strips, but have been greatly reduced. Efficiencies are generally above 98\%.}
\label{fig:efficiency2} 
\end{figure}

An improved cluster finding algorithm to 
account for common-mode noise and channel-by-channel pedestal subtraction 
has been implemented, and the effect on chamber efficiencies has been studied
in test beamtimes in December 2014 and summer 2015. The structures in the 
efficiency maps as seen e.g. in Fig.~\ref{fig:efficiency}
have largely disappeared with the 
improved analysis scheme. As depicted in Fig.~\ref{fig:efficiency2}, 
efficiencies for each element are now above 98\% with inefficiency patterns 
greatly reduced.

In summer 2015 the GEM telescope modification for the final
configuration began with reduced gaps between GEM elements, which should be 
about 8 cm to optimize the balance between spatial resolution requirements and
minimizing the effects of multiple scattering. In order to reduce the 
distance between two elements to less than 12 cm, it was required to re-arrange
the APV cards such that they would be oriented within the plane of each GEM.
At the same time, cabling was modified to accomodate the new geometry,
and the length of the cables was reduced to less than 10 m from previously
25 m. Studies are ongoing to assess and optimize the noise situation.

Alongside modification of the digital and analog cables, we decided 
to upgrade the previous MPD v3 to v4. The v4 has standard HDMI-A front panel
connectors for the analog cables instead of the previous, very uncommon, HDMI-B
connectors. 
The migration to the new hardware has been challenging, as the operation of the
new MPD version and firmware with the existing DAQ frontend software was
unsatisfactory. In particular, the I2C addressing of the APVs was problematic, 
which delayed the successful re-commissioning of the telescope in the 2016 
beamtimes. In spring 2017 we decided to retire the previous DAQ software and
to migrate to the meanwhile mature INFN/JLab version of the code, which ran
successfully at JLab in fall 2016. At the June 2017 PSI test 
beamtime, work is now in progress to meet the milestone of
low-noise operation and reproduction of high efficiency. The new frontend 
software offers improved control features for the MPD and APV operation. 

\paragraph{Path to completion:}
In the OLYMPUS experiment, the readout rate of the telescopes was 
$\approx$100 Hz.
During the test beamtimes in 2014
and 2015 with one telescope, 
GEM readout required $\approx$0.8 ms, which would limit the DAQ rate
to $\approx$1 kHz.
In order to achieve a readout rate of 
order 2 kHz at less than 20\%
deadtime, an order of magnitude increase in readout speed is required. 

With the hardware and software upgrades we also switched to a faster VME 
controller (XVB-601 replacing the V7768 model) and are going to establish in 
steps 32-bit and 64-bit block transfer, which together is expected to decrease
the readout time by at least a factor 10, sufficient for MUSE operation. 
With the new controller, the VXS extensions of VME systems are also 
supported by all hardware elements, potentially allowing a significant further
speed increase.
The new DAQ software allows multi-frame readout of the APV cards, to be tested
in the June 2017 beamtime. With multiple samples the event timing can be 
extracted. This can be used to suppress accidental hits occurring 
predominantly at high rates.
For operation at the highest rates, we plan to migrate to MPD firmware version 4
Extraction of the data from the MPD via optical link would then be possible. 
Following the
scheme presently being implemented for the SBS setup at JLab, a Sub-System 
Processor (SSP) with a large FPGA collects the data from the MPD via optical 
link and performs real-time data reduction by applying common-mode correction 
and pedestal subtraction, zero suppression, and timing analysis in multi-sample
readout mode to suppress accidental hits.

The system will be optimized in 2017 for its final configuration and operation.

\subsubsection{Beam Veto}
\label{sec:beamveto}
\paragraph{Purpose:}
The beam veto detector is used to reduce trigger rate, by vetoing 
scattering or beam particle decay events upstream of the scattering 
chamber. 

\paragraph{Requirements:}
The beam veto detector uses the same technology as the scattered particle
scintillators of
Section~\ref{sec:scint}, but with a modified geometry and only
single-ended readout.
The requirements are presented in Table~\ref{table:veto_requirements}.

\begin{table}
\caption{Beam veto detector requirements}
\label{table:veto_requirements}
\begin{tabular}{|c|c|c|}
\hline 
Parameter & Performance Requirement & Achieved\tabularnewline
\hline 
\hline 
Time Resolution & 1~ns / plane & not attempted; easy\tabularnewline
\hline 
Efficiency & 99\%  & not attempted; easy \tabularnewline
\hline 
Positioning & $\approx$1~mm, $\approx$1~mr & not attempted; easy\tabularnewline
\hline 
Rate Capability & 1~MHz / plane & not attempted; easy\tabularnewline
\hline 
\end{tabular}
\end{table}

\begin{figure}[h]
\centerline{
\includegraphics[height=2.5in]{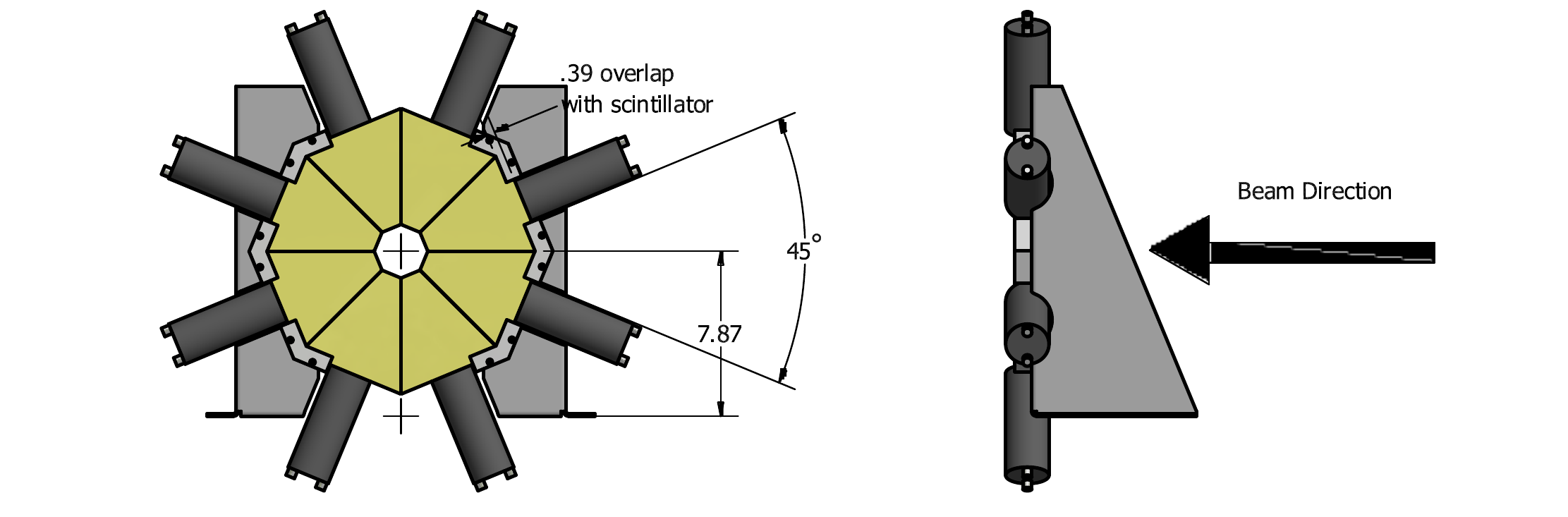}}
\caption{
Design of the beam veto detector and its frame. Dimensions are in inches.
}
\label{fig:BeamVeto}
\end{figure}

\paragraph{Detector design:}
The veto detector design is shown in Fig.~\ref{fig:BeamVeto}.
The veto detector is an approximately annular detector that surrrounds
the beam. 
It consists of 8 trapezoidal scintillators read out with Hamamatsu
R13435 PMTs.
The 3-cm inner radius matches the target vacuum chamber thin entrance
window.
Simulations were used to determine the veto detector coverage, leading
to an outer radius of about 16 cm.

\paragraph{Current status:}
The veto detector has been designed using the same technology as
the scattered particle scintillators, but has more modest performance
requirements.
It has not been prototyped or tested in the planned veto geometry.

\paragraph{Path to completion:}
We plan to build the veto detector in late summer / early fall 2017,
so that it is available for the dress rehearsal run.

\subsubsection{Beam Monitor}
\label{sec:beamscint}
\paragraph{Purpose:}
The beam monitor provides a high-precision particle time 
measurement and a flux determination of beam particles
downstream of the target.
For scattered particle events, it provides a determination of the
time and particle type of randomly coincident unscattered particles,
to monitor beam stability.
The time of flight from the beam monitor determines particle type 
independently from the RF time, and also determines muon and pion 
momenta.
For Moller / Bhabha scattering events that generate triggers, the
beam monitor detects the forward-going, high-momentum 
electron / positron.

\paragraph{Requirements:}
The beam monitor comprises a central hodoscope similar to the beam
hodoscope of Section~\ref{sec:beamsipm} and an outer hodoscope
similar to the scattered particle scintillators of
Section~\ref{sec:scint}.
In both cases the underlying technology and the requirements are the
same; the detector geometry is different.

\paragraph{Detector design:}
The central hodoscope of the beam monitor comprises two 
offset planes of 16-paddles, each similar to the beam hodoscope 
shown in Fig.~\ref{fig:SiPMProto}.
We use BC-404 scintillator paddles with dimensions 
30 cm long $\times$ 12 mm wide $\times$ 3 mm thick.
The long length is intended to limit radiation damage to the SiPMs.
We use three Hamamatsu S13360-3075PE SiPMs in series
to read out each end of each paddle.
The same readout electronics is used as for the beam hodoscope.

The outer hodoscope consists of 4 paddles identical in technology
to the scattered particle scintillator paddles, but of dimensions 
30 cm long $\times$ 6 cm wide $\times$ 6 cm thick.
For these paddles we use constant-fraction discriminators to maintain
good timing for randomly coincident beam particles, for which we
will not have pulse size information and thus cannot do walk
corrections.

\paragraph{Current status:}
Full-size prototype paddles for the central hodoscope have been 
constructed and will be tested in June 2017, to verify performance.
We have successfully tested 12-mm wide prototypes previously, but only
up to 16-cm long.
All components needed to construct the detector itself are on hand,
except for the detector frames.
Prototype scintillator paddles similar to the outer hodoscope paddles, 
but 50 cm long, have excellent performance characteristics in
extensive testing in PiM1.
The shorter outer hodoscope paddles should have superior 
resolution and efficiency.

\paragraph{Path to completion:}

The first beam monitor central hodoscope plane will be completed
in summer 2017 for use in the fall 2017 dress rehearsal run.
The second central hodoscope plane and the outer hodoscope 
paddles will be completed in fall 2017.

\subsection{Target}
\label{sec:target}

\subsubsection{Purpose}

The MUSE experiment requires a liquid hydrogen target of very stable density, and sufficient cooling power to minimize uncertainty in target length. The target will also be used for tracking tests and background subtraction measurements, which require multiple targets in addition to the full liquid hydrogen cell. 

\subsubsection{Requirements}

Although liquid hydrogen (LH$_2$) targets have been fabricated and widely used at various laboratories, special precautions have to be taken for each individual target to ensure safe handling of hydrogen which is a flammable gas. The MUSE LH$_2$ cryotarget system is further complicated by the physics need 
for a target ladder that will allow the beam to strike either a LH$_2$ target, a dummy target for background study, a carbon target for detector alignment, or no target at all -- an empty target position -- and 
by the experimental requirement for large vacuum windows. The latter requirement arises from the very large solid angle subtended by the detectors.
Thus, the MUSE cryotarget system requires appropriate engineering and safety considerations. An external company, Creare Inc.\footnote{Creare website: www.creare.com}, has been selected to design and potentially fabricate the system. Part of the design effort was to build and test prototypes of the most critical components, such as the target cells and the entrance and exit windows. We discuss here the major sub-systems of the cryotarget system, the engineering design, and safety measures and tests for safe operation of the target at PSI. An outline of the main physics requirements for the target can be found in table \ref{table-target-requirements}.
\par

\begin{table}

\caption{Target system requirements.}
 \label{table-target-requirements}

\begin{tabular}{|c|c|c|}
\hline 
Parameter & Performance Requirement & Achieved?\tabularnewline
\hline 
Liquid hydrogen & maintain liquid hydrogen-filled & not attempted;  \tabularnewline
& cell at T$\approx$19 k and P$\geq$1 atm  & moderate \tabularnewline
\hline 
Cool down time & $<$ 3 days & not attempted; \tabularnewline
 &  & moderate \tabularnewline
\hline 
Beam entrance window & $>$6~cm & not attempted; \tabularnewline
& & easy \tabularnewline
\hline 
Exit window(s) & $20^{\circ}<\theta<100^{\circ}$; & prototyping  \tabularnewline
(One continuous or two & $\phi = 0^{\circ}\pm45^{\circ}$ at $\theta = 60^{\circ}$ & underway; \tabularnewline
symmetric on beam & beam up-down and & challenging \tabularnewline
 left and beam right) & beam left-right symmetry & \tabularnewline
\hline

\end{tabular}
\end{table}

\subsubsection{Target Design: Major Sub-Systems}
\label{sssec:sub_sys}
The MUSE target system consists of many sub-systems. One of them is the target ladder. It consists of two approximately $300$ ml cylindrical cells ($6$~cm inner diameter (I.D.), $14$~cm height including the end caps) mounted coaxially, one above the other, and a carbon target ($6~\times~6$~cm$^2$ area). Each cell consists of a Kapton cylinder with a wall thickness of about $0.12$~mm and two copper end caps. The pipes of the ladder are made of copper. The upper cell is filled with LH$_2$ and serves as the cryogenic target. It is operated at a temperature of $20$~K 
and a pressure of $1$ bar. The lower cell is empty and serves as the dummy target to measure the background. The carbon target is used for detector alignment.\par

A cryocooler is utilized to cool a copper condenser and the copper structure of the target ladder. The cryocooler is bolted to the condenser to ensure good thermal contact. The CH110LT single-stage cryocooler from Sumitomo Industries has been selected, partially because similar cryocoolers have been used at PSI and because they have a service center in Darmstadt, Germany. The total power load on the LH$_2$ target  at $20$~K is only a few $\mu$W from the beam and a few $100$~mW (taking into consideration a few layers of aluminized Mylar superinsulation wrapped around the target) due to radiation. Thus, the CH110LT cryocooler, which has $25$~W cooling power at $20$~K and $50$~Hz, has sufficient power to liquefy hydrogen and maintain it at $20$~K.\par 

The target is controlled and monitored via a FPGA based PLC target control system. The target system has an external pressure sensor and heaters on the condenser to regulate the hydrogen pressure and thus the temperature of the LH$_2$ cell. Close to $20$~K, a change of $0.1$~K in LH$_2$ temperature results in a change of $24$~Torr in pressure~\footnote{See {NIST website: webbook.nist.gov/cgi/fluid.cgi?ID=C1333740$\&$Action=Page}}. Hence, a more precise temperature regulation is achieved by having a feedback loop on hydrogen pressure instead of temperature. Additionally, temperature sensors are installed on the copper condenser to support a backup temperature regulation system. Each target cell has one temperature sensor, one level sensor, and two heaters, which are all monitored by the slow control system.\par
\begin{figure}[ht!]
\centering
  \includegraphics[height=4.5in]{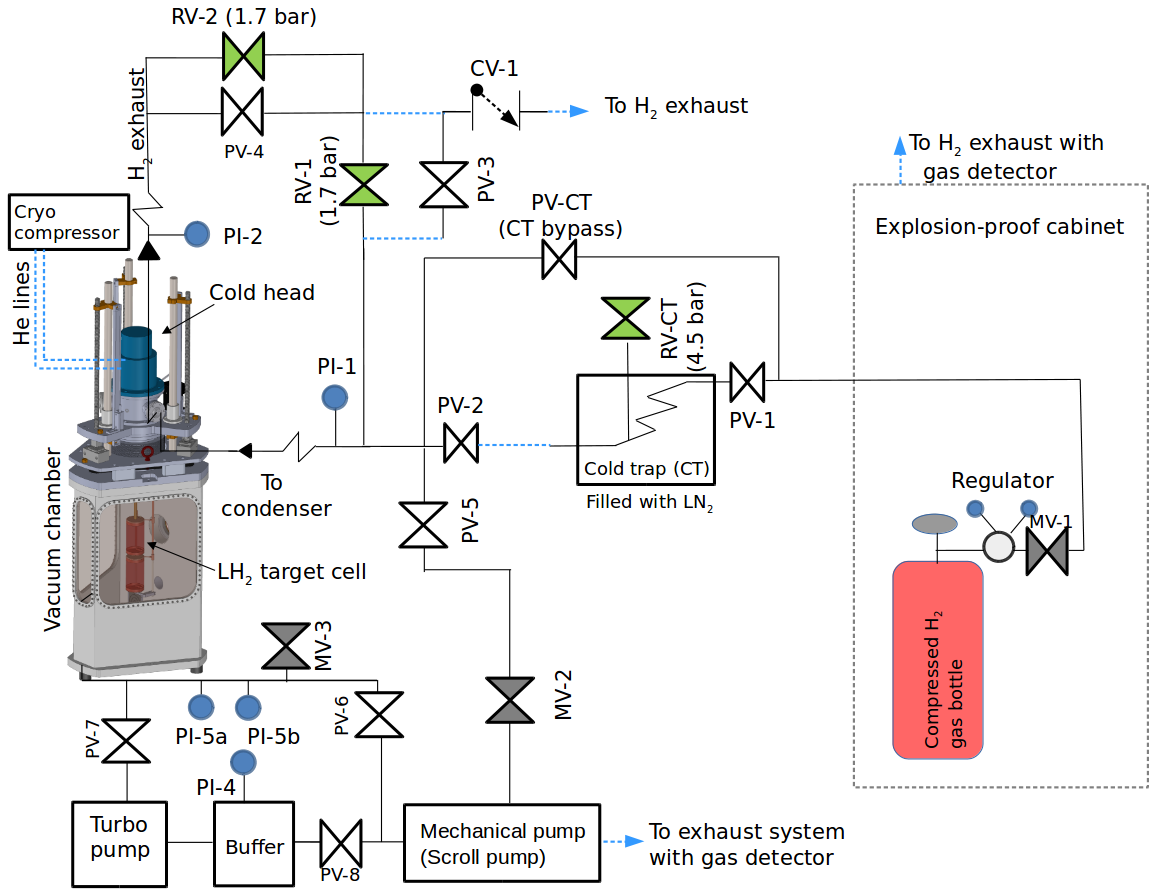}\hfill
  \\[2 ex]
 \includegraphics[height=0.9in]{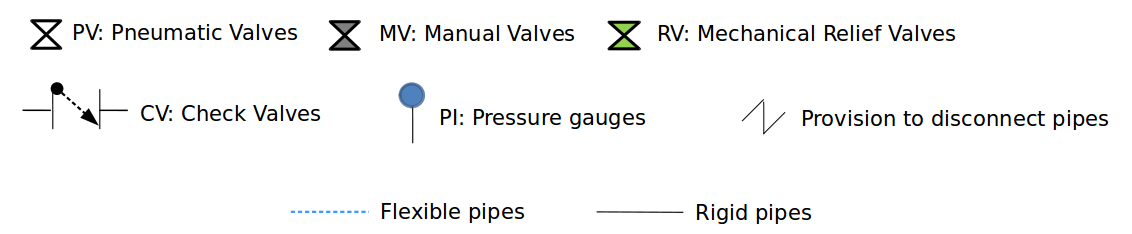} 

\caption{The schematic layout of the gas system. Note that flexible pipes are required to allow vertical movement of the target ladder and the condenser. 
}
\label{fig:gas_sys}
\end{figure}

The target ladder and the condenser are housed inside a vacuum chamber to provide good thermal insulation. The outer diameter (O.D.) cannot exceed $49$~cm so that it does not interfere with the rest of the experimental set up. The design is discussed in details in Section~\ref{sssec:eng_design}. The chamber has a bellows which allows the cryocooler, the condenser and the targets to move vertically together. The chamber requires two windows: a beam entrance window made of Kapton with approximately $7$~cm I.D. and thickness less than $0.2$~mm, and an exit window(s) for the beam and scattered particles, covering a $\theta$ range of $[20, 100]^{\circ}$ on each side of the beamline, and $\phi$ range of $[-45, 45]^{\circ}$ from the target center at $\theta=60^{\circ}$. It is highly preferable to have a continuous window with a thickness of up to $285$~g/m$^2$ covering $\theta$ region of $[0, 110]^{\circ}$ on both sides of the beamline, if possible. The window must be capable of withstanding the pressure difference between atmospheric pressure and vacuum. Each window assembly consists of a frame on which the window sheet is glued. The sheet seals onto the vacuum chamber with an O-ring.\\
Another important sub-system is the vacuum system. It is provided by PSI. It consists of mechanical and turbo pumps, valves and a PSI-built control system. The layout of the pumps and valves is shown in Figure~\ref{fig:gas_sys}. The fore-vacuum side of the turbo pump is connected to a scroll pump through a buffer to increase the lifetime of the scroll pump. The same mechanical pump is also used to pump out the supply line, the target cell and the cold trap for purging. The exhaust of the mechanical pump goes into a dedicated hydrogen exhaust pipe in the PiM1 area.\\
A gas system is used to control hydrogen flow. Figure~\ref{fig:gas_sys} shows the schematic layout of the gas system that is fabricated by PSI. Hydrogen gas for the target is supplied from a compressed hydrogen gas tank. The tank is placed outside the PiM1 experimental area (in an explosion-proof cabinet). Approximately $0.3$~m$^3$ of H$_2$ gas is needed. Safe disposal of hydrogen gas leaving the target cell is achieved by the direct release of the hydrogen gas into a dedicated hydrogen exhaust pipe in the PiM1 area. The pneumatic and mechanical overpressure relief valves are configured such that the pressure in the LH$_2$ target, the hydrogen input and exhaust lines always remains below $2$~bar. 

\subsubsection{Target Design: Engineering Design by Creare}
\label{sssec:eng_design}
 
 \begin{figure}[tbh]
\centerline{
  \includegraphics[height=3.8in]{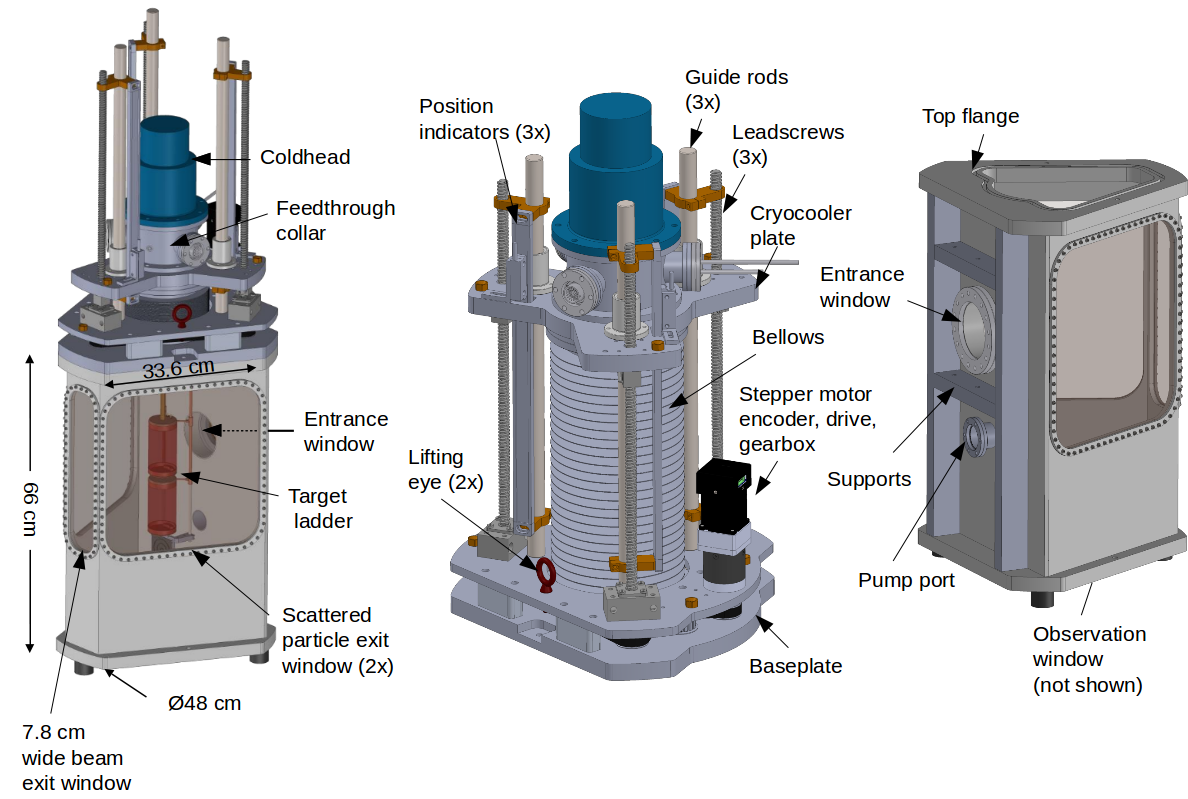}
  }
\caption{The vacuum chamber design by Creare. (Left) A full view of the chamber with the lift system. (Center) The details of the lift system implemented on the top of the chamber. (Right) A back view of the chamber showing the chamber supports, entrance window and pump port.}

\label{fig:chamber}
\end{figure}

A CAD rendering of the vacuum chamber with entrance and exit windows, bellows, cryocooler and target ladder is shown in Fig.~\ref{fig:chamber}. The chamber material is chosen to be stainless steel. The vacuum chamber has an O.D. that fits within $49$~cm, and it has a height of $66$~cm. This does not include the lift system. The chamber wall thickness was determined to be $9.5$~mm, based on chamber stress analysis. This thickness provides a safety factor of at least a factor of two from the yield point.\par

A large edge-welded bellows allows the vertical movement of the cryocooler, the condenser and the target ladder. The bellows has an I.D. of $15.2$~cm and is capable of providing the required vertical movement of $34$~cm to attain all four target positions. Such bellows are commercially available. The vertical movement of the cryocooler plate (shown in Figure~\ref{fig:chamber} (center)) and the target ladder is achieved by using three leadscrews which are driven by a cogged timing belt and motor system. Three guide rods keep the cryocooler plate centered and prevent movement in the horizontal plane. As a precautionary measure, a magnetic position scale is attached to each guide rod to provide the true position of the cryocooler plate at three equally spaced radial positions.\par

Figure~\ref{fig:chamber} also shows a feedthrough collar with three conflat flanges connected to the chamber top. Two of them serve as electrical and fluid feedthroughs while the third is a spare. The condenser has two copies of a heater circuit ($25$~$\Omega$, $200$~W each) and a Lakeshore Cernox temperature sensor, one for regular operation and the other as a backup. The target cell and the empty cell each have a $50$~$\Omega$, $50$~W heater for regular operation and a second as a backup, one Lakeshore Cernox temperature sensor and one Allen Bradley level sensor attached inside the top end cap. The electrical feedthrough has sufficient pins to accommodate all electrical connections and also provide spare pins.
\par
The vacuum port is placed below the entrance window such that it lies above the detector support table. The bottom of the chamber has a view port where a camera can be attached to survey the X and Z coordinates of the target. The design has provisions for bolting the chamber to a stand which, in turn, can be bolted to the raised platform of the experimental hall, thus providing a very stable configuration. The stand has translation mechanisms to adjust the chamber position in all three directions. The height of the stand is chosen such that the adjustment screws lie above the detector support table.
\par

The beam entrance window has a $7.5$~cm O.D. clear aperture in accordance with the specified experimental requirements. Creare has successfully vacuum tested a $50$~$\mu$m thick Kapton sheet for the entrance window using a test assembly fabricated from a $6$~inch diameter conflat nipple. Pressure tests using the same configuration confirmed that the sheet can withstand a pressure difference of $4$~bar, thus providing a safety factor of four with respect to operating conditions. 

It is preferred that the exit window for the beam and the scattered particles is a single, continuous window covering the $\theta$ region of $[0, 110]^{\circ}$ on both sides of the beamline and the $\phi$ region of $[-45, 45]^{\circ}$ at $\theta=60^{\circ}$. This requires a $64.1\,\times\,33.7$~cm$^2$ clear aperture. There are several other constraints imposed by the experimental requirements on the design. It should be homogeneous and there should be a minimum distance of $10$~cm between the center of the target and the window along the beam direction to allow the study of background and vertex reconstructions. It should be able to not only routinely operate under atmospheric pressure, but also have a pressure safety margin of $3$ times the operating pressure. Furthermore, the window material should be chosen such that the angular resolution in the scattering angle, $\Delta\,\theta$, is $19$~mrad or better for a precise determination of the proton's electric radius. \par

Building an exit window which satisfies all these constraints has proved to be very challenging. Several types of materials have been tested for the exit window.  Among those were a $125$~$\mu$m thick Kapton (the thickest Kapton that is commercially available) and various sailcloth fabrics~\footnote{These sailcloth fabrics were manufactured by Dimension-Polyant. Website: http://www.dimension-polyant.com/} which are Mylar laminated on Kevlar-like fabric. Due to the cylindrical shape and large size, all these windows deflect inside to varying degrees and form pleats when vacuum is established inside the chamber. 
Unfortunately, these pleats lead to inhomogeneities which are difficult to simulate and they lead to large multiple scattering for particles traversing through them. \par 
  
We have settled on a trapezoidal design for the chamber which would use three separate, flat windows made of a $120$~$\mu$m thick Kapton and a sailcloth for the beam exit window and the two scattered particle windows, respectively. The use of three small-sized flat windows instead of one large-sized cylindrical window has eliminated the problem of pleats without violating the hard requirement of covering a $\theta$ range of $[20, 100]^{\circ}$ from the target center on both sides of the beamline. The analysis will be more complicated due to background events generated in the two support strips for the window frames at the very forward angles, which can generate additional event triggers. We are studying with Geant4 simulations the backgrounds, background rejection in the analysis, and the use of small veto scintillators just outside the two support strips to tag particles which scatter from these strips into the acceptance. Preliminary results are very positive.\par
\begin{figure}[tbh]
\centerline{
  \includegraphics[height=4.4in]{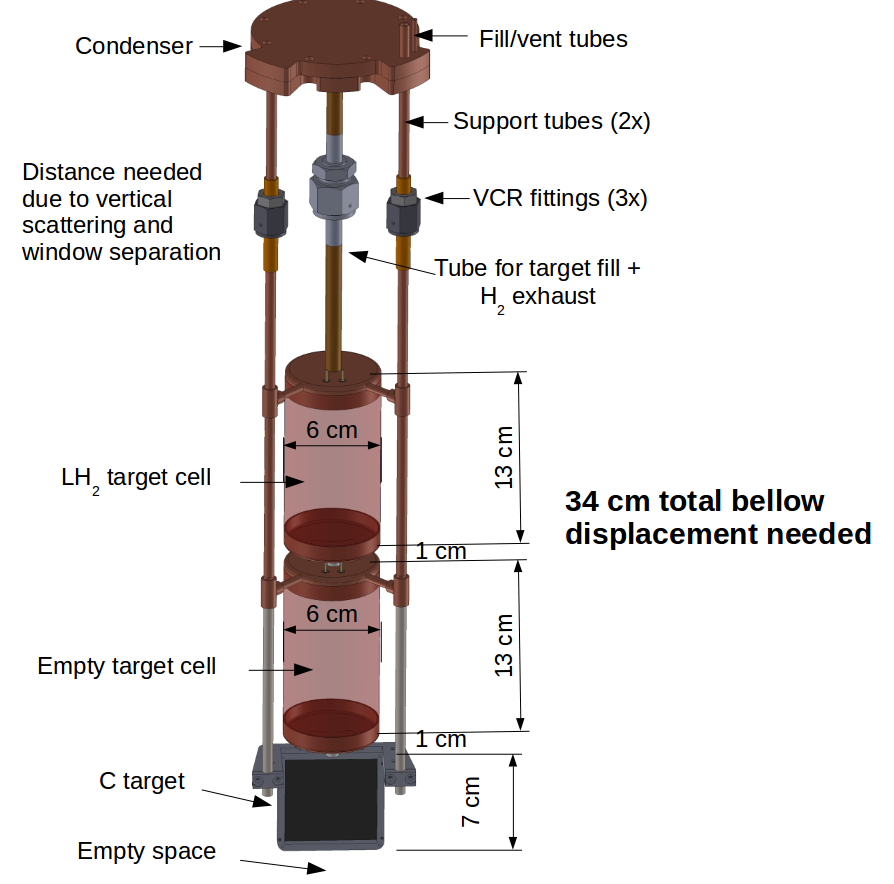}
}
\caption{A schematic view of the target ladder.}   

\label{fig:ladder}
\end{figure}
Figure~\ref{fig:ladder} shows a schematic view of the target ladder along with the dimensions of each cell. In this configuration, a single copper tube with $1$~cm O.D. connects the inside of the condenser to the top end cap of the LH$_2$ cell. It serves as the LH$_2$ inlet tube as well as H$_2$ exhaust outlet tube for the target cell. The bottom end cap of the target cell is connected only to the Kapton cylindrical wall of the cell. Laboratory tests performed on target ladder prototypes at the University of Michigan (U-M) have shown that this configuration gives a higher safety factor for the LH$_2$ target cell than the configuration in which a LH$_2$ inlet tube is connected to the bottom end cap  and a separate tube is connected to the top end cap to remove the hydrogen exhaust from the target cell. This may be because the latter configuration over-constrains the target's Kapton wall.\par

Two support tubes ($0.63$~cm O.D.) shown in Figure~\ref{fig:ladder} connect the condenser to the three targets via radial stubs ($0.32$ cm O.D.). These support structures provide additional mechanical stability to the ladder. The condenser and the entire pipe structure, starting from the bottom of the condenser up to the top end cap of the empty target cell, are made of copper to establish thermal equilibrium between these components. The support tubes are placed to the sides and rear (upstream side) of the target ladder, away from the interaction area and in the shadow of the supporting chamber walls, avoiding additional sources of background.\par
\begin{figure}[tbh]
\centerline{
  \includegraphics[height=3.1in]{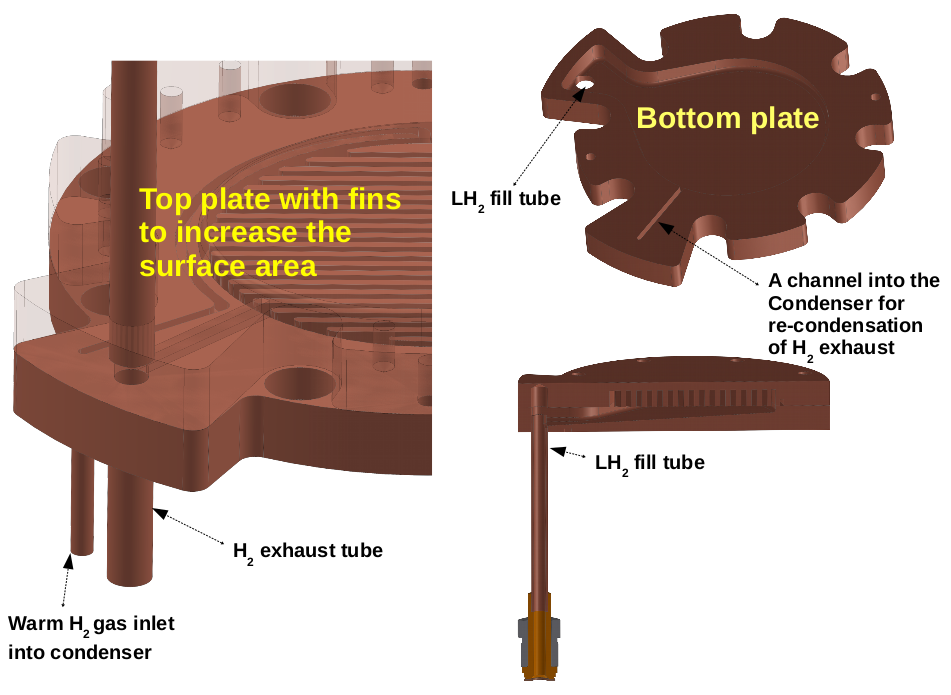}
}
\caption{SolidWorks drawings of the condenser. 
}
\label{fig:condenser}
\end{figure}
Figure~\ref{fig:condenser} shows SolidWorks drawings of the condenser. The top plate consists of a cavity which contains a series of fins to provide enough surface area for condensing hydrogen. The internal floor of the bottom plate is tilted towards the liquid hydrogen fill pipe that is connected to the target. A channel milled into the bottom plate allows the hydrogen exhaust from the target to re-enter the condenser and get liquefied. The condenser weighs $1.27$~kg and has an internal condensing surface area of $350$~cm$^2$.\par

\subsubsection{Target Design: Target Cell Design and Fabrication}

\begin{figure}[tbh]
\centerline{
  \includegraphics[height=1.8in]{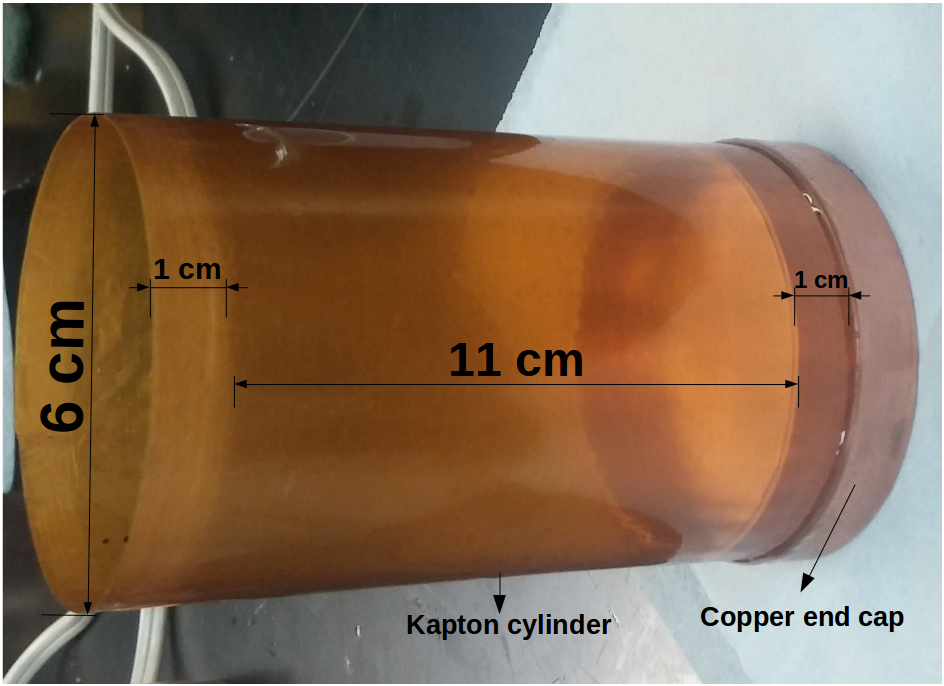}\hfill
  \includegraphics[height=2.8in, angle=90]{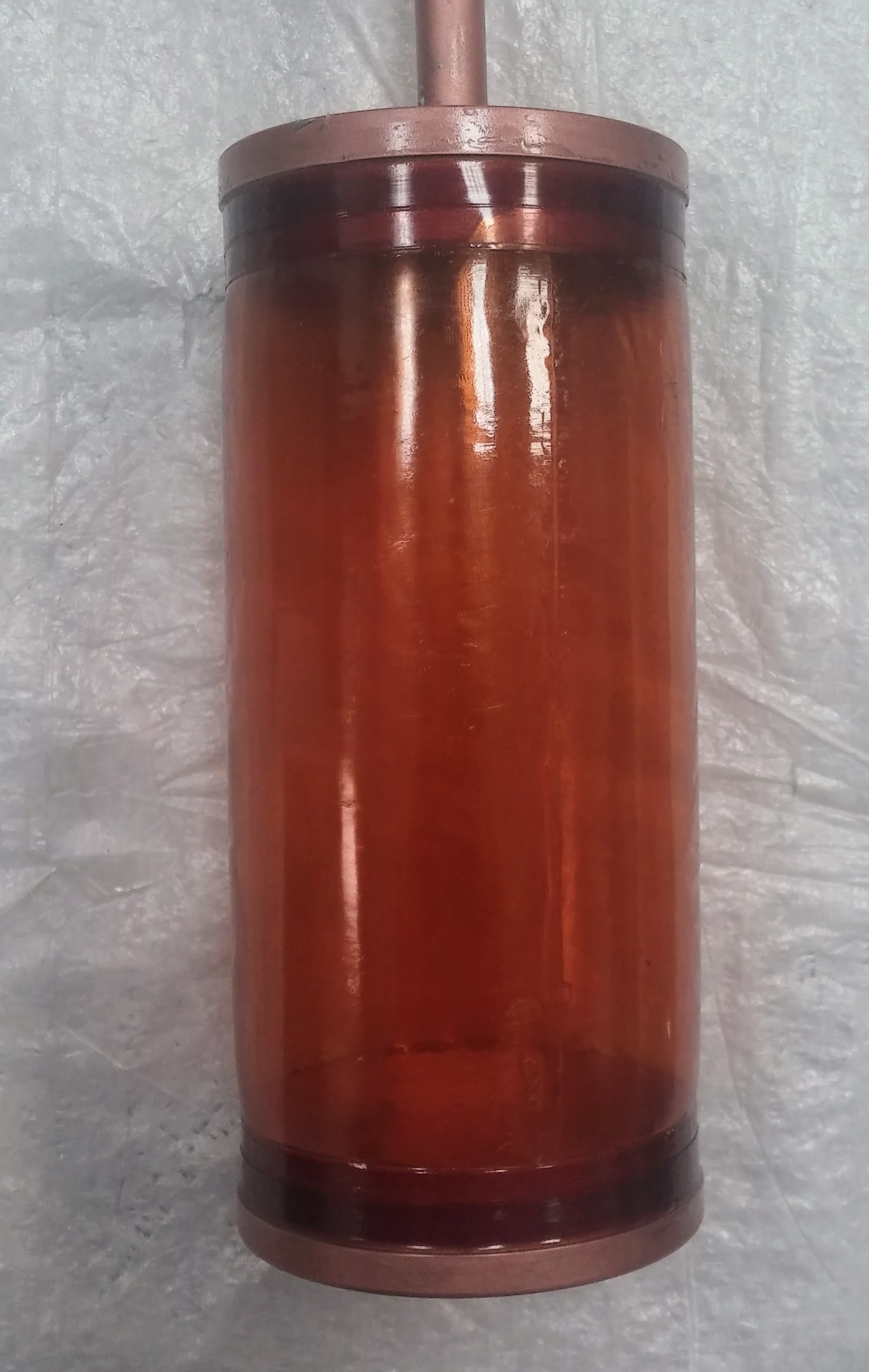}\hfill
}
\caption{Photographs of a target cell prototype built at U-M. (Left) The Kapton cylinder is made from three wraps of a $25$~$\mu$m thick Kapton sheet glued together using Stycast $1266$. (Right) Two copper end caps are glued to the Kapton cylinder using Stycast $1266$. The glue joints between the cylinder and the end caps are further reinforced by gluing few layers of Kapton strips.
}
\label{fig:target_U-M}
\end{figure}

Various target cells have been fabricated and tested at U-M to fine-tune the fabrication technique. These cells have cylindrical shapes with walls that are made with three wraps of a $25$~$\mu$m thick Kapton sheet (see Figure~\ref{fig:target_U-M}). The three wraps are glued together using Stycast $1266$. The thickness of the glue is controlled to achieve a total thickness of about $0.12$~mm for the cell wall. This method leads to a strong and robust wall. The height of the cell is consistent with the experimental requirement ($11$~cm without including the end caps). The Kapton cylinder has an extra $1$~cm length at both ends to glue the end caps to the cylinder. Cell tests were performed at a differential pressure of $2$~bar at liquid nitrogen temperature over several temperature cycles. Destruction tests have consistently shown that the cells survive a differential pressures of $3.8$~bar at about $300$~K, thus providing a higher safety factor than the factor of three recommended by PSI. 
\par

\subsubsection{Target Design: Safety Measures and Tests} 
\label{ssec:safety}

Full details for the safe operation of the cryotarget are provided in the 
MUSE Hydrogen Target Safety and Operating Procedures Report, which has to be approved by PSI before MUSE can operate the target in the PiM1 area. The main points are discussed here.\par

The pneumatic valves in the gas system controlled by the slow control system protect the target cell from high pressure. For additional safety, these valves are normally open-type valves. As 
backups, the mechanical over-pressure relief valves, operating independently of 
the slow control system, guarantee protection of the target from high pressure (i.e. above $1.7$~bar). Furthermore, gas detectors and safety alarms are placed at various places to detect hydrogen leaks and 
issue warnings.\par

If there is a loss of power, valves to the vacuum pumps will close and thus maintain reasonable vacuum inside the chamber. Pneumatic valves in the inlet and exhaust lines will open. LH$_2$ will slowly warm up and the evaporated H$_2$ gas will be directed to a dedicated H$_2$ exhaust pipe in the PiM1 area.\par

In case there is a slow loss of vacuum due to a leak, the valves to the pumps
 will close. The evaporation process will be slow and the evaporated H$_2$ will be directed to the 
 dedicated H$_2$ exhaust pipe in the PiM1 area.\par
 
If there is a rapid loss of vacuum, such as from a window rupture, the shock wave may rupture the target cell.  
In that situation, hydrogen will mix with air and possibly lead to an explosion (if the mixture of H$_2$:O$_2$
reaches a ratio of $10$:$1$). Note that the target cell will evaporate only about 
$0.25$ m$^3$ of H$_2$ into the area, which has the significantly greater volume of about $200$ m$^3$. 
Furthermore, the target will be wrapped in multilayer insulation, so the shock wave
should be significantly mitigated. To protect the vacuum chamber windows, protective plastic shields are used to cover the windows whenever there is access to the area.
 
Lastly, one should consider the possibility of a target rupture. The released liquid will evaporate and expand quickly in the chamber. The outcome will depend on the extent of failure. If the target cell develops a slow leak, the windows will most likely remain intact and the evaporated hydrogen will be pumped out by the vacuum system into the H$_2$ exhaust system. On the other hand, if the target cell breaks releasing a large amount of LH$_2$ in a short period of time into the chamber, this may break the chamber window(s) and lead to an explosion as discussed above. 

The cryotarget system will undergo multiple stages of safety tests before the production run begins in the summer of 2018. The overall procedure is as follows. The pipe work of the gas system will be thoroughly inspected for leaks. It will also be checked whether the pipes can successfully hold pressure up to the relief pressure of $1.7$~bar. Prototypes of windows will be vacuum tested and pressure tested at $3$~bar. Testing the target will involve multiple steps. First, it will be pressurized to $2$~bar differential pressure using helium gas. At this pressure, the target will undergo several temperature cycles using liquid nitrogen and will be thoroughly checked for leaks after each cycle. Next, a cooldown test with neon, which has a similar boiling point to hydrogen but is not explosive, will be performed using the final cryotarget system in a staging area. The target will be pressure tested up to the set pressure of the mechanical relief valves. If successful, a cooldown test with a few milliliters of LH$_2$ will be performed in the staging area. Finally, a complete integration test at the PiM1 beam line will be performed to fully test all components, including the slow control system and safety procedures before the production run begins.

\subsubsection{Current status and path to completion}
The target design process started in mid-December of 2016, and was essentially completed in June 2017, with: {a successful completion of engineering design review of the chamber and target ladder by PSI; constructed cell prototypes meeting design requirements; a fully designed gas system; safety documentation under review; and tests of safety factors of the large exit windows well underway. We and Creare have both estimated that, given the current status and rate of progress, the complete target system can be ready for beam in summer 2018.  

\subsubsection{Operation and Maintenance of the Target}

Once commissioning is completed and operations begin, target experts along with shift workers will be responsible for normal
monitoring and operation of the target.
The local PSI target group will assist as needed, such as for maintenance and/or repair work. 
To that effect, European standards are used to construct the
system; all designs, off the shelf and manufactured components are metric
as requested by the local group.

\section{Scattered Particle Spectrometer}

\subsection{Straw Chambers}
\label{sec:chambers}
\paragraph{Purpose:}
The Straw Tube Tracker provides high resolution and high efficiency
tracking of the scattered particles from the target. 

\paragraph{Requirements:}
The design requirements are given in
Table~\ref{table:WBS4_requirements}.  
The MUSE straw tube tracker is based on recent developments 
in straw chamber design~\cite{pandasttdesign} 
being implemented for the PANDA experiment~\cite{pandaexperiment};
this design meets our requirements, but for a different geometry chamber
in a different accelerator environment. 

\begin{table}
\caption{Straw Tube Tracker requirements}
\label{table:WBS4_requirements}
\begin{tabular}{|c|c|c|}
\hline 
Parameter & Performance Requirement & Achieved\tabularnewline
\hline 
\hline 
Position Resolution & 150~$\mu$m & $\checkmark$ $<$120~$\mu$m\tabularnewline
\hline 
Efficiency & 99.8\% tracking & $\approx$ 99\% in prototype; moderate\tabularnewline
\hline 
Positioning & $\approx$0.1~mm, 0.2 mr in $\theta$ & not attempted; moderate\tabularnewline
\hline 
Positioning & $\approx$0.5~mr pitch, yaw, roll & not attempted; moderate\tabularnewline
\hline 
Positioning & 50~$\mu$m wire spacing & not attempted; moderate\tabularnewline
\hline 
Rate Capability & 0.5~MHz & not attempted; easy\tabularnewline
\hline 
\end{tabular}
\end{table}

\paragraph{Detector design:}
The detector design is a combination of individual straw design, which
is based on the PANDA straw chamber~\cite{pandasttdesign}, with a
geometry appropriate for the MUSE experiment.

The PANDA design uses thin-walled, over-pressured straws, allowing for 
significantly less straw material while providing mechanical
stability.
We adopt the same straws, wires, end pieces, and feed throughs from this design.
The PANDA chambers have operated successfully at rates exceeding 
8~kHz/cm, significantly higher than the MUSE rates.
The PANDA chambers have also achieved a position resolution of 
$\approx$150 $\mu$m.

MUSE requires symmetric scattered particle detector systems to beam left and right.
We use 2 chambers on each side of the beam, each chamber with 5 vertical
planes and 5 horizontal planes, to achieve high tracking efficiency. 
In order to provide better resolution on the scattering angle, the 
vertical straw planes will be placed closer to the target.
In order to break symmetries that make certain trajectories hard
to track, the straws in the front and rear planes will be offset by
$\approx$2~mm from each other.
Straw spacing is 1.01 cm, and adjacent offset straw planes are 
centered 0.87 cm apart. 

Table~\ref{tab:chambers}  summarizes chamber geometry
and the number of straws per chamber.  
The spacing between the chambers will be about 6 cm from the back of
one chamber to the front of the next chamber.  
The front chambers each have 275 60-cm long vertical straws 
and 300 55-cm long horizontal straws.
The rear chambers each have 400 90-cm long vertical straws 
and 450 80-cm long horizontal straws.
The total number of straws in the system is 2850.
No stereo planes are needed because the low beam flux reduces 
the likelihood of multiple particle tracks on the same side of the
beam, and the scattered particle scintillators
provide a precise time and approximate position for any second tracks.

\begin{table}[htb]
\caption{\label{tab:chambers} Straw chamber parameters including the distance
from the pivot, chamber active area, and the number of straws.}
\begin{tabular}{|l|c|c|c|} \hline
Chamber & Distance & Active Area & Number of Straws  \\
        & (cm)     & $\left(\rm{cm}^2\right)$ & per Chamber   \\
\hline
\hline
Front  & 30 & 60 $\times$ 55 & 575  \\
Back   & 45 & 90 $\times$ 80 &  850 \\ \hline
\end{tabular}
\end{table}

The chambers will be operated using a mixture of 90\% Ar $+$ 10\% CO$_2$
at a pressure of 2 bar, using pressure-control 
transducers and mass flow controllers (Bronkhorst Ltd.).
Each of the 4 chambers will be provided with an independent gas supply system.
The gas mixture will run in continuous-flow mode, completely replacing 
the full gas load in the chambers every 12 hours, which will allow us
to run without bubblers. 

The straws operate at 1700V.
Voltage is applied and readout connected using dedicated HV / readout
cards, into which the PADIWA frontend cards plug directly
(see Section \ref{sec:triggerdaq}).
The PADIWAs are in turn read out by TRB3 TDCs. 

The full design for the chamber frames, including the straws, gas
lines, and readout electronics, has been realized in CAD format, to
ensure compatibility with other MUSE systems at the design level.
Figure~\ref{fig:chambercad} shows the CAD drawing.
 
\begin{figure}[h]
\centering{
\includegraphics[width=0.6\textwidth]{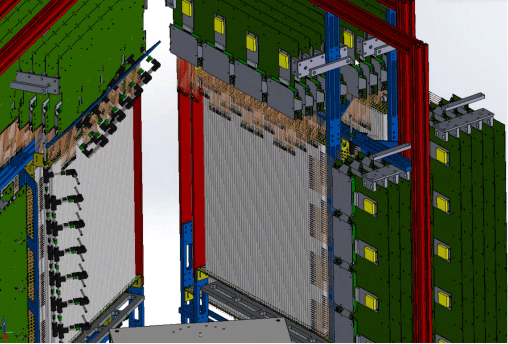}
\caption{\label{fig:chambercad} Snapshot of CAD drawing of the four STT chambers 
mounted on the table. Straws (light gray), HV/readout cards (dark
gray) with PADIWAs (yellow on gray), frames (blue and red), and parts
of the gas system plumbing (black) and other components can be seen.}
}
\end{figure}

\paragraph{Current status:}

\begin{figure}[h]
\centering{
\includegraphics[width=0.3\textwidth]{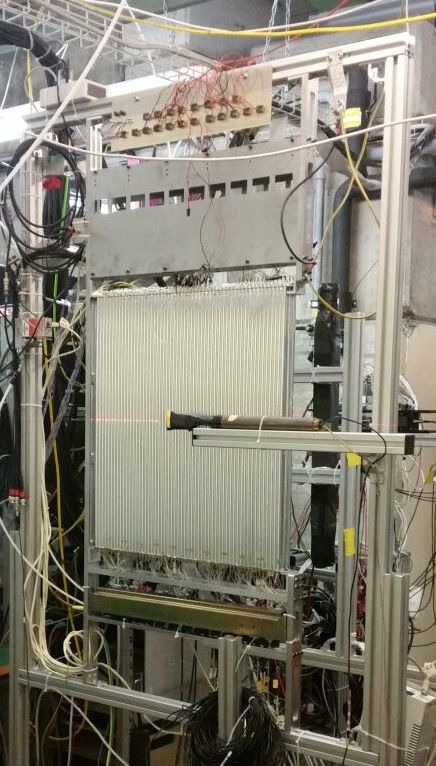}
\caption{\label{fig:chamber_const} A prototype half-chamber
  mounted at PSI.
}}
\end{figure}

\begin{figure}[h]
\centering{q
\includegraphics[width=0.5\textwidth]{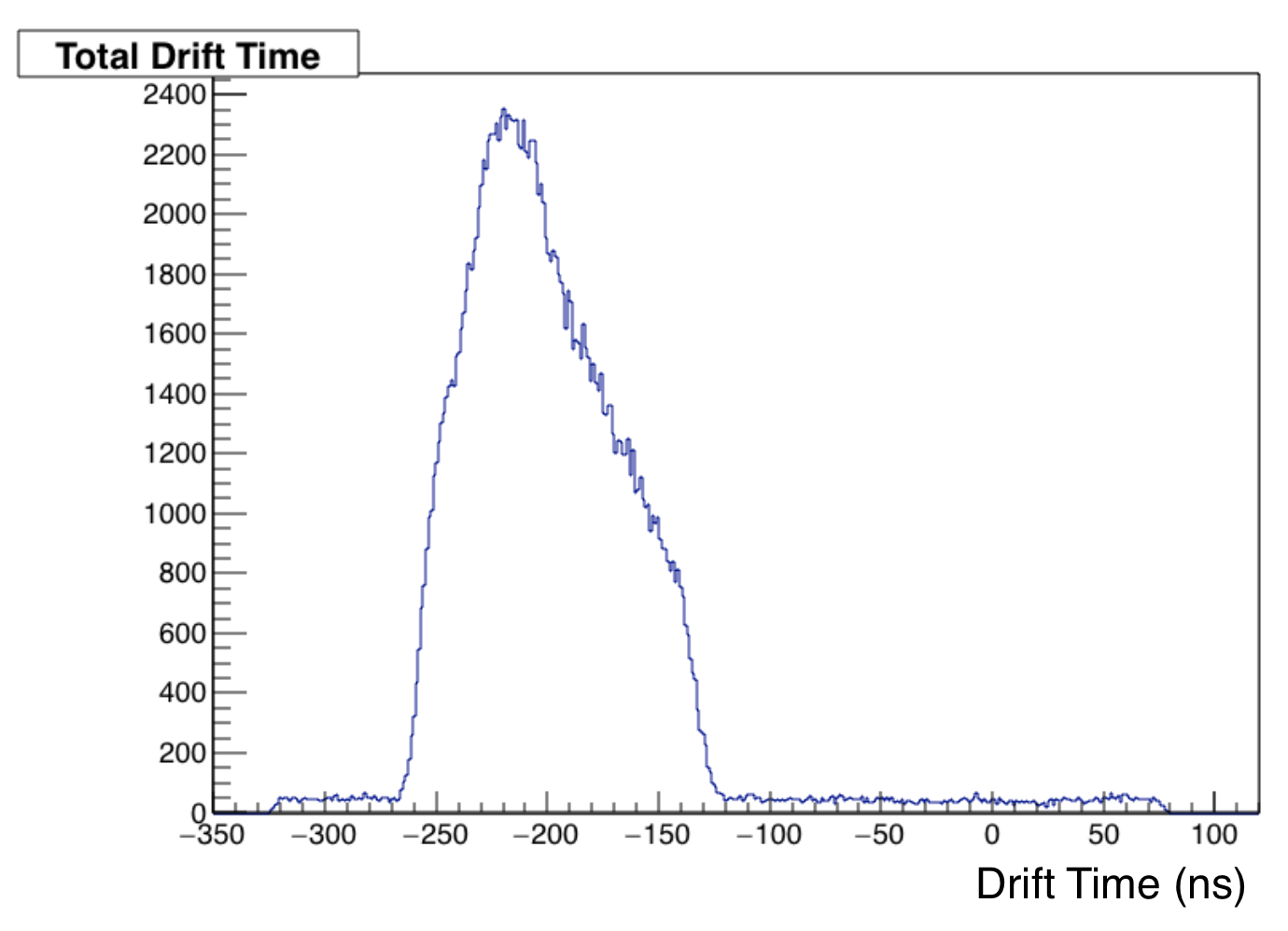}
\caption{\label{fig:chamber_sharkfin}A typical drift time histogram
  from the chamber tested during the Dec 2016 test period, showing the
  expected 'shark fin' shape.}
}
\end{figure}

\begin{figure}[h]
\centering{
\includegraphics[width=0.5\textwidth]{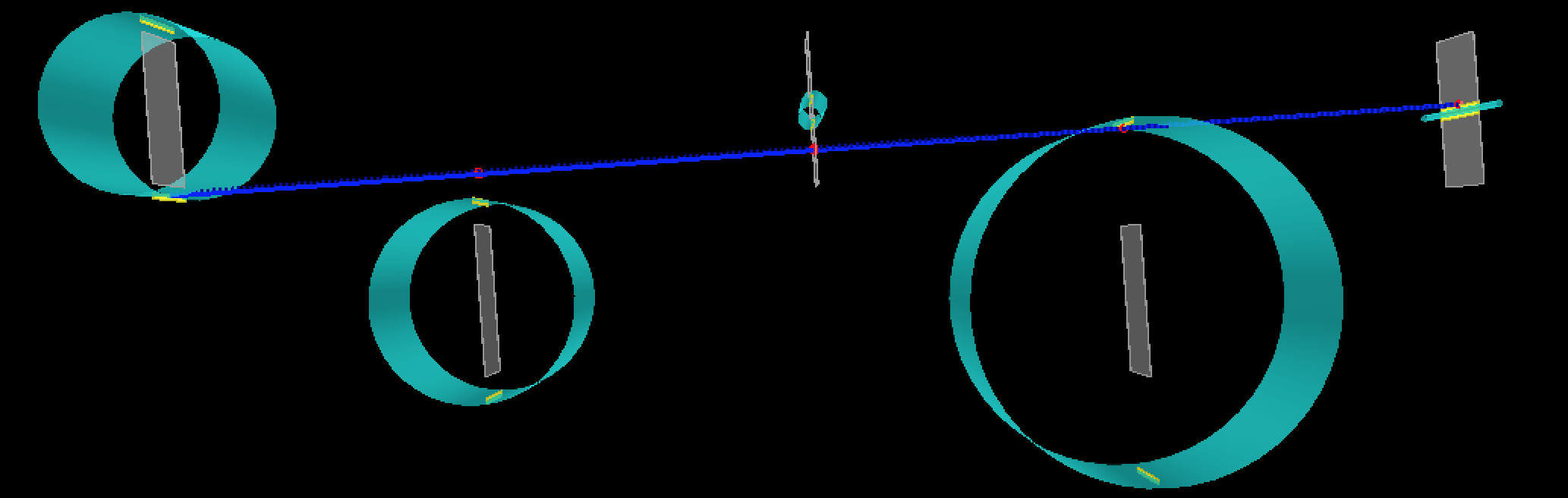}
\caption{\label{fig:chamber_track}A track generated by the track fitting code used for the straw tube tracker.}
}
\end{figure}

\begin{figure}[h]
\centering{
\includegraphics[width=0.5\textwidth]{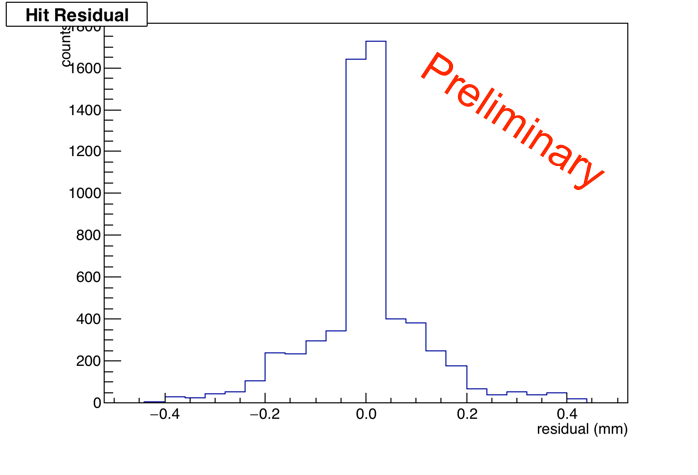}
\caption{\label{fig:chamber_resolution}A plot of the track residuals
  from a genfit analysis of beam test data.}
}
\end{figure}

Figure~\ref{fig:chamber_const} shows a prototype half-chamber, 
constructed at Hebrew University, and partially operated at PSI in in December
2016.
In these test beam conditions the straws were shown to operate
reliably with approximately 90\% efficiency, which yields 99\% tracking
efficiency for 5 planes.
Fig. \ref{fig:chamber_sharkfin} shows a typical 'shark fin' diagram of 
the straw time distribution, above a flat, random noise background. 

Importantly, during the same beam test, tracking information was 
obtained from a half chamber (vertical straws only) and a tracking 
algorithm was implemented. 
Figure~\ref{fig:chamber_track} shows a track obtained from 
the tracking code. 
A preliminary analysis of the chamber resolution using a small 
calibration dataset shows a resolution of approximately 115 $\mu$m;
Fig.~\ref{fig:chamber_resolution} shows the residuals of the track 
distances.
It is clear that the design resolution can be achieved.

\paragraph{Path to completion:}
In order to avoid straw  damage during transport to PSI, the
construction plan has individual straws being built at Hebrew University 
and shipped to PSI, where the chambers will be assembled and tested.
The first chamber straws are at PSI, being assembled into a chamber.
The chambers are assembled by gluing the individual straws
together using a precision machined jig. 
The straw packages are mounted between printed circuit board end
plates, which are mounted into the detector frames.
Gas, HV, and readout are then added.

Straw construction will continue at the Hebrew University with the
second chamber straws shipped to PSI for the December 2017 beam time,
and the straws for the final chambers being shipped in early 2018.

New HV / readout cards are being built, incorporating a gain
modification to improve the signal to noise ratio. This will be tested
during the June 2017 beam time. 

The assembled chambers are flushed, and the wire positions determined
using a well-collimated radioactive source and trigger scintillator, 
moved by a precision stepper motor across the chamber.
The drift time measurements allow the wire to be located.
This technique has previously been used with precision $\approx$50~$\mu$m.
The first chamber will be scanned with this technique in August 2017.

\setcounter{paragraph}{0}
\subsection{Scattered-Particle Scintillators}
\label{sec:scint}
\paragraph{Purpose:}
The scattered-particle scintillators are part of the event trigger and
help with the particle separation via time-of-flight (TOF)
measurements. 

\paragraph{Requirements:}

This purpose requires high detection efficiency for the
particles of interest and excellent timing resolution.  
The requirements for the scattered-particle scintillators are laid out
in Table~\ref{table:WBS7_SC_requirements}.

\begin{table}
\caption{Scattered-particle scintillation-detector requirements}
\label{table:WBS7_SC_requirements}
\begin{tabular}{|c|p{5.5cm}|p{5.5cm}|}
\hline 
Parameter & \centering{Performance Requirement} & \centering{Achieved}\tabularnewline
\hline 
\hline 
Time Resolution & \centering{$\approx$60~ps / plane} & \centering{$\checkmark$ 55~ps}\tabularnewline
\hline 
Efficiency & \centering{99\%, $\ll 1$\% paddle to paddle uncertainty}  &
                                                             \centering{$\checkmark$
                                                             99\%,
                                                             paddle to
                                                             paddle
                                                             not
                                                             attempted,
                                                             moderate}\tabularnewline
\hline 
Positioning & \centering{$\approx$1~mm, $\approx$1~mr} & \centering{not attempted; easy} \tabularnewline
\hline 
Rate Capability & \centering{0.5~MHz / paddle} & \centering{$\checkmark$ 1~MHz}\tabularnewline
\hline 
\end{tabular}
\end{table}

\paragraph{Detector design:}

The Experimental Nuclear Physics Group at USC is building
the scattered-particle scintillators for the MUSE experiment. 

The group previously designed, prototyped, and built the new
FToF12 detector for the upgraded CLAS12 at Jefferson Lab. The design
and construction procedures of the MUSE detectors follow that of the
FToF12 detector.  The detector will be made of Eljen Technology EJ-204
plastic scintillators, which have a high light output and fast rise
time.  Each end of the long scintillator bars is fitted with black
tape, which masks the corners while leaving a circular window that
extends one millimeter into the area that will be covered by the
photocathode. The corner blocking reduces the amount of reflected
light contributing to the leading edge of the PMT signal.  Hamamatsu
R13435 PMTs are then glued to each end of the scintillator.  The bare
counter is wrapped with precision-cut aluminized mylar and
DuPont$^{\rm TM}$ Tedlar. The Tedlar film extends beyond each PMT onto
the anode, dynode, and high-voltage cables, providing a single
light-tight casing for the entire counter.  Pairs of scintillator bars
are mounted on aluminum sheets that serve as backing structure.  These
units will be mounted in a frame.  Details about the construction
process and system tests for quality assurance can be found in
Ref.~\cite{gothe2009}.

Table~\ref{tab:scwall} lists the design parameters for the
scintillator walls. The front wall is approximately square and covers at least a
horizontal angular range from $20^\circ$ to $100^\circ$ from all
points within the target.  The back wall has an
increased angular acceptance to account for particles which scatter in
the front wall material.

\begin{table}[h!]
\begin{center}
\renewcommand{\tabcolsep}{5mm}
\renewcommand{\arraystretch}{1.1}
\caption{Design parameters for the scintillator walls.
  \label{tab:scwall}}
\begin{tabular}{lrr}
  \hline\hline
 & Front wall & Back wall \\
\hline
Number of scintillator bars &  18 &  28 \\
Scintillator cross section & 6~cm~$\times$~3~cm &  6~cm~$\times$~6~cm \\ 
Scintillator length & 120~cm &  220~cm \\ 
Target to front-face distance & 52~cm & 74~cm \\
Gap between scintillator bars & 0.3~mm & 0.3~mm \\
Scintillation material & EJ--204 & EJ--204\\
Photomultiplier & Hamamatsu	
  R13435 & Hamamatsu	
  R13435\\
\hline\hline
\end{tabular}
\end{center}
\end{table}

\paragraph{Current status:}

We are presently building the 18 short and 28 long detectors for the
left TOF wall.  All photomultiplier tubes have been tested and
characterized; almost all photomultiplier tubes have been glued to the
scintillation bars; most detectors are built.  We continue to measure
position-dependent time resolutions in cosmic-ray tests before we
mount pairs of detectors on their backing structure.  Average time
resolutions of $\sigma_{avg} = 55$~ps for the 220-cm long bars and
below 50~ps for the 110-cm long bars were obtained; see
Fig.~\ref{fig:sc_time_resolution}.
\begin{figure}[ht]
\centerline{\includegraphics[width=3.5in]{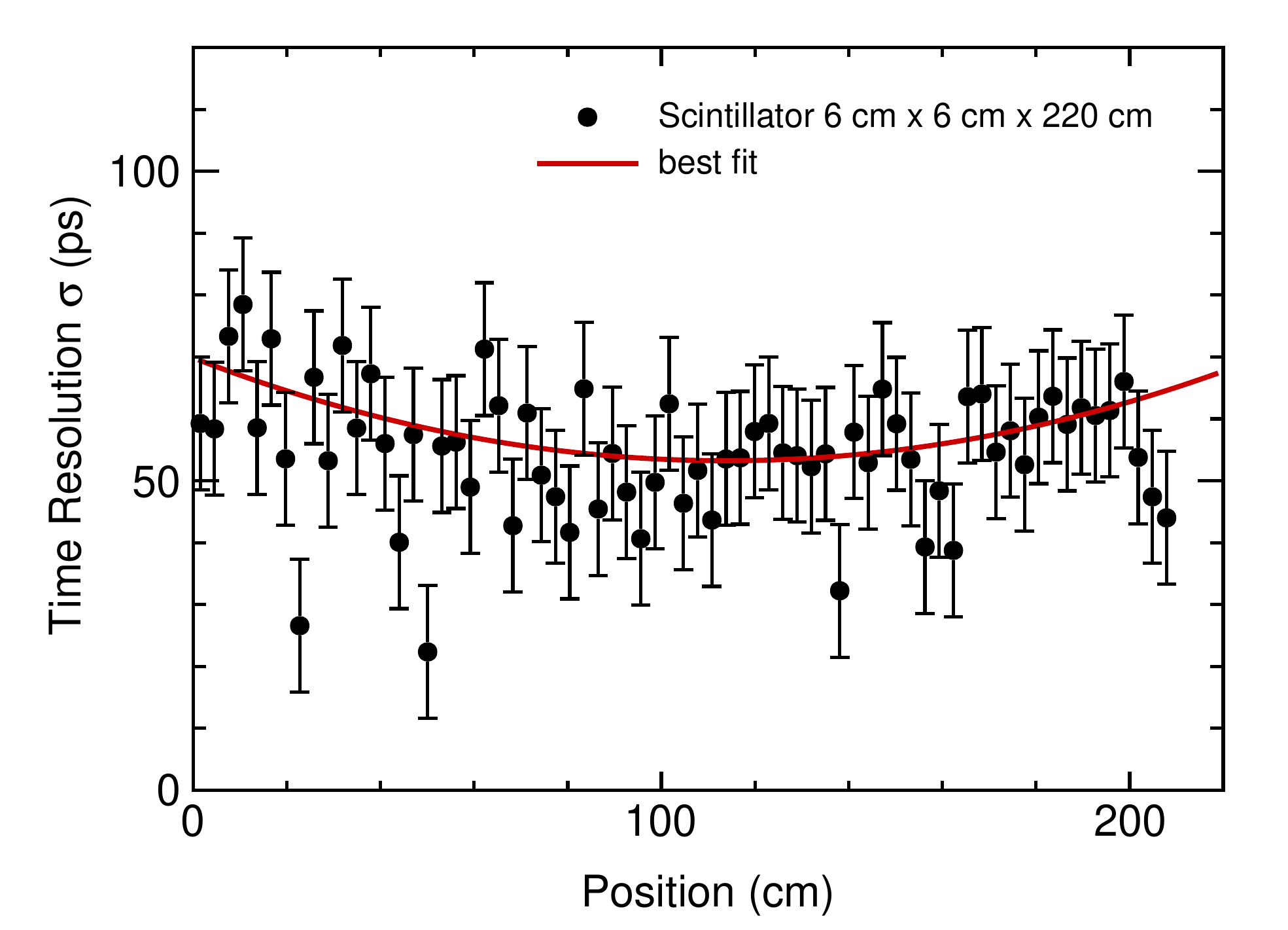}}
\caption{Position-dependent time resolution for a prototype MUSE 220-cm 
  long scintillation bar after calibration, event selection, and time-walk
  correction. The average time resolution is $\sigma_{avg} = 55$~ps.} 
\label{fig:sc_time_resolution}
\end{figure}

We have studied the performance of the proposed scattered-particle
scintillators with Geant4 simulations of the planned setup.  The
particle interactions and their energy deposition within the
scintillators have been calculated.  Figure~\ref{fig:sc_edep_ftof}
shows the distribution of deposited energy in a 5~cm $\times$ 5~cm
scintillator which was used in the summer 2013 test measurement at
$\pi$M1.  The incident particles were 153~MeV/$c$ muons.  The
simulated energy distribution agrees nicely with the measured data.
The energy deposited by particles whose paths do not traverse at least
the full thickness of the scintillator is lower than the energy of the
lower edge of the Landau-like portion of the energy distribution.
\begin{figure}[hbt]
\centerline{\includegraphics[width=4.0in]{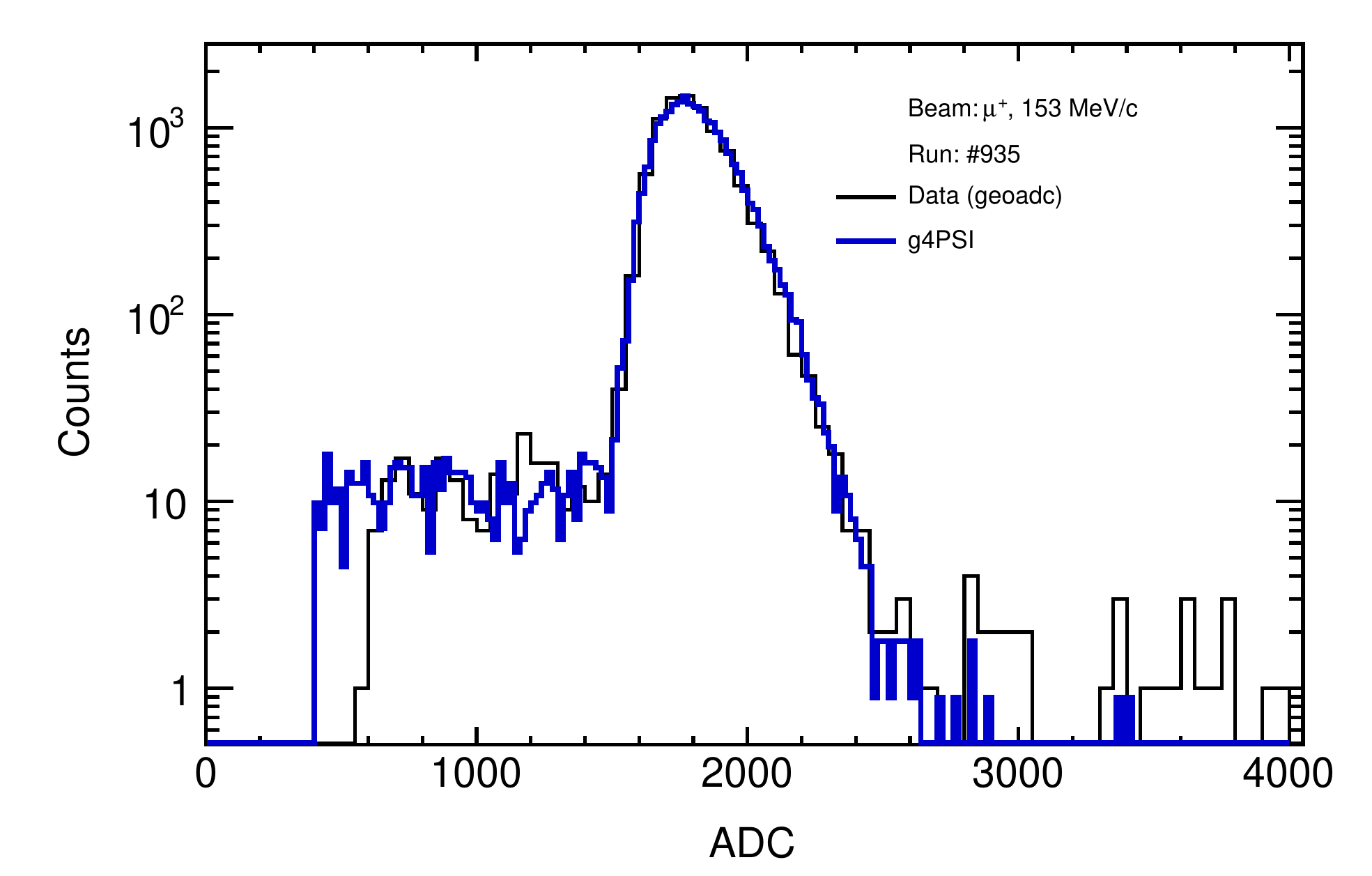}}
\caption{Deposited energy of muons passing through a 5~cm $\times$
  5~cm scintillator bar.  The data are from the summer 2013 test
   measurement at $\pi M1$.  The blue histogram shows the result of the
  Geant4 simulation.}
\label{fig:sc_edep_ftof}
\end{figure}

Simulated energy distributions for the 6~cm $\times$ 3~cm and 6~cm
$\times$ 6~cm scintillator bars are shown in Fig.~\ref{fig:sc_edep}
for scattered electrons (left panel) and muons (right panel) at various
beam momenta.  
\begin{figure}[t]
\centerline{
\includegraphics[width=3.1in]{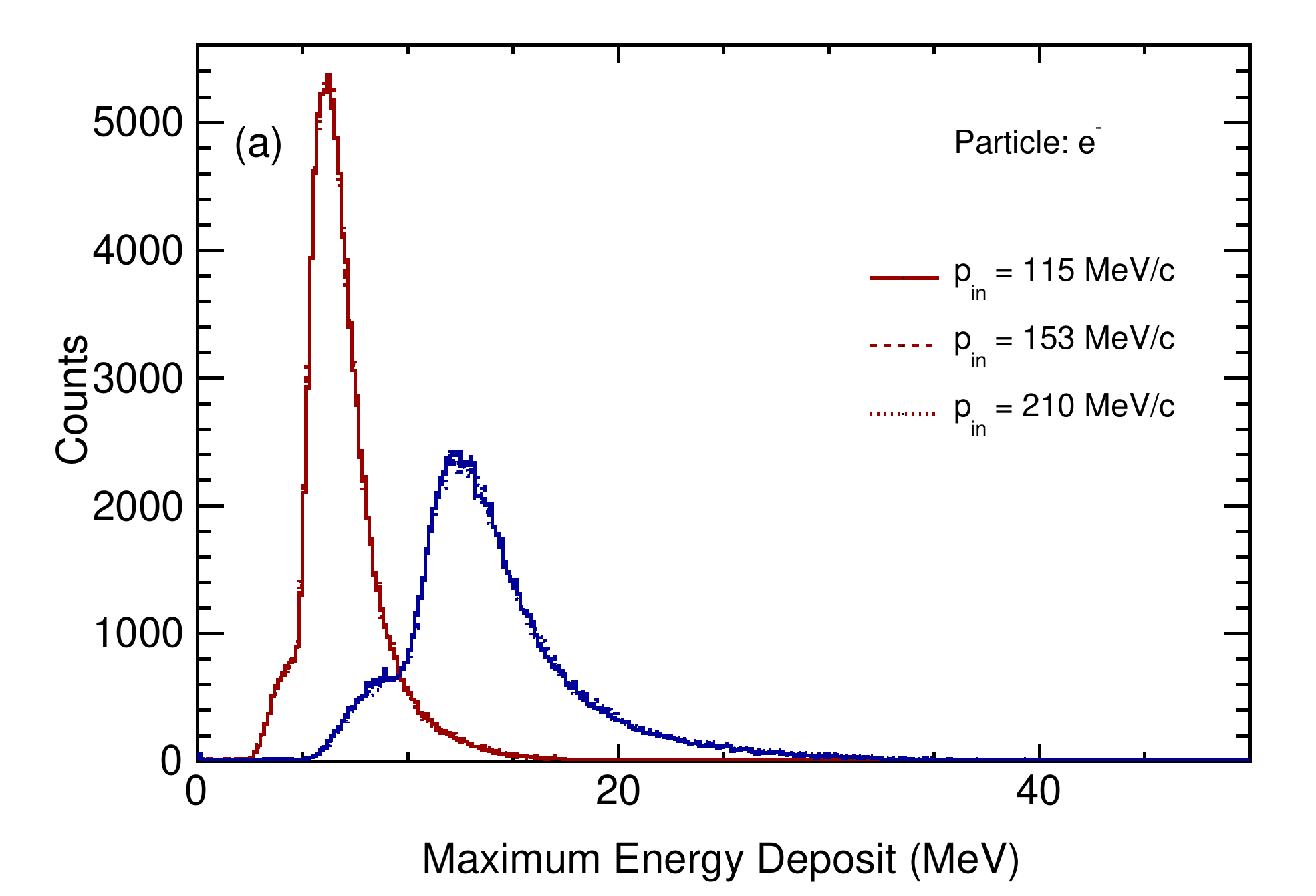}
\includegraphics[width=3.1in]{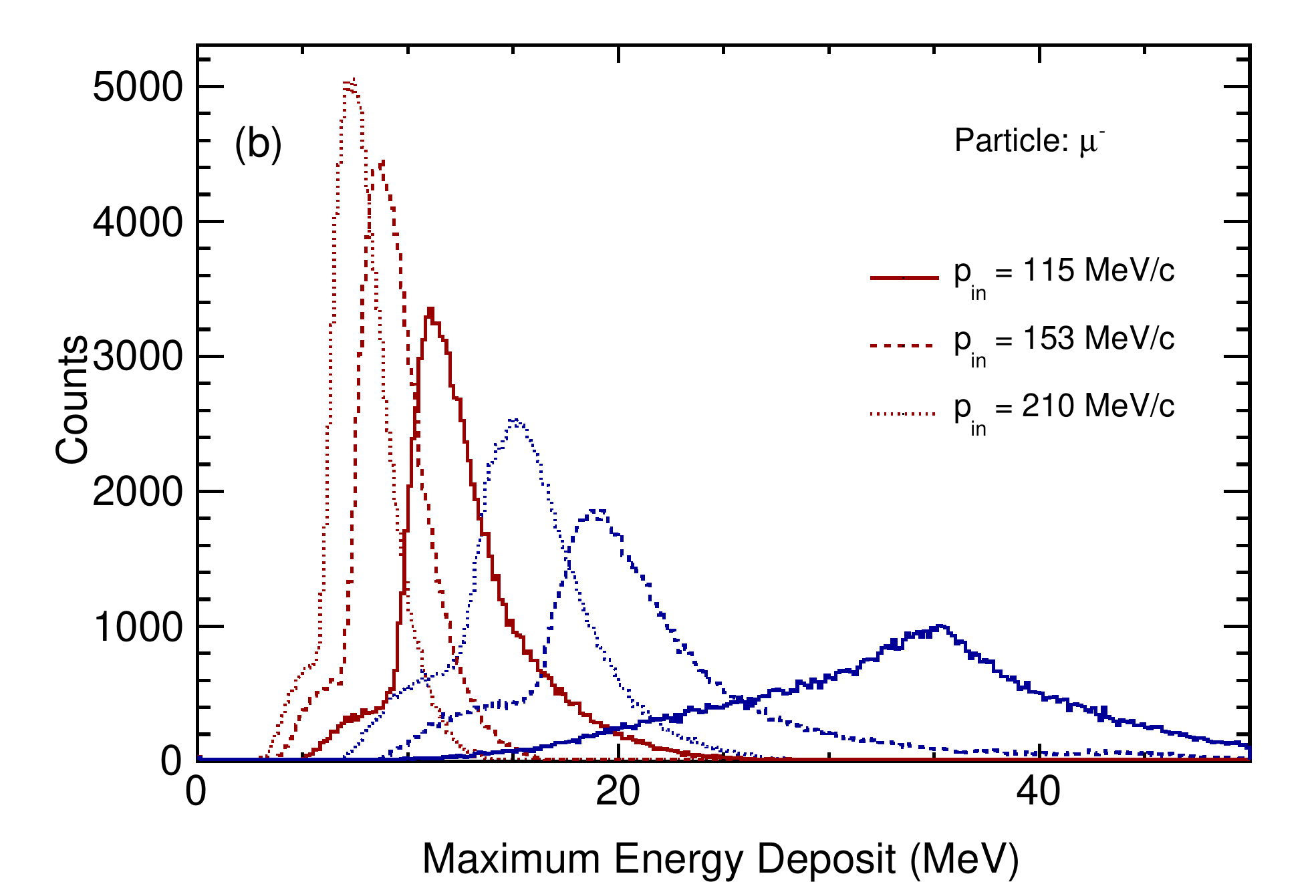}
}
\caption{Simulated energy deposition for scattered electrons (left) and
  muons (right), traversing the 6~cm $\times$ 3~cm bars of the front and
  6~cm $\times$ 6~cm bars of the back
  scattered-particle scintillator wall.  The simulation recorded for
  each event the maximum energy deposition in a scintillator of a
  given plane. }
\label{fig:sc_edep}
\end{figure}
The set of curves with low energy deposition (red)
is for the front wall; the set of curves with high energy deposition (blue)
is for the thicker back wall.  In the studied range, the energy
depositions for $e^{\pm}$ are independent of the beam momentum.  The
simulation shows for each event the maximum energy deposition in any
front- or back-wall bar.  Very nearly all events have energy
depositions above threshold, $E_{th} = 2$~MeV, in (at least) one bar.
The detection efficiency is indeed very high.

A detailed view of the particle detection efficiencies for the
scattered-particle scintillator walls at 115 MeV/$c$ is shown in
Fig.~\ref{fig:sc_trig_eff} for electrons and positrons as a function
of the particle scattering angle.  All panels are for the same
detection threshold of $E_{th} = 2$~MeV.  The solid dots give the
ratio of events with an above-threshold hit in the front plane per
incident particle.  Particles were incident on the ``active'' area of
the scintillator plane; the physical size of the plane is slightly
larger.  The average acceptance is higher for higher momenta and it is
higher for $\mu^\pm$ than for $e^-$.
\begin{figure}[!ht]
\centerline{
\includegraphics[width=3.1in]{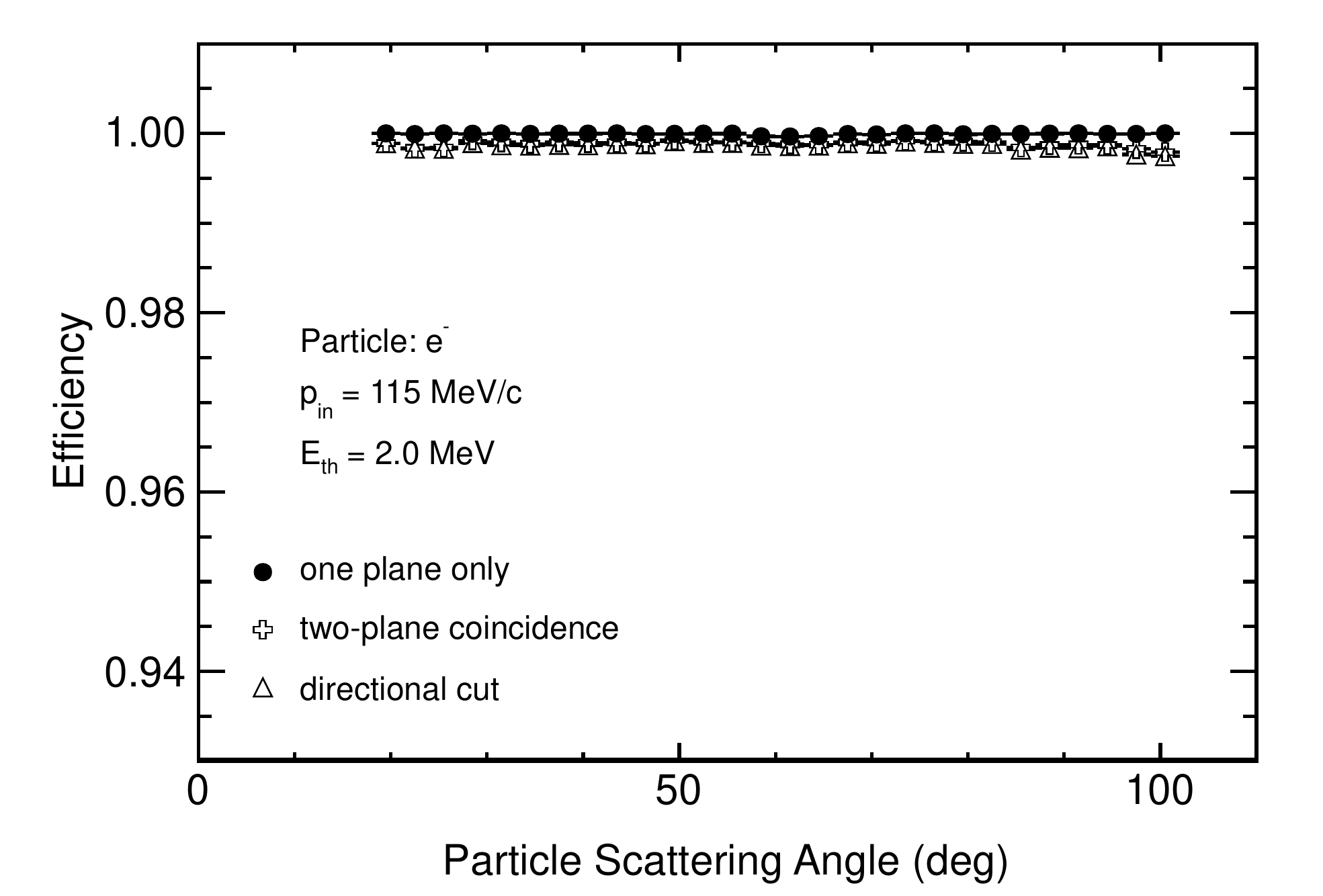}
\includegraphics[width=3.1in]{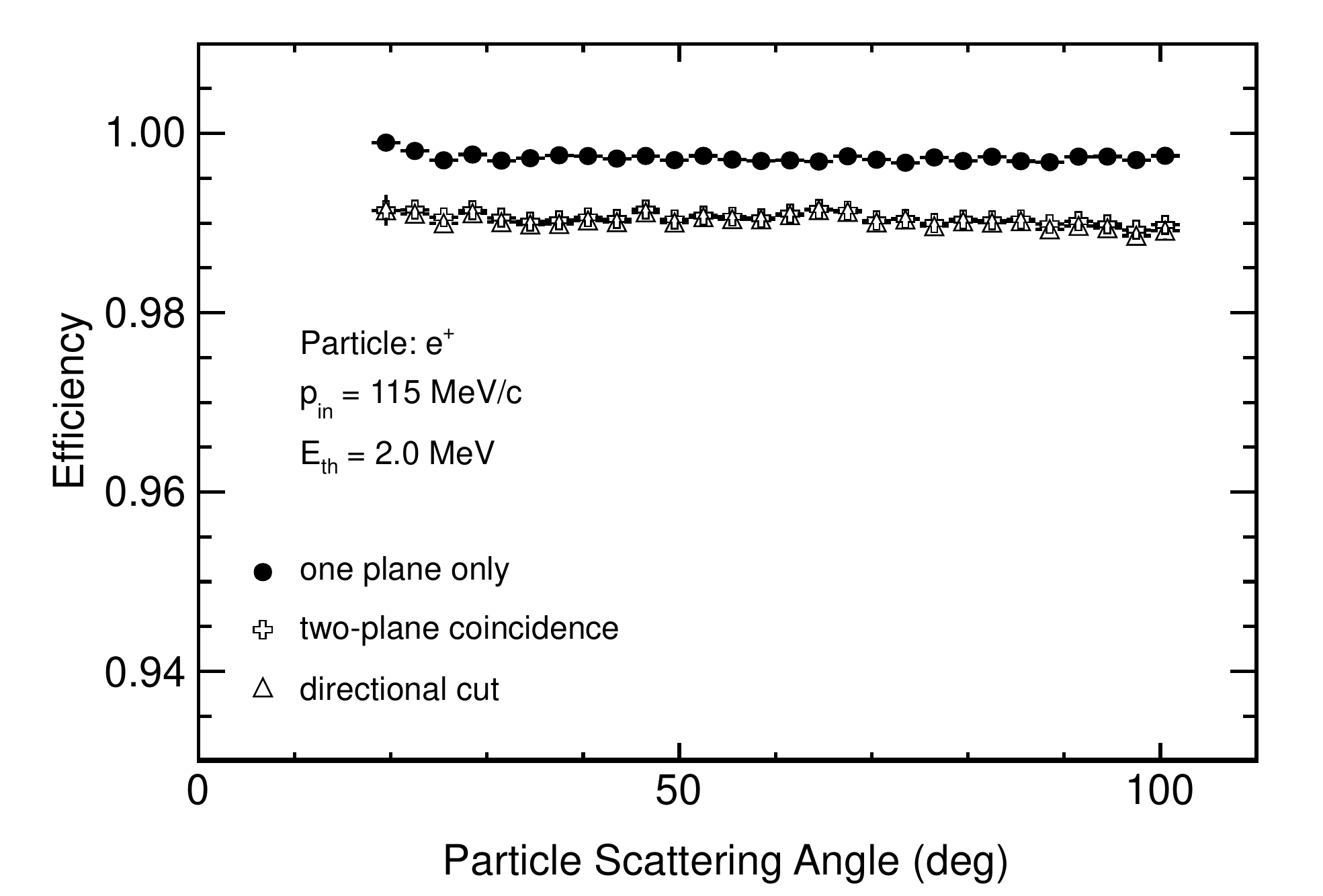}}
\caption{Estimated detection efficiency as a function of particle
  scattering angle for $e^+$ and $e^-$ at beam momenta of 115~MeV/$c$.
  The change of momentum of the scattered particle with scattering
  angle is taken into account.}
\label{fig:sc_trig_eff}
\end{figure}
The one-plane efficiency is practically 100\%.  The two-plane
coincidence (plus symbol in Fig.~\ref{fig:sc_trig_eff})
requires above-threshold hits in both the front and back
planes.  It is in all cases well above 99.5\%, except for $e^+$.  The
``directional cut'' (triangle points in Fig.~\ref{fig:sc_trig_eff})
utilizes the fact that scattered particles, which
originate in the target, deposit energy mostly in certain combinations
of front- and back-wall scintillators.  For an event to pass this cut,
each hit in a scintillator bar of the back wall must coincide with
hits in up to three corresponding neighboring scintillators in the
front wall.  This directional cut does not affect the efficiency much but
helps to suppress triggers from background events which do not
originate within the target.  Figure~\ref{fig:sc_paddle} illustrates
this correlation of scintillator-bar numbers for muons with different
momenta originating in the target volume.
\begin{figure}[!ht]
\centerline{
\includegraphics[width=3.1in]{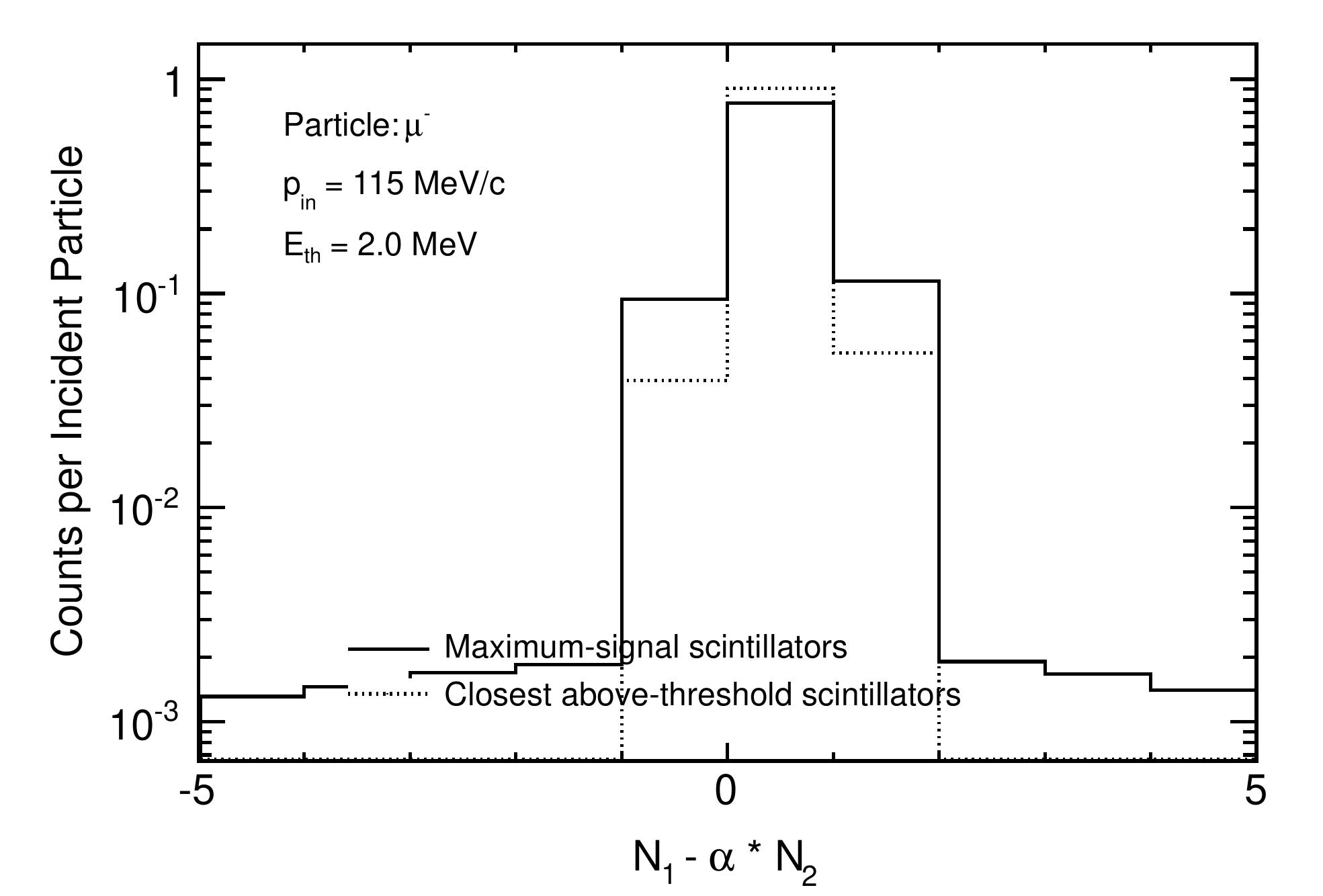}
\includegraphics[width=3.1in]{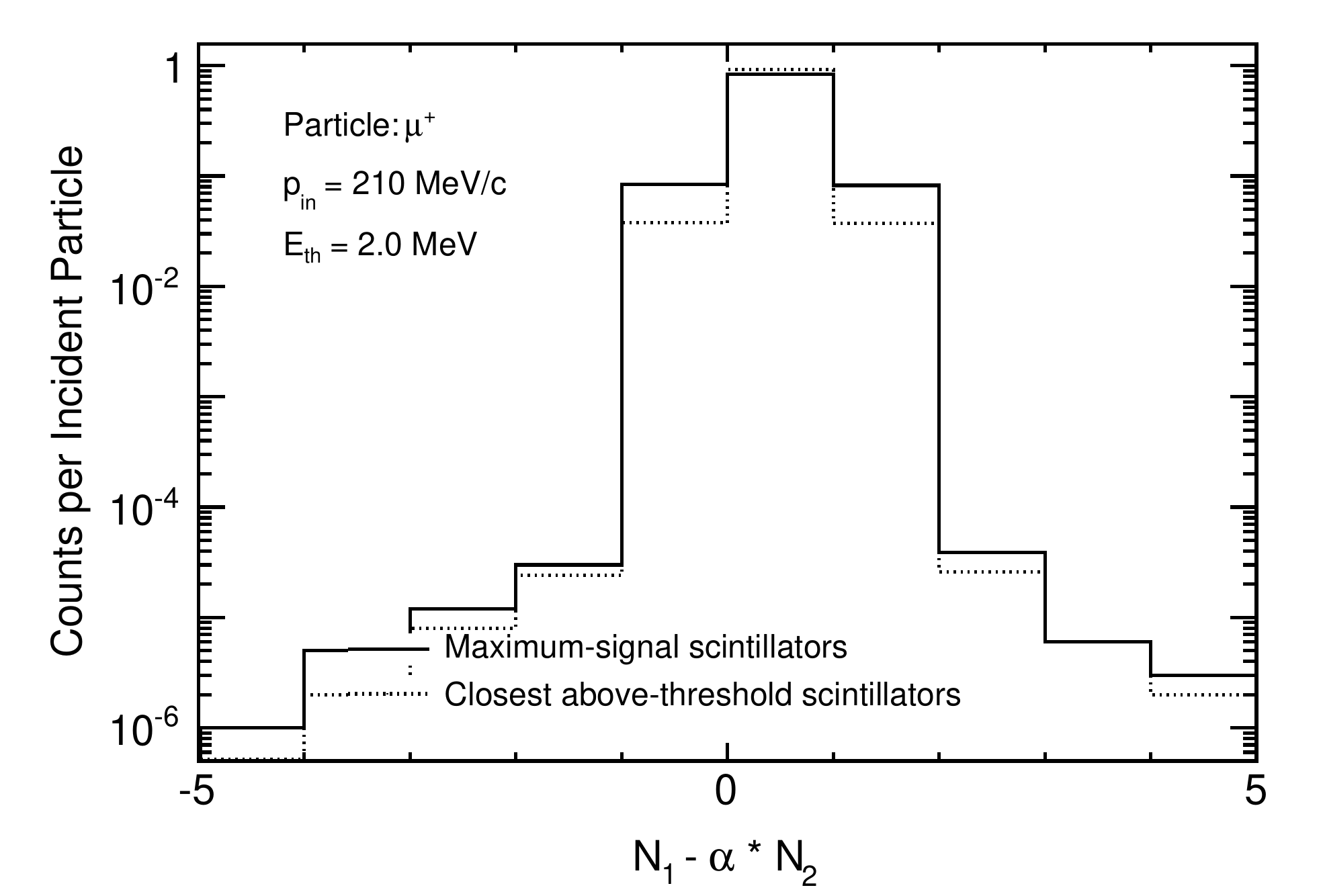}}
\caption{Typical paddle-number correlations between paddle numbers $N_1$ and
  $N_2$ from the front- and back-wall scintillators, respectively.
  The factor $\alpha$ is the ratio of the distances from the target
  to the front and back scintillator-wall mid-planes, respectively.}
\label{fig:sc_paddle}
\end{figure}

If uncorrected, detection inefficiencies in the scattered-particle
detector will lead to errors in the measured cross sections. The
average corrections for detector inefficiencies are of the order of
0.1\% for $\mu^\pm$ and $e^-$; and 0.4\% to 0.9\%
for $e^+$.
These values require a threshold
of $E_{th} = 2$~MeV.  The positron efficiency is reduced due to
possible annihilation processes.  The detector inefficiencies show
some angular dependence at low scattered particle momentum (backward
angles at 115~MeV/$c$ beam momentum); see
Fig.~\ref{fig:sc_trig_eff}.  After correction for these effects, we
expect the contribution from the scattered-particle detector to the
systematic uncertainties of the absolute cross section to be less than
0.1\%.  The uncertainty is larger for $e^\pm$ cross sections if the
threshold can not be kept stable.  Because of their very similar
detector response, we expect the contributions to the systematic
uncertainties of relative cross sections for $\mu^+$ and $\mu^-$ to be
negligible.  Also, the $\mu^\pm$ and $e^-$ relative cross section
uncertainties should be much smaller than 0.1\%.

\paragraph{Path to completion:}

The larger part of the detectors for the left wall have been built and
we expect to complete its construction on schedule by the end of
summer 2017.  Thereafter, as soon as the remaining construction
funding becomes available, we will continue building detectors for the
right wall.

\setcounter{paragraph}{0}
\section{DAQ}
\label{sec:daq}

\subsection{Electronics and Readout}
\label{sec:readout}

\paragraph{Purpose:}
The Data Acquisition System (DAQ) reads out the fast event data from
the detectors, times and pulse sizes, for event analysis and detector
calibration.
This sub-system also reads out slow controls, and includes the event trigger.

\paragraph{Requirements:}
The requirements for the DAQ hardware are listed in table \ref{table-DAQrequirements}.  
The readout channels necessary for the experiment are listed in table
\ref{table-DAQsummary},
except that we treat the GEM DAQ as part of the GEM system, discussed
in Section~\ref{sec:gems}.

\begin{table}

\caption{Data acquisition system hardware requirements.}
 \label{table-DAQrequirements}

\begin{tabular}{|c|c|c|}
\hline 
Parameter & Performance Requirement & Achieved?\tabularnewline
\hline 
TDC Resolution & $\leq$ 40~ps & $\checkmark$ $<$ 30~ps \tabularnewline
\hline 
QDC Resolution & $>$ 10~bit & $\checkmark$ 12~bit \tabularnewline
\hline 
Readout Rate & 100 $\mu$s per event & 400 $\mu$s per event to \tabularnewline
& & date, moderate \tabularnewline
\hline 

\end{tabular}
\end{table}

\begin{table}

\caption{Summary of detector needs. The Gate Source is either
``Exp't'' for the full experiment trigger, which requires delaying
signals, or ``Fast'' for an earlier Level 1 trigger.}

\label{table-DAQsummary}

\begin{tabular}{|c|c|c|c|c|c|c|}
\hline 
Detector & No. & TDC & Disc. & QDC & Gate & Trigger\tabularnewline
 & Chan's &  &  &  & Source & Input\tabularnewline
\hline 
\multicolumn{7}{|c|}{Upstream Beam Line Detectors:}\tabularnewline
\hline 
SiPM Hodo. & 64-128 & $\checkmark$ & MCFD & $\checkmark$ & Fast & $\checkmark$ \tabularnewline
\hline 
Veto & 8 & $\checkmark$ & PADIWA & $\checkmark$ & Exp't & $\checkmark$ \tabularnewline
\hline 
\multicolumn{7}{|c|}{Beam Monitor:}\tabularnewline
\hline 
SiPM & 64 & $\checkmark$ & MCFD & $\checkmark$ & Fast & $\checkmark$ \tabularnewline
\hline 
Paddles & 8 & $\checkmark$ & MCFD & $\checkmark$ & Fast & $\checkmark$ \tabularnewline
\hline 
\multicolumn{7}{|c|}{Scattered Particle Detectors:}\tabularnewline
\hline 
STT & 2850 & $\checkmark$ & PADIWA & $\times$ & N/A & $\times$ \tabularnewline
\hline 
SPS & 184 & $\checkmark$ & PADIWA & $\checkmark$ & Exp't & $\checkmark$ \tabularnewline
\hline 
\end{tabular}
\end{table}

\paragraph{Data acquisition system design}

The basic components of the MUSE DAQ are outlined in the following paragraphs.

The analog detector signals are digitized by discriminators of two kinds:
PADIWA level discriminators, or Mesytec MCFDs.

\textbf{PADIWAs} were custom-designed at GSI to provide
a fast, compact and cost-effective readout for FAIR experiments.
They are 16-channel level discriminators. 
The customized MUSE PADIWA design removes the standard
 $\times$10 input amplification.
PADIWAs have
an independent threshold for each channel, set via
TRB3 board control lines, using either a script
or the TRB3 web interface. 
The PADIWAs send LVDS signals to the TRB3 TDC.
Tests have shown 
that for fast scintillators, level discriminators used
in combination with QDC-signals for time-walk
corrections provide better time resolution
than constant fraction discriminators.

\textbf{Mesytec MCFDs} provide constant fraction discrimination
for all detectors read out by SiPMs, and for the beam monitor
paddles where we will want to precisely time random coincidence 
signals with no corresponding pulse size measurement.
The MUSE MCFDs have been customized with LVDS outputs for
direct input to the TRB3 system, instead of externally converting the
standard ECL output.
Each board discriminates 16 channels, and provides two outputs
per channel: a copy of the analog signal for input to the QDC,
and the discriminated logic pulse.
The MCFD board also
provides a single 16OR LEMO output, which can be used
as a fast gate of the MQDC-32.  
MCFDs are controlled via usb interface.
 
\textbf{TRB3} boards were designed by GSI as a customizable, programable,
scalable, versatile building block for modern data acquisition systems.  
Each TRB3 has a central FPGA for control, communication, and
triggering,
plus four peripheral FPGAs programmed as high-resolution -- 11 ps --
TDCs plus scalers.
Each peripheral FPGA controls and reads out 
three PADIWAs or 48 channels.
Thus, one TRB3 can control and read out a maximum of
12 PADIWAs or 192 channels.
The TRB3s are powered by a 48 V supply, and are independently 
controlled and read out over gigabit ethernet.  
They require no VME crate. 
The TRB3s will be distributed
throughout the experimental equipment, leading to shorter cabling
and better timing.

 \textbf{Mesytec-MQDC-32s} provide signal size information,
for pulse-height corrections to improve timing, 
and for monitoring detector response, to check for gain shifts.
The MQDC-32 accepts both positive and negative inputs in the same 
hardware module, through the use of jumpers, and reads out all
channels in 250 ns.
The MQDC-32s can be gated with a fast trigger or with the experimental
trigger. For the fast trigger, a final read-out trigger or fast clear
needs to then
be provided by the experiment trigger. 
This eliminates the need for long delay cables.
The MQDC-32 modules are housed in VME crates.

\textbf{VME} infrastructure is provided via standard
VME crates, read out via a \textbf{CAEN v2718 VME-PCI bridge} in
each crate over a \textbf{CAEN A3818C - PCIe 4-link optical
bridge}. In order to synchronize the data provided by 
each of the crates and that provided by each TRB3,
an event number, generated in the TRB3 triggering
system, is fed into the VME systems using a 
\textbf{VULOM 4b} board, designed and produced
by GSI. This VME board has an FPGA, and a series
of inputs and outputs which will received the trigger 
number, and other trigger header information. This
information will be copied into the VME buffer and
sent with the data to ensure synchronization
between systems. 

\begin{figure}[h]
\centerline{
\includegraphics[height=2.5in]{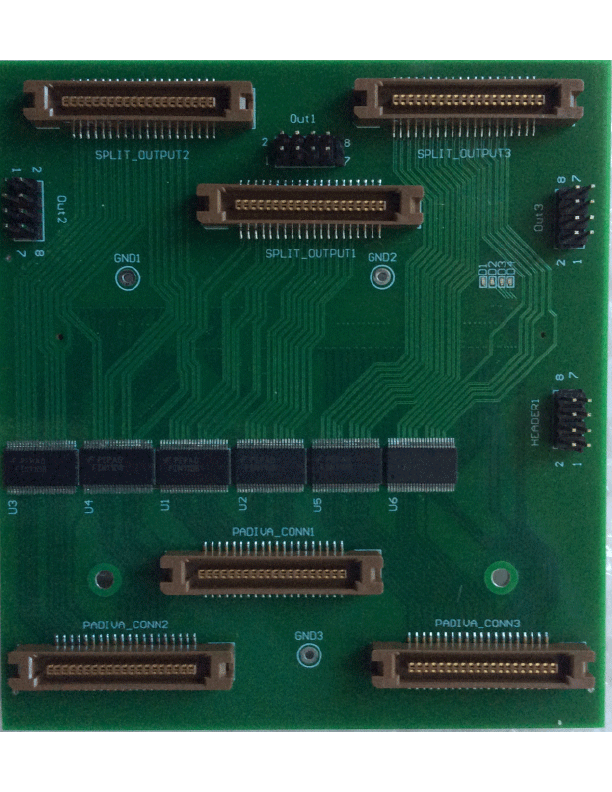}
}
\caption{
A prototype splitter board. The board mounts directly onto the TRB3
peripheral FPGA, in place of the usual board used to couple in TDC
input signals.
}
\label{fig:SplitterProto}
\end{figure}

\textbf{A custom splitter} copies logic signals going to 
the TRB3 TDCs for use in trigger logic and QDC gating.
The prototype splitter board shown in Fig.~\ref{fig:SplitterProto} was built by 
Marcin Kajetanowicz, who works with the group of Piotr Salabura 
of Jagiellonian University of Krakow, Poland.
It directly couples the logic signals from trigger detectors into the
 high precision TDC FPGAs, and copies the signal to a second
trigger FPGA.
Test data show that timing resolution of the directly coupled signal
is basically unaffected, while the copied signal timing has a total
jitter increased by $\approx$10 ps -- see Fig.~\ref{fig:SplitterProtoData} --
which is negligible at the trigger level.

\begin{figure}[h]
\centerline{
\includegraphics[height=3.5in]{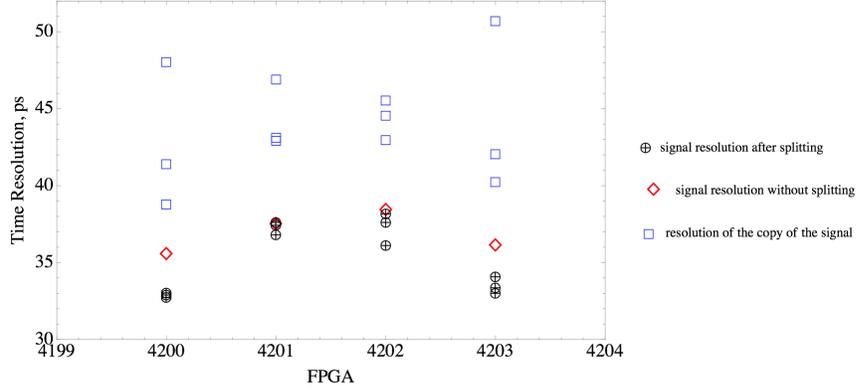}
}
\caption{
Measured TRB3 timing resolution for signals in $\pi$M1 without 
and with the prototype splitter board.
}
\label{fig:SplitterProtoData}
\end{figure}

For diagrams of the systems planed for each 
detector read out, see Appendix~\ref{DAQ-appendix}. 

\paragraph{Current status:} 
All major DAQ components 
have been designed, prototyped and investigated during beam tests,
all components function as required.
The modified PADIWA design to date has only been tested by modifying
standard PADIWAs, as we await delivery of the modified design.
Also, two simple circuit boards to adjust cable form factors are being 
designed and built by Basel.
One plugs into the Mesytec 34-pin output and converts to a 
TRB3 cable form factor for TRB3 input.
The second takes scintillator signal coax cable inputs to a
pin connector on which a PADIWA mounts.

One specification still to be met is the readout time,
which is currently limited by the GEM readout, discussed elsewhere.

POs have been issued for nearly all procurement of DAQ components
needed in 2017.
Many cable components have been ordered, but final cable
procurement and construction awaits final cryotarget design, 
review, and acceptance, to ensure procurement of the appropriate cable lengths.

\paragraph{Path to completion:}
The remaining electronics will be purchased as
soon as the second year of MUSE funding is approved.
All electronics systems will then be assembled, with the 
cable construction and testing, at PSI, in 
time to allow all electronics systems to be 
fully tested before needed for data taking.

\setcounter{paragraph}{0}
\subsection{Trigger}
\label{sec:triggerdaq}

\paragraph{Purpose:}
The trigger identifies events of interest for the fast DAQ, and initiates 
the read out of detector signal times and sizes.

\paragraph{Requirements:}

The trigger needs to efficiently identify muon and electron scattering
events, suppress pion scattering events, and provide multiple 
potentially prescaled trigger types corresponding to the different
types of physics and / or background events of interest.
The trigger needs to be $\approx$99\% efficient at
identifying $e$ and $\mu$ scattering events to maintain statistical
precision and $\approx$99\% efficient at identifying and not
triggering from pion induced background events to keep dead time low. 
Trigger angle-dependent efficiency variation should be small and known
at the  $\approx$0.1\% level.
Finally, the trigger must be generated within several hundred ns of
the event.

\paragraph{Trigger system design}

We use a two level trigger consisting of several level 1 trigger
components feeding into a level 2 master trigger.
The components include the following:
\begin{itemize}
\item {\it The beam PID system} uses RF timing of signals from 
the beam hodoscope described in Section~\ref{sec:beamsipm} to 
identify $e$'s, $\mu$'s, and $\pi$'s.
\item{\it The veto system} detects particles that will enter the vacuum
chamber through the veto scintillators and the thick metal 
vacuum chamber wall behind them. These particles arise from the beam tails, 
from particle decays, or from scattering in the beam line detectors.
\item{\it The scattered particle system} uses hit patterns
in the scattered particle scintillators to identify possible
scattering events from the cryotarget region.
\item{\it The beam monitor system} identifies forward-going 
particles after the target. Due to the
potential of this information to change the shape of the angular
distribution, this system is not planned to be used in production data taking.
It is intended for calibration data such as trigger verification, 
beam momentum measurements, and determination of backgrounds in the 
scattered particle detectors when they do not have a level 1 trigger.
\item{\it The random system} generates a random in time (with respect
to beam interactions) signal to help measure unbiased detector
backgrounds.
\item{\it The NIM system} generates a trigger independently via NIM
logic modules. We have used NIM trigger generation in test times to
date, and the NIM system will continue to be used as the FPGA
trigger is developed, as an aid to its verification.
\item{\it The level 2 master trigger} takes the outputs of the
aforementioned level 1 systems to determine if any trigger 
conditions are satisfied. It includes different prescale factors for
the different trigger types and a latch function.
\end{itemize}

Typically the various detector logic signals are sent to a high
precision TDC on a TRB3 FPGA through a splitter board, with the copied
signal sent to a second TRB3 FPGA programmed with the appropriate
trigger logic.

The beam PID system logic uses individual channel logic signals plus
the RF signal as a clock to determine the RF phase of each logic
signal. 
All 32 signals from each beam hodoscope plane are fed into the beam 
PID system; each plane is analyzed in a separate FPGA.
The different particle types arrive in the beam line detectors about
every 20 ns, with $\approx$3 - 6 ns separation between 
particle types and $\approx$0.3 ns rms width of the particle bunches
in time.
We require both ends of a paddle to fire to identify particle type, 
to suppress noise.
The system can identify multiple particle types at the ``same'' time
from the 16 paddles in a plane.

The veto system is a simple logical OR of all 8 PMTs in the veto
detector.

The scattered particle system uses logical ORs and ANDs to identify
hit patterns of scintillator paddles correlated between the front and
rear walls. Due the the number of channels, the system is divided 
among four FPGAs, for top left, lower left, top right, and lower right
PMTs. Each FPGA has inputs from 18 front $+$ 28 read PMTs.

The beam monitor system comprises three FPGAs, using ORs and
ANDs to analyze that both ends of one of the 16 hodoscope paddles 
fire in either the front or rear hodoscope planes, or that both ends
of one of the 4 large flanking paddles fire.

The random system will be a logic signal generated from noise in a PMT
away from the beam line. 
The rate will be kept low by setting an appropriate discriminator threshold.

The level 2 master trigger then generates triggers, including the
following\footnote{ We indicate a logical NOT with the symbol ``!''
  and a logical OR with the symbol ``$\vee$''.}:
\begin{itemize}
\item {\it Scattered electron:} Beam PID system indicates $e$, !$\mu$,
  !$\pi$; !VETO; Scattered particle system indicates  TOPLEFT $\vee$
  TOPRIGHT $\vee$ BOTTOMLEFT $\vee$ BOTTOMRIGHT; beam monitor ignored,
  random ignored, NIM ignored.
\item  {\it Scattered muon:} Beam PID system indicates $\mu$, !$e$,
  !$\pi$; !VETO; Scattered particle system indicates  TOPLEFT $\vee$
  TOPRIGHT $\vee$ BOTTOMLEFT $\vee$ BOTTOMRIGHT; beam monitor ignored,
  random ignored, NIM ignored.
\item {\it Scattered pion:} Beam PID system indicates $\pi$, !$e$,
  !$\mu$; !VETO; Scattered particle system indicates  TOPLEFT $\vee$
  TOPRIGHT $\vee$ BOTTOMLEFT $\vee$ BOTTOMRIGHT; beam monitor ignored,
  random ignored, NIM ignored.
\item {\it VETO:} Beam PID system ignored; VETO; 
  Scattered particle system ignored; beam monitor ignored,
  random ignored, NIM ignored.
\item {\it Beam particle:} Beam PID system indicates $e$ $\vee$$\mu$
  $\vee$ $\pi$; !VETO; Scattered particle system ignored; beam monitor TRUE,
  random ignored, NIM ignored.
\item {\it Random:} Beam PID system ignored; VETO ignored; 
  Scattered particle system ignored; beam monitor ignored,
  random TRUE, NIM ignored. 
\item {\it NIM:} Beam PID system ignored; VETO ignored; 
  Scattered particle system ignored; beam monitor ignored,
  random ignored, NIM TRUE.
\end{itemize}

\paragraph{Trigger system current status}

Trigger programming has been ongoing and all needed functions of the
trigger have been demonstrated in prototypes.

To test the veto, scattered particle, and beam monitor systems, we
programmed initially OR and AND logic for testing.
Different operations were encoded for different input channels to test
the various options needed.
The code generated logic pulses of widths 20, 30, and 40~ns from the
inputs, and performed the same logic on all pulse widths.
This is a standard technique to provide information about trigger
efficiency when signal times or widths can vary, and we plan to use
this technique during the experiment.
We found that the TRB3 FPGA generates these logic outputs within 
$\approx$40~ns.

A much harder task is the TDC function, using the RF clock to
determine particle identification. 
The TDC was encoded with similar, but lower resolution, techniques to
the TRB3 high precision TDC code, to speed up processing.
This still requires detailed placement of the logic gates and logic 
routing through the FPGA.
We coded for 48 input channels, with the beam RF time divided into
32 0.6-ns wide TDC bins.\footnote{With 0.6-ns wide bins and 
measured beam and detector time resolutions, 
particle identification gates in the trigger
can be set on the RF time that are $\approx$99.99\% efficient for 
both accepting electrons and muons and rejecting pions. 
Actual performance will depend on the bin widths in the time region 
between the peaks.}
By varying the phase of the input signal relative to the clock, we
determined that the bins width variation is sufficiently small, with
bin widths ranging from about 0.3 - 1.0 ns.
The phase determined is compared to separate $e$, $\mu$, and 
$\pi$ arrays to determine the particle type - this allows a single
particle in a detector to be identified as multiple types in that
detector, if the two types are close enough in time.
We found that the TRB3 FPGA generates these TDC outputs within
$\approx$80~ns.

\paragraph{Path to completion}

The TDC trigger code will be beam tested in the June 2017 beam time.
Two main types of work remain to develop the trigger.
The first, more straightforward, development is to adjust the generic code
we have developed to the specific input logic combinations needed
for the various level 1 triggers and the level 2 trigger.
The second, less straightforward, development is to understand
the TRBnet code implemented in the fast TDCs, and include it in
our FPGA codes. Implementing TRBnet will allow monitoring and 
controlling - for example, adjusting the TDC bins that correspond
to each particle type - the level 1 triggers through the network,
rather than through the JTAG connectors on the boards.

\subsection{Software Systems}
\label{sec:daqsystem}

MUSE has been using and will continue to use the PSI MIDAS system for data
acquisition.
MIDAS supports the slow controls and standard data acquisition
modules, and we have extended it to successfully read out all
electronics used in the experiment: the GEM electronics, the TRB3s,
and the Mesytec QDCs. 
The read out system has been incrementally extended as we have obtained 
additional modules and crates to read out, with ongoing performance tests.
MIDAS has been operated at rates up to 2.4 kHz, with the limit
determined by the number (0, 1, or 2) of GEM telescopes we read out, 
since the GEMs have been the slowest component, requiring nearly 
(1~ms / GEM telescope) in our original implementation.
Block readout times for other electronics are $\approx$0.1 ms, but
were not implemented in the original rate tests.

The DESY OLYMPUS analysis framework has been adopted into the experiment
as ``MUSECOOKER,'' for data analysis.
The analysis also uses  CERN ROOT and GENFIT, a general wire chamber tracking
program.

With a $\approx$2 kHz trigger rate, data rates are about 12 MB/s, 
with nearly all of the raw data coming from the GEMs.
We have about 30 TB of local storage for experimental data,
sufficient for about 1 month of full rate raw data,
along with 90 TB of long term archival storage being arranged
through PSI. 
We plan to reduce long term storage needs by zero-suppressing 
the GEM data, rather than retaining the full set of ADC values for each GEM plane.

\section{Tests, Commissioning, Calibrations, Runs}
\label{sec:commissioning}
\subsection{Installation and Commissioning}
\label{startcommissioning}

The MUSE equipment is currently planned to be partially installed in
late 2017 for a dress rehearsal run at the end of the year, with full 
installation completed in winter/spring 2018 for production data
taking starting mid 2018.
The 2017 dress rehearsal run will include most of the beam line
detectors and one spectrometer arm.
The beam line detectors for 2017 will be the first 2 of the 4 beam hodoscope
planes, the GEM telescope, the veto detector, and the central
hodoscope of the beam monitor. 
Also, only a solid target will be used in 2017.

Installation and commissioning of the detectors will follow typical procedures.
Detector installation first requires assembly and installation of the 
corresponding support structures, the beam line detector table,
the scattered particle frames, and the beam monitor frame. 
Also, the appropriate electronics, high and low voltage supplies, and
gas systems must be available.
From late summer to early fall 2017 we expect to install and survey
the beam line hodoscope planes, GEM chambers, and veto detector 
on the beam line detector table,
the left front STT on the beam line detector table,
the left scintillators walls onto their support frame,
and the central beam monitor hodoscope onto its frame.

Approximate operating conditions are known for all of these 
detectors from previous experience, including prototyping and testing
after construction.
Initial detector checkout / commissioning will be done with cosmic
rays and sources, and ultimately with beam.
Techniques are standard. 

\subsection{Calibrations}

MUSE requires several nonstandard calibration procedures, which we
address here.

A special mount has been built that allows the GEM telescope
to be tilted at various precisely known angles relative to the beam.
We can use these data sets provide a check of the consistency of the 
internal GEM coordinate system, since the beam is not moving.

The STT wire spacing is determined with a precision moving collimated 
source in a special calibration setup.
Also, to limit experimental systematic uncertainties, we require knowledge
of measured scattering angle offsets at the sub-mr level. 
The experimental table is designed so that the STT can be rotated into the beam 
behind the GEMs, into several precisely determined angle settings.
These measurements allow the STT wire
spacing, position, and orientation to be checked, and also
determine the STT orientation relative to the beam in the data
taking setting. 
This can be achieved since the GEMs determine angles to
$\approx$ 1 mr (100$\mu$ spatial resolution over 16.8 cm) 
the STTs determine angles to 
$\approx$ 5 mr (150$\mu$ spatial resolution over 4 cm) for one STT
or $\approx$ 1 mr for the STT pair,
and multiple scattering can be limited to 2-3 mr at higher channel
momenta for alignment determination.

The GEMs and STT combined can be used to obtain data for
multiple scattering from target and detector elements to verify
simulations.
We leave the STT in the beam line and use a beam trigger, but we
add various materials between the GEMs and the STT. 
Note that the GEM telescope + STT setup verifies calculated
multiple scattering in the GEMs, and the GEM telescope with
2 STTs verifies multiple scattering in the STTs.
Of course, each of the measurements convolutes resolutions
and multiple scattering.

The level 1 beam PID trigger needs to be calibrated for each momentum.
If system electronics and cables and the accelerator setup are kept
constant, the electron RF time is constant, whereas the pion and muon 
times vary with momentum.
The beam PID will be calibrated with a beam trigger at a low beam flux 
of a several kHz, where the accidental rate is negligible. 
Once the trigger PID cuts are determined, the trigger will be operated 
with increasing beam flux, up to the planned 3.3 MHz flux, to verify
that its efficiency for accepting events of interest and for rejecting
background events, as well as the effects of accidentals, are understood.
Note that the scattered particle trigger requires additional study.
The timing of the scattered particle trigger needs to be determined,
and the prescale factors for the various trigger types need to be
adjusted to optimize in particular the muon scattering uncertainties.

The table and beam monitor support are designed so that the 
beam hodoscope and beam monitor positions along the beam line 
can be varied by precisely known distances. 
Multiple measurements at different positions allows the muon and
pion momenta to be determined.\footnote{We have obtained 0.2\% --
  0.3\% measurements with 50-cm distance changes in PiM1. We have
designed for 2-m changes in the experiment.}
We also use the downstream beam line to analyze the relative momentum
of beam particles, by turning off the quadrupoles after the second
dipole, and comparing positions of pions, muons, and electrons
at the target. 
This measurement is performed for multiple positions of a narrow 
collimator at the channel intermediate focus, where the beam
dispersion is 7 cm / \%.
TURTLE calculations indicate that the dispersion at the target 
is about the same, providing a $\approx$10$^{-4}$ measurement 
of the relative momentum.\footnote{This technique is being tested 
in the June 2017 test run.}
The same measurement also gives information on energy loss in
materials placed at the IFP, for validating simulations.

\subsection{Dress Rehearsal Run Plan}

After the commissioning and calibrations are completed, 
the dress rehearsal run can start. 
The aim of a dress rehearsal run is to take a $\approx$10\%
sample of the intended full data set in near production data
conditions, to be subjected to analysis so that any issues
can be identified and corrected before starting the
production data taking.

The dress rehearsal and production data taking run plan for an individual
beam momentum are similar. In production data taking we will:
\begin{enumerate}
\item Determine the beam momentum.
\item Calibrate the trigger and determine its performance. Optimize
prescale factors for different trigger types.
\item Take data in the no-target setting to measure pion and muon decay
and other backgrounds.
\item Take data with solid calibration target to verify position
  reconstruction.
\item Cycle through full hydrogen cell $+$ empty cell settings to
  determine signal $+$ backgrounds and backgrounds.
\item Repeat for opposite polarity.
\end{enumerate}
The cryotarget will not be ready in time for the dress rehearsal run,
so for \#5 we will instead measure with solid CH$_2$ and carbon targets.

\subsection{Run Plan}
\label{sec:runplan}

Based on estimated beam fluxes, cross sections, and efficiencies,
the experiment requires 12 months of primary production beam time
for a 5$\sigma$ measurement of whether the proton radius is the
smaller muonic hydrogen or larger electron value.
To date we have assumed an even distribution of 2 months beam time 
for each of the 6 momentum / beam polarity combinations.
We have not tried to optimize this division since the optimization 
depends in part on actual system performance and in part on physics
motivation. 
Also, the systematic and statistical uncertainties are roughly
matched, reducing the effects of redistributing time.
Since the muon scattering rates are higher for positive charge than
negative, but lower in each case than the electron scattering rates,
the division of time between the beam charges can improve the
statistical uncertainties for the radius extraction by spending more
time at positive polarity, or can improve the statistics for the
two-photon measurement by spending more time at negative polarity.

Figure~\ref{fig:elmurates} in appendix~\ref{sec:physreacback} shows
that the ratio of hydrogen elastic scattering to carbon elastic
scattering ranges from a factor of 3 - 4 at small angles to an order
of magnitude or more at large angles. Muon decay backgrounds
are smaller than elastic scattering by an order of magnitude or more,
once timing cuts are applied to the data. 
For each kinematic setting the time needs to be split between
measuring the signal plus backgrounds with target full , and
measuring the backgrounds with target empty.
Figure~\ref{fig:bgsubunc} shows that the final statistical uncertainty
on the background subtracted cross sections is not very sensitive 
to the division of time between the two measurements.
In our case it appears the signal to background varies with angle
from about 3 - 10, and choosing to spend about 75\% of the time
on signal plus background will not cause the background subtracted
statistical uncertainty at any angle to be more than a few percent
worse than it would be if the division of time were optimized for that
angle. 
Since changing between full and empty targets takes only a few
minutes, we expect to cycle from target full to empty
and back every several hours. 

\begin{figure}[h]
\centerline{\includegraphics[width=0.4\textwidth]{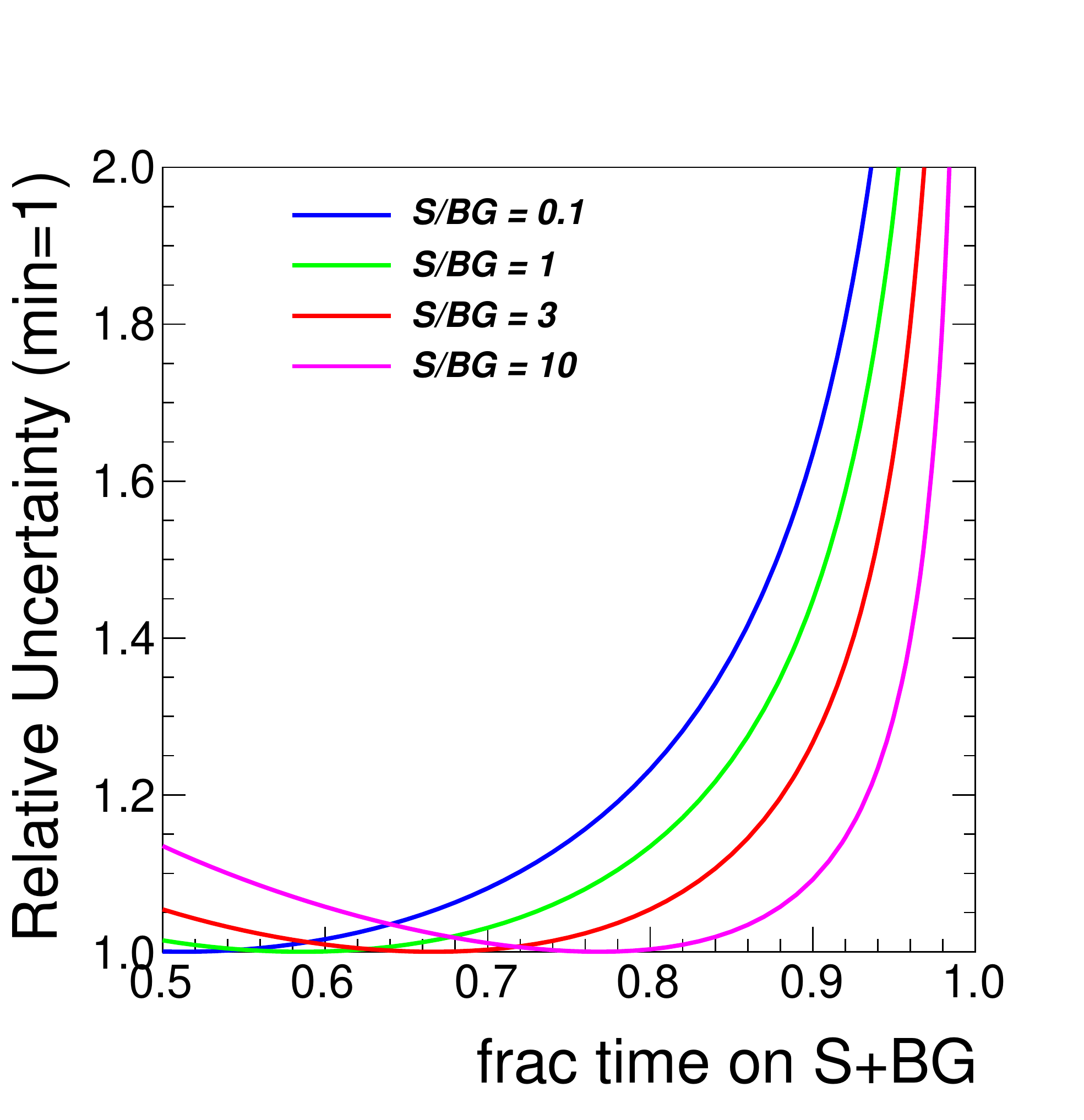}}
\caption{The relative statistical uncertainty for the background
  subtracted cross section as a function of the fraction of time
  spent measuring the signal $+$ background - the remaining
  fraction of time would be spent measuring the background.}
\label{fig:bgsubunc}
\end{figure}

To avoid being affected by long term issues, and as a check of
stability, it is desirable to regularly change between momentum
settings. 
However, changing momenta runs into two potential issues.
First, momentum settings may not be reproducible with sufficient
accuracy. 
We have established that if we cycle the
channel, setting the channel momentum to at least 300 MeV/$c$ until
the Hall probe stabilizes before reducing the momentum to the desired
value, then the RF time distribution and apparently the beam spot
are reproduced, but we have not proven this with sufficient precision
to consider it routine;
we also have not tested this for channel polarity changes.
Second, we plan to optimize the data by adjusting the number of
beam hodoscope planes from 2 at 115 MeV/$c$ to
3 at 153 MeV/$c$ to 4 at 210 MeV/$c$.
This requires working on the detector, near the cryotarget,
to remove or reinstall beam hodoscope planes. 
We prefer to avoid this activity due to the potential safety
hazard of working near the cryotarget and due to a desire to
minimize handling the detector, with the consequent risk of
damage to some equipment.
We cannot commit to more frequent momentum changes until we have
more confidence in the procedure.

Once calibrations are finished and normal production 
operations are established, we expect up to a few target changes daily
and momentum / polarity much less frequently.

We are considering two further checks on systematics, and our ability
to extract the same form factors from different settings,
beyond the three primary momentum settings,
two identical spectrometers, and two beam polarities.
We can obtain data with the straw tube trackers slightly rotated in
angle, rather than symmetric, and run some time
at a beam momentum off by a few MeV/$c$ from one of our 3 planned
settings. Each of these would require about half of the time of one of
the already planned settings.

The detectors are built to have high efficiency, and are monitored
through redundant planes, scaler rates, efficiency triggers, pulse size spectra,
and slow controls.
We expect to not need special efficiency calibration runs, after the
initial setup.
We also note that random coincidence beam particles provide an 
unbiased measurement of beam parameters and beam line detector 
performance and stability.

\subsection{Personnel for Production Runs}

For MUSE production running, we plan to run single person shifts.
We plan to have 3 - 4 experts -- MUSE postdocs and Ph.D.\ students -- 
on site during production data taking, to act as experts 
on call or in case a second person is needed for beam area access.
The experts also need to be at PSI for additional
time for checkout of the apparatus before production data taking
starts.
This plan leads to about 1400 expert and 1200 non-expert 
shifts for the two years of production data.
Table~\ref{tab:shiftcommit}  shows the shift commitments of
core MUSE institutions involved in the construction project
that were presented to the PSI BV48 meeting in February 2017.
We are slightly under-committed for non-expert shifts, but strongly
over-committed for expert shifts.

\begin{table}[h]
\caption{\label{tab:shiftcommit} 
MUSE shift commitments.
}
\begin{tabular}{|l|r|r|}
\hline
Institution & Expert shifts & non-expert shifts  \\
\hline
GW          & 360 & 168 \\
Hampton & 120 & 56 \\
Hebrew   & 180 & 90 \\
Michigan & 308 & 96  \\
Montgomery & 0 & 112 \\
Rutgers   & 720 & 112 \\
South Carolina & 120 & 196 \\
Tel Aviv  & 0 & 196 \\
Temple   & 360 & 90 \\
\hline
TOTAL & 2168 & 1116 \\
\hline
\end{tabular}
\end{table}

\section{Analysis, Corrections, Systematics, Results}
\subsection{Data Analysis}
\label{sec:dataanalysis}

\subsubsection{Determination of Yields}
\label{sec:getyields}

Here we present various standard steps in the data analysis leading to the
determination of yields.
The event-data analysis inputs are QDC and TDC
signals from the detectors along with trigger information.
\begin{itemize}
\item QDC spectra will be monitored to check for stability of
detector gains, threshold setting, and consistency with simulations.
\item Timing of fast scintillators will be improved with QDC walk
  corrections. 
All paddles (except veto paddles) have double-ended readout, 
so mean times will be determined.
\item A raw time of flight is calculated from the beam hodoscope
  to the scattered particle scintillators. We average over multiple
  detector planes.
\item GEM hit positions are determined from GEM clusters.
\item Straw drift times are converted to drift distances.
\item We use GENFIT for both GEMs and straws to determine 
  tracks and  residuals / resolutions. Tracking also determines 
  efficiencies, independently from hit spectra.
\item The GEM track is compared to the beam hodoscope hit
  position for consistency. We use the GEM track projected to the 
  target to put a fiducial cut on the target to decrease the number
  of background, side-wall events.
\item The straw track is compared to the scintillator paddle hit for consistency.
\item The GEM and straw tracks together determine 
  an interaction vertex, the quality of the reconstructed vertex,
  and the scattering and azimuthal angles.
  Simulations indicate few mm position resolutions, growing to the cm
  level for $z$ position in cases of forward angle scattering, and
  typically 10 -- 15 mr angle resolution, due to multiple scattering.
  We use a loose $z_{target}$ cut as we prefer to remove background
  through full cell / dummy cell subtractions.
  See Fig.~\ref{fig:ztarget210}.
\item GEM and straw tracks determine a path length between the beam
  hodoscope and scattered particle scintillators, leading to a
  corrected time of flight and $\beta$.
\item The RF time and the corrected  time of flight or $\beta$
determine the the reaction type, scattering or decay of
some particle. We have also found in Monte Carlo studies that a neural 
network analysis approach, discussed below, is very efficient.
\item Fiducial cuts are applied to the straw chambers, so that we do
not use particles near the edge of the acceptance to determine
cross sections.
\item The data will be analyzed for accidental coincidences with other
 beam particles. Additional beam particles that might interfere with
an unambiguous analysis will cause the event to be thrown out.
Clearly additional beam particles provide unbiased measurements
of beam properties (trajectories in the target, and perhaps momentum 
from time fo flight from beam hodoscope to beam monitor).
\end{itemize}

\begin{figure}
\centerline{\includegraphics[height=3.5in]{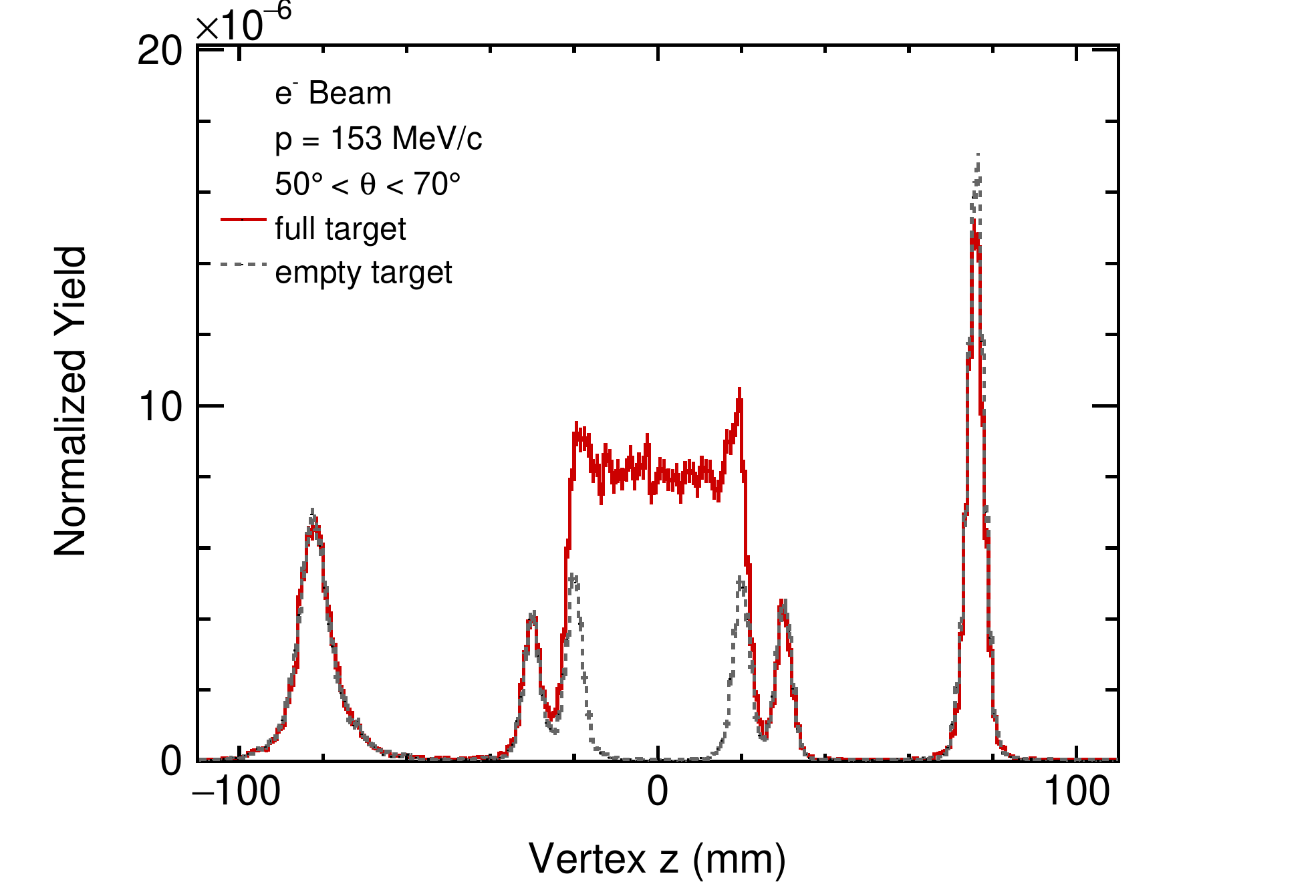}
}
\caption{Geant4 simulation of reconstructed interaction position along the beam line
for electrons, with a full target (solid) and an empty target (dotted). The resolution along the beam line is
a few mm; see also Fig.~\ref{fig:vertex}.  The structures in the
empty-target distribution are (from the outside to the inside):
scattering-chamber windows, super insulation, and target entrance/exit
walls.
}
\label{fig:ztarget210}
\end{figure}

\subsubsection{ Backgrounds}

All reactions other than elastic scattering of electrons and muons
from protons are background reactions.
There is only a small probability that beam particles scatter or decay
upstream and produce hits in the scattered particle detectors without
giving hits in the beam line detectors -- which would lead to throwing
out the event dur to multiple beam particles.
As a result, we only consider reactions possibly leading to triggers
that might be mistaken for electron or muon elastic scattering.

Pion events are removed by RF time in the beam line detectors.
These events would also have a long corrected time of flight / 
small $\beta$.
We do not consider them further.

\begin{figure}[ht]
\centerline{\includegraphics[width=\linewidth]{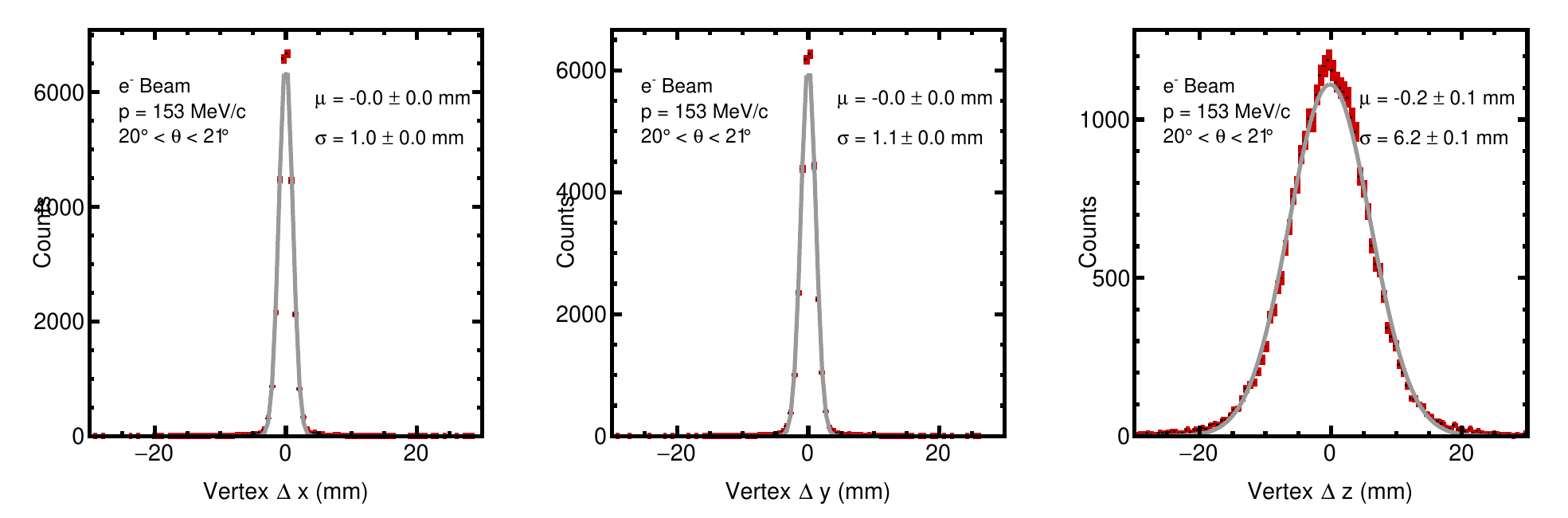}}
\centerline{\includegraphics[width=\linewidth]{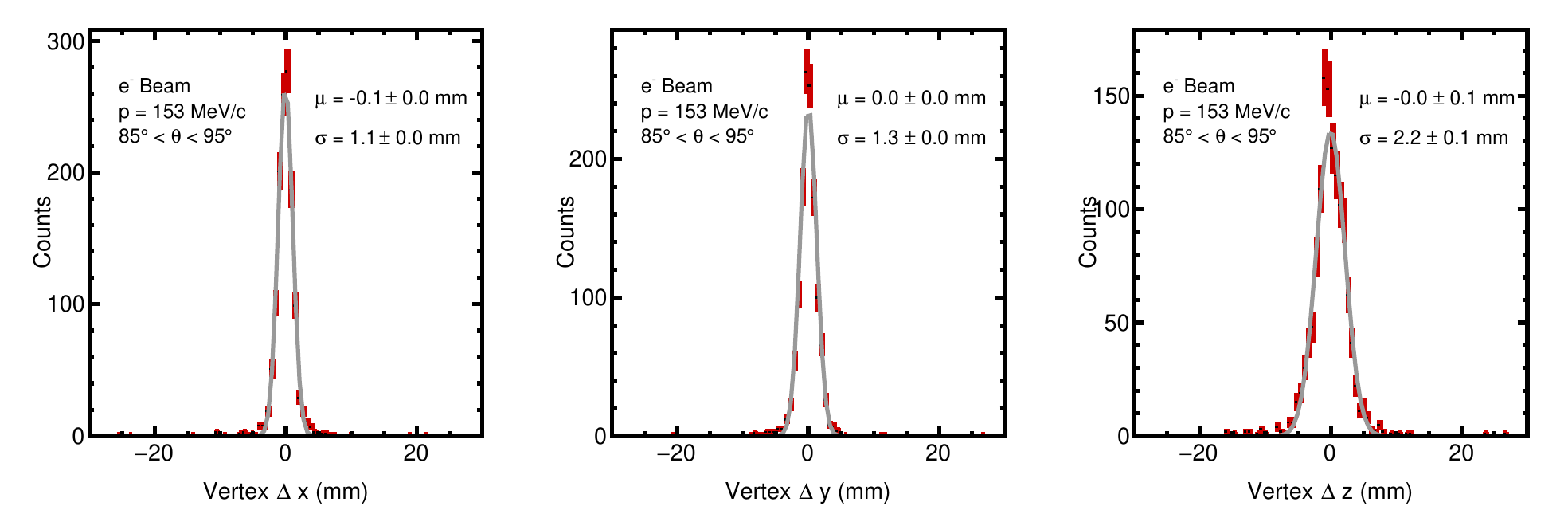}}
\caption{Examples of differences between reconstructed and actual
  scattering reaction vertex from simulations for electrons at
  forward angles (upper panels) and perpendicular to the beam line
  (lower panels) and a beam momentum of 153~MeV/$c$.}
\label{fig:vertex}
\end{figure}

Figure~\ref{fig:ztarget210} shows a Geant4
simulated reconstructed image of the target along the beam line.
The simulation included effects of the various detector and
target materials on the particles. 
The resolution of the vertex $z$ coordinate varies as a function of angle due to a
$1/\sin\theta$ geometric effect.
The reaction vertex is determined by the position of the closest
approach between the incident-particle track measured by the GEM chambers
and the scattered-particle track measured by the straw chambers.
The good reconstruction with few mm resolution is evident, which
allows rejection of all events that do not result from scattering or
decay in the immediate vicinty of the target cell; see also Fig.~\ref{fig:vertex}.
We use the reconstructed vertex information to remove
scattering backgrounds from vacuum chamber windows, which are far from
the target.
Scattering from the target cell walls must be subtracted.
Monte Carlo simulations show that 
LH$_2$ cell - empty cell subtractions work well, but care must be taken
for the downstream windows, due to the slight loss of flux, energy, and change
in beam position distribution caused by interactions in the 
liquid hydrogen.

The shape of the target leads to an increasing ratio of wall events to
liquid hydrogen events for trajectories further from the beam axis.
We apply target fiducial cuts using the GEM detector tracks to reduce
scattering from the sides of the target cell.

M\o{}ller and Bhabha scattering generate a low-energy forward-angle background.
The recoil electrons with enough energy (10 -- 20 MeV) to trigger our system all go
to smaller angles than the detectors, but can multiple scatter out
into the detectors and generate events that look like forward angle scattering,
$<$25$^{\circ}$.
The beam monitor is an efficient veto detector for these M\o{}ller
scattering background events in the offline analysis, as they have a forward going ``high''
momentum beam particle that continues into the high-precision
beam scintillators after the target.

\label{sec:sim_muon_decay}
Muon decay with the electron or positron detected has similar kinematics to muon
scattering, but a faster outgoing particle.
We have studied the muon-decay-in-flight background with a trained
neural network based on data from our Geant4 simulation of the full
detector setup.  The data included five parameters: time-of-flight
(SiPM to front TOF SC wall and SiPM to rear TOF SC wall) information,
energy deposition (in front- and rear SC bars), and front-to-rear SC
bar correlations.  We assumed overall 100~ps, 92~ps, and 89~ps timing
resolutions for the system at 115, 153, and 210~MeV/$c$, respectively,
as we plan to use more beam hodoscope planes at the higher momenta.
Results for the muon scattering signal and for the muon-decay
background for these observables are shown in Fig.~\ref{fig:nn_obs}
for a beam momentum of 153~MeV/$c$.
\begin{figure}[ht]
\centerline{
\includegraphics[width=3.1in]{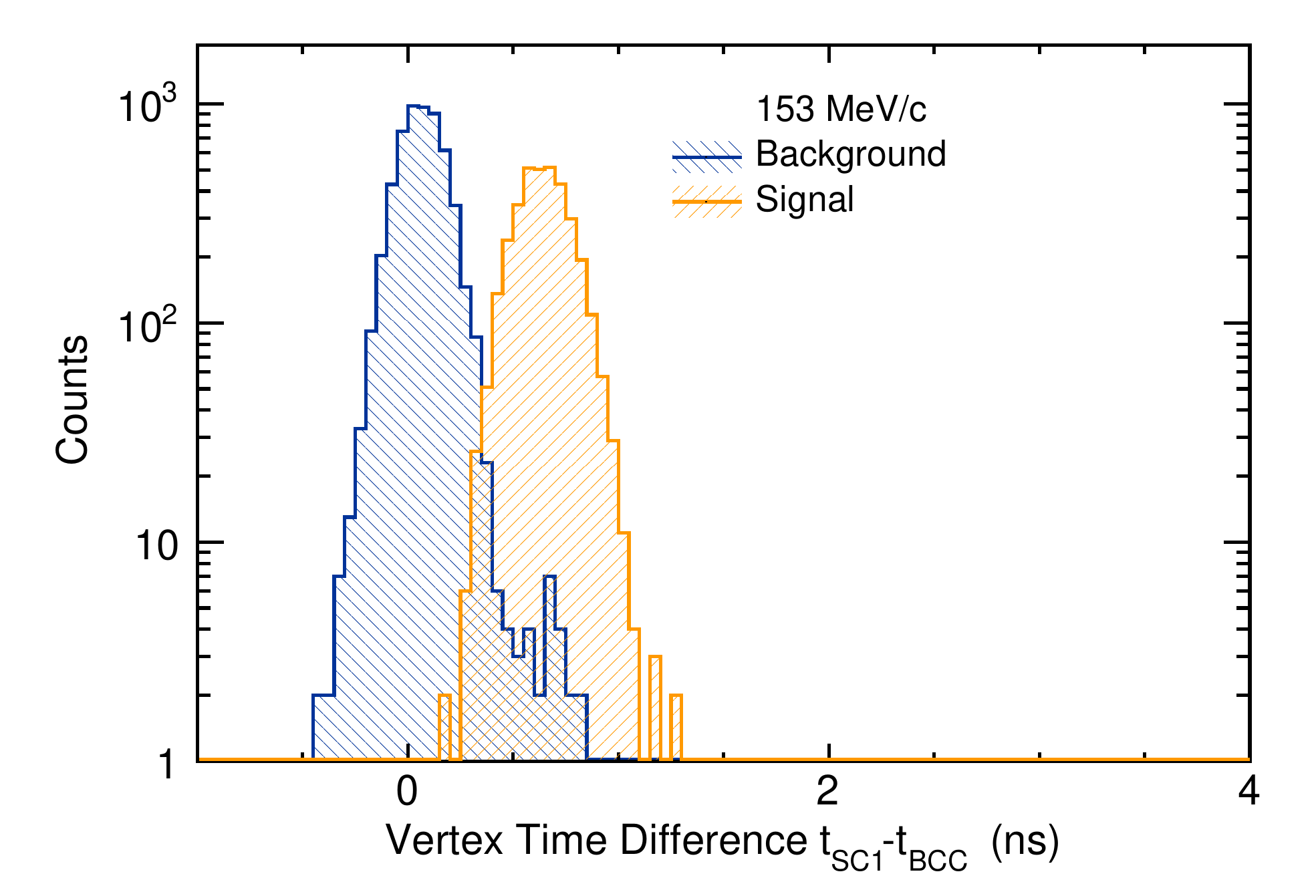}\hfill
\includegraphics[width=3.1in]{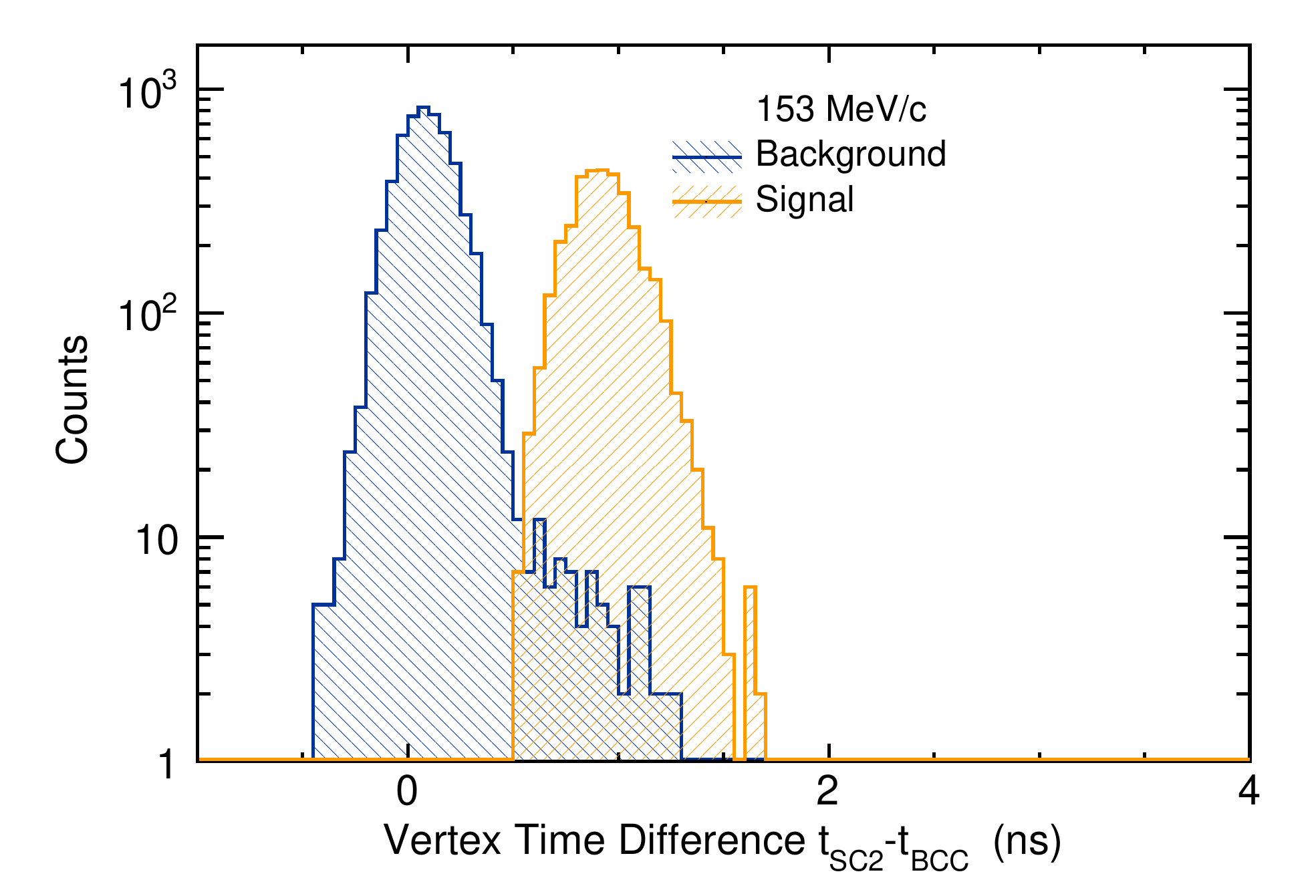}
}
\centerline{
\includegraphics[width=3.1in]{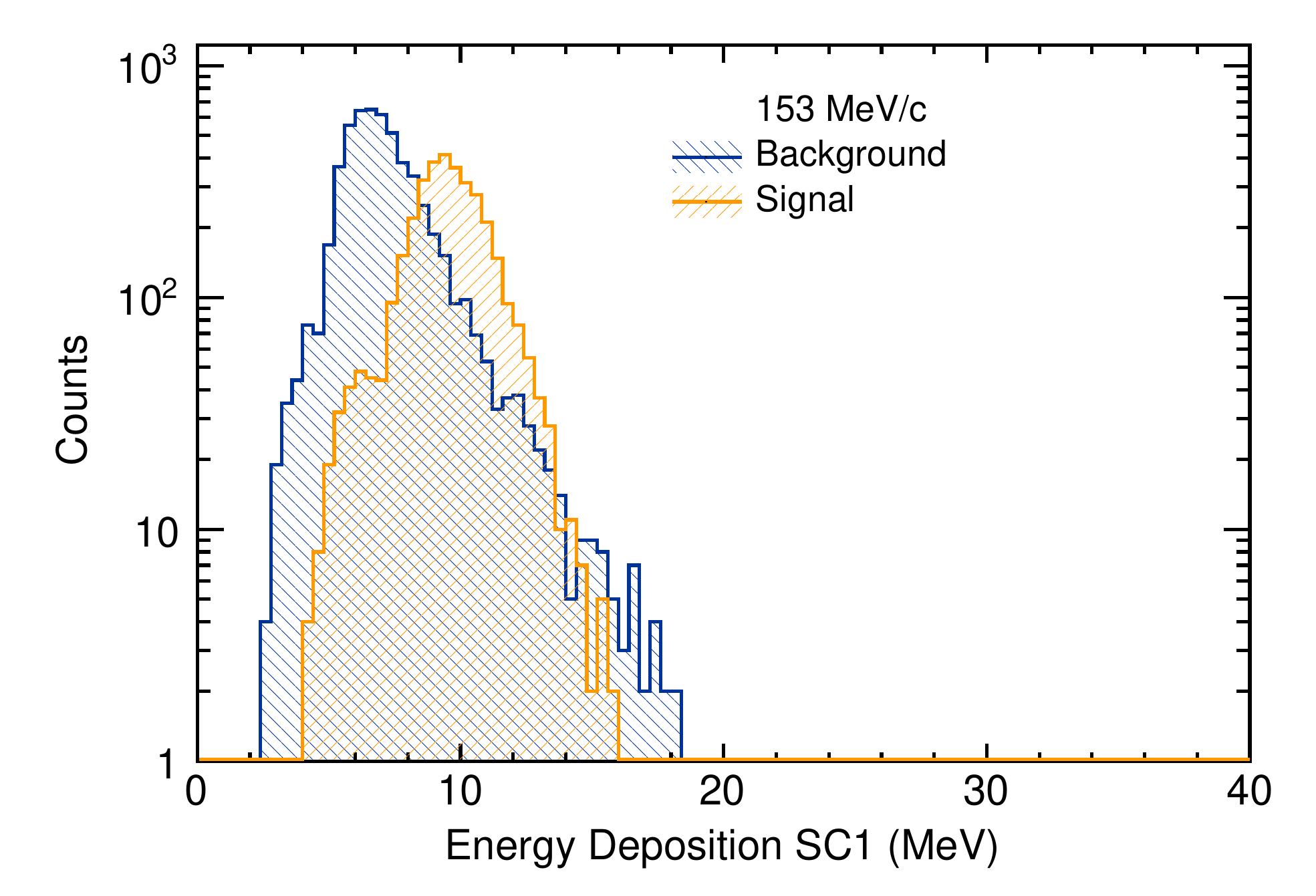}\hfill
\includegraphics[width=3.1in]{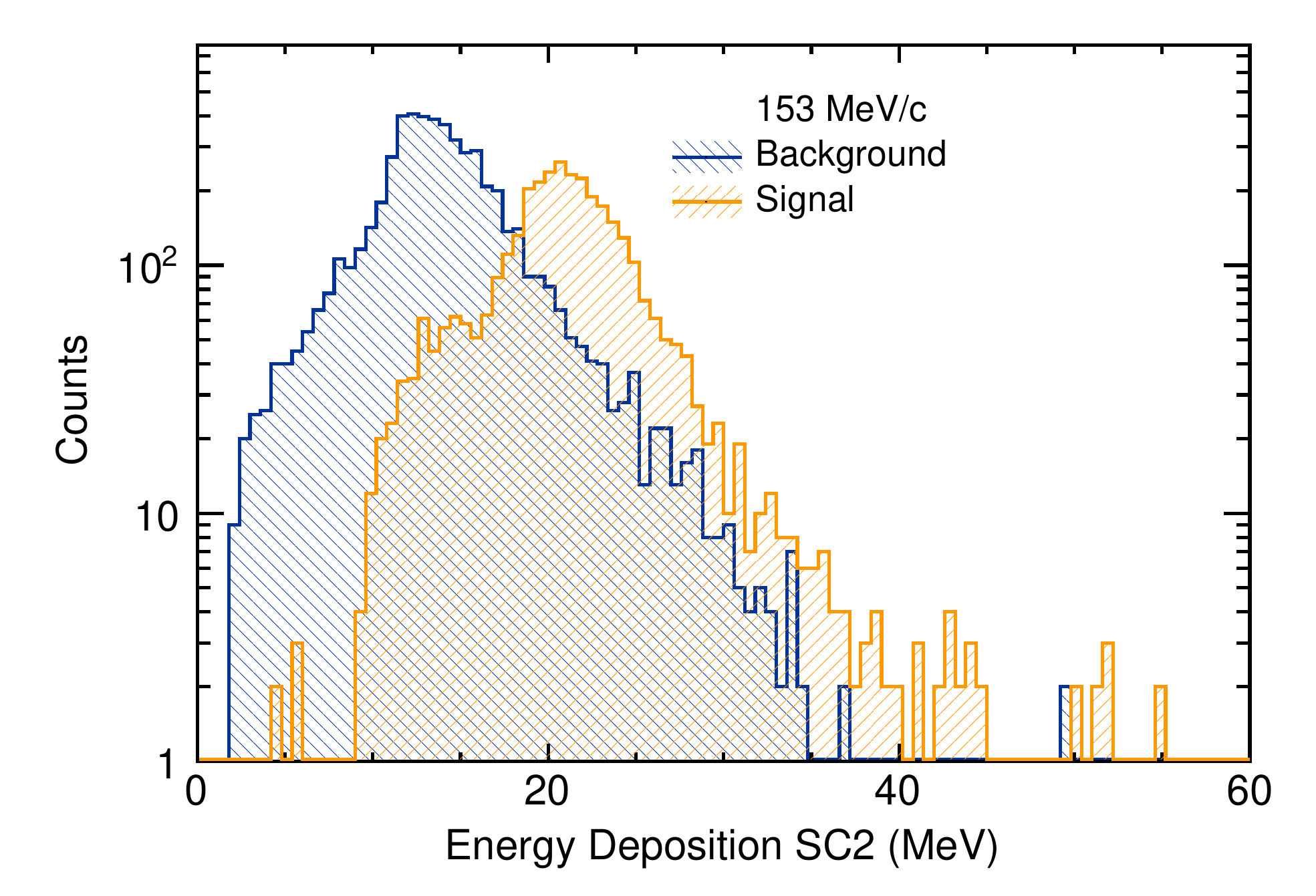}
}
\includegraphics[width=3.1in]{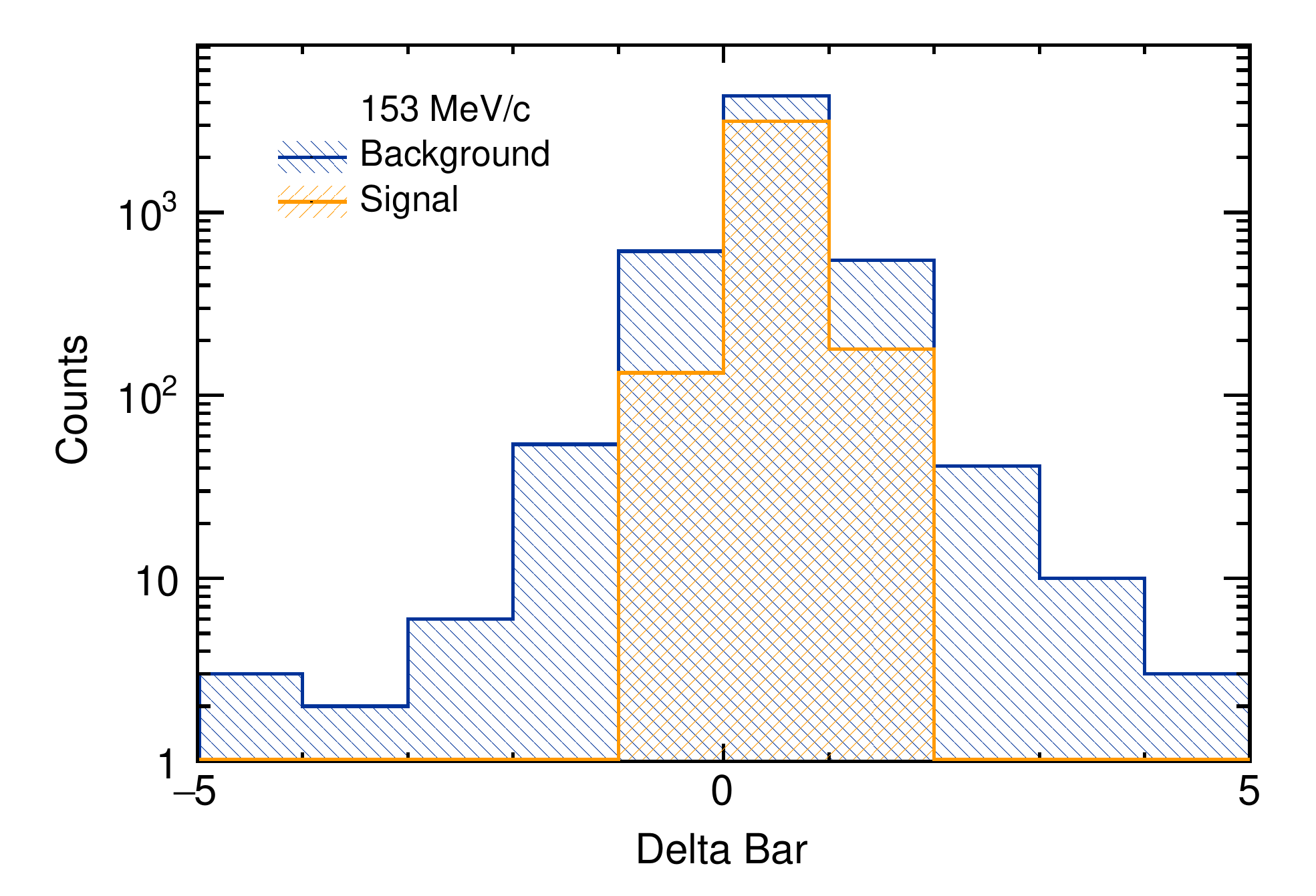}\hfill

\caption{Simulated timing, energy, and detector correlations for
  muon-scattering signal and muon-decay background and a beam momentum
of 153~MeV/$c$.}
\label{fig:nn_obs}
\end{figure}
Particularly the simultaneous use of time-of-flight information from
both scintillator planes proved crucial for a good discrimination of
the muon-scattering signal from the decay background.  Results for 115
and 210 MeV/$c$ momenta have also been obtained, but are not shown
here in detail.  

The output of the trained neural network for an event is related to
the probability of the event being signal or background.  Depending on
the cut on the output variable larger and larger fractions of signal
events can be selected at the expense of also accepting larger
fractions of background events.  This is shown in
Fig.~\ref{fig:nn_result}.
\begin{figure}[ht]
\includegraphics[width=3.5in]{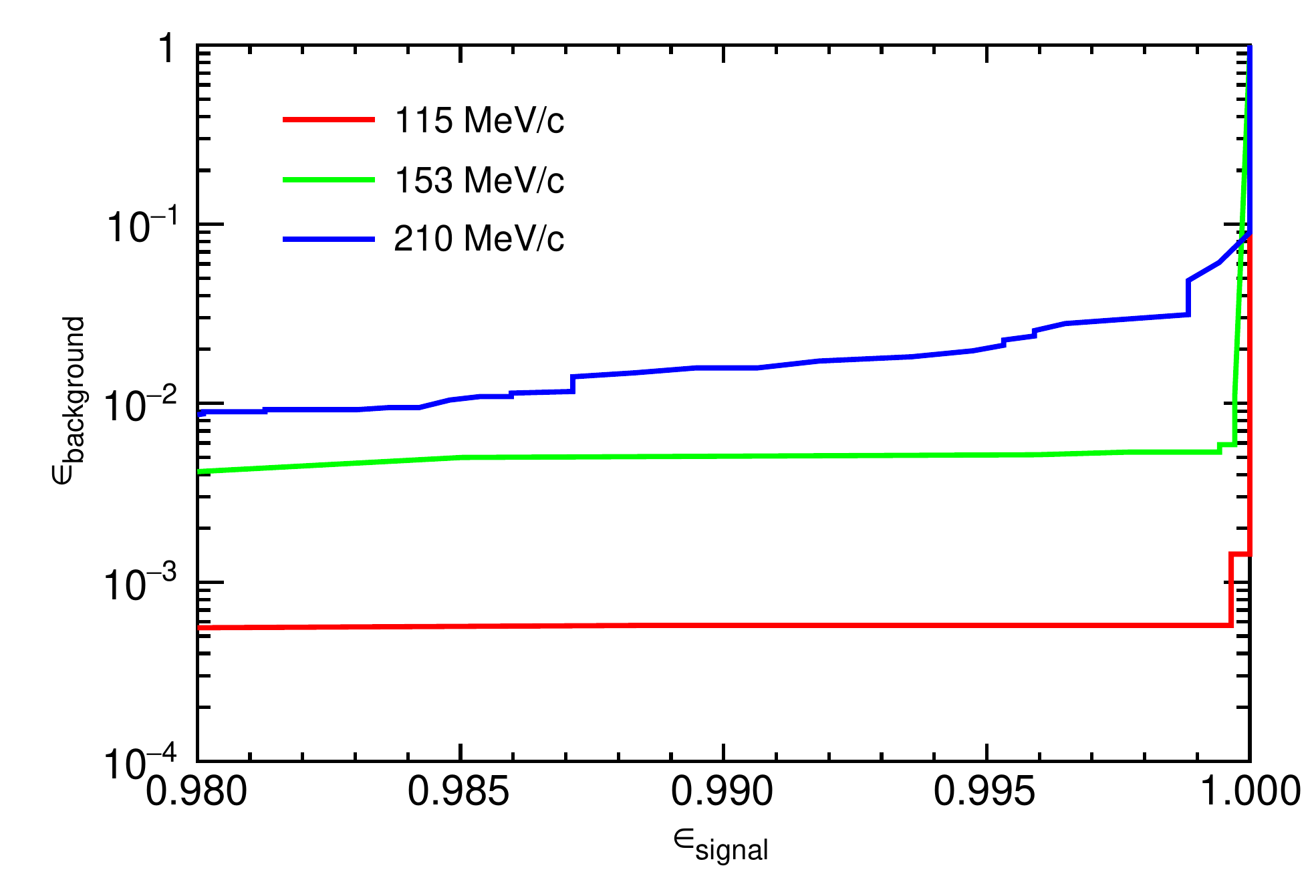}

\caption{Neural-network result for the discrimination of
  muon-scattering signal from muon-decay background at beam momenta of
  115, 153, and 210~MeV/$c$.  The red curve for example indicates that
  one can obtain near 100\% efficiency identifying scattering
  particles while only misidentifying $<0.1$\% of decay particles.  In
  the worst case, at 210 MeV/$c$, one can reduce the decay events by
  99\% while retaining over 98\% of the scattering events. There is a
  large relative uncertainty in the 115 MeV/$c$ and 153 MeV/$c$
  results due to a lack of statistics because the remaining background
  after the neural net cut is very small.}
\label{fig:nn_result}
\end{figure}
As expected, the separation of peak and background
events is much better at the lower 115~MeV/$c$ beam momentum and worse
for 210~MeV/$c$. We estimate that a close to 100\% signal efficiency can be obtained with less
than 0.1\% background for the 115 MeV/$c$ setting and less than 0.5\%
background for the 153 MeV/$c$ settings.  The time-of-flight
distributions and pulse-height distributions of the simulation can be
carefully validated in these settings.  For the 210 MeV/$c$ setting we
estimate that an acceptance of 98\% of the signal would include 1\% of
the background events.  These effects can be corrected for with data
from the simulation (after validation) and/or with background
measurements.

\subsubsection{Solid Angle}

Determination of the cross section from the yields requires knowing
the solid angle. 
The MUSE beam size is sufficiently large that a realistic beam profile 
needs to be taken into account in calculating the solid angle.
We have studied the solid angle in Monte Carlo.
Figure~\ref{fig:sc_geom_acc} shows the estimated geometrical acceptance.
We have also studied the effects of offsets in the relative positions
of the target, beam, and detectors on the solid angle - the
systematic uncertainty.
Generally the solid angle is not much changed except at the edges of
the acceptance, and we are most sensitive to offsets along the
beamline, since offsets transverse to the beamline cancel.
These studies have led to a specification that the relative positions
be determined at the 100$\mu$m level.

\begin{figure}[ht]
\centerline{
  %%%   sc_geom_acceptance.C with input file:
  %%%   "/Volumes/Work/g4PSI/results_scwall/proposal13/scwall_mp_210MeV.root";
\includegraphics[width=3.6in]{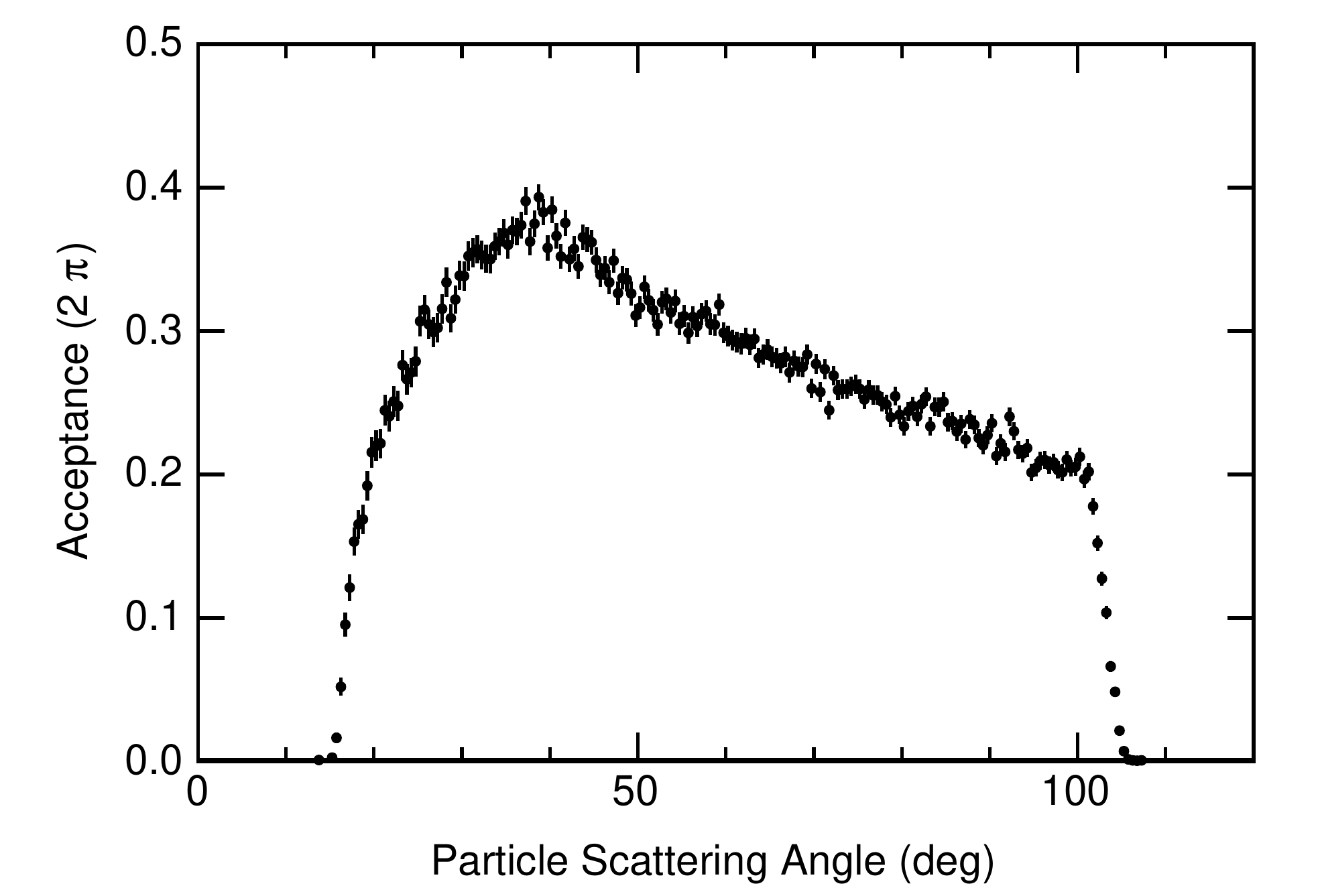}
}
\caption{Estimate of the geometrical acceptance of one scintillator
 wall as the fraction of high-energy muons originating uniformly 
distributed from the
target which hit the straw chambers and front and back scintillator walls.  }
\label{fig:sc_geom_acc}
\end{figure}

\subsubsection{Projected Data with Statistical Uncertainties}
\label{sec:dataprojections}

Using the information given above, it is possible to take the run
plan, make rate estimates for these processes, take into account
various inefficiencies and the increases in uncertainties from
measuring and subtracting backgrounds, and work out the 
resulting statistical uncertainties.
This is shown in Fig.~\ref{fig:simdatastats} for the positive
polarity $\mu^+ p$ scattering and Fig.~\ref{fig:simdatastatsm}
for  $\mu^- p$  scattering.
For both polarities $ep$ uncertainties are a factor of a couple times smaller due to
the greater particle flux and the lack of a decay background to be cut
and subtracted.
Some momenta shown have generally larger uncertainties due to a
smaller $\mu$ flux than at other momenta.
Uncertainties generally increase with $Q^2$ or angle due to the
decreasing cross section.
The statistical uncertainties are below the 1\% level for most of the
data set, but grow to about 2\% in the least precise cases.
Thus the form factor will be determined to better than about 1\%
in all cases.
Negative polarity beams have more electrons and fewer muons,
leading to increased uncertainties for the muons when measuring for
the same integrated luminosity.

\begin{figure}
\centerline{\includegraphics[width=4.5in]{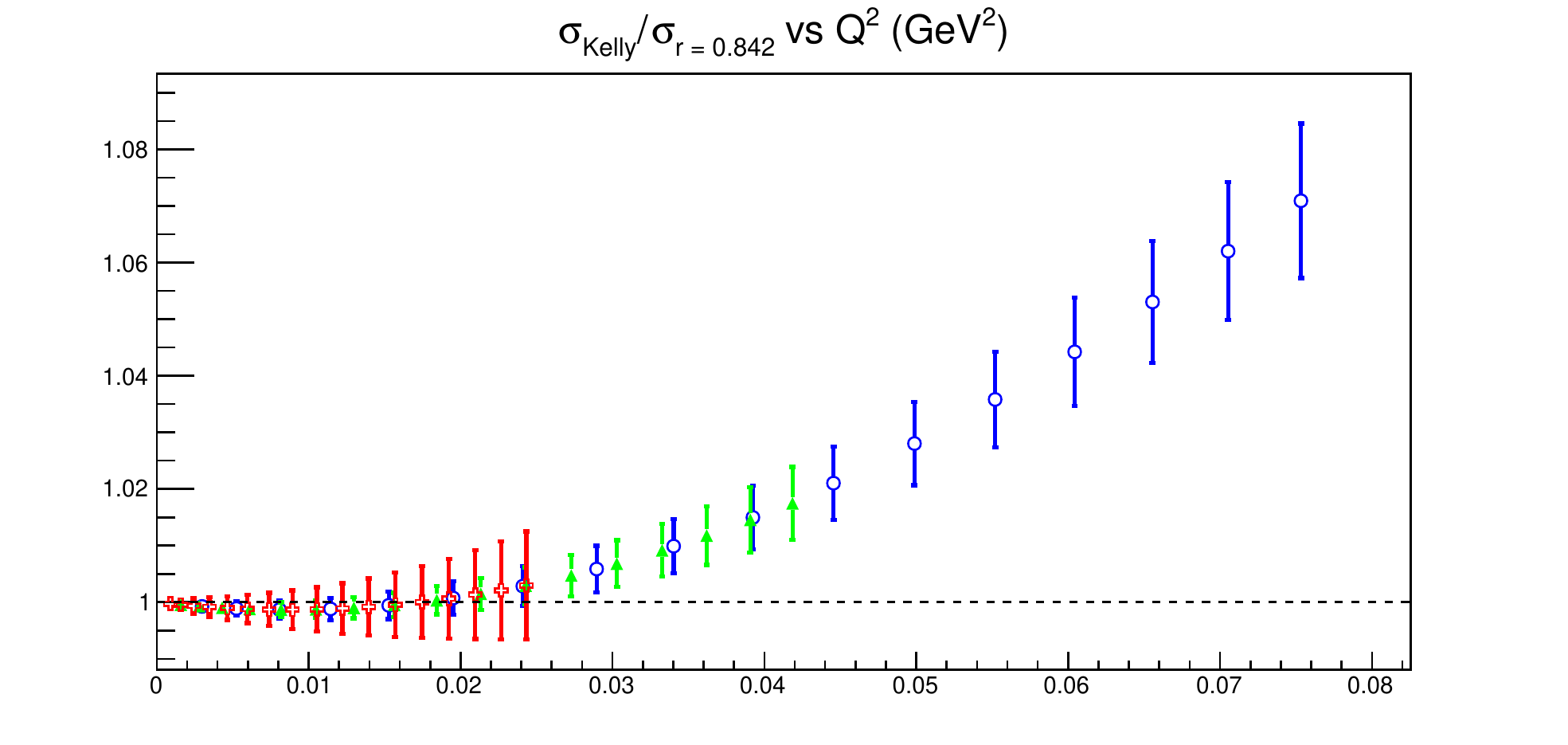}}
\caption{Estimated statistical uncertainties for $\mu^+ p$ elastic
  scattering cross sections, after background subtraction, and
  including experimental inefficiencies. 
  Each point corresponds to a 5$^{\circ}$ bin in
  scattering angle.  }
\label{fig:simdatastats}
\end{figure}

\begin{figure}
\centerline{\includegraphics[width=4.5in]{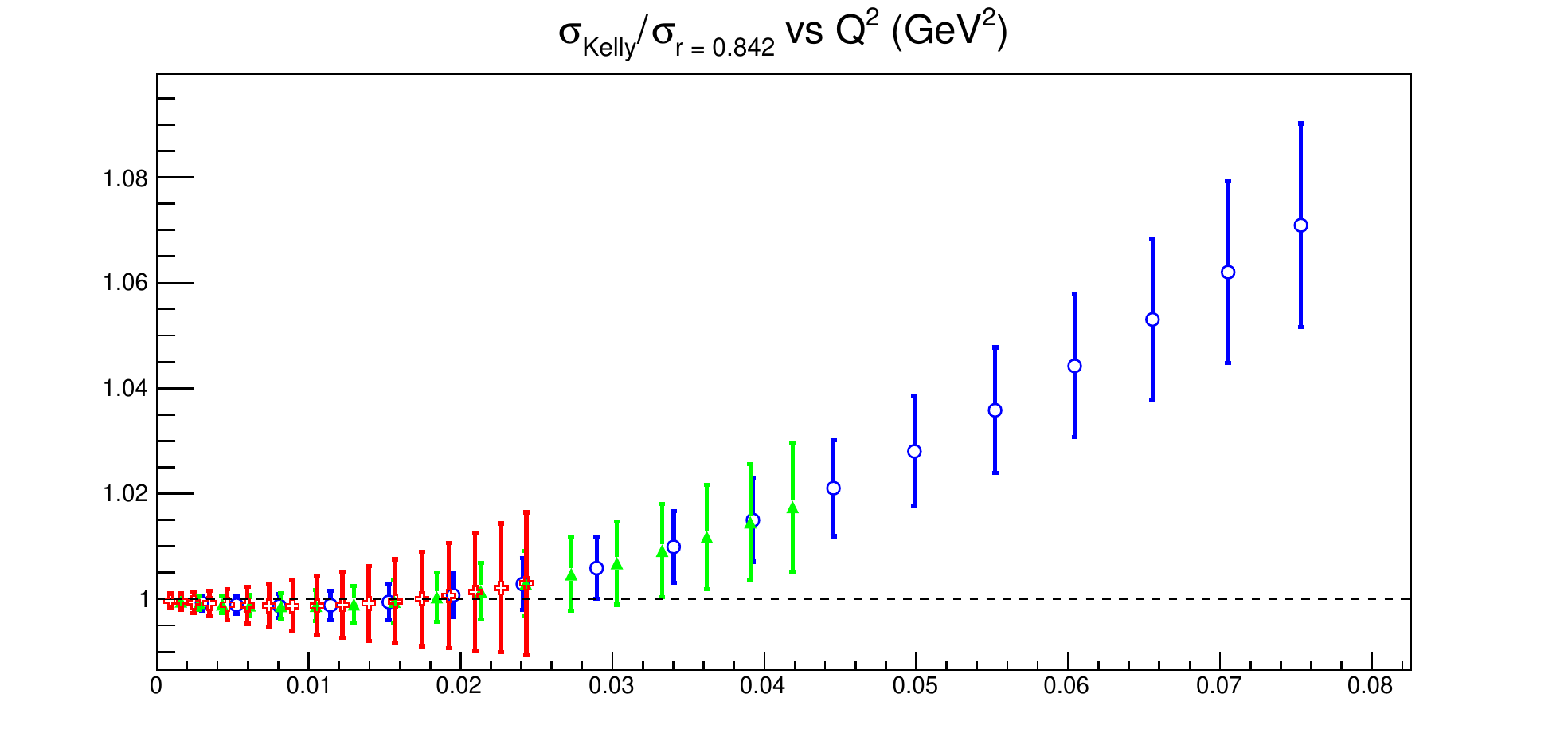}}
\caption{Estimated statistical uncertainties for $\mu^- p$ elastic
  scattering cross sections, after background subtraction, and
  including experimental inefficiencies. 
  Each point corresponds to a 5$^{\circ}$ bin in
  scattering angle.  }
\label{fig:simdatastatsm}
\end{figure}

\subsubsection{Derived Data with Statistical Uncertainties}

From the cross sections shown in Figs.~\ref{fig:simdatastats} and 
\ref{fig:simdatastatsm}, we will construct ratios to determine the
consistency of the ``electron'' and ``muon'' form factors, and
the size of two-photon effects.

\begin{figure}
\centerline{\includegraphics[width=2.8in]{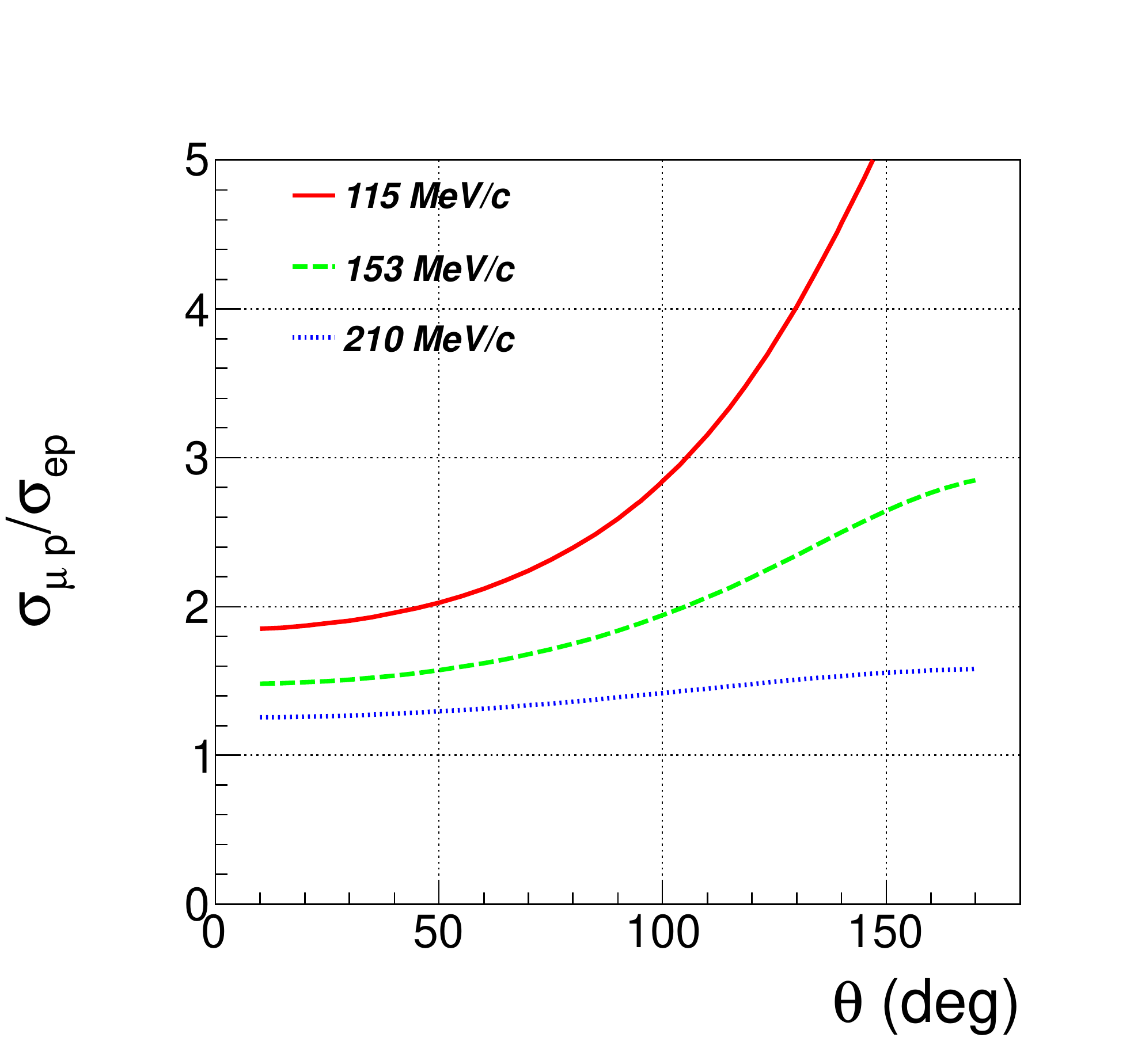}}
\caption{Calculated ratio of $\mu p$ to $ep$ elastic cross sections
 at the same angle. 
}
\label{fig:mup2epcsrat}
\end{figure} 

\begin{figure}
\centerline{\includegraphics[width=6.0in]{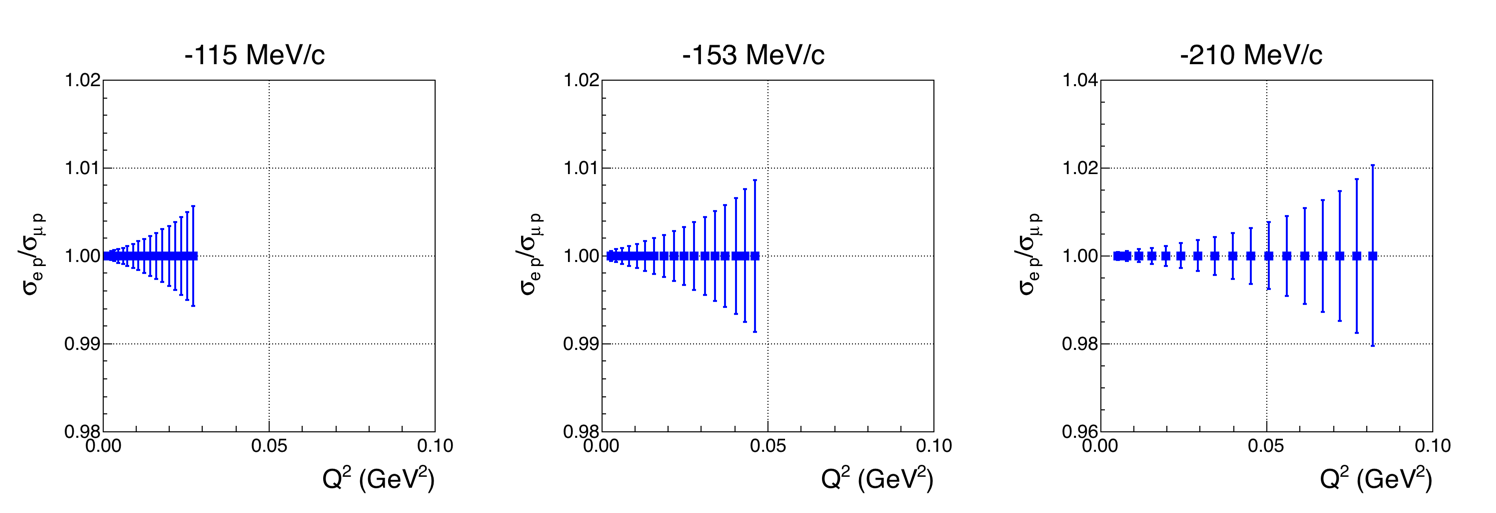}}
\caption{Relative uncertainty in the ratio of $\mu p$ to $ep$ elastic cross sections.
The relative statistical uncertainties in  the form factors are half as
large, since $d\sigma/d\Omega \propto G^2$.
}
\label{fig:csratstat}
\end{figure}

Figure~\ref{fig:mup2epcsrat} shows the ratio of $\mu p$ to $ep$
elastic cross sections. 
The ratio is not unity due to terms in the full cross
section formula proportional to $m/E$ and $m/M_p$, which are
about 0 for the electron. 
We neglect this to show the relative
statistical uncertainty in the cross section ratio, in Fig.~\ref{fig:csratstat}.
The statistical uncertainties are dominated by
the $\mu p$ uncertainties which are a few times larger. 
Ultimately, we will want to compare the electric form factor at the 
same $Q^2$.
For the form factor, the statistical uncertainty is reduced a factor
of 2,  compared to the cross section ratio, as $d\sigma/d\Omega
\propto G^2$,
but there are additional systematic uncertainties, 
from comparing the cross sections in the same angle bin vs.\
the form factors in the same $Q^2$ bin. 
It can be seen that the form factor ratio will have statistical
uncertainties generally below 1\%.

\begin{figure}
\centerline{\includegraphics[width=6.0in]{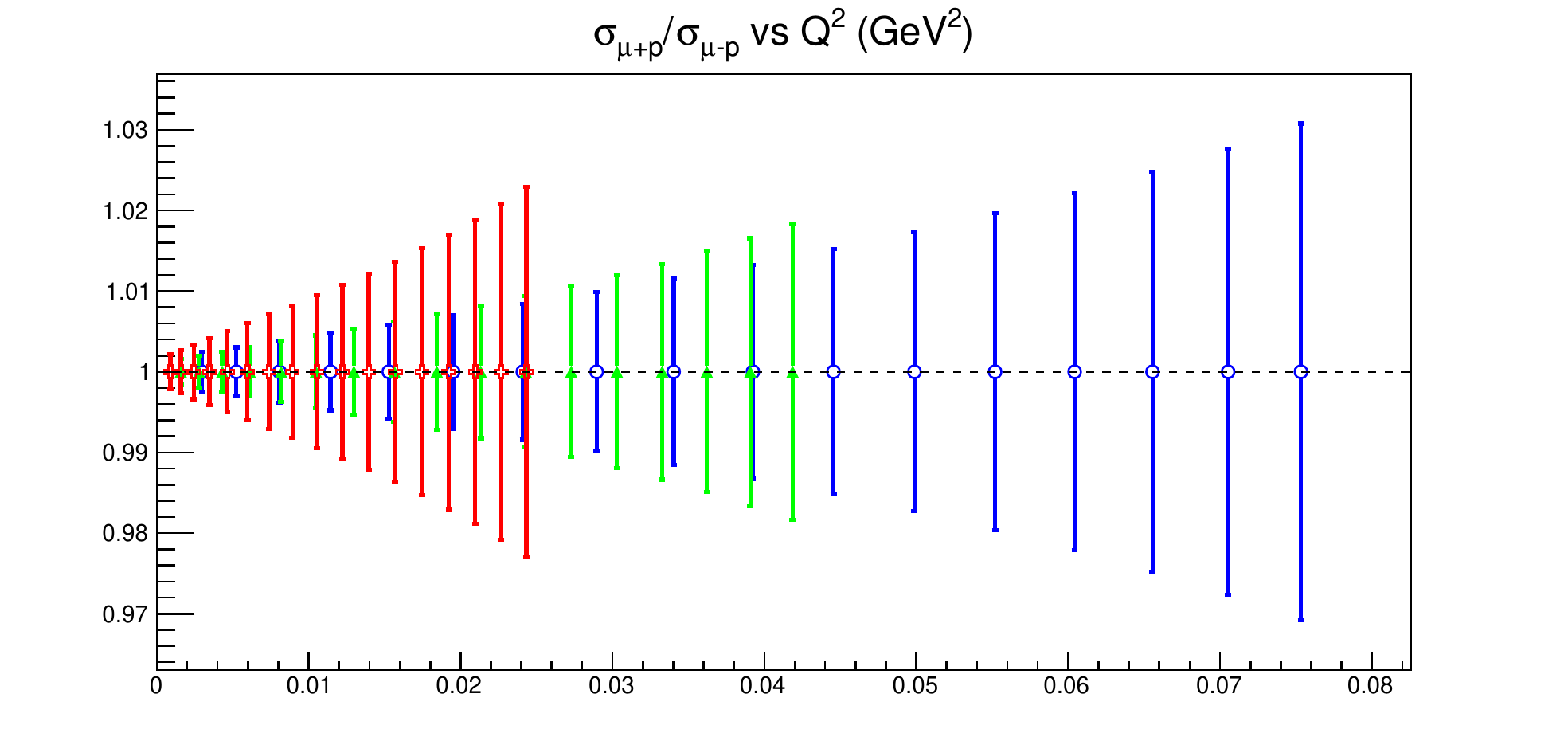}}
\caption{Relative uncertainty in the ratio of $\mu^+ p$ to $\mu^-p$ elastic cross sections.
}
\label{fig:twogamuncmuon}
\end{figure} 

The ratios of positive polarity cross section to negative polarity
ones differs from unity due to the two-photon exchange contribution,
which reverses sign between positive and negative polarities.
The size of the effect varies with $Q^2$, and is typically not more
than a few percent, but calculations disagree with the best existing
data at the percent level.
The relative statistical uncertainties for the ratio of positive to
negative polarity cross sections for muons are
shown in Fig.~\ref{fig:twogamuncmuon}; uncertainties for the electron
ratios are a few times better.
Some theoretical estimates are in Section~\ref{sec:corrections}.
There are no existing data in this $Q^2$ range; there are
two $\approx$4\% points at $Q^2$ $\approx$ 0.15 GeV$^2$, to be 
compared with the 48 often much more precise projected data
shown here.

\subsection{Corrections}
\label{sec:corrections}

The $ep$ and $\mu p$ cross sections determined from the background-subtracted
yields must be corrected for a number of experimental and theoretical
effects.
The experimental efficiency corrections included in
Eq.~(\ref{eq:cseq}) are all standard, so we do not discuss them further.
Resolutions and multiple scattering lead to an angle averaging, which
warps the cross section as shown in Fig.~\ref{fig:msanglesyst} in
Appendix~\ref{sec:uncertaintiesoverview}.
This effect can be corrected by either calculation or Monte Carlo,
and we do not discuss it further.

There are three types of theoretical corrections we discuss here:
magnetic and $Q^2$ corrections,
radiative corrections, and
two-photon exchange corrections.

\subsubsection{Magnetic and $Q^2$ Correction}

\begin{figure}[h]
\centerline{\includegraphics[width=6.0in]{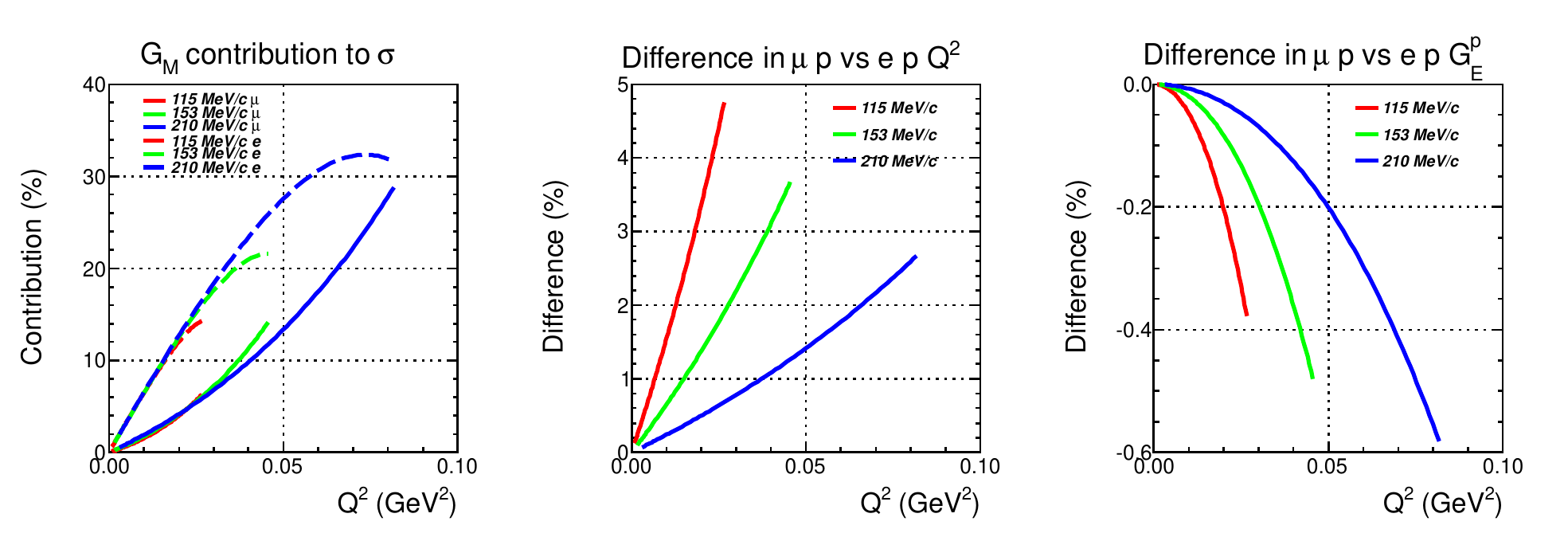}}
\caption{Corrections related to extracting the electric form factor
  from the cross section for comparison at constant $Q^2$ rather
  that constant scattering angle.
  Left: Portion of the cross section coming from the magnetic
  form factor $G_M$. Solid (dashed) lines for for $\mu p$ ($ep$)
  elastic scattering.
  Middle: Percentage difference in $Q^2$ between $\mu p$ and $ep$
  elastic scattering at the same beam momentum and angle.
  Right: The difference in the electric form factor arising from the
  different $Q^2$. All estimates use the Kelly form factors.
}
\label{fig:geunc}
\end{figure}

To extract and compare the proton electric form factors and {\em electric}
radius with $ep$ and $\mu p$,
corrections have to be made for the magnetic contribution to the
cross section and for the $Q^2$ difference between $\mu p$ and $ep$
elastic scattering at the same beam momentum and angle.
Figure~\ref{fig:geunc} shows factors related to this determination.
The magnetic contribution ranges up to about 30\%
at the largest angles.
In the range of the MUSE experiment, fits of the Bernauer data 
suggest that the uncertainty in the 
magnetic form factor is $\leq$0.3\%. 
The uncertainty of the magnetic contribution to the cross section
is then $\leq$ 30\% $\times$ 0.3\% $\approx$ 0.1\%.
The changes in $Q^2$ are purely kinematic.

\subsubsection{Radiative Corrections}
\label{sec:radcorr}

Radiative corrections procedures for electron-proton scattering are well
established, and numerous codes exist.
The precision of the corrections is limited by approximations in the
codes, the precision with which the experiment is known (acceptance, 
bin kinematics, etc.), and the knowledge of the form factors.
The important difference for $\mu p$ scattering vs.\ $ep$ scattering is that the larger
muon mass suppresses the emission of bremsstrahlung radiation.
However, most older codes assume the peaking approximation and / or the ultra-relativistic
approximation ($m/E \to 0$).
Afanasev has accounted for these effects and provided us with an exact
calculation of the muon bremsstrahlung correction in MUSE kinematics. 
The correction is near zero at $\theta = 0^{\circ}$, and grows with
angle and beam momentum, becoming as large as 3\%
for $\theta = 100^{\circ}$ at a beam momentum of 210 MeV/$c$.
Afanasev estimates the uncertainty in the correction to be over 
an order of magnitude smaller than the correction, around the 0.1\%
level.
The correction for $ep$ scattering is about 5 times larger and
similarly less precise.
These estimates need to be re-evaluated when the experiment is
assembled and the actual uncertainties can be evaluated.

\subsubsection{Radiative Corrections and Beam Momentum}
\label{sec:radcorrbeamp}

The beam momentum at the interaction point is degraded from the 
momentum out of the channel due to interactions with detectors 
before the target.
A major part of this energy loss is external bremsstrahlung, part of
the radiative corrections.
Figure~\ref{fig:realbeamenergymu} shows a Geant4 calculation of the 
beam momentum entering and exiting the target, starting from a flat channel
momentum distribution $\pm$1.5\%-wide.\footnote{The distributions will
also be {\em measured} as part of our simulation calibration program.}
The energy shifts are similar at the three beam settings, leading to
larger fractional momentum shifts at the lowest beam momentum setting.
We can use these calculated momentum distributions to study
corrections from the spectrum shape.
We compare the average of the cross sections from the calculated
spectrum to the cross section at the central momentum of the spectrum.
The right panel of Fig.~\ref{fig:realbeamenergymu} shows that this
is about 0.05\% -- 0.1\%, and the variation with angle is about 0.01\%.
Thus knowing the average momentum is sufficent to make this correction
quite small.

\begin{figure}
\centerline{\includegraphics[width=2.6in]{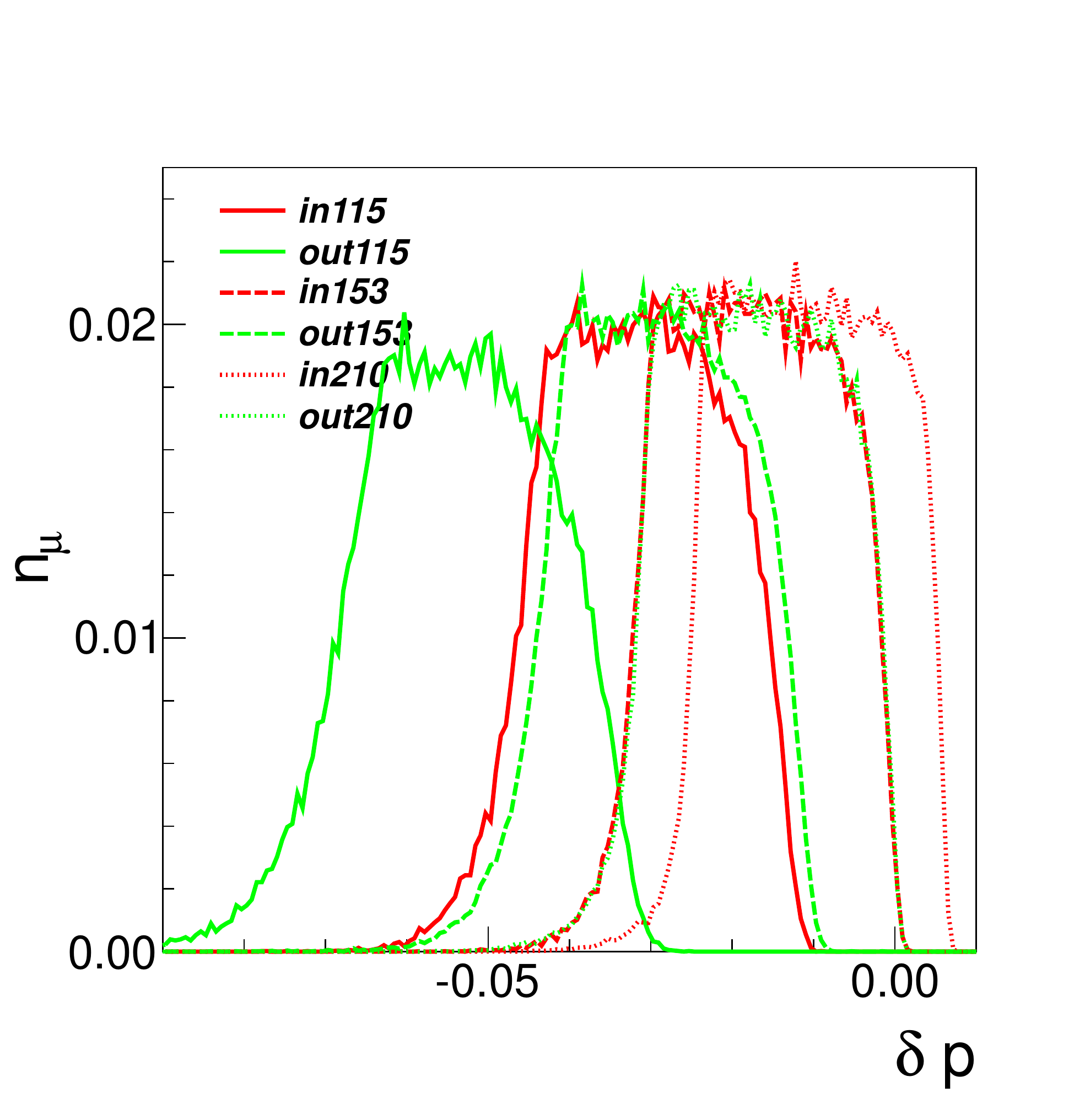}
                           \includegraphics[width=2.6in]{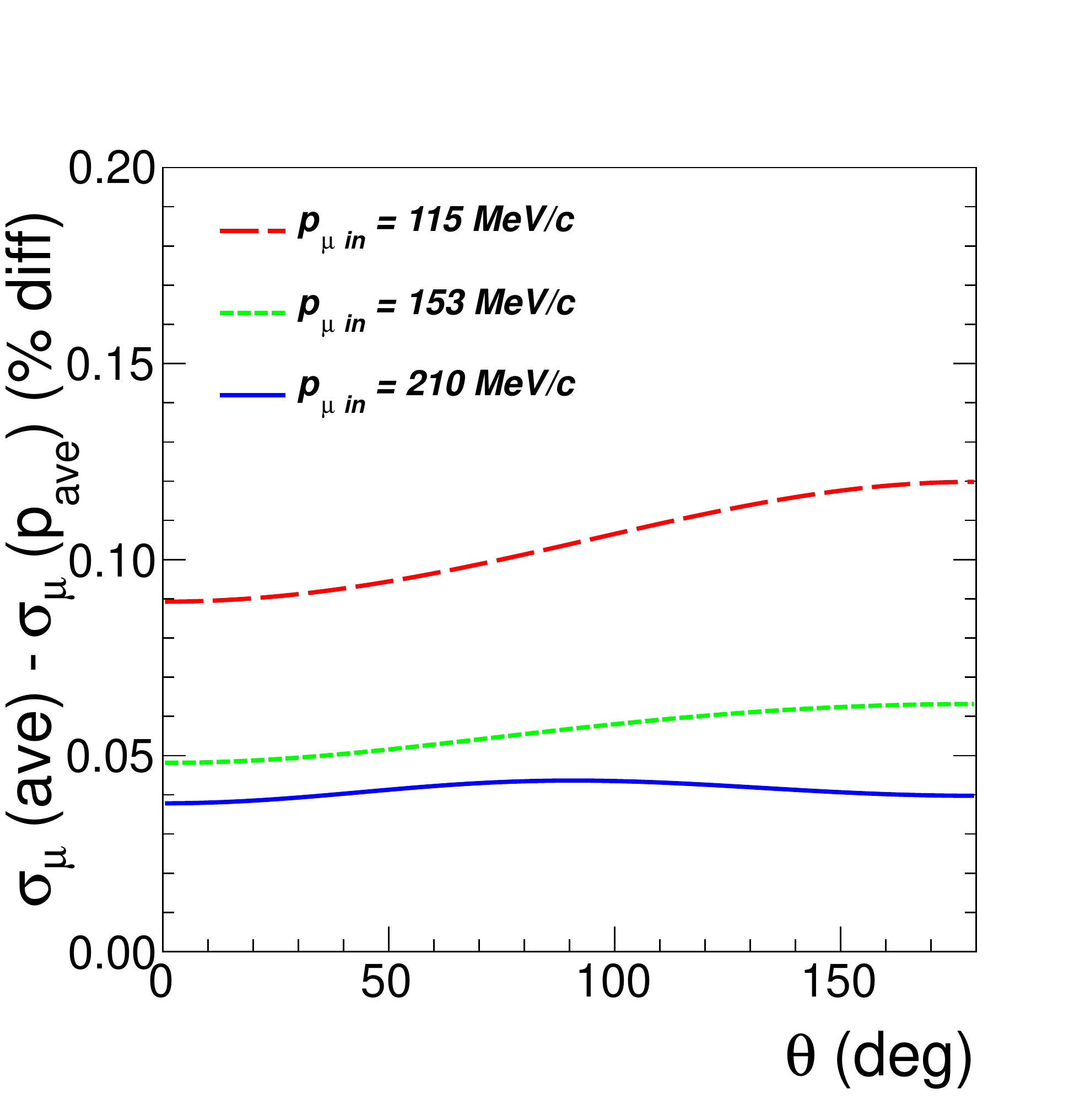}}
\caption{Left: Momentum spectrum for beam muons entering and
  exiting the target. A fractional number of events per bin is plotted
vs.\ the momentum relative to the central momentum of the $\pi$M1 channel.
Right: Difference between the cross section calculated for the average beam
momentum and the cross section averaged over the incident spectra shown to the left.
}
\label{fig:realbeamenergymu}
\end{figure}

\subsubsection{Two-photon exchange}

At very low $Q^2$, calculations of two-photon exchange (TPE) within a hadronic
framework~\cite{Blunden:2005ew,Blunden:2005jv,Arrington:2011dn} are typically 
expected to be reliable, and are in good agreement with a low $Q^2$ TPE
expansion~\cite{Borisyuk:2006uq}, which is expected to be valid up to
$Q^2=0.1$~GeV$^2$, therefore covering our entire $Q^2$ range.  
The recent generation of higher precision (TPE) measurements comparing
cross sections for $e^+p$ and $e^-p$ shows that TPE is small, at the
percent level, but at this level it is also significantly different
from calculations.
For $\mu p$, the calculations have additional uncertainties from the 
knowledge of helicity-flip amplitudes.
The calculations also depend on what intermediate states are included.
There have been two calculations of conventional, or soft, TPE calculations
for MUSE, both predicting small effects on the MUSE cross sections.  
These calculations can be benchmarked to low-$Q^2$ measurements of the 
imaginary part of TPE, for example the upcoming results from 
the measurement of the beam normal single spin asymmetry for the Qweak experiment.

Afanasev's calculations (see Fig.~\ref{fig:tpe}, left) of TPE show an
effect that approaches 
zero at forward angles and increases with scattering angle.
The effect is no more than about 0.2\% in MUSE kinematics, 
with little difference seen between the correction for muons and for
electrons, and estimated uncertainties about half of the correction.
Tomalak \& Vanderhaeghen \cite{tomalak2014} calculate the expected corrections
for $\mu^- p$ and find it varies between 0.25\% - 0.5\% (see Fig.~\ref{fig:tpe}, right).

\begin{figure}[h]
\centerline{\includegraphics[width=0.55\textwidth]{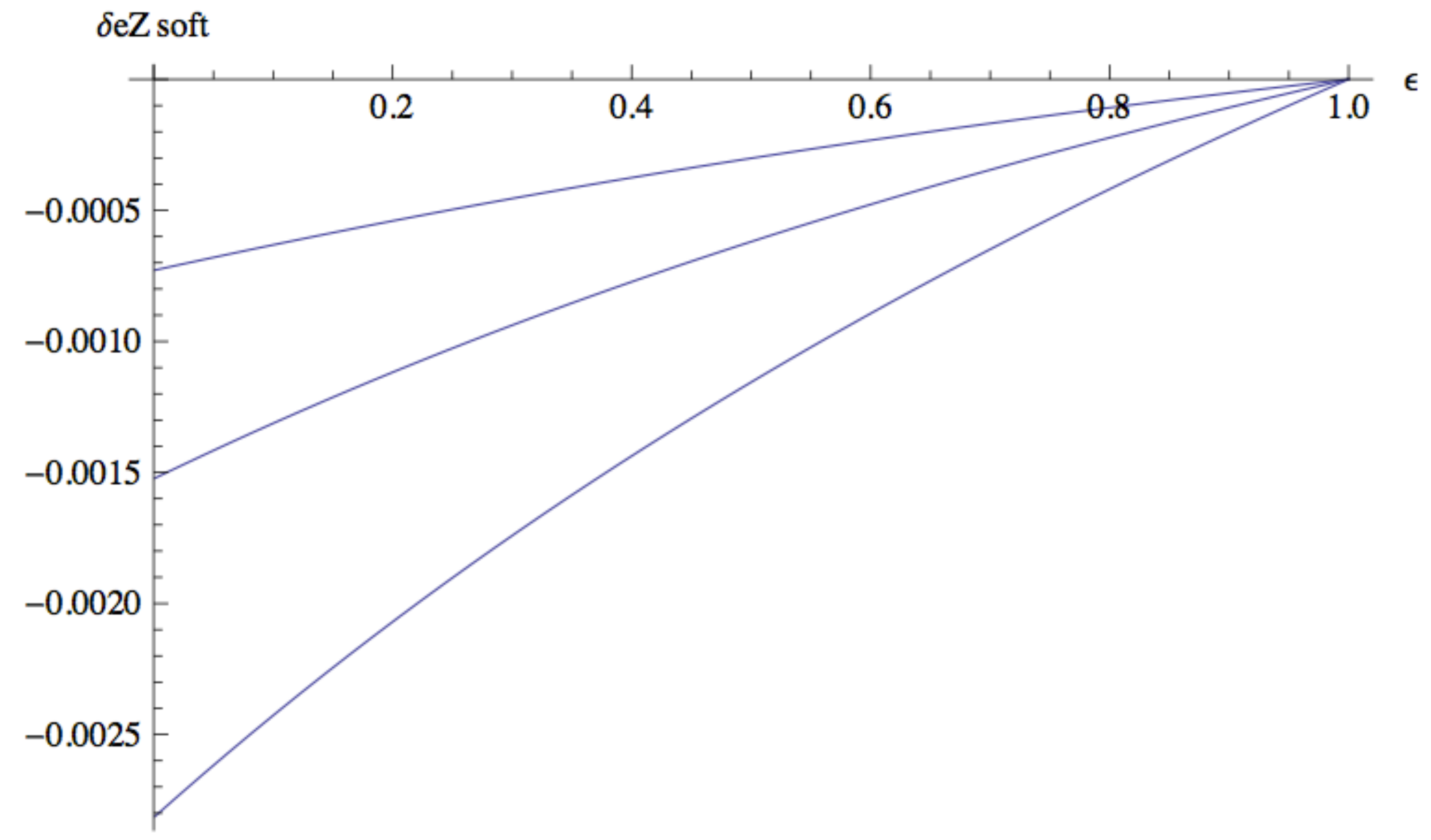}
\includegraphics[width=0.44\textwidth]{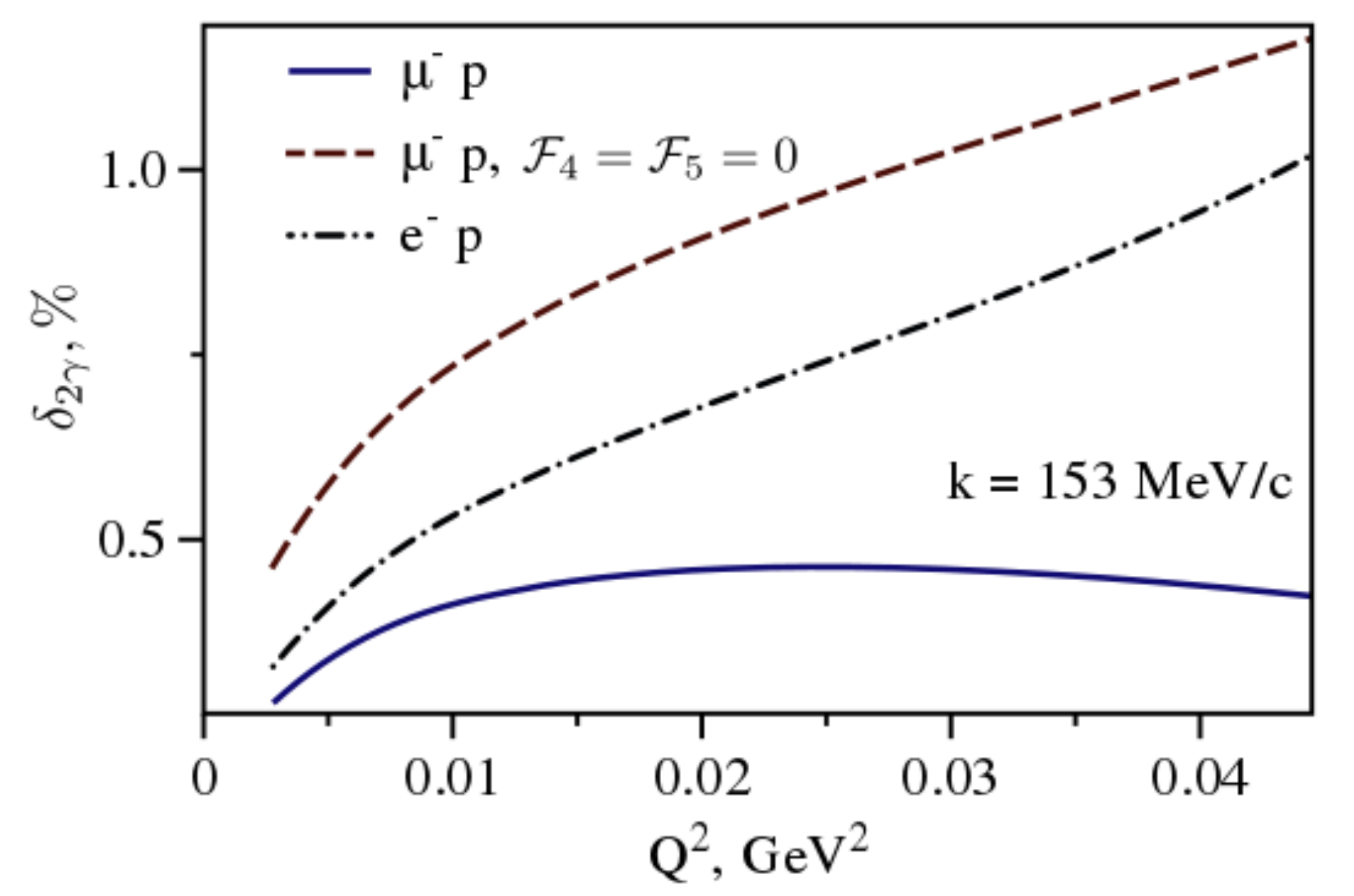}}
\caption{Calculations of TPE for $\mu p$ cross sections.
  Left: $ep$ TPE corrections for the three MUSE beam energies as a
  function of $\epsilon$ calculated by Afanasev. The correction grows with energy.
  The muon calculations are very similar.
  Right: $\mu^- p$ and $e^- p$ TPE corrections calculated by Tomalak \cite{tomalak2014}. The red-dashed curve shows the muon correction without accounting for the helicity-flip amplitude. 
}
\label{fig:tpe}
\end{figure}

The approach in MUSE is to directly measure the TPE effect in $ep$ and
$\mu p$ scattering by comparing the cross sections for positive and negative polarity
beams, rather than relying on calculations, since these measurements
directly test physics that might help explain the proton radius puzzle.

\subsection{Systematics}
\label{sec:systematics}

\begin{table}[h]
\caption{\label{tab:osystsummary} 
Estimated MUSE relative systematic cross section uncertainties
for the shape of angular distributions, the ratio of
muon and electron scattering cross sections, and the ratio
of $+$ charge to  $-$ charge cross sections.
}
\begin{tabular}{|l|c|c|c|}
\hline
Uncertainty &  angular distribution & $\mu$/$e$ & $+$/$-$ \\
     &   (\%)  & (\%) & (\%) \\
\hline
Detector efficiencies                         & 0.1 &0.1 & 0.1 \\
Solid angle                                        & 0.1 & small & small \\
Luminosity                                        & small & small & small \\
Scattering angle offset                      & 0.2 & small & small \\
Multiple scattering correction           & 0.15 & small & small \\
Beam momentum offset                      & 0.1 & 0.1 & 0.1 \\
Radiative correction                             & 0.1 ($\mu$), 0.5 ($e$) & 0.5 & 1$\gamma$small \\
Magnetic contribution                          & 0.15 & small & small\\
Subtraction of $\mu$ decay from $\mu p$ & 0.1 & 0.1 & small \\
Subtraction of target walls                    & 0.3 & small & small\\
Subtraction of pion-induced events     & small & small & small \\
Beam PID / reaction misidentification  & 0.1 & 0.1 & 0.1 \\
Subtraction of $\mu$ decay from $e p$ & small & small & small \\
Subtraction of $ee$ from $e p$ & small & small & small \\
\hline
TOTAL & 0.5 ($\mu$), 0.7 ($e$) & 0.5 & 0.2 \\
\hline 
\end{tabular}
\end{table}

Estimated systematic uncertainties 
are shown in Table~\ref{tab:osystsummary}, based on the analysis in
Appendix~\ref{sec:uncertaintiesoverview}.
We focus on relative uncertainties here since the absolute uncertainties cannot be
determined precisely enough, at the few tenths of a percent level, so
that the data can be used without a normalization factor.
Instead we cross normalize the angular distributions to each other and
to the $Q^2 = 0$ form factors. 
We summarize here various aspects of the systematics.  

\subsubsection{Beam Detector Related Systematics}

It is a good approximation that beam detectors do
not change the angular distribution of scattered particles,
so the related systematic uncertainty vanishes.
There is a small probability of mis-identified reaction types 
that do change the shape of the angular distribution.
By measuring the shape of these background angular
distributions and quantifying how they are misidentified and mixed
into the data, their effects can be subtracted.

\subsubsection{Beam Momentum Determination Systematics}
\label{sec:beammomentum}

We showed that beam momentum offsets and averaging over beam momentum
distributions lead essentially to a renormalization of the
data with small angle-dependent variations, at or below 0.1\%.
We plan to control this systematic with beam momentum and
energy loss measurements, using a combination of time of flight
and channel dispersion techniques.
We have demonstrated absolute muon and pion momentum measurements
at the needed 0.2\% -- 0.3\% level, and estimated from TURTLE simulations
that the relative momentum of all particle types can be determined to $\ll$ 0.1\%.

\subsubsection{Target Systematics}

The target thickness various with trajectory, and multiple scattering
in the upstream detectors leads to slightly different beam spots for
different particles and beam momenta. 
However we measure the particle dependent trajectories into the
target, so corrections can be made through the simulation.

\subsubsection{Scattered Particle Detector Systematics}

The scintillators and straw chambers are designed so that we have
very high detection and tracking efficiencies, near 100\%, so that 
there is little room for the angular distribution shape to be changed.
There is only a $\approx$99\% efficiency for $e^+$ due to positron annihilation, 
for which we estimate a 0.1\% relative systematic
uncertainty.

While the rate varies across the detector angular range, the rates are
sufficiently low that even for the most forward
scintillator bars the dead time will only be $\approx$0.1\%.
The dead time can be deduced from scaler readings.

The experiment is sensitive to offsets in the measured angles and
multiple scattering that averages the cross section, due to the
sharply falling electromagnetic cross section.
We have a program of calibration measurements to control both
effects.
These uncertainties strongly increase towards the forward angles,
much smaller impact at larger angles.

\subsubsection{Solid Angle}

The determination of the solid angle is controlled through a combination
of careful design and construction of equipment, temperature
control, and survey and dedicated calibration measurements
of equipment positions.
The left/right and up/down symmetry of our systems provide
checks on offsets in those directions that might affect the
solid angle.
All the offsets have the largest effect on cross sections at our
most forward and most backward scattering angles, where the 
acceptance is getting small.

\subsubsection{Electronics, Trigger, and Computer Live Time Uncertainties}

We use standard techniques to control electronics uncertainties,
including scalers, measuring read out times, performing logic with
multiple width pulses, using a random trigger, and taking data at
multiple beam fluxes.
We plan to calibrate the trigger at low beam rates and study its
performance as a function of rate, monitoring in the fast DAQ the
trigger inputs, intermediate stages, and final trigger.

\subsubsection{Uncertainties in Theoretical Corrections} 

We do not consider uncertainties in the TPE calculations since MUSE
will measure the TPE contribution directly.
Radiative corrections for the muon, an example shown in Fig.~\ref{fig:aaradcorr}, 
are estimated to be $<$3\%,
with overall uncertainties about one-tenth of the correction, or 0.3\%
and angle-dependent variations smaller, about 0.1\%.
Electron radiative corrections and uncertainties are about 5 times larger.
In the case of the muon, the radiative tail is quite small, whereas for
the electron, the radiative tail is long, so the correction
averages over a wider range of vertex kinematics from
pre-radiation.

\begin{figure}[h]
\centerline{\includegraphics[height=2.0in]{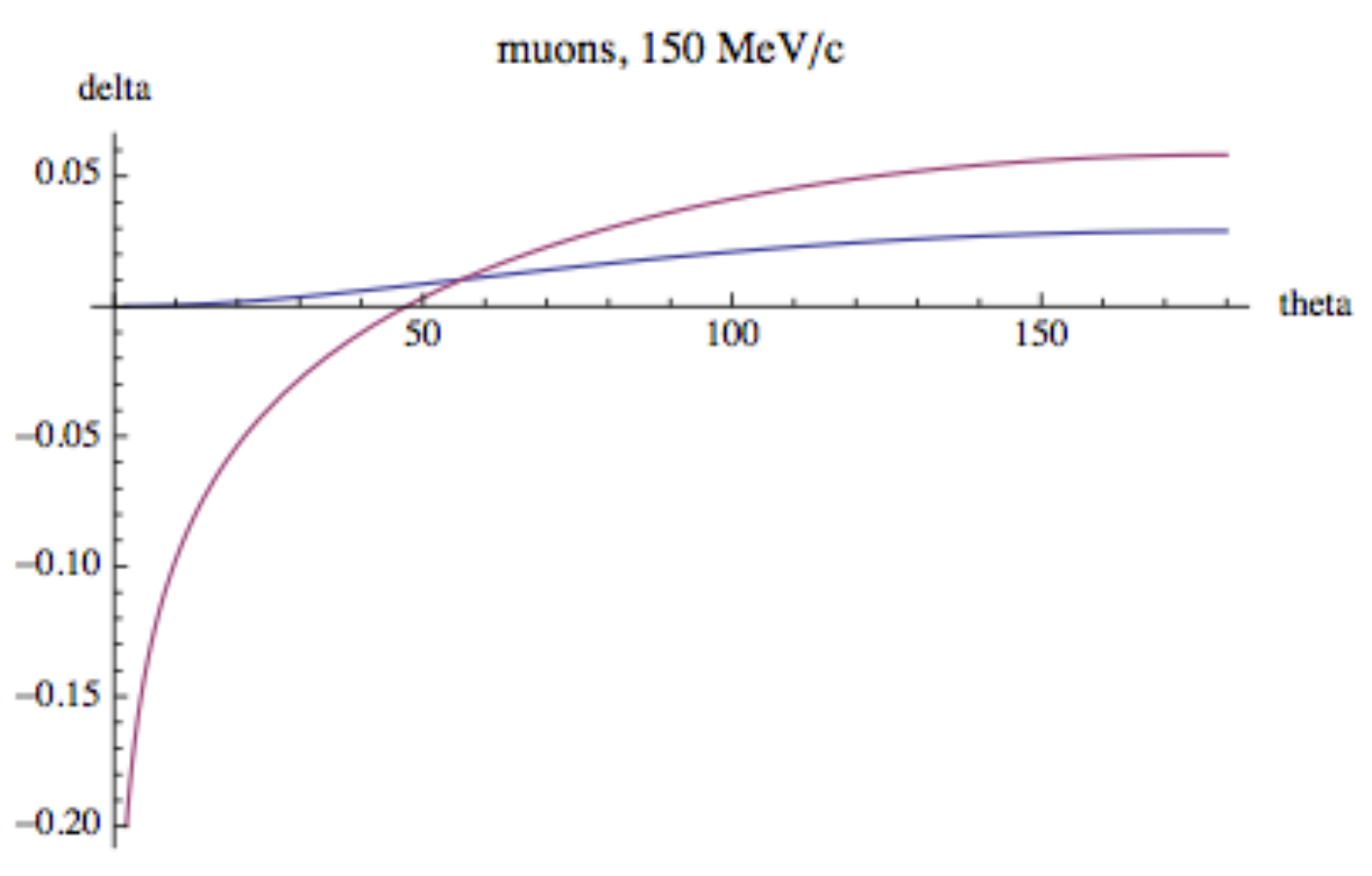}}
\caption{Radiative correction calculation from Afanasev showing the 
  muon radiative corrections at 150 MeV/$c$ for MUSE.  The blue (red)
  curve is the full (approximate) calculation.
}
\label{fig:aaradcorr}
\end{figure}

\subsubsection{Analysis  and Instability Uncertainties}

Additional systematic uncertainties might arise from the analysis
procedures or instabilities in the data. 
For example, the cross section varies over time or with a range of reasonable
cuts by some amount, with no good explanation as to why.
The variation presumably reflects some unidentified systematic effect, 
and can be used to estimate an uncertainty from the unknown effect.
It is not possible at this time to assign an uncertainty here.

\subsubsection{Systematic Uncertainties for Ratios}

As shown in Table~\ref{tab:osystsummary}, certain factors
are the same or very close to the same when comparing $ep$
to $\mu p$ or $+$ to $-$ cross sections, which results in
some cancellations of the systematic uncertainties.
The mass effects limit the cancellations in $ep$ vs.\ $\mu p$.

\subsection{Radius Extraction}
\label{sec:radiusextraction}

The direct comparison of $\mu p$ and $ep$ cross sections, corrected for the
effects of the muon mass, or the extracted electric form factor, will
make it clear in the MUSE experiment whether the two particles have
different interactions or not.
If there is a difference, the comparison of + and - charge cross
sections should make it clear if two-photon exchange might be
responsible for the difference.
But the radius, extracted from the slope of the form factor
at $Q^2$ = 0, directly connects to the atomic physics measurements.
Since we do not measure at $Q^2$ = 0, the radius relies on fitting 
the measured data with a form factor parameterization. 
In this section we very briefly discuss the radius extraction and 
the resulting uncertainties.

\begin{figure}
\centerline{\includegraphics[width=0.6\textwidth]{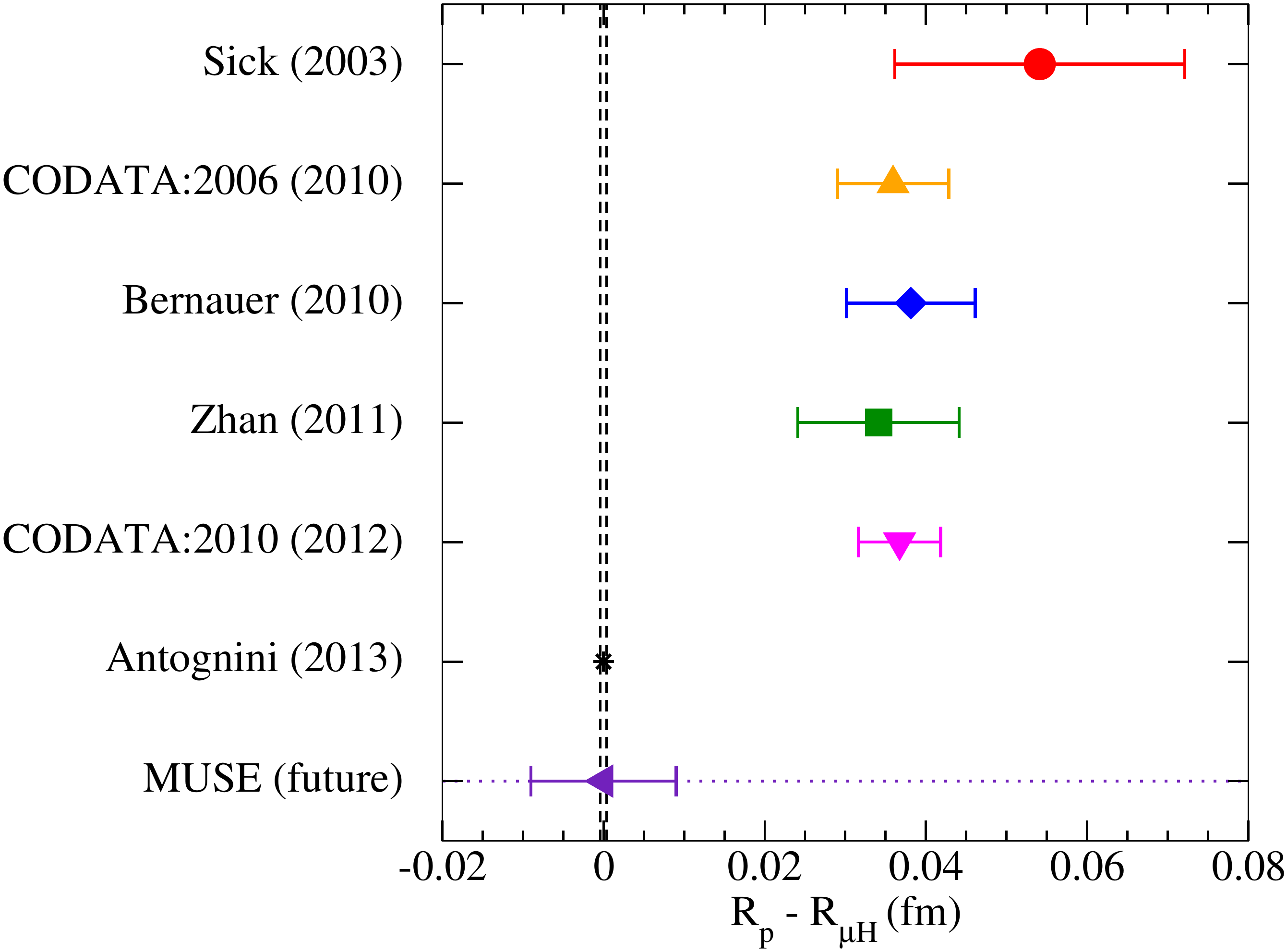}}
\caption{
A summary of some recent proton charge radius
  determinations, relative to the muonic hydrogen result, including
Sick \cite{Sick:2003gm}, 
CODATA 2006 \cite{Mohr:2008fa}, 
Bernauer {\it el al.} \cite{Bernauer:2010wm}, 
Zhan {\it et al.} \cite{Zhan:2011ji},
CODATA 2010 \cite{2012arXiv1203.5425M}, 
and
Antognini {\it el al.} \cite{Antognini:2013}, 
which includes Pohl {\it et al.} \cite{Pohl:2010zza},
along with our expected MUSE result, arbitrarily placed at 0.
}
\label{fig:radiusresults}
\end{figure}

We consider here the standard approaches of doing simple 
polynomial fits -- note the criticism of
this approach in \cite{PhysRevC.90.045206} -- 
and of doing inverse polynomial fit to the data sets. 
Pade fits have too many parameters for the MUSE data
range, and $z$-expansion fits yield similar results to
what we present.
We followed the fitting procedures of \cite{PhysRevC.90.045206},
and have consistent results for our polynomial fits.

We fit pseudodata, with the
data generated according to various world parameterizations and
with our expected uncertainties. We used the same generating
functions for $ep$ and $\mu p$ and fit the same way, and compared
the results and the differences in the results.
Table~\ref{tab:fittsummary} summarizes our results.
In comparing $ep$ to $\mu p$, fitting issues largely go away
and the difference is reliable -- if as assumed here the form factors
are the same in both cases.
In first order fits, the truncation error is larger than the
uncertainty, especially for polynomial fits, so the absolute radius
is unreliable.
Going to a second order fit increases the uncertainty and reduces
the truncation error, making the extracted radius more reliable.
Going from polynomial to inverse polynomial fits reduces the
truncation error while leaving the uncertainty approximately unaffected.
Thus we advocate for a first-order inverse polynomial fit to compare
$ep$ to $\mu p$ most precisely, and, if the extracted radii are
consistent, a joint fit of the $ep$ and $\mu p$ data sets with a
second order inverse polynomial fit to obtain the absolute radius.
The difference fit gives an $\approx$8$\sigma$ determination
of the possible 0.04 fm difference, 
while the joint fit gives an $\approx$5$\sigma$ determination
of the absolute radius.

\begin{table}[h]
\caption{\label{tab:fittsummary} 
Typical results from fitting estimated MUSE data for several
generating functions.
}
\begin{tabular}{|l|c|c|c|c|c|}
\hline
 & 
\multicolumn{2}{|c|}{Truncation Error} &
 \multicolumn{2}{|c|}{Uncertainty}  &
 $ep$ vs.\ $\mu p$\\
\hline
 & $ep$ & $\mu p$ &   $ep$ & $\mu p$ &   difference \\
 & (fm) & (fm) & (fm) & (fm) & (fm) \\
\hline
First order polynomial                       & -0.050 & -0.048 & 0.003  & 0.003 & -0.002 \\
Second order polynomial                    & -0.005 & -0.005 & 0.009  & 0.007 & 0.001 \\
First order inverse polynomial            & 0.007 & 0.007 & 0.004  & 0.003 & 0.001 \\
Second order inverse polynomial         & -0.001 & -0.001 & 0.01  & 0.008 & 0 \\
\hline 
\end{tabular}
\end{table}

\section{Project Management}
\label{sec:management}
\begin{table}[h]
\caption{MUSE project WBS tasks and responsible people.}
\begin{tabular}{|c|c|c|}
\hline 
\textbf{Task/WBS} & \textbf{Responsible Persons} & \textbf{Institution}\tabularnewline
\hline 
\hline 
Management & Gilman/Cioffi/Reimer/Briscoe   & Rutgers/GW/Argonne/GW\tabularnewline
\hline 
Frames and Design WBS 1 & Reimer & Argonne\tabularnewline
\hline
Scintillator/Silicon photodector WBS 2 & Piasetzky & Tel Aviv Univ. \tabularnewline
\hline 
Cerenkov detector WBS 3 & Gilman & Rutgers \tabularnewline
\hline
Wire Chambers WBS 4& Ron & Hebrew \tabularnewline 
\hline 
Cryogenic Target WBS 5 & Lorenzon & University of Michigan \tabularnewline
\hline 
 Electronics and DAQ WBS 6 & Downie  & George Washington \tabularnewline
\hline 
 Scintillators WBS 7& Strauch &  South Carolina \tabularnewline
\hline 
GEM chambers WBS 8& Kohl  & Hampton   \tabularnewline
\hline
Installation WBS 9 & Briscoe & GW \tabularnewline
\hline
Software WBS 10 & Golossanov & GW \tabularnewline
\hline
Data Taking WBS 10 & Downie & GW \tabularnewline
\hline
\hline
\end{tabular}
\label{tab:WBS}
\end{table}

The MUSE construction project has put in place a formal 
project management structure due to the extent and cost
of the construction project.  
The project manager team consists of:

\begin{itemize}
\item R. Gilman (Rutgers) Project Manager.  Professor Gilman 
 stepped down as MUSE spokesperson to take over project
 management for the experiment, as he was the best fit for
 the position.
\item D. F. Cioffi (George Washington) Deputy Project Manager.
  Professor Cioffi taught project management at GW for more than ten
  years, during much of that time he was director of the master's
  Project Management Program.  He has extensive
  federal government experience, including being a National Science
  Foundation Associate Program Director (1991).  A faculty member of
  the Decision Sciences Department, 
  he currently serves as Senior Advisor to the Dean of the Business School, 
  as well as retaining an appointment in the Physics Department.
\item P. Reimer (Argonne National Laboratory) Deputy Project Manager.
  Dr. Reimer has played a lead role in a number of experiments, and
  most recently served as deputy project manager for the Fermilab
  SeaQuest experiment.  
\item W. J. Briscoe (George Washington) Deputy Project Manager.
  Professor Briscoe is currently Chair of the Department of Physics at
  George Washington University.  He has managed several large
  experimental efforts.
\end{itemize}

The project management is described in detail in the Project Execution
Plan; the final version was submitted to NSF on June 1, 2017.  The
Plan covers numerous topics including the Work Breakdown Structure,
responsible people, key performance parameters, milestones, risks and
risk management, contingencies, project oversight and communication,
and the funding profile.

This material is based upon work supported by the National Science
Foundation under Grant Numbers OISE-1150594, PHY-1205782, PHY-1207672,
PHY-1306126, PHY-1309130, PHY-1314148, OISE-1358175, PHY-1401974,
PHY-1404271, PHY-1404342, PHY- 1505458, PHY-1505615, PHY-1505934,
PHY-1506000, PHY-1506061, PHY-1506160, PHY-1612495, PHY-1614456,
PHY-1614773, PHY-1614850, PHY-1614938, PHY-1649873, HRD-1649909, and
PHY-1714833; the Department of Energy under Grant Numbers
DE-SC0012485, DE-0012589, DE-AC02-06CH11357, and DE-SC0013941; the
Schweizerischer Nationalfonds under grant numbers 200020-156983, and
2015.0594; and the Binational Science Foundation under Grant Number
2015618, 2015625, and 2014709; and the Paul Scherrer Institute.

\subsection{Collaboration}
\label{sec:collaboration}

The MUSE collaboration is comprised primarily of people with
experience in electron scattering experiments, some of whom have
worked together for over 20 years.
The collaboration lead spokesperson is E.\ Downie (GW), with deputy
spokespeople G.\ Ron (HUJI) and S.\ Strauch (SC).
The collaboration has experience with experiments of the size and
scale of MUSE, primarily electron and photon scattering experiments.
The core of the collaboration comprises the
institutions constructing the
experimental equipment.
A summary of some commitments beyond what is reported in Table~\ref{tab:WBS}
on the WBS structure 
is shown in Table~\ref{tab:musesplit}.
The collaboration formalized its
structure by adopting a charter at its January 2014 meeting,
and has regularly held collaboration meetings since.

\begin{table}[h]
\caption{Some MUSE collaboration responsibilites not covered by the
  WBS structure.}
\begin{tabular}{|c|c|c|}
\hline 
\textbf{Responsibility} & \textbf{Person} & \textbf{Institution}\tabularnewline
\hline 
\hline 
$\pi$M1 Channel & K. Deiters   & PSI\tabularnewline
\hline 
Trigger & R. Gilman (Project Manager) & Rutgers \tabularnewline
\hline 
Radiative Corrections & A. Afanasev & George Washington \tabularnewline
\hline 
Analysis Software & J. Bernauer & MIT\tabularnewline
\hline 
Simulations & S. Strauch & South Carolina \tabularnewline
\hline
\end{tabular}
\label{tab:musesplit}
\end{table}

\newpage
\appendix

\section{Backgrounds}
\label{app:backgrounds}
\label{sec:physreacback}

\begin{figure}[h]
\centerline{\includegraphics[width=1.0\textwidth]{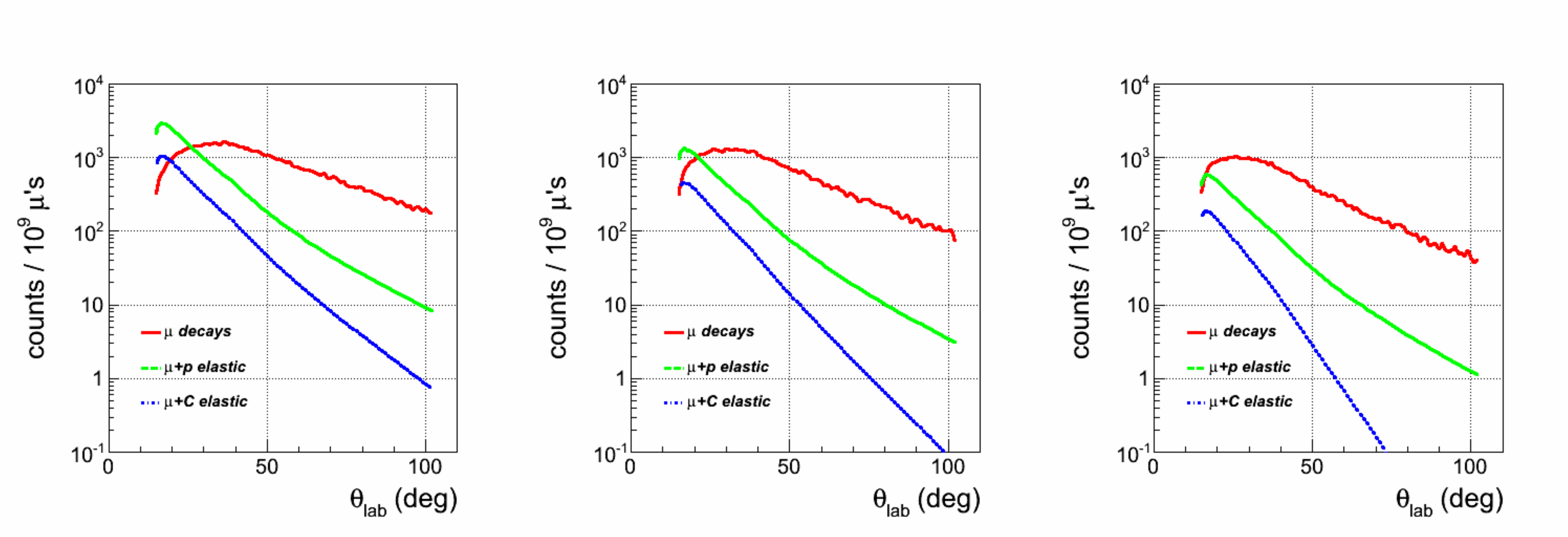}}
\caption{
Estimated rates at 115 MeV/$c$ (left), 153  MeV/$c$ (middle), and 210  MeV/$c$ (right)
as a function of angle for muon elastic scattering
from protons and from carbon in the target end caps, and for electrons
from muon decays in flight in a 10 cm region near the target.
The counts shown are based on the detector geometry,
1$^{\circ}$ angle bins, and 
10$^9$ incident $\mu$'s,
corresponding to about 1 hour of data.
Elastic scattering rates were directly calculated,
while in-flight muon decay rates are from a numerical simulation.
}
\label{fig:elmurates}
\end{figure}

\begin{figure}
\centerline{\includegraphics[width=1\textwidth]{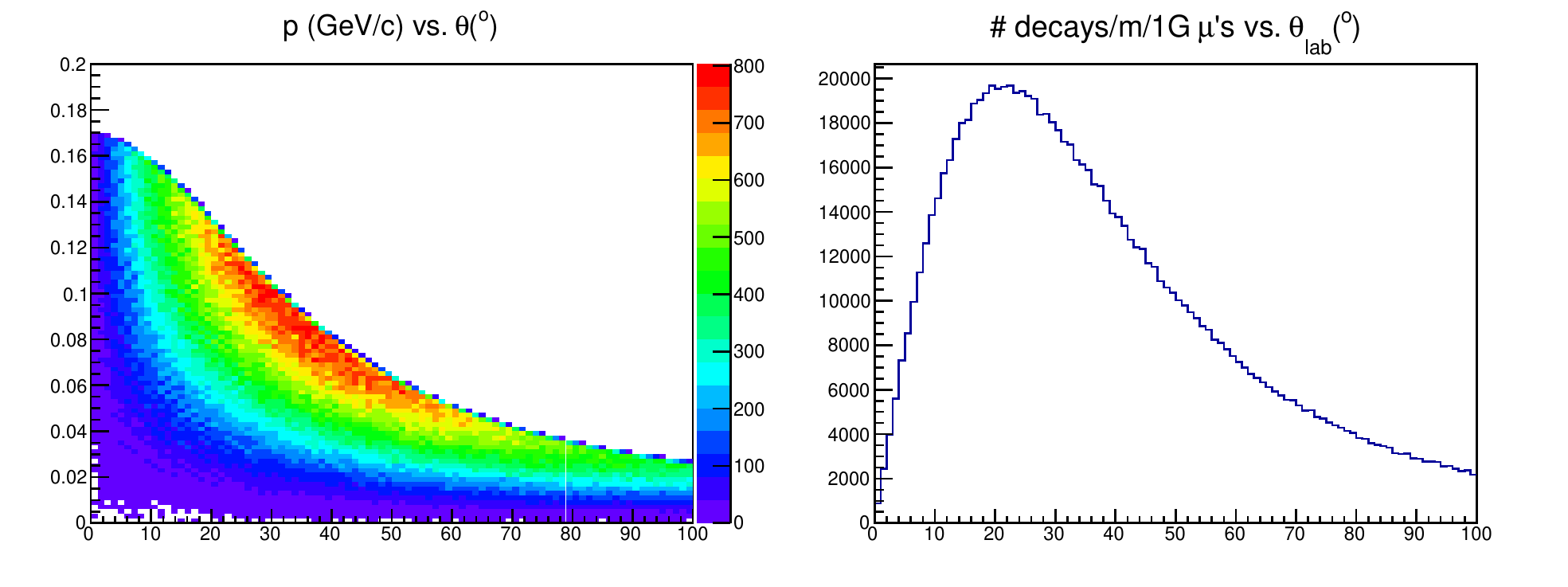}}
\caption{Left: Simulation of $e$ momentum vs.\ angle from
decays in flight of 153 MeV/$c$ unpolarized $\mu$'s.
Right: A projection showing the angular distribution of the electrons
from muon decays.
The distribution shifts slightly to smaller or larger angles depending
on muon polarization direction.
The numbers of electrons are per meter of flight path and per 
10$^9$ incident $\mu$'s.
}
\label{fig:beammudkad}
\end{figure}

In this appendix we enumerate many of the physics backgrounds in the
MUSE experiment.
The MUSE experiment goal is to measure precise $ep$ and $\mu p$
elastic scattering cross sections.
Backgrounds can increase uncertainties, due to cuts, subtractions,
or DAQ dead times, as well as modify cross sections if not correctly
handled.

Background processes include the following:
\begin{itemize}
\item For incident $\mu$'s:
scattering from the target end windows,
decaying in flight, and knocking out $\delta$'s from the target.
The rates for elastic muon scattering and in-flight muon decay
are shown in Fig.~\ref{fig:elmurates}.
The $e$'s from $\mu$ decay have a wide range of angles due to the 
3-body nature of the decay, as shown in Fig.~\ref{fig:beammudkad},
and, at the trigger level, resemble scattering events 
for decays near the target.
An example of such an event is shown in the left panel of
Fig.~\ref{fig:GeantBackgroundEvents}.
Simulations (Section~\ref{sec:sim_muon_decay})
show that the muon decay background can be suppressed by at least 3
orders of magnitude at the lower momenta, and 2 orders of magnitude at
210 MeV/$c$, while maintaining high elastic scattering identification,
by time-of-flight cuts in a neural network approach.
Cell end-cap backgrounds are removed mainly by empty-target subtractions.

\begin{figure}[tbh]
\centerline{\includegraphics[height=2.5in]{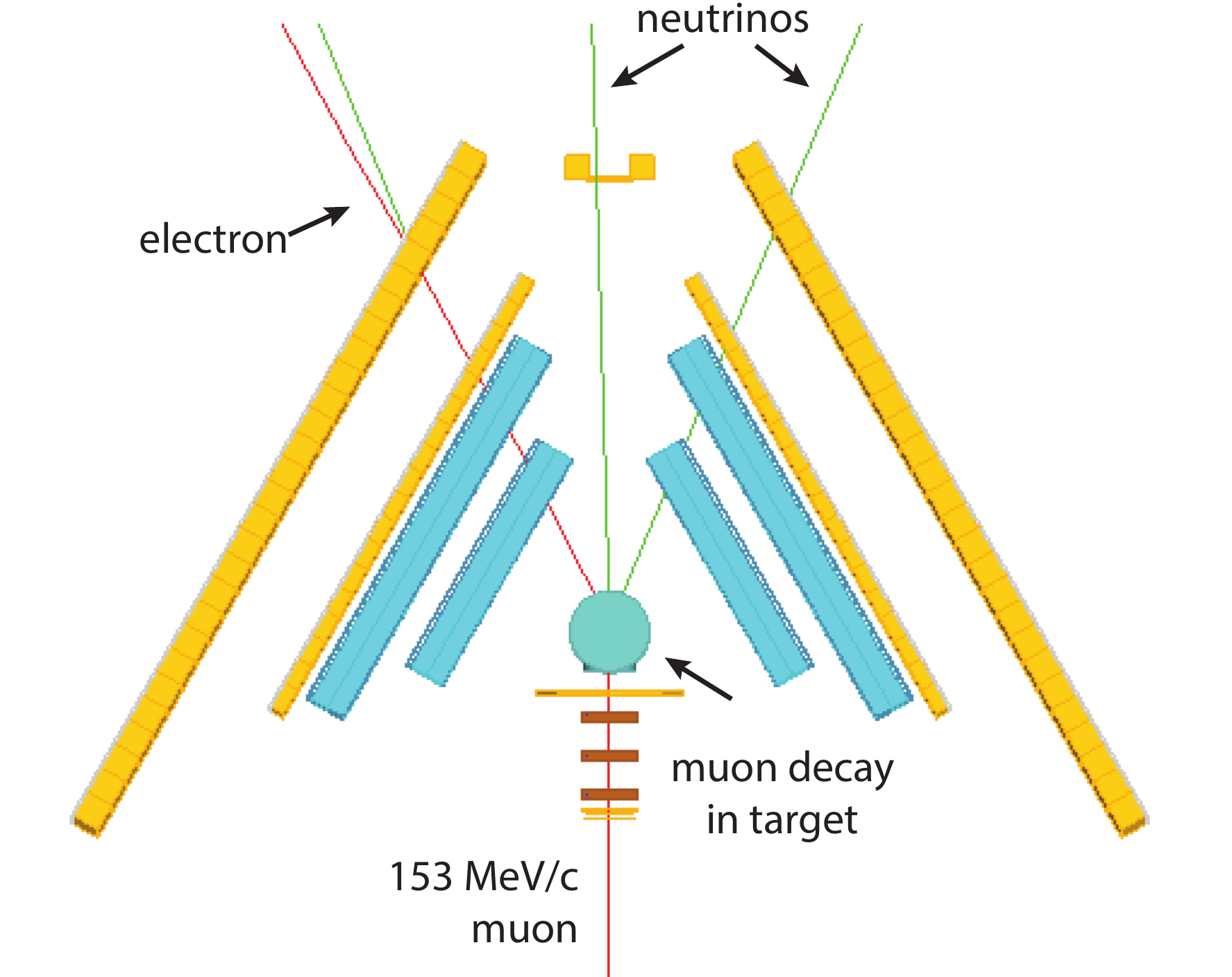}
\includegraphics[height=2.5in]{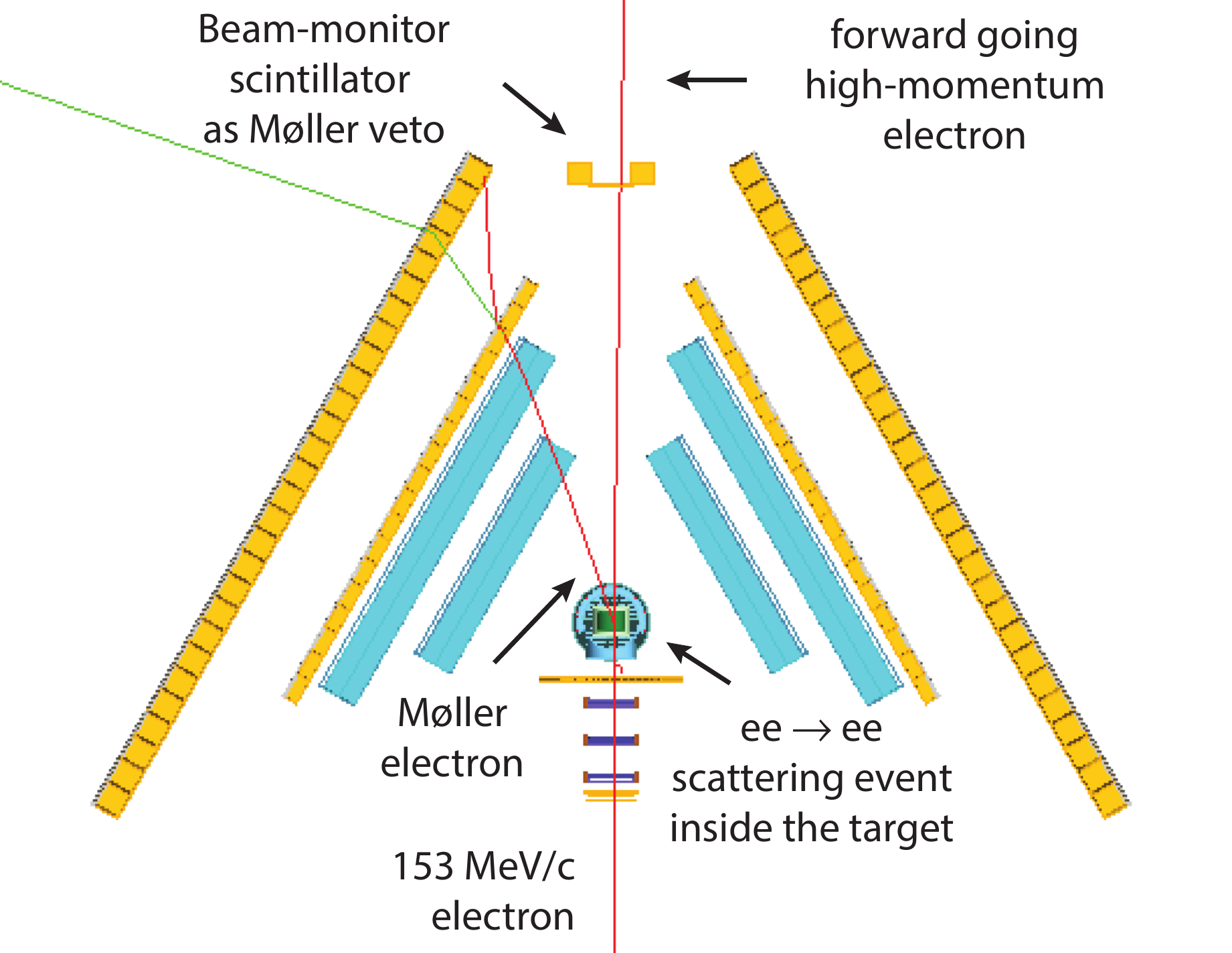}}
\caption{Geant4 simulation showing two background events. Red tracks
  are charged particles, while green tracks are neutrals.
Left: an apparent muon scattering event is actually a muon decay
within the target. At the trigger level the outgoing electron forms a
coincidence with the incoming muon. The electron knocks out several 
neutrals from the rear plane.
Right: An incident electron M\o ller scatters in the target. 
The high-energy forward going electron is detected in the beam monitor,
which can be used to greatly suppresses the M\o ller scattering.
The scattered electron is low energy and multiple scatters in the
front scintillator plane so that it ends up a few paddles from the
expected position in the rear scintillator plane, but still causes a trigger.
}
\label{fig:GeantBackgroundEvents}
\end{figure}

\item For incident $e$'s:
scattering from the target end windows, 
M\o ller and Bhabha scattering from atomic electrons,
and positron annihilation.
Empty target subtraction removes the end window background.
M\o ller and Bhabha electrons and positrons are suppressed
because the large-angle, low-momentum particles do not
efficiently trigger our system, and because the beam monitor
scintillator allows further suppression at the analysis stage.
A sample M\o ller event is shown in the right panel in Fig.~\ref{fig:GeantBackgroundEvents}.
Positron annihilation at the target generates photons that we 
are inefficient at detecting, while annihilation of scattered
positrons represents a potential inefficiency that can be understood
and corrected through simulations.
Note that radiative corrections are much more important 
for electrons and positrons than for muons.

\item For incident $\pi$'s: all processes are backgrounds.
Beam particle identification (PID) with beam line detector timing
suppresses pion events at the trigger and analysis level.

\item Beam background events:
in-flight decays of $\pi$'s and $\mu$'s in the beam generate
a halo of particles with large emittance as well as a
momentum halo for the occasional $\mu$ decay in the near-0$^{\circ}$
direction. Simulations and test data have shown that:
\begin{itemize}
\item Decays of pions in the first few m of the channel, before
magnetic elements, can lead to muons with timing in the tails
of the muon RF timing peak.
\item Decays within the magnetic elements of the channel
are largely swept out of the acceptance, because the decay 
product has a momentum outside the channel acceptance.
\item Decays after the magnetic elements of the channel are mostly at
large angles to the beam direction, and, if detected in the beam line
detector elements, add tails to the angle, position and time distributions.
\end{itemize}
\end{itemize}
Section~\ref{sec:dataanalysis} describes in more detail estimated
uncertainties resulting from removing backgrounds through
cuts and subtractions.

\section{Systematic Uncertainties}
\label{sec:uncertaintiesoverview}
In this appendix we present details of some 
experimental systematics that lead to the experiment performance
requirements.
At the smallest MUSE $Q^2$ setting, 0.0016 GeV$^2$, the elastic cross section
differs from that of a point particle by $\approx$1\%.
Absolute uncertainties will almost certainly be at least twice as large, so the
data need to be normalized to the $Q^2$ = 0 point, and it is the
relative systematic uncertainties that are important.

The relative systematic uncertainties are typically at the few-tenths of a percent
level, and can be divided into two categories.
First, the factors in Eq.~(\ref{eq:cseq}) can be systematically
different from angle to angle.
Second, the cross section depends strongly on $Q^2$, 
so offsets in determining the kinematics can change the extracted
form factors.
Overall systematics, such as the factors $N_{beam}$ and 
$(x\rho)_{target}$, and beam line detector efficiencies, are the same independent
of scattering angle, so the relative uncertainty vanishes.
Our goal is to measure
the relative cross section with systematic uncertainties of 0.4\% for muons and 0.6\% for electrons.

The solid angle $\Delta\Omega$ can be calculated by integrating
$\Delta\Omega = dA/r^2 = (dxdy)/r^2$.
For MUSE, we have to integrate not only over the detector, but also
over the target volume including the spatial distribution of the beam.
We have also studied the systematic uncertainty in the solid angle arising
from offsets in the chamber position and orientation relative to the
beam and target.
The essential finding is that the relative solid angle uncertainty is greatest
at the edges of the acceptance, the most forward and most backward
angles.
As long as the chamber provides ``full'' $\phi$ coverage --
the arc of constant $\theta$ reaches from the bottom to the top of the
chamber, rather than ending on the sides -- chamber offsets do not 
affect the $\phi$ coverage much.
Examing the numerical change in the solid angle vs offsets, we set a
positioning knowledge requirement of 0.1 mm to keep cross section
systematic uncertainties close to the desired 0.1\% level.
However, offsets in the directions transverse to the beam largely
cancel due to the up/down and left/right symmetry of the experiment;
analysis can be used to search for such offsets.

When binning data, the concern is how well the solid angle is known
for a bin, which is determined by the precision of the straw chambers.
Since reconstruction resolutions do not change the solid angle, 
what is important is that the chamber wire positions need to be
determined precisely. 
Standard machining techniques should allow determination of
wire positions at the $\approx$25 $\mu$m level\footnote{At this level
  of precision, temperature control is required.},
compared to $\approx$2.5 cm wide bins, and a $\approx$35 cm 
distance to the pivot, so that relative solid angle uncertainties 
will be about 0.14\%. 
We will use a traveling collimated source technique to measure wire
positions at the $\approx$50 $\mu$m level.
But note that bin to bin fluctuations are correlated -- if one bin is
too wide the neighbor is too narrow, as long as there is no overall
scale issue in the chamber.
In effect, these are fluctuations that add noise to the data.

The detector system is designed for high detection efficiency, 
so that the angle-to-angle variation is below the tenths of a percent level. 
(At larger angles statistics will limit our ability to know the
efficiency this well.)
Scintillator thresholds will be calibrated with QDC spectra checked
against simulations, with resulting uncertainties estimated to be $<$0.1\%.
Trigger programming will have to be carefully studied to ensure that it
does not introduce paddle-dependent efficiencies.
Straw chamber detection and tracking efficiencies should be
$\approx$100\% 
due to the use of redundant straw chamber planes.

Not evident in Eq.~(\ref{eq:cseq}) is that the cross section varies with
beam momentum and scattering angle, so offsets in these can lead to
changes in the cross sections that vary with angle.
Offsets in beam energy, $E$, change the scattering kinematics and factors of
$E$, $E^{\prime}$, and $Q^2$ that go into the cross section formula,
and lead to the form factor being determined at the wrong $Q^2$.
Multiple scattering and tracking detector resolution move events from
bin to bin, with multiple scattering having a much larger effect.

\begin{figure}
\centerline{\includegraphics[width=2.4in]{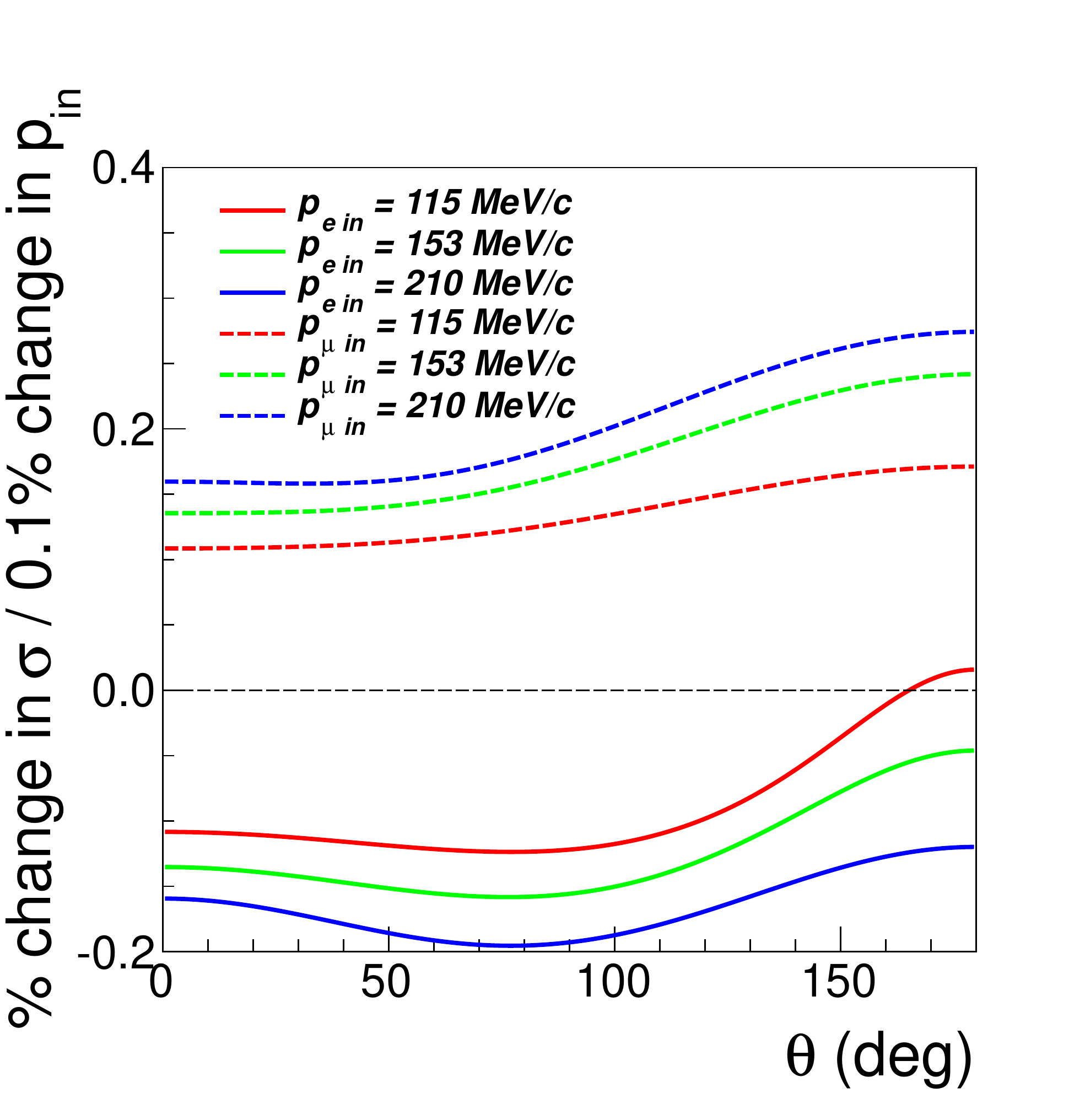}
\includegraphics[width=2.4in]{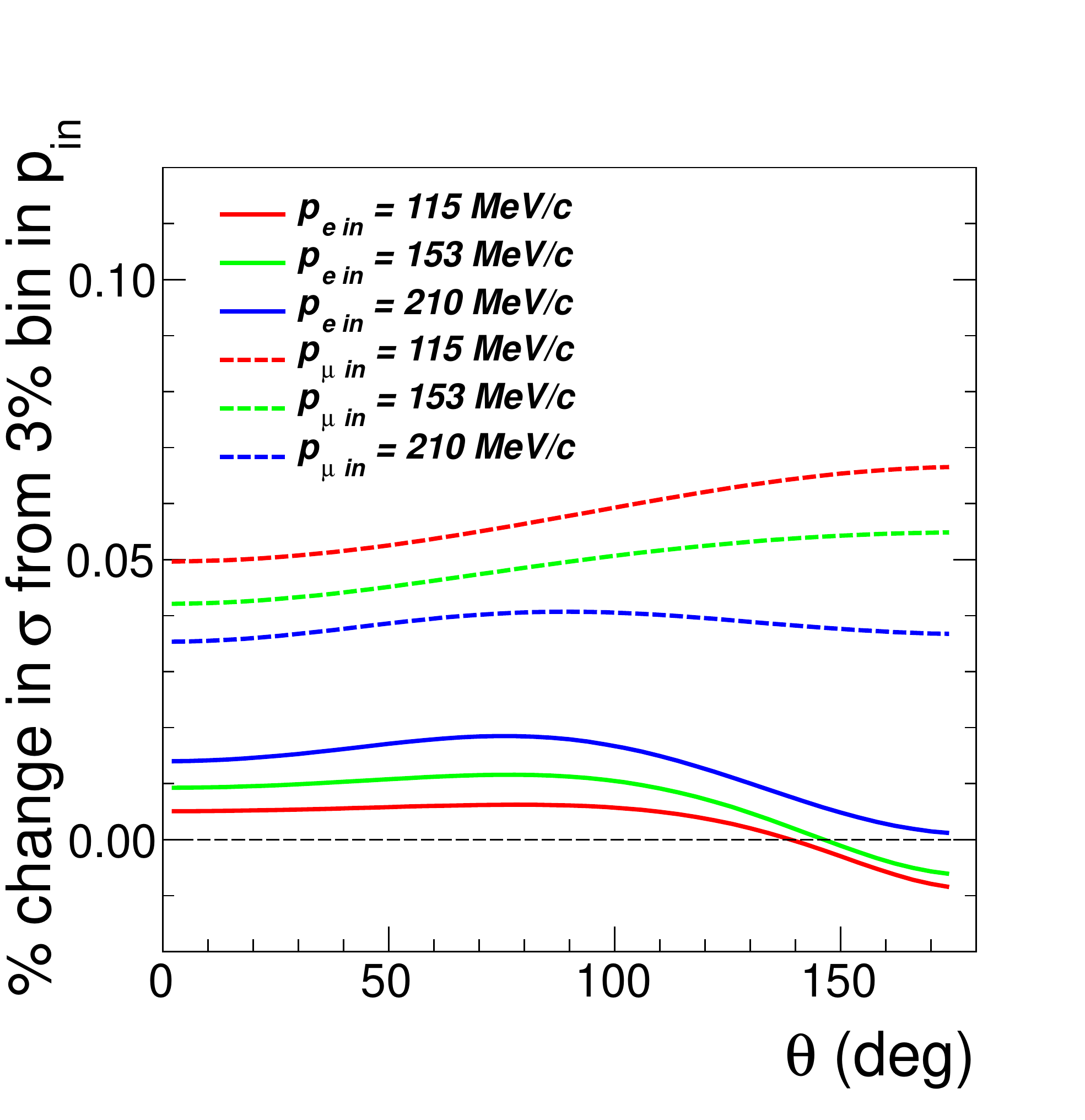}}
\caption{Left: Change in cross section in percent for a 0.1\%
change in the beam momentum. 
Right: Change in cross section in percent when averaging over a $\pm$1.5\%
bin in the beam momentum. 
We assumed a uniform distribution in incident momentum, and
evaluated the average cross section for the full momentum bin 
compared to the cross section for a mono-energetic beam at 
the central momentum.
Both studies used the Kelly form factor parameterization \cite{Kelly:2004hm}.
}
\label{fig:energysensitivity}
\end{figure}

Figure~\ref{fig:energysensitivity} shows the sensitivity of the
measured cross sections to offsets in the beam energy
and to averaging over the $\pi$M1 momentum acceptance.
For the planned kinematic coverage of 20$^{\circ}$ -- 100$^{\circ}$,
both effects act roughly as overall normalization offsets.
As the data will be renormalized in the end, the important issue is
the angle-to-angle variations of the curves in
Fig.~\ref{fig:energysensitivity}, 
not the offsets from 0.

The beam momentum distribution will be measured
by monitoring the flux with beam-line detectors as the IFP collimator
is set to a thin slit and moved across the acceptance. (It is also
possible to measure with an active detector at the IFP.)

The central beam momentum is more difficult.
Knowing the beam momentum at the $\approx$0.3\% level will keep
this systematic to below 0.1\% (0.05\%) on the cross section ($G_E$)
and thus it will be nearly insignificant.
We have in test measurements determined the muon beam 
momentum at a level of $\approx$0.2\% or 0.3\%
through time-of-flight measurements, using multiple positions of the
thin target scintillators and measuring time of flight to the beam
monitor.
We have also verified the channel dispersion and checked the
consistency of the pion and muon momenta.

The electron beam momentum cannot be measured with time of flight.
We have a proof of principle demonstration from TURTLE simulations
of the channel that we can turn of the downstream quadrupoles after
the final dipole, to obtain a dispersion of about 7 cm / \% at the
PiM1 target. By combining RF time $+$time of flight measurements to identify
particles with position measurements, we should be able to again
vary collimator position at the central dispersion point of the
channel and check the consistency of electron, muon, and pion momenta
at close to the 0.01\% level.
A first test measurement with this technique is planned for June 2017.
We note that this technique also allows us to verify simulations of
energy loss in materials.

We conclude that the beam momentum distribution can be determined well
enough that the residual systematic uncertainty on the cross sections
will be no more than about 0.1\%, with the uncertainty on the electric
form factor about 0.05\%.
Section~\ref{sec:beammomentum} provides more detail, including
a more realistic (Geant simulated) beam momentum spectrum, but the
conclusions are unchanged.

\begin{figure}
\centerline{\includegraphics[height=2.4in]{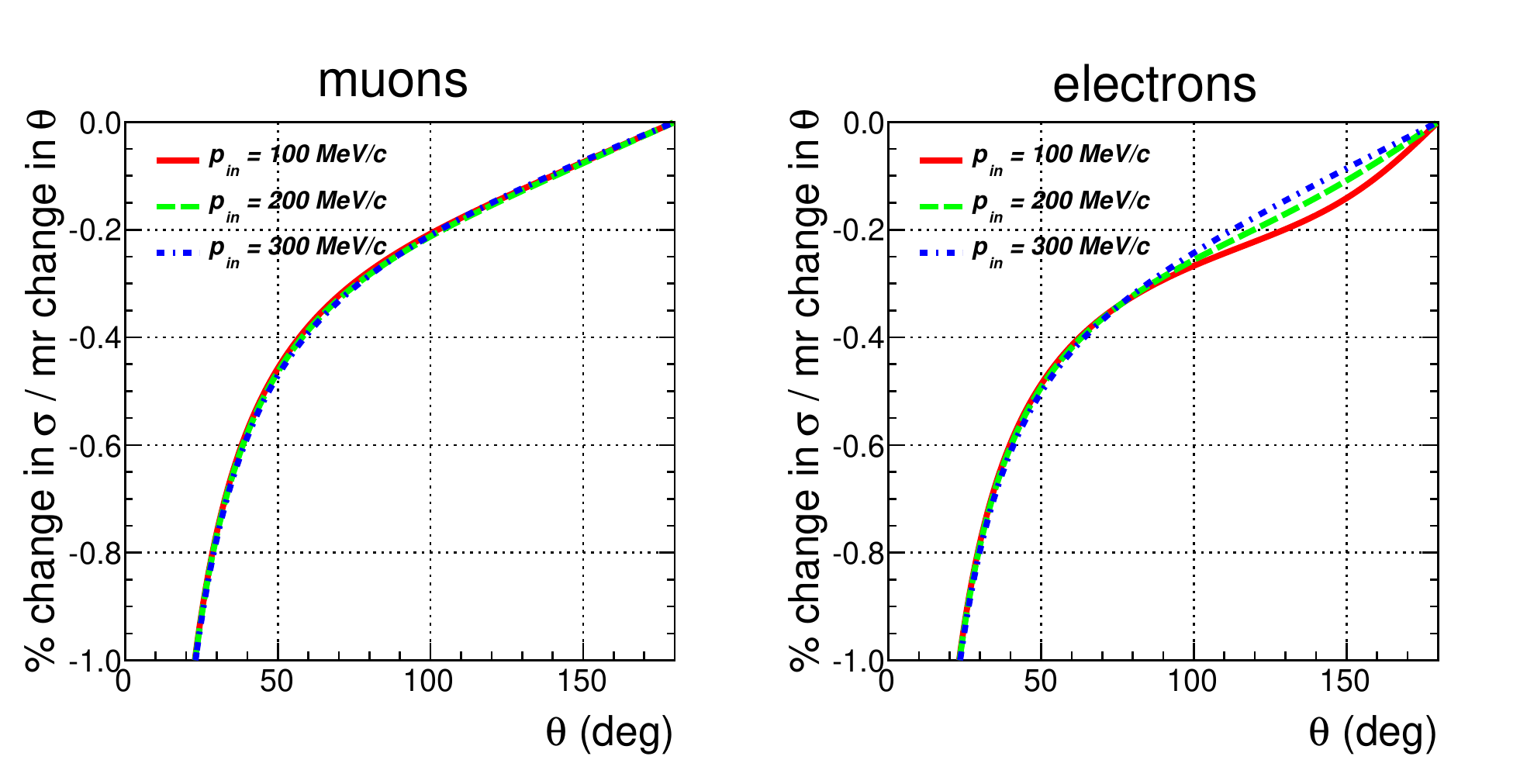}}
\caption{Change in cross section from a +1 mr offset in the scattering
 angle. 
Estimates were done with the Kelly form factor parameterization
\cite{Kelly:2004hm}.
}
\label{fig:anglesyst}
\end{figure}

\begin{figure}
\centerline{\includegraphics[height=2.4in]{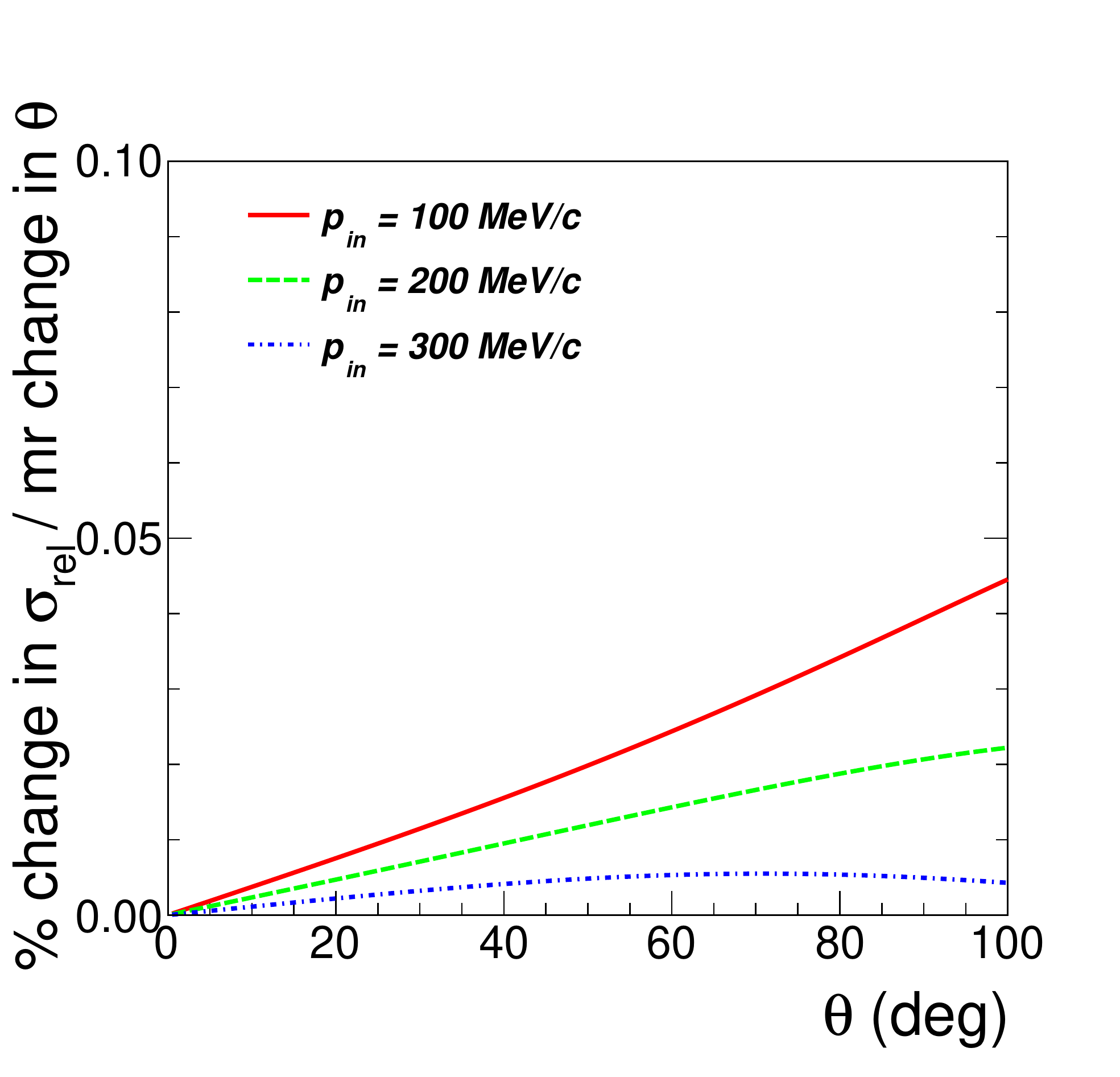}}
\caption{Relative change in cross section -- 
the change in the ratio of cross sections of $\mu p$ to $ep$ scattering -- 
from a +1 mr offset in the scattering angle. 
Estimates were done with the Kelly form factor parameterization
\cite{Kelly:2004hm}.
}
\label{fig:anglesystrelrel}
\end{figure}

Figure~\ref{fig:anglesyst} shows that offsets in the scattering angle
change the cross section, which changes the slope of the form factor 
vs.\ $Q^2$ and the radius. 
For determining a precise absolute radius, an absolute scattering angle
uncertainty at the 1 mr level leads to cross section uncertainties at the
1\% level, close to the allowed limits.
Section~\ref{sec:commissioning} describes a dedicated calibration
measurement
to determine the spectrometer angles to $\approx$0.2 mr
with dedicated calibration data.
We precisely rotate the straw
chambers into the beam behind the GEMs, and use high energy particles
to limit multiple scattering.
Thus the angle offset effect should be below 0.2\%, with a relative
uncertainty at the 0.1\% level.

\begin{figure}
\centerline{\includegraphics[height=2.4in]{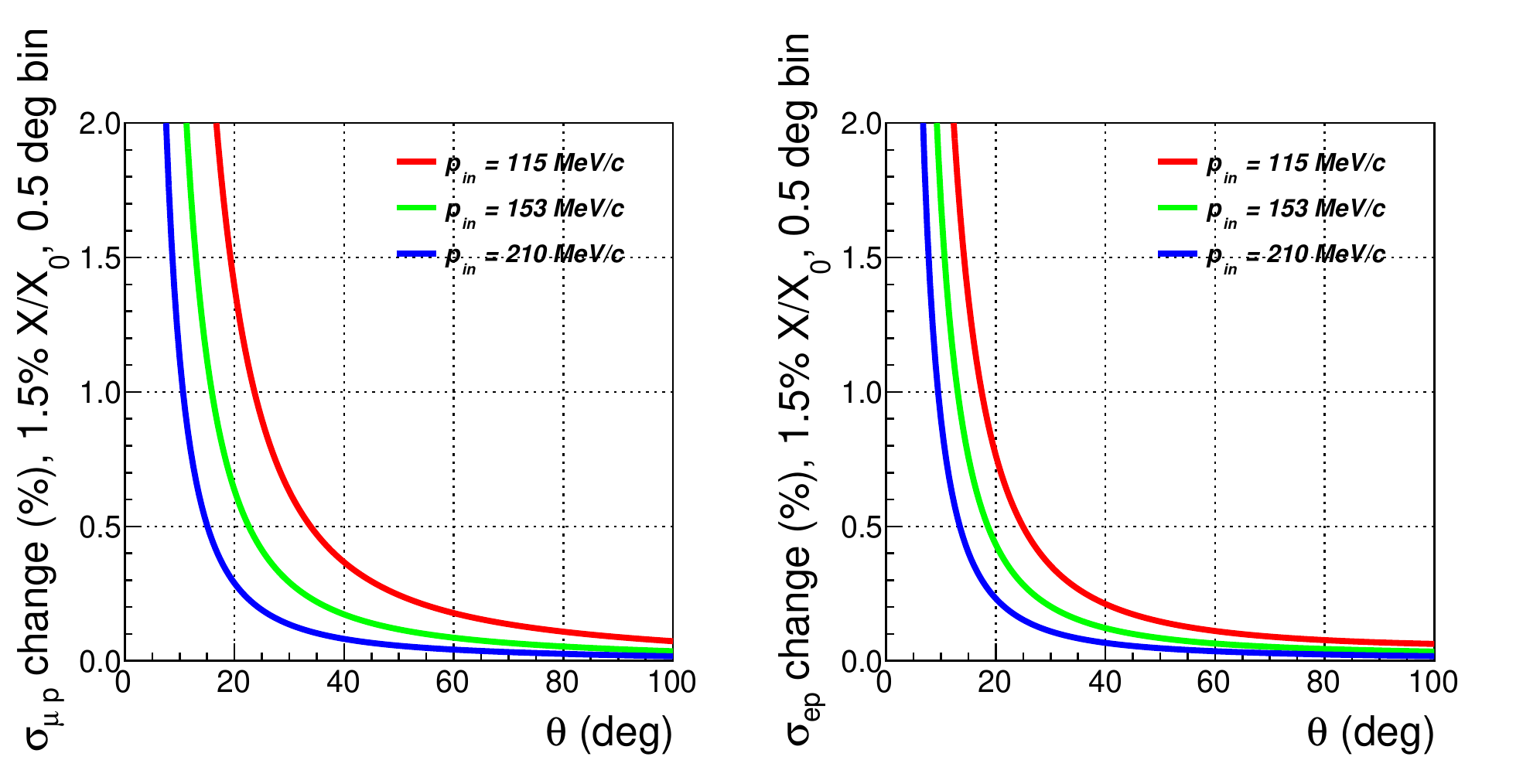}}
\centerline{\includegraphics[height=2.4in]{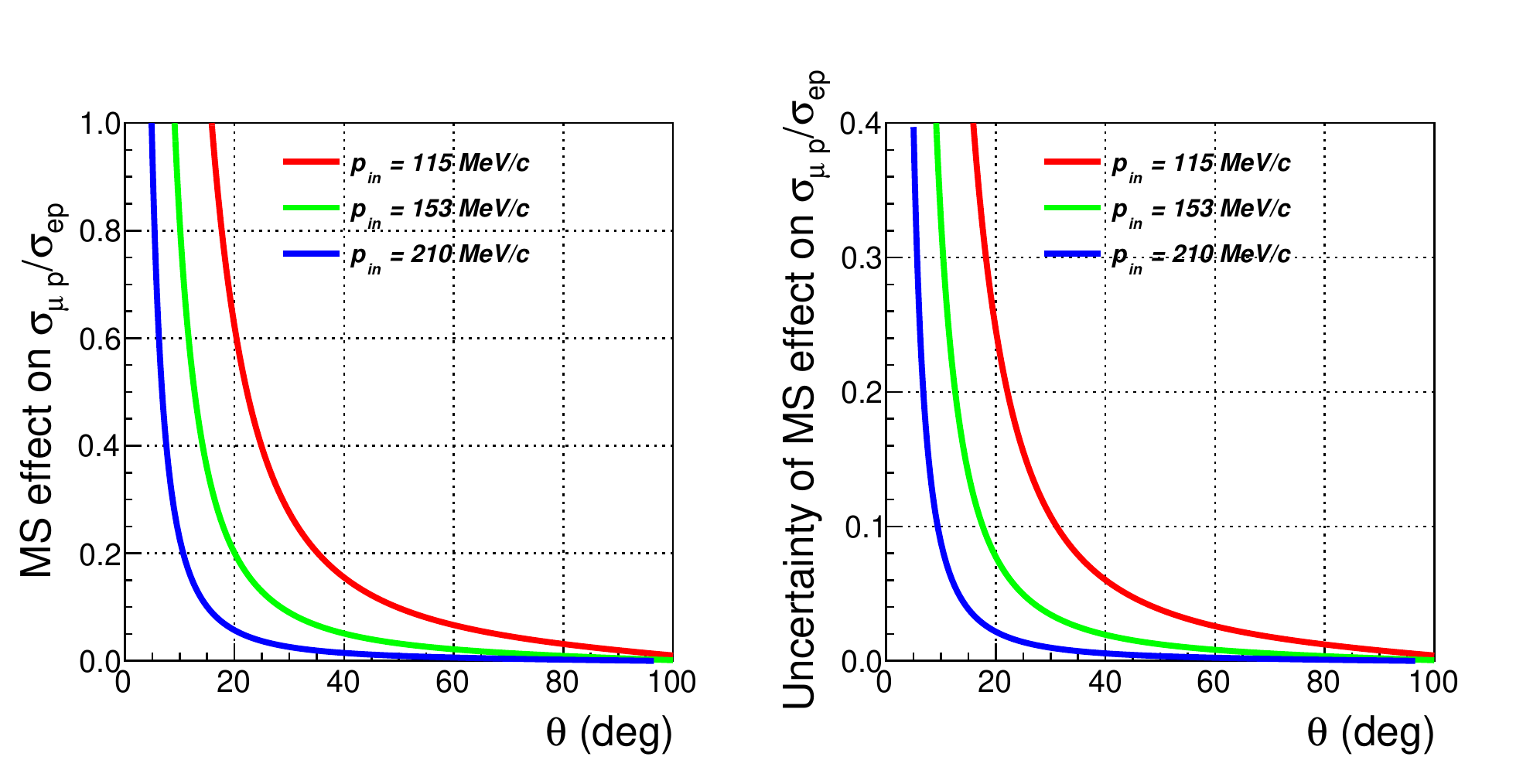}}
\caption{Top: Change in the cross section from multiple scattering.
Bottom left: Same calculation applied to the ratio of $\mu p$ to $ep$
cross sections.
Bottom right: Estimated uncertainty in the Gaussian approximation.
We assume the approximation is exact, and calculate the difference
in the multiple scattering correction for the best estimate of 
$X/X_0$ = 1.5\% and $X/X_0$ = 1.67\%, about 11\% more material.
}
\label{fig:msanglesyst}
\end{figure}

Figure~\ref{fig:msanglesyst} shows the effect of multiple scattering, 
which averages over scattering angles.
The estimates use the simple Gaussian approximation, 
the Kelly form factor parameterization \cite{Kelly:2004hm},
and material corresponding to $X/X_0$ = 1.5\% based on the 
thicknesses of the last GEM chamber, the first straw chamber,
and the cryotarget windows and liquid hydrogen cell.
The multiple scattering effect is similar in
shape at all beam momenta, but decreases in magnitude with
momentum due to a $1/p\beta$ dependence.
The upper panels show that the multiple scattering effect
quickly decreases with angle, with about 1\% effects at forward angles.
The lower left panel shows that the effect is partially canceled
in considering the ratio of $\mu p$ to $ep$ scattering cross
sections.
Since the effect is known it can be corrected for.
The lower right panel estimates how accurate the correction might
be, comparing multiple scattering for two different amounts of
material.
The correction should be able to reduce the uncertainty from multiple
scattering to be a factor of 2 or so smaller than the effect of
multiple scattering.
A simple estimate is that the uncertainty from multiple scattering
is about 0.3 $\pm$ 0.3\% for the individual cross sections but only
0.1 $\pm$ 0.1\% for the ratio of cross sections. The uncertainty is
half as large for the electric form factor.

In Geant4 simulations we have compared the reconstructed
scattering angle with the actual scattering angle in the $ep$ and
$\mu p$ reactions; see Figure~\ref{fig:scat_angle}.
\begin{figure}[ht]
\centerline{\includegraphics[width=3.2in]{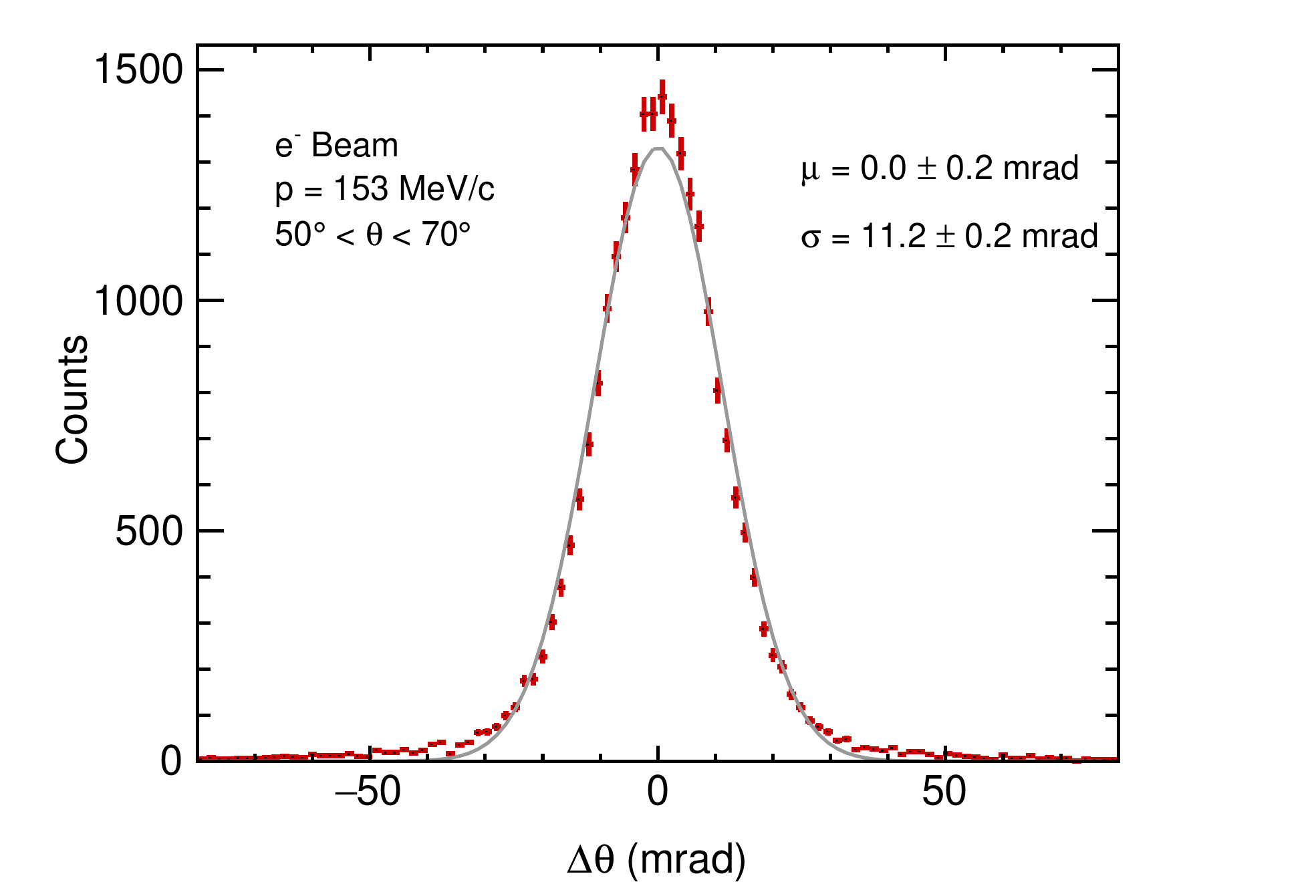}\hfill
\includegraphics[width=3.2in]{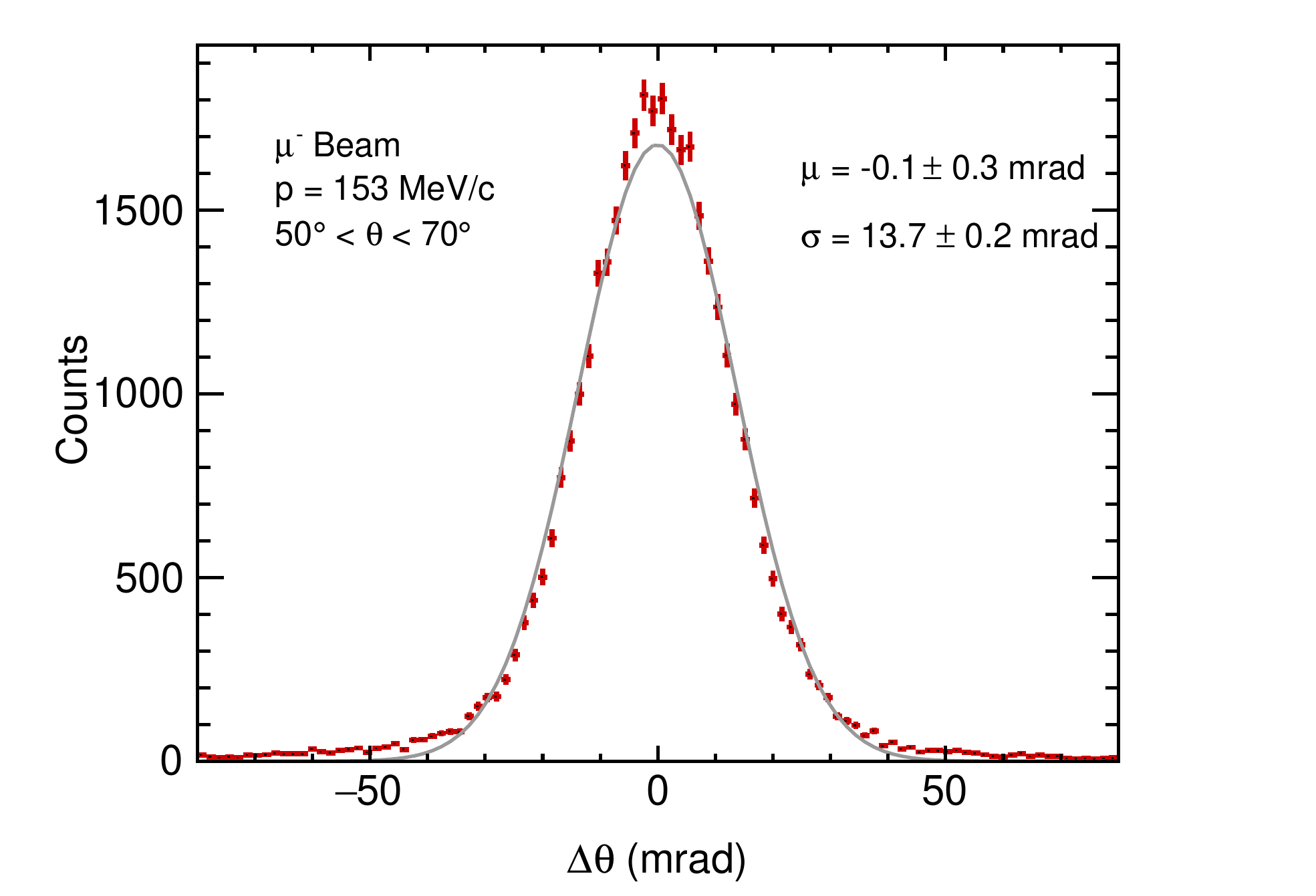}}
\caption{Examples of differences between reconstructed and actual
  reaction scattering angle from simulations for electrons (left
  panel) and muons (right panel) with a beam momentum of 153~MeV/$c$
  and an average scattering angle of about $60^\circ$.}
\label{fig:scat_angle}
\end{figure}
From similar simulations, multiple scattering corrections will be
calculated for the data, and the final systematic uncertainty might be
much smaller if we can show in high precision calibration data that
the simulations accurately model the effects of multiple scattering.
To ensure accurate corrections can be done, we will measure multiple
scatter from experiment components to verify the simulations.
The uncertainty ultimately will depend on the accuracy of the
simulations,
and the mutliple scattering uncertainty given here is only a rough estimate.

The systematic uncertainties will be verified by
measuring multiple cross sections with 6 primary experiment settings --
two beam polarities $\times$ three beam momenta.
These primary settings will be supplemented with additional
measurements at offset angles and momenta.
The quality of the overlap of these data provide a check on the
estimated systematics.

\newpage

\section{Electronics Layout}
\label{DAQ-appendix}
\begin{figure}[h]

\includegraphics[width=0.4\columnwidth]{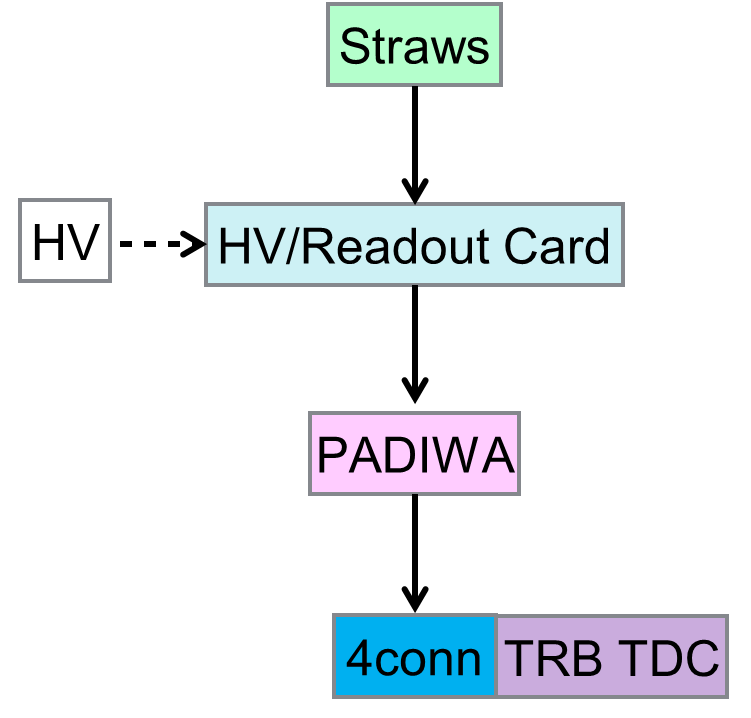}\includegraphics[width=0.55\columnwidth]{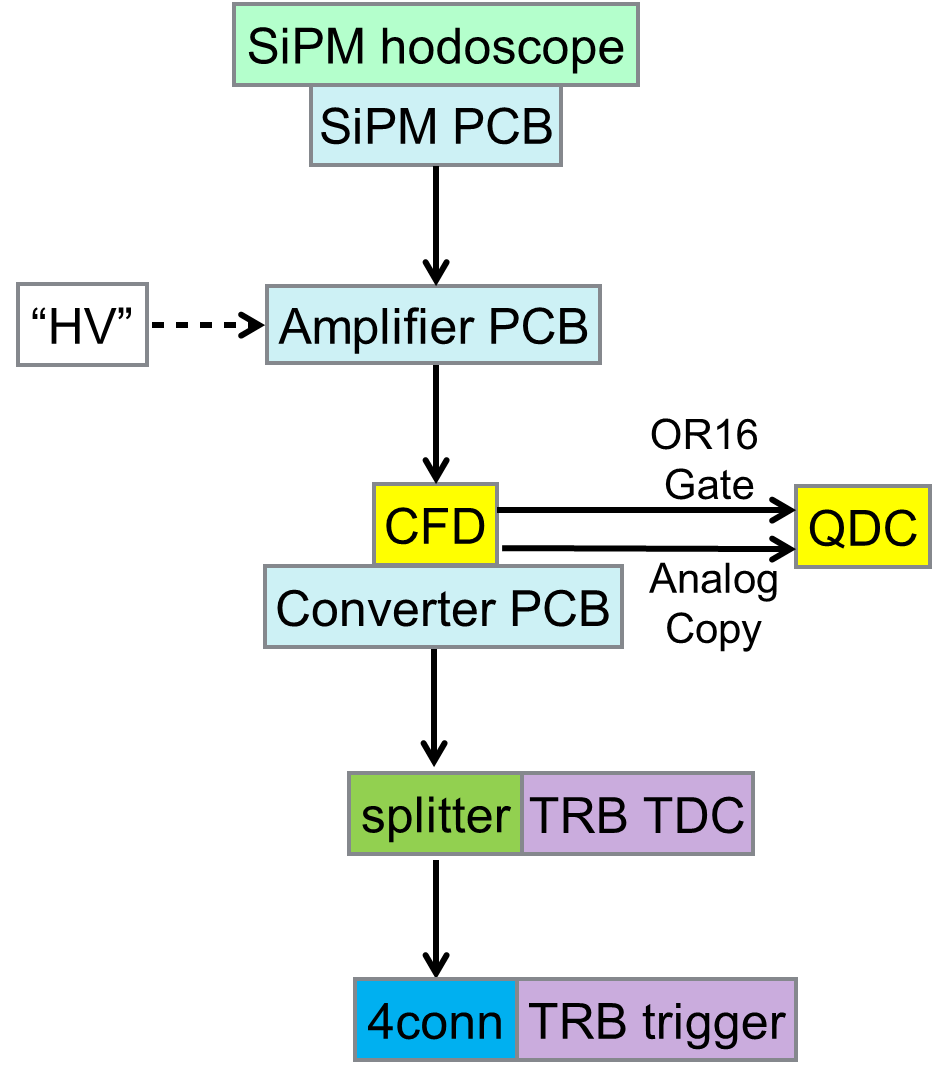}

\caption{Left: Electronics layout of the Straw Tube Tracker. Right: Electronics
layout of all SiPM detectors incuding beam hodoscope and beam monitor
SiPM detectors.}

\end{figure}

\newpage

\begin{figure}
\includegraphics[width=0.49\columnwidth]{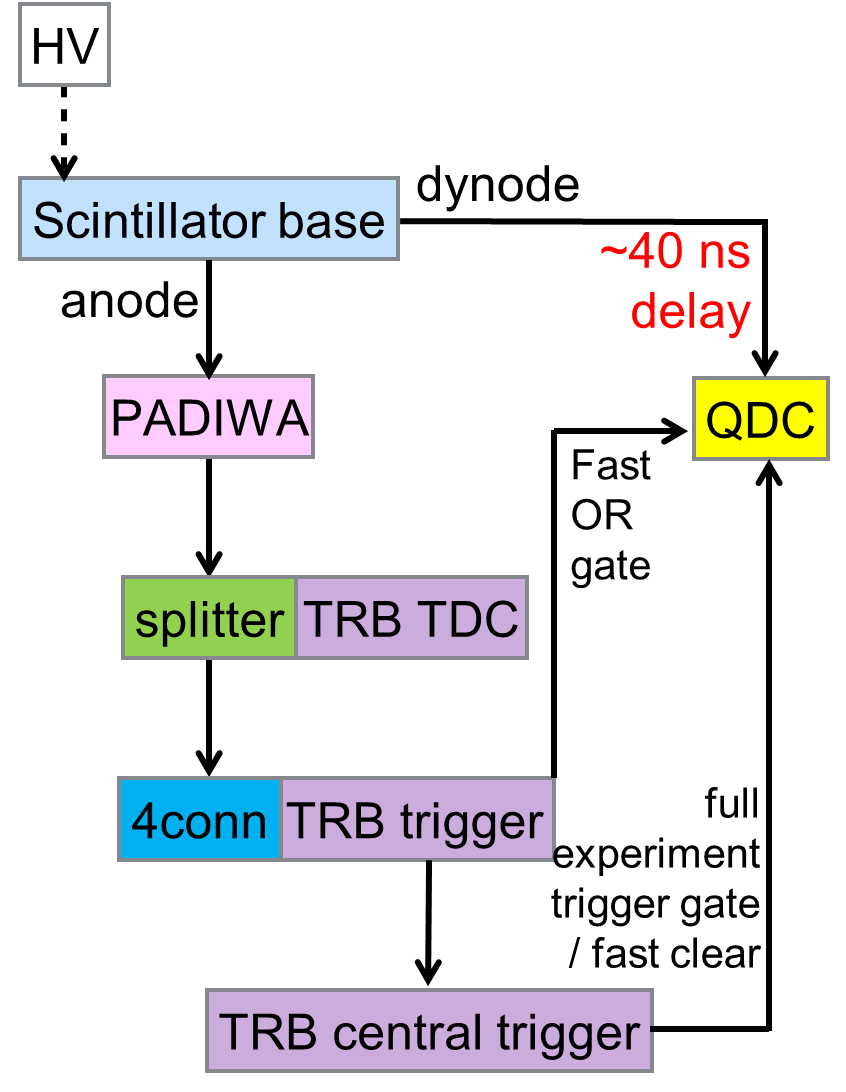}\includegraphics[width=0.49\columnwidth]{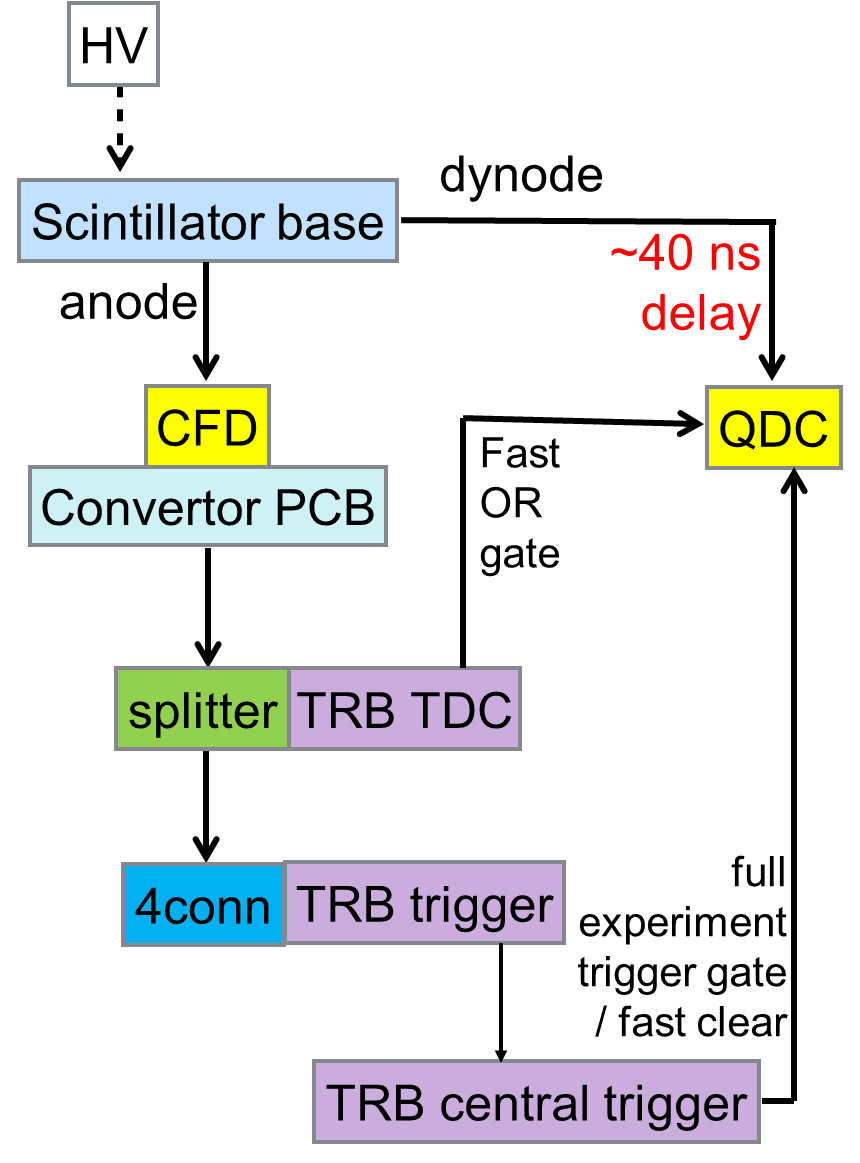}

\protect\caption{Left: Electronics layout of the Scattered Particle Scintillators.
Right: Electronics layout of the beam monitor scintillator paddles.
Note the exchange of the PADIWA for an MCFD in the case of the beam
monitor paddles in order to get good timing for out-of-time background
particles for which there will be no good QCD signal. }
\end{figure}

\newpage \null \newpage

\bibliography{muse_tdr_mediumfont}

\end{document}